\documentclass[a4paper,11pt]{book}

\usepackage[a4paper]{geometry}
\newenvironment{dedication}{ 
  \cleardoublepage
  \thispagestyle{empty}
  \vspace*{\stretch{1}}
  \itshape
  \raggedleft
 }
 {
  \par
  \vspace{\stretch{3}}
  \cleardoublepage
}
\usepackage{emptypage} 
\usepackage{fancyhdr} 
\usepackage{titlesec} 
\titleformat{\chapter}[display]{\filright\normalfont\Huge\bfseries}{\thechapter}{15pt}{} 
\usepackage{appendix} 
\usepackage[nottoc,notlot,notlof]{tocbibind} 
\usepackage{etoolbox} 
\pretocmd{\frontmatter}{\pdfbookmark[-1]{Front matter}{frontmatter}}{}{} 
\pretocmd{\backmatter}{\pdfbookmark[-1]{Back matter}{backmatter}}{}{} 
\patchcmd{\part}{\vfil}{\vfil\thispagestyle{empty}}{}{} 
\pretocmd{\appendix}{\cleardoublepage}{}{} 
\makeatletter 
\patchcmd{\@chap@pppage}{\thispagestyle{plain}}{\thispagestyle{empty}}{}{} 
\makeatother 
\patchcmd{\addappheadtotoc}{\addcontentsline{toc}{chapter}}{\addcontentsline{toc}{part}}{}{} 
\patchcmd{\addappheadtotoc}{\contentsline{toc}{chapter}}{\contentsline{toc}{part}}{}{} 

\usepackage{amsmath}
\usepackage{amssymb}
\usepackage{bbm}
\usepackage{braket}
\usepackage{float}
\usepackage[T1]{fontenc}
\usepackage{graphicx}
\usepackage{import}
\usepackage[utf8]{inputenc}
\usepackage{lipsum}
\usepackage{mathrsfs}
\usepackage{pdfpages}
\usepackage{slashed}
\usepackage{xcolor}

\newcommand{\BE}{\begin{equation}}
\newcommand{\EE}{\end{equation}}
\newcommand{\avg}[1]{\left\langle#1\right\rangle}
\newcommand{\Tr}[1]{\text{Tr}\left\{#1\right\}}
\newcommand{\cbar}{\overline{c}}
\newcommand{\mc}[1]{\mathcal{#1}}
\newcommand{\mf}[1]{\mathfrak{#1}}
\newcommand{\psibar}{\overline{\psi}}
\newcommand{\ELL}{\mathcal{L}}
\newcommand{\one}{\mathbbm{1}}
\newcommand{\Tavg}[1]{\left\langle T\left\{#1\right\}\right\rangle}
\newcommand{\MSbar}{$\overline{\text{MS}}$}
\newcommand{\mubar}{\overline{\mu}}
\newcommand{\cleannlnp}{\setlength\parfillskip{\leftskip}\newpage
\setlength\parfillskip{0.0pt plus 1.0fil}\noindent}

\usepackage[linktocpage,bookmarksopen,bookmarksopenlevel=0]{hyperref}
\hypersetup{
  pdfpagemode=UseNone,
  pdftitle={Perturbative methods in non-perturbative Quantum Chromodynamics},
  pdfauthor={Giorgio Comitini},
  pdfdisplaydoctitle,
  pdflang=en-US,
  pdfpagelayout=TwoColumnRight
}

\begin{document}
\frontmatter
\begin{titlepage}
\begin{center}
        \vspace*{1cm}
        \Huge
        \textbf{Perturbative methods in non-perturbative Quantum Chromodynamics}
        \LARGE
       \\
        \vspace{0.8cm}
        \textbf{Giorgio Comitini}
        \\
        \vspace{0.8cm}
 	\Large
        Dissertation presented in partial fulfillment of the requirements for the degree of Doctor of Physics/Doctor of Science (PhD) in Physics
        \vfill
        \textbf{Supervisors}: Prof. Dr. F. Siringo and Prof. Dr. D. Dudal
        \vfill
        \vspace{0.5cm}
        \vfill
        \includegraphics[width=0.25\textwidth]{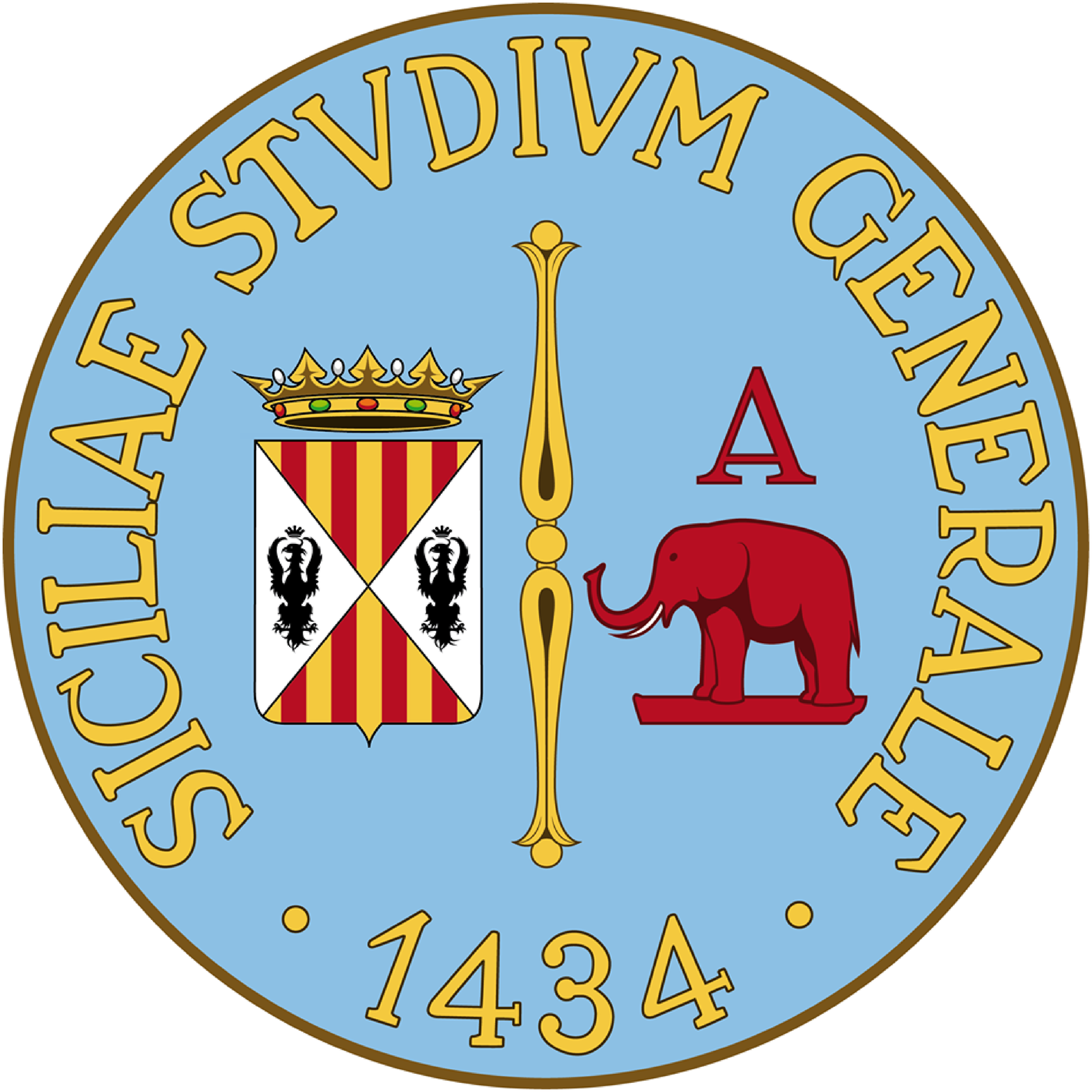}
        \vspace{0.5cm}
        
        Department of Physics and Astronomy ``E. Majorana''\\
        Universit\`{a} degli Studi di Catania\\
        Italy
       
        \vspace{1cm}
        \includegraphics[width=0.32\textwidth]{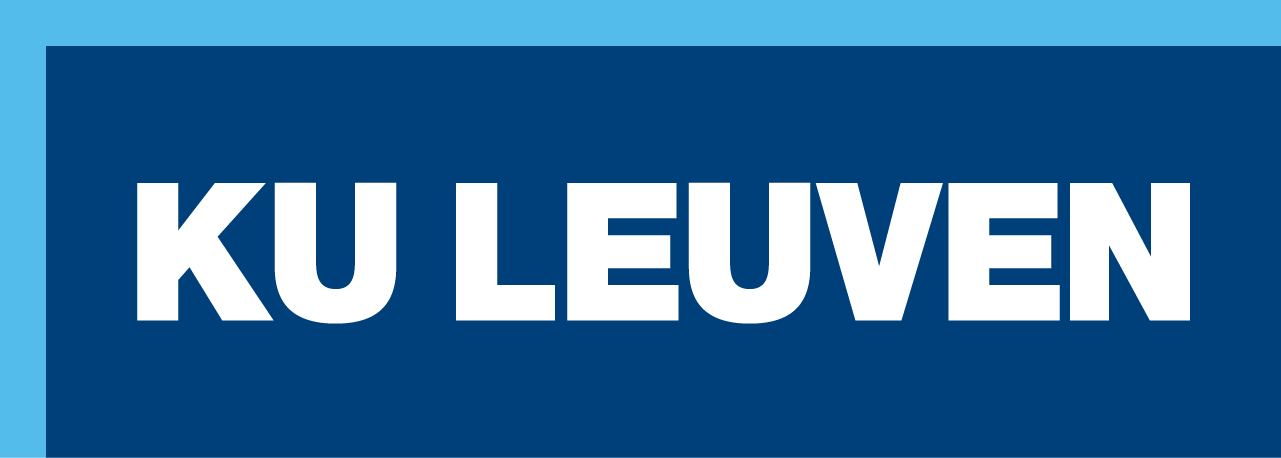}
        \vspace{0.5cm}
        
        Department of Physics and Astronomy, Faculty of Science\\
        KU Leuven -- Campus Kortrijk\\
        Belgium
        
        \vfill
        \vspace{0.5cm}
        January 2023
        \vspace{1cm} 
    \end{center}
\end{titlepage}

\begin{dedication}
A Cinzia
\end{dedication}

\tableofcontents
\renewcommand{\leftmark}{Contents}
\renewcommand{\rightmark}{Contents}
\cleardoublepage

\chapter{Abstract}
\renewcommand{\leftmark}{Abstract}
\renewcommand{\rightmark}{Abstract}

The objective of this thesis is to present two new perturbative frameworks for the study of low-energy Quantum Chromodynamics (QCD), termed the Screened Massive Expansion and the Dynamical Model. Both the frameworks paint a picture of the infrared regime of QCD which is consistent with the current knowledge provided by the lattice calculations and by other non-perturbative methods, displaying dynamical mass generation in the gluon sector and a massless ghost propagator. The Screened Massive Expansion achieves this by operating a shift of the QCD perturbative series, performed by adding a mass term for the transverse gluons in the kinetic part of the Faddeev-Popov Lagrangian and subtracting it back from its interaction part so that the total action remains unchanged. The Dynamical Model, on the other hand, interprets the generation of a dynamical mass for the gluons as being triggered by a non-vanishing condensate of the form $\langle (A^{h})^{2}\rangle$, where $A^{h}$ is a gauge- and BRST-invariant non-local version of the gluon field, and explores the consequences of the inclusion of the former in the partition function of the theory. Since the main focus of this thesis is on the gauge sector of QCD, most of our calculations will be carried out in the context of pure Yang-Mills theory. There we will show that the gluon and the ghost propagator derived by making use of the two frameworks are in good agreement with the Euclidean Landau-gauge lattice data, within the limits of a one-loop approximation. During the course of the thesis we will address topics such as the first-principles status of the two methods, the absence of Landau poles from the strong running coupling constant and the extension of the Screened Massive Expansion to finite temperatures and to full QCD. Future research prospects are discussed in the Conclusions.

\chapter{Abstract (Italian)}
\renewcommand{\leftmark}{Abstract}
\renewcommand{\rightmark}{Abstract}

L'obiettivo di questa tesi \`{e} presentare due nuovi framework perturbativi per lo studio della Cromodinamica Quantistica (QCD) alle basse energie, denominati Sviluppo Perturbativo Massivo (Screened Massive Expansion) e Modello Dinamico (Dynamical Model). Entrambi i framework forniscono un quadro del regime infrarosso della QCD in accordo con le conoscenze attuali ottenute grazie a calcoli su reticolo e ad altri metodi non perturbativi, mostrando generazione dinamica di massa nel settore gluonico e un propagatore ghost non massivo. Lo Sviluppo Perturbativo Massivo perviene a tale risultato attraverso una modifica della serie perturbativa della QCD, operata aggiungendo un termine di massa per i gluoni trasversali nella parte cinetica della Lagrangiana di Faddeev-Popov e sottraendo lo stesso termine dalla parte di interazione, in modo che l'azione totale rimanga inalterata. Il Modello Dinamico, per contro, interpreta la generazione di una massa dinamica per i gluoni come innescata da un condensato non nullo della forma $\langle (A^{h})^{2}\rangle$, dove $A^{h}$ \`{e} una versione non-locale gauge e BRST invariante del campo gluonico, ed esplora le conseguenze della sua introduzione nella funzione di partizione della teoria. Poich\'{e} il focus di questa tesi \`{e} sul settore di gauge della QCD, la maggior parte dei nostri calcoli saranno condotti nel contesto della teoria di Yang-Mills pura. In esso mostreremo che i propagatori gluonico e ghost derivati nell'ambito dei due framework sono in buon accordo con i dati euclidei sul reticolo nella gauge di Landau, entro i limiti di un'approssimazione a one loop. Nel corso della tesi affronteremo argomenti quali lo status da princ\`{i}pi primi dei due metodi, l'assenza di poli di Landau nella costante di accoppiamento forte e l'estensione dello Sviluppo Perturbativo Massivo a temperature finite e alla QCD completa. Nelle Conclusioni verranno discusse alcune prospettive di ricerca future.

\chapter{Abstract (Dutch)}
\renewcommand{\leftmark}{Abstract}
\renewcommand{\rightmark}{Abstract}

Het doel van deze thesis is om 2 nieuwe perturbatieve raamwerken voor te stellen om lage energie Kwantumchromodynamica (QCD) te bestuderen: de Gescreende Massieve Expansie (Screened Massive Expansion) en het Dynamisch Model (Dynamical Model). Beide raamwerken schetsen een beeld van het infrarood regime van QCD dat consistent is met onze huidige kennis zoals aangeleverd door roostersimulaties en andere niet-perturbatieve methodes: dynamische massageneratie in de gluonsector en een massaloos spookdeeltje. De Gescreende Massieve Expansie bekomt dit door een gepaste shift van de QCD perturbatiereeks, meerbepaald door een nieuwe massaterm toe te voegen in de kinetische term voor de transversale gluonenn op het niveau van de Faddeev Popov Lagrangiaan, waarbij deze nieuwe term dan weer wordt afgetrokken in het interactiegedeelte. Daardoor blijft de totale actie wel onveranderd. Het Dynamisch Model daarentegen bekomt een gluonmassageneratie als een gevolg van een niet-verdwijnend massacondensaat van de vorm $\langle (A^{h})^{2}\rangle$, waarbij $A^{h}$ een ijk- en BRST-invariante niet-lokale versie van het gluonveld is. Verschillende niet-triviale consequenties van het toevoegen van deze laatste aan de partitiefunctie van de theorie worden uitvoerig besproken. Vermits het hoofddoel van deze thesis de ijksector van QCD is, zullen we de meeste berekeningen in de context van pure ijktheorieën uitvoeren. We zullen daarbij aantonen dat de gluon- en spookpropagator, zoals deze kunnen bepaald worden vanuit beide raamwerken, in goede overeenkomst zijn met de Euclidische Landau-ijk roosterdata en dat binnen de beperkingen van een één-lus benadering. In de loop van de thesis zullen we verschillende onderwerpen bespreken zoals daar zijn de ab initio status van beide methodieken, het ontbreken van een Landau-pool in de sterke koppelingsconstante en de veralgemening van de Gescreende Massieve Expansie naar eindige temperatuur en naar volledige QCD.  Toekomstige onderzoeksuitbreidingen bespreken we tenslotte in de Conclusies.

\chapter{Introduction}
\renewcommand{\leftmark}{Introduction}
\renewcommand{\rightmark}{Introduction}

\section*{Quantum Chromodynamics as the theory of the strong interactions}
\renewcommand{\rightmark}{Quantum Chromodynamics as the theory of the strong interactions}
\label{chpt:intro}

Quantum Chromodynamics was born in 1973 with the publication of three seminal papers by D.~J.~Gross and F.~Wilczek \cite{GW73}, H.~D.~Politzer \cite{Pol73}, and H.~Fritzsch, M.~Gell-Mann and H.~Leutwyler \cite{FGL73}. During the late '60s and early '70s, evidence had begun accumulating \cite{Pan68,BCD69,BFK69,Tay69,MP71,FK72,MBB72,Per72,BBB75} that the \textit{quarks}, fermionic degrees of freedom originally devised as a mathematically convenient tool for explaining the observed hadron spectrum \cite{GM61,Ne61,Gre64,HN65,GM64,Zwe64a,Zwe64b}, might have more physical significance than was initially attributed to them. Experiments on deep inelastic electron-proton scattering carried out at SLAC \cite{Pan68,BCD69,BFK69,Tay69,FK72,MBB72}, together with later experiments on neutrinos \cite{MP71,Per72,BBB75}, painted a picture of the nucleon structure which was in general agreement with the theoretical predictions obtained by J. D. Bjorken, R. P. Feynman and others \cite{Bjo69,BP69,Fey69a,Fey69b} using the \textit{parton} model. The latter regarded the nucleons as loosely bound conglomerates of more elementary components -- the partons -- unable to exchange large momenta via their reciprocal non-electromagnetic interactions.

While the scientific community started to get accustomed with the idea that the quarks might in fact exist as elementary particles, the proponents of the quark model maintained a more abstract, algebraic point of view \cite{FG71,FG72}. The reason for this was the complete lack of evidence for the existence of free quarks, combined with the fact that no explanation had yet been given for the curious ``switching-off'' of the strong interactions at large momentum transfers. It is in this spirit of abstraction that in 1973 Fritzsch, Gell-Mann and Leutwyler advocated that the strong interactions inside the hadrons could be modeled by an octet of massless \textit{gluon} fields carrying color charge \cite{FGL73}. In their paper, they argued that the coloredness of the gluons -- along with the established postulate that any physical state be colorless -- might explain why the gluons were not observed as free particles, just like the quarks were not. This property of the strong interactions is today known as \textit{confinement}. The color octet gluon picture would also lead to other physically meaningful consequences, such as the fact that quark-antiquark bound pairs are preferably created in colorless states (in compliance with the aforementioned principle of color-neutrality for physical states) and the existence of eight instead of nine massless pseudoscalar mesons in the limit of zero mass for the constituent quarks. These would be the charged and neutral pions and kaons plus the lighter neutral eta meson, if nature had not decided to go its own way and provide the quarks with a mass.

In the meantime, the solution to the problem of the switching-off of the strong interactions at high energies had been given by Gross and Wilczek and Politzer \cite{GW73,Pol73}. Using the Renormalization Group (RG) approach of Gell-Mann, F. E. Low, C. G. Callan and K. Symanzik \cite{GML54,Cal70,Sym70}, Gross, Wilczek and Politzer showed that the non-abelian gauge theory formulated by C. N. Yang and R. Mills in 1954 \cite{YM54} possesses the property of \textit{asymptotic freedom}: in the limit of high energies -- provided that the number of fermions coupled to the gauge bosons is not too large -- the running coupling constant of the Yang-Mills (YM) theory tends to zero, thus making the theory effectively free at sufficiently large energies. In particular, modeling the strong interactions as a Yang-Mills theory with gauge group SU(3) -- in which the gluons were to be identified with the massless (color-charged) gauge bosons -- would be sufficient to explain the success that the parton model had in describing the scaling properties exhibited by the deep inelastic scattering cross-sections. Such an approach also predicted violations to scaling, which would be subsequently observed in the experiments \cite{BDD78,dGHH79a,dGHH79b,dGHH79c}.

With the theoretical machinery in place for turning ideas into numbers, the following years were spent verifying the hypothesis that Quantum Chromodynamics was the right theory of the strong interactions. By 1975, little doubt was left that quarks were true dynamical degrees of freedom of the hadrons. In addition to the deep inelastic scattering data, this was confirmed by the first measurements of the hadronic cross-section in $e^{+}e^{-}$ collisions \cite{Ric74,SBB75} and by the discovery of 2-jet events at SLAC \cite{HAB75}. The latter were interpreted as the product of the hadronization of a quark-antiquark pair, created by a single virtual photon in the process $e^{+}e^{-}\to (\gamma)\to q\overline{q}\to 2\,\text{jets}$.

The discovery of the gluon, on the other hand, had to wait until the end of the decade. The first indirect evidence for the existence of the gluon had been obtained in 1970-1971 by measuring the structure function of the nucleons \cite{LS70,KW71,LS71}. Then it was observed that the quarks and antiquarks inside the nucleons did not exhaust the momentum sum rules of the structure functions, which would therefore also need to receive contributions from flavorless partons yet to be seen. The obvious candidate for the fulfillment of the sum rules was, of course, the gluon. Conclusive proof of its existence, however, only came in 1979, when four different collaborations -- MARK-J, JADE, PLUTO and TASSO -- working at the PETRA electron-positron collider detected the occurrence of 3-jet events in the hadronic channel of $e^{+}e^{-}$ annihilation \cite{MARKJ79,TASSO79,PLUTO79,JADE80}. Since the quarks were fermions, the third jet in the event could not possibly stem from the hadronization of a quark. Instead, it had to originate from a boson. Interpreting the third jet as due to gluon bremsstrahlung in the QED/QCD process $e^{+}e^{-}\to(\gamma)\to q\overline{q}g\to3\,\text{jets}$ was sufficient (albeit far from trivial in terms of the model employed for jet formation) to match the experimental data on the cross section of the channel and on the momentum distribution of the decay products. Soon enough, analyses of the angular distribution of the three jets confirmed the spin-1 nature of the gluon \cite{TASSO80,CELLO80,PLUTO80}.\\

Since the '60s and '70s, the amount of evidence in favor of QCD being the true theory of the strong interactions has multiplied to the point that nobody today questions the validity of the model. From a mathematical perspective, QCD is a non-abelian gauge theory of Yang-Mills type with gauge group SU(3). The global charges associated to the local SU(3) symmetry are identified with the color charge carried by the gluons and quarks, the latter taken to be Dirac fields living in the fundamental representation of the gauge group.

Thanks to the asymptotic freedom typical of non-abelian gauge theories, the high-energy regime of QCD has been tested to an astonishing degree of precision using ordinary methods of perturbation theory. Theoretical results have been derived up to fifth order in the strong coupling constant $\alpha_{s}$ \cite{vRVL97,Cza05,LMM16,BCK17,CFH17,HRU17}, and the fundamental parameters of the theory -- that is, the coupling constant itself and the quark masses -- have been measured extensively \cite{HRZ22,LMQ22,MLB22}.\newpage

Unfortunately, the other side of the coin of ultraviolet (UV) asymptotic freedom is the unbounded increase of the value of the strong coupling the infrared (IR). Since the beta-function coefficients computed in perturbative QCD (pQCD) turn out to be negative up to the current reaches of the perturbative calculations \cite{HRU17}, perturbation theory predicts that, at low energies, the strong coupling constant grows to infinity at a finite, non-zero scale, thus developing an \textit{infrared Landau pole}. While a strongly-coupled IR regime is perfectly consistent with the experimental observations, the fact that pQCD -- whose applicability rests precisely on the smallness of $\alpha_{s}$~-- yields an infinite IR coupling marks the breakdown of the method at low energies. In particular, the existence itself of the Landau pole cannot be trusted, being derived in a domain in which the assumptions of perturbation theory are invalid.

In order to extract predictions from low-energy QCD, one has to resort to \textit{non-pertur\-bative} methods, the most common of which are lattice QCD, the Dyson-Schwinger Equations and the Operator Product Expansion and Gribov-Zwanziger approaches. In the next section we will give a brief introduction to these techniques.

\section*{Non-perturbative methods in Quantum Chromodynamics}
\renewcommand{\rightmark}{Non-perturbative methods in Quantum Chromodynamics}

The term ``non-perturbative'', in general, can be understood to have two meanings. First of all, it can mean \textit{methodologically} non-perturbative -- that is, not making use of any form of perturbative expansion. Second, it can mean \textit{intrinsically} non-perturbative -- i.e., able to incorporate features which cannot be described at any finite order in ordinary perturbation theory. Needless to say, calculational techniques which are methodologically non-perturbative are usually employed to study features of the theory which are intrinsically non-perturbative. Broadly speaking, lattice QCD and the Dyson-Schwinger equation approach are methodologically non-perturbative techniques, whereas the Operator Product Expansion and Gri\-bov-Zwanziger approaches are intrinsically non-perturbative techniques.\\

In lattice QCD (LQCD) \cite{Cre85,IM97,DD06,GL10,LM15,HSL22}, the fundamental fields of the theory -- that is, the gluon and quark fields~-- are defined on a discrete lattice of finite volume. Ordinary (continuum) QCD is then recovered by extrapolating the lattice results towards the limit of zero lattice spacing and infinite volume. Since the number of sites in the lattice is finite, the number of degrees of freedom of LQCD is also finite. As a result, the Green functions of the theory can be computed numerically by averaging over the values of finitely many variables.

For a typical state-of-the-art lattice calculation, the number of lattice sites can be as large as $128^{4} \sim 3\cdot 10^{8}$. Since the gluon has $4\times 8$ degrees of freedom per site, while a single quark has $4$, performing a LQCD calculation requires to evaluate integrals with as many as $\sim10^{10}$ variables of integration. Clearly, this can only be done on extremely powerful supercomputers using Monte Carlo techniques.

In the intermediate- to high-energy regime, LQCD provides us with an independent determination of the values of the strong coupling constant $\alpha_{s}$ \cite{ABC22,HRZ22} and of the quark masses \cite{ABC22,MLB22} which is in excellent agreement with the results of perturbation theory. At low energy, amongst the most notable achievements of the lattice approach, we mention the calculation of the decay constants of the pseudoscalar mesons \cite{BBB15,FIK15,CDK16,GLT18,DCMG19,ABC22,RSV22}~-- which, in addition to being significant in its own right, is also essential for measuring the elements of the Cabibbo-Kobayashi-Maskawa (CKM) quark-mixing matrix \cite{CLS22}~-- and the (partial) determination of the hadron spectrum \cite{ABD04,DFF08,AII09,BTB10,CDI10,LLO10,BBG11,BDDP11,DEJ11,GDK11,MW11,BLB12,DDH12,GIR12,MOU13,NAI13,ADJ14,BDM14,PEM14,PRCB15,AK17,DKL19,ADK22}. The fact that the lattice calculations are able to predict the lighter hadron masses within an error of a few percent from their experimentally measured values is arguably the most compelling proof that QCD truly provides a complete description of the strong interactions, from the TeV scales reached at the hadron colliders, down to the MeV scales typical of low-energy hadronic processes.\\

In contrast to lattice QCD, the Gribov-Zwanziger (GZ) approach is a continuum method whose main concern is to address the existence of \textit{Gribov copies} in the configurations of the gluon field. In order to fully contextualize the method, we must first take a step back and discuss some of the issues that arise when quantizing a gauge theory.

The local gauge invariance that characterizes the theories like QCD causes some of the degrees of freedom of theory to be redundant, in the sense that field configurations which are related to one another via a gauge transformation describe the very same underlying physics. In order to extract physical predictions from a gauge theory, one must first dispose of such a redundancy by \textit{fixing a gauge} -- that is, by choosing a gauge in which to carry out the calculations.

In continuum quantum field theories, the gauge is usually fixed by employing a procedure devised by L. D. Faddeev and V. Popov (FP) \cite{FP67}. The FP procedure consists in integrating out the redundant degrees of freedom from the partition function of the theory while introducing fictitious \textit{ghost fields} whose role is to remove any leftover unphysical contribution from the computed gauge-invariant quantities. The resulting FP action is no longer gauge invariant, but possesses instead a fermionic global symmetry known as BRST symmetry from the names of their discoverers, C. Becchi, A. Rouet and R. Stora \cite{BRS75,BRS76} and I. V. Tyutin \cite{Tyu75}. Being realized through global transformations, BRST symmetry does not pose any obstacle to the proper calculation of physical quantities. On the contrary, it is nowadays used as the customary starting point for proving a large number of properties of the gauge theories, such as their perturbative renormalizability \cite{Wei96}.\\

In 1978, V. N. Gribov \cite{GRI78} observed that, at the non-perturbative level, the FP procedure fails to fully fix the gauge of the non-abelian theories due to the existence of zero modes of the so-called \textit{Faddeev-Popov operator} $-\partial^{\mu}D_{\mu}$ \cite{FP67}. These zero modes can be used to construct gauge transformations which relate distinct field configurations of the FP partition function -- the Gribov copies -- to one another. As a result, the FP procedure is invalidated.

In order to solve this issue, Gribov proposed to restrict the Faddeev-Popov partition function to the configurations belonging to the domain since known as the \textit{Gribov region} \cite{GRI78}, defined by the requirement that their associated Faddeev-Popov operator be positive. A local and renormalizable action capable of implementing the Gribov constraint was discovered in 1989 by D. Zwanziger \cite{ZWA89b}, paving the way for the systematic study of the gauge sector of QCD under the lens of the Gribov hypothesis.

Since the eigenvalues of the Faddeev-Popov operator are strictly positive for small enough values of the gauge fields, the Gribov copies have no effect on the perturbative (UV) regime of QCD. In the deep infrared, on the other hand, the restriction of the fields to the Gribov region turns out to considerably alter the dynamics of the gluons: in \cite{GRI78,ZWA89b} it was shown that, within the GZ approach, instead of growing to infinity as is typical of massless fields, the zero-order gluon propagator vanishes at zero momentum.\newpage

Nowadays, thanks to relatively recent lattice calculations, a consensus has been reached that the IR behavior displayed by the standard GZ gluon propagator is not the correct one (we shall have more to say on this topic in the following section). Nonetheless, extensions of the GZ framework that take into account the non-perturbative effects brought by the \textit{vacuum condensates}, like the Refined Gribov-Zwanziger approach of \cite{DGSV08,DSVV08,DOV10,DSV11}, do manage to reproduce the exact low-energy dynamics of the theory. These extensions shed light on key aspects of QCD such as the analytical structure of the propagators, and provide us with important benchmarks for the quantitative study of its infrared regime.\\

Vacuum condensates -- that is, vacuum expectation values of products of operators evaluated at the same spacetime point -- play a central role in the approach known as the Operator Product Expansion (OPE). First proposed by K. G. Wilson in 1969 \cite{Wil69} and put on firm mathematical grounds by W. Zimmermann in 1970 \cite{Zim70}, the OPE allows us to compute the first non-perturbative corrections to the behavior of the Green functions due to the non-vanishing of the condensates. Such corrections have been calculated for quantities like the strong coupling constant $\alpha_{s}$, the heavy quark-antiquark effective potential and numerous cross-sections -- see e.g. \cite{PS95,IFL10} for an overview.

While strictly valid only at intermediate- to high-energy scales, the OPE can be used at all scales as a tool to prove that terms which -- often for dimensional reasons -- would be forbidden to enter the perturbative series of a Green function can nonetheless emerge from non-perturbative contributions. A classic example of this is the appearance of a mass term in the quark propagator due to the non-vanishing of the quark condensate $\avg{\psibar\psi}$ even in the limit of zero quark mass, where such a term could never arise by plain perturbation theory.\\

Within the functional approach to the quantization of the field theories, it is possible to derive integral equations that describe the exact behavior of the $n$-point Green functions in terms of higher-point Green functions. Such equations are known as the Dyson-Schwinger Equations (DSE) from the names of their discoverers, F. J. Dyson and J. Schwinger \cite{Dys49,Sch51}, and are customarily used to investigate the non-perturbative behavior of QCD.

In order to solve the DSE, one has to truncate the infinite tower of equations by making assumptions on the form of the higher-point Green functions. A solution is then searched for in a self-consistent way, by improving the accuracy of the approximation step-by-step in the calculation until convergence is achieved.

One specific instance of a DSE, the Bethe-Salpeter equation \cite{SB51}, is the standard tool for the study of bound states in relativistic quantum field theory.\\

Lattice QCD, the Schwinger-Dyson Equations and the Gribov-Zwanziger approach all predict that, in the deep infrared, the dynamics of the gluons is substantially different from what is expected from the calculations carried out in ordinary perturbation theory. The low-energy behavior of the gluons in Quantum Chromodynamics is the subject of the next section.

\section*{The mass of the gluon}
\renewcommand{\rightmark}{The mass of the gluon}

While at high energies the experimental observations are consistent with the ordinary perturbative picture of gluons as massless particles, it has been suggested in the literature that a non-vanishing gluon mass could help explain some of the data gathered on the low energy behavior of QCD. Studies have been carried out on processes such as the decay and formation of the pseudoscalar and vector mesons \cite{PP80,JA90,CF94,LW96,CF97,MN00,Fie02,Nat09}, $e^{+}e^{-}$ annihilation into hadrons \cite{Fie94,LW96} and $pp$ and $p\overline{p}$ scattering \cite{HKN93,LMM05,Nat09}, all of which show that the IR data are better fitted by assuming that the gluon possesses a mass in the range $\approx 500$-$1000$~MeV. The gluon mass was included in the theoretical predictions by making use of a multitude of techniques, ranging from the calculation of phase-space effects \cite{PP80,CF94,LW96,CF97,MN00,Fie02}, to the implementation of the solutions of the Dyson-Schwinger Equations \cite{Nat09}, to the calculation of non-perturbative corrections due to the quadratic $\avg{F^{2}}$ gluon condensate \cite{JA90,LW96} which in papers such as \cite{Cor82} were linked to dynamical mass generation in the gluon sector.

At the turn of the century, numerical simulations performed on larger and larger lattices \cite{LSWP98a,LSWP98b,BBLW00,BBLW01} made it possible to explore the deep infrared regime of pure Yang-Mills theory -- that is, QCD in the absence of quarks. The lattice data clearly showed that, in the limit of vanishing momentum, the gluon propagator does not grow to infinity as would be expected from a massless field, but saturates instead to a finite, non-zero value, just like the propagator of a massive particle. This was not completely unexpected, as approaches like that of Gribov and Zwanziger \cite{GRI78,ZWA89b} or the discovery of the so-called \textit{scaling} solutions of the Dyson-Schwinger Equations \cite{VHA97,AB98} had already pointed out that non-perturbative effects could lead to the strong suppression of the gluon propagator in the IR; moreover, the possibility that the gluons might acquire a mass due to the strong interactions had already been investigated in studies such as \cite{Cor82} and in some of the previously mentioned phenomenological analyses. Nonetheless, the results of the lattice calculations marked a turning point in the field of low-energy QCD both by providing the first clear evidence of the occurrence of dynamical mass generation for the gluons and by revealing that the zero-momentum limit of the gluon propagator is in fact finite, instead of vanishing, as had been predicted within the GZ and DSE frameworks. The massiveness and the zero-momentum finiteness of the gluon propagator have since been confirmed by a number of lattice studies carried out both in pure Yang-Mills theory \cite{SIMS05,CM08b,BIMS09,ISI09,BMM10,BLLM12,OS12,BBCO15,DOS16} and in full QCD \cite{BHLP04,BHLP07,IMSS07,SO10,ABBC12}, and by the discovery of the so-called \textit{decoupling} solutions of the DSE \cite{AN04,AP06,ABP08,AP08,HV13}, to the point that, today, they are regarded as established facts by the low-energy QCD community.

The occurrence of dynamical mass generation (DMG) in the gluon sector of QCD has far-reaching implications both on the phenomenology and on the theoretical investigation of the strong interactions in the infrared regime. From a phenomenological perspective, as shown e.g. by the aforementioned \cite{PP80,JA90,HKN93,CF94,Fie94,LW96,CF97,MN00,Fie02,LMM05,Nat09}, it is clear that a non-vanishing gluon mass does indeed affect the outcome of the experiments carried out at low energies. Nonetheless, we should remark that the relation of the lattice/DSE findings to the empirical data is far from clear at present: since the gluon propagator is a gauge-dependent quantity which must necessarily enter the physical predictions in a gauge-invariant way, it is not straightforward to spell out the influence of the saturation of the propagator on the QCD observables in the absence of a complete theory of the gluon mass.

From a theoretical perspective, on the other hand, dynamical mass generation is crucial to our understanding of the strong interactions, given that ordinary perturbation theory forbids the gluons to acquire a mass to any finite order in the coupling constant: it can be shown that the radiative corrections to the zero-momentum limit of the\cleannlnp gluon propagator vanish in pQCD, so that a singular gluon polarization \textit{á la} Schwinger \cite{Sch62a,Sch62b,ABP16,ADFP22} yielding a finite propagator can never be obtained by ordinary perturbative methods. Of course, it could be argued that, since pQCD breaks down in the infrared regime, it makes little sense to try to extract predictions on the low-energy behavior of the strong interactions by making use of perturbation theory. And indeed, we will see that the issue of gluon DMG and that of the formation of a Landau pole in the strong coupling constant are deeply related, to the extent that succeeding in describing the first also manages to solve the second. Nonetheless, the fact remains that the failure of standard pQCD to account for DMG in the gluon sector leaves us with little to no fully analytical tools to explore the correct low-energy limit of QCD starting from its ordinary formulation, and creates the need to look for alternative computational methods.\\

At the beginning of the last decade, M. Tissier and N. Wschebor \cite{TW10,TW11} showed that, by adding a mass term for the gluons in the Landau-gauge Faddeev-Popov Lagrangian of pure Yang-Mills theory, one could perturbatively derive a gluon and a ghost propagator that accurately reproduced the infrared lattice data already to one loop, while yielding a strong coupling constant with no Landau poles. Since the gluon mass term breaks the BRST invariance of the FP action\footnote{Although the action still possesses a generalized, non-nilpotent BRST symmetry that can be exploited to prove the renormalizability of the corresponding quantum theory -- see \cite{CF76}.}, their \textit{Curci-Ferrari} (CF) \textit{model} -- so named after its original proponents G. Curci and R. Ferrari \cite{CF76} -- was to be regarded as an effective description of the strong interactions. Following the publication of \cite{TW10,TW11}, the CF model was used to compute the three-point gauge vertices \cite{PTW13}, extended to full QCD \cite{PTW14,PTW15,PRST17,RSTT17,PRST21b} and to finite temperatures \cite{RSTW14,RST15,RSTW15a,RSTW15b,RSTW16}, worked out to two loops \cite{GPRT19,BPRW20,BGPR21} and employed to study the analytical structure of the propagators \cite{HK19,HK20} -- see also \cite{RSTW17,PRST21a}. In all cases, the CF technique provided essential insights both into the viability of perturbative techniques in Quantum Chromodynamics and into the IR behavior of QCD itself. The results obtained by making use of the model showed a remarkable agreement with the available lattice data, which only improved by going to higher order in perturbation theory.

The success of the Curci-Ferrari model in describing the low-energy regime of the strong interactions suggested that treating the gluons as massive at tree level could be sufficient to restore the validity of perturbation theory in the infrared, yielding a perturbative series that correctly displays dynamical mass generation in the gluon sector while at the same time remaining self-consistent thanks to the absence of Landau poles in the coupling constant. This suggestion prompted the research of further perturbative techniques by which a gluon mass term could be introduced in the expansion of the Green functions, without however changing the content of the Faddeev-Popov Lagrangian. The Screened Massive Expansion and the Dynamical Model are two examples of such techniques.

\section*{Massive perturbative formulations of Quantum Chromodynamics and the outline of this thesis}
\renewcommand{\rightmark}{Massive perturbative formulations of QCD and the outline of this thesis}

The objective of this thesis is to present the main results obtained by making use of two new perturbative frameworks for Quantum Chromodynamics, termed the Screened Massive Expansion and the Dynamical Model. In this section we give a brief overview of the two methods and of the contents of the thesis.\newpage

The Screened Massive Expansion (SME) was formulated in 2015 by F. Siringo \cite{SIR15a,SIR15b,SIR16b} in the context of pure Yang-Mills theory with the aim of providing a massive perturbation theory for QCD \textit{á la} Curci-Ferrari without modifying the overall Faddeev-Popov action. This was achieved by adding a transverse mass term for the gluons in the kinetic part of the Lagrangian and subtracting it back from its interaction part, so that the transverse gluons would propagate as massive at order zero in the perturbative expansion while preserving the total action of the theory. As a result of the subtraction of the mass term from the interaction Lagrangian, a new two-point interaction vertex -- termed the \textit{gluon mass counterterm} -- must be included in the Feynman rules of the expansion, giving rise to new Feynman diagrams which are not present in the perturbative series of the Curci-Ferrari model. To any order in the gluon mass counterterm, these diagrams can be shown to be equal to derivatives of corresponding Curci-Ferrari diagrams with respect to the gluon mass parameter. When all the new diagrams are resummed, the ordinary, massless perturbative series of QCD is recovered, proving that the SME is indeed perturbatively equivalent to pQCD. For obvious reasons, such a resummation is not performed in practice.

The Screened Massive Expansion neglects the existence of Gribov copies in the configuration space of QCD. The rationale for this is that the massiveness of the gluon -- which is already taken care of by the SME -- suppresses the large field configurations, so that the dynamical effects of the copies are expected to be suppressed not only in the UV -- where the SME reproduces the results of ordinary perturbation theory -- but also in the IR.

The gluon and ghost propagators computed within the SME are found to be in excellent agreement with the lattice data already at one loop \cite{SIR15a,SIR15b,SIR16b,SIR17d}, displaying mass generation in the gluon sector. Within the SME, the latter occurs in a non-trivial way: the tree-level mass term introduced in the gluon propagator by shifting the expansion point of perturbation theory cancels with an opposite term in the gluon polarization, so that the gluon mass only survives inside the loops of the expansion. In other words, the saturation of the gluon propagator at zero momentum is a truly dynamical effect of the interactions.

The SME was extended to the chiral limit of full QCD and used to study the analytic structure of the propagators in \cite{SIR16b,SIR17c,SIR17a}. There it was shown that the one-loop gluon propagators possesses a pair of complex-conjugate poles and a spectral function which violates the positivity conditions that must hold for physical particles. This can be interpreted as evidence for gluon confinement. In the quark sector, an analogous finding was made for the quark spectral functions, pointing to quark confinement, but a single real quark pole was observed instead. Recent calculations, first presented in \cite{CRBS21} and carried out with the aid of more accurate lattice data, show that, on the contrary, the poles of the quark propagator are actually complex conjugate like in the gluon sector. The quark mass functions computed in \cite{CRBS21} turn out to be in very good agreement with the lattice, whereas the quark $Z$-functions display the wrong behavior due to the limitations of the one-loop approximation.

In \cite{SIR17b,SC21} the Screened Massive Expansion of pure Yang-Mills theory was extended to non-zero temperatures with the aim of studying the temperature-dependence of the gluon propagator and of deriving dispersion relations for the gluon quasi-particles. A comparison with the lattice data yielded good results in the (spatially) transverse sector and mixed results in the (spatially) longitudinal sector, the latter accounted for by the fact that a $4$-dimensionally transverse gluon mass term for the gluons might be sub-optimal at high temperatures. The finite-temperature behavior of YM theory was\cleannlnp also investigated under the lens of the Gaussian Effective Potential in \cite{CS18}, where it was shown that a discontinuity in the optimal value of the SME gluon mass parameter produces a corresponding discontinuity in the entropy density, marking the occurrence of the deconfinement phase transition.

The topic of the predictiveness of the Screened Massive Expansion was addressed in \cite{SC18}, where an optimization procedure based on the Nielsen identities -- see also \cite{SC22b}~-- was formulated with the aim of reducing the number of free parameters of the expansion. By enforcing the gauge-parameter independence of the position of the poles and of the phases of the residues of the gluon propagator, it was possible to obtain an expression for the propagator which -- modulo multiplicative renormalization -- only depends on the value of the gluon mass parameter. When compared to the lattice results, it was found that the optimized propagator was indistinguishable from one obtained by a full fit of the lattice data, thus demonstrating the soundness of the method. The optimization of the two-point sector of pure Yang-Mills theory was completed in \cite{SIR19a,SIR19b} with the determination of the parameters of the ghost propagator. The results of \cite{SC18} were used as a starting point for the studies carried out in \cite{CRBS21,SC21}.

In \cite{CS20} the one-loop pure Yang-Mills gluon and ghost propagators were improved by making use of Renormalization Group methods. For not-too-large initial values of the coupling constant, the Taylor-scheme running coupling was shown to be free of Landau poles and to remain moderately small at all energy scales, thus confirming that the SME is self-consistent in the infrared. As in most massive models of QCD, the finiteness of the coupling is made possible by the fact that the gluon mass parameter provides the beta function with a scale at which the RG flow is allowed to slow down. While the RG-improved propagators display a good agreement with the lattice data at intermediate- to high-energy energy scales, essentially reducing to their ordinary pQCD analogues in the deep UV, the SME running coupling turns out to be too large at its maximum for the one-loop approximation to be sufficiently accurate in the deep IR, below momenta of approximately $500$~MeV. At such low energies, the optimized fixed-scale results of \cite{SC18} still constitute our best estimate of the behavior of the Yang-Mills propagators.\\

While the Screened Massive Expansion does not explicitly address the origin of the gluon mass, the Dynamical Model (DM) -- born from studies carried out in the framework of the Gribov-Zwanziger approach \cite{CDFG15,CDFG16a,CDFG16b,CDPF17,CFPS17,CDGP18,MPPS19,DFPR19} --, advances the hypothesis that dynamical mass generation might be triggered by a non-vanishing BRST-invariant quadratic gluon condensate of the form $\avg{(A^{h})^{2}}$, where $A^{h}$ is a gauge-invariant version of the gluon field $A$. The formation of such a condensate can be proved to be energetically favored in pure Yang-Mills theory by making use of Local Composite Operator methods, which allow us to include the operator $(A^{h})^{2}$ in the Faddeev-Popov Lagrangian from first principles, without changing the physical content of the theory. An effective potential for the condensate can then be derived and minimized to provide the on-shell value of $\avg{(A^{h})^{2}}$, which is found to be different from zero.

In the process of deriving the effective potential, successive transformations of the Faddeev-Popov action yield a new action $I$ in which the condensate is coupled to the quadratic operator $(A^{h})^{2}$. Since to lowest order in perturbation theory $(A^{h})^{2}$ reduces to the square $A^{2}$ of the gluon field, the non-vanishing condensate generates a mass term for the gluons, with a mass parameter proportional to the condensate itself. We remark that the action $I$ and the Faddeev-Popov action are dynamically equivalent on the shell of the gap equation -- that is, on the minima of the effective potential. $I$ is taken to be the\cleannlnp defining action of the Dynamical Model.

The renormalizability of the Dynamical Model was proved in \cite{CFGM16,CVPG18}. The first preliminary results on the DM gluon and ghost propagators in the Landau gauge, on the other hand, were obtained in \cite{Dem20}, where it was shown that the propagators have the same expressions as in the Curci-Ferrari model, with the notable exception that the tree-level mass term in the gluon propagator disappears once the gap equation is enforced. This feature shows that the description of the IR dynamics of Yang-Mills theory provided by the DM lies somewhere in between the Curci-Ferrari model and the Screened Massive Expansion. Like in the latter, DMG in the Dynamical Model is a result of the radiative corrections brought by the interactions alone.

A new renormalization scheme for the RG analysis of the Dynamical Model in the Landau gauge, termed the Dynamically Infrared-Safe (DIS) scheme, is presented in this thesis. Within the DIS scheme, it is possible to derive a finite running coupling and one-loop RG-improved propagators which display a very good agreement with the lattice data over a wide range of momenta, only failing below approximately $500$~MeV just like in the SME.

As a final note, we should mention that the Dynamical Model was recently extended to finite temperature in \cite{DVRV22} with the aim of probing the deconfinement transition of pure Yang-Mills theory.\\

In this thesis we will give a theoretical overview of the Screened Massive Expansion and of the Dynamical Model and present the main results that have been obtained within the two frameworks. In detail, its contents are as follows. In Chapter~\ref{chpt:stdfor} we review the formalism of Quantum Chromodynamics and its ordinary perturbative formulation and discuss the breakdown of the latter in the infrared. In Chapter~\ref{chpt:npmet} we review some of the non-perturbative results obtained by lattice QCD, the Operator Product Expansion and Gribov-Zwanziger approaches and the Curci-Ferrari model, upon which we will rely during the rest of the thesis. In Chapter~\ref{chpt:sme} we discuss the set-up of the Screened Massive Expansion of pure Yang-Mills theory, report explicit expressions for the one-loop gluon and ghost SME propagators, describe the optimization procedure by which the spurious free parameters of the expansion are fixed from principles of gauge invariance and perform the RG improvement of the propagators. In Chapter~\ref{chpt:smeapp} we present two applications of the Screened Massive Expansion -- namely, its extension to finite temperature and to the quark sector of full QCD. In Chapter~\ref{chpt:dynmod} we define the gauge-invariant gluon field $A^{h}$, derive the one-loop effective potential for its quadratic condensate $\avg{(A^{h})^{2}}$ and the action $I$ of the Dynamical Model, report expressions for the one-loop DM gluon and ghost propagators and perform their RG improvement in the DIS scheme. The results of the lattice calculations are used throughout Chapters \ref{chpt:sme} to \ref{chpt:dynmod} as a benchmark for the validity of our calculations. In Chapter~\ref{chpt:concl} we present our conclusions and discuss potential future developments of the Screened Massive Expansion and of the Dynamical Model.

\mainmatter

\part{Quantum Chromodynamics and its infrared regime}

\chapter{The standard formulation of QCD}
\renewcommand{\leftmark}{\thechapter\ \ \ The standard formulation of QCD}
\renewcommand{\rightmark}{\thechapter\ \ \ The standard formulation of QCD}
\label{chpt:stdfor}

In this first chapter we start by reviewing the definition of Quantum Chromodynamics. The main objective of our review is to fix the notation and highlight some of the properties of QCD which are assumed to be valid at all scales. We will discuss the symmetries of the theory, its quantization in the functional formalism and the issue of gauge fixing. Thanks to the BRST symmetry which survives the fixing of the gauge, we will be able to derive the general form of the well-known Slavnov-Taylor identities and that of the lesser-known Nielsen identities. The latter of these will play a fundamental role in obtaining some of the results presented in Chapter \ref{chpt:sme}.

Next, we move on to the standard perturbative formulation of QCD. Although not suitable for studying the infrared dynamics of the strong interactions, standard perturbation theory remains the most important benchmark for any analytical treatment of QCD. Going through the derivation of the perturbative series for an arbitrary Green function will allow us to introduce the Feynman rules of standard perturbation theory, to discuss the validity of the approximation and to show how the formalism leaves the doors open to possible modifications of the series. By making use of the Renormalization Group, we will address the asymptotic freedom typical of the non-abelian gauge theories and illustrate the breakdown of the method at low energy.

\section{Action functionals for QCD and their symmetries}
\renewcommand{\rightmark}{\thesection\ \ \ Action functionals for QCD and their symmetries}

\subsection{The classical action}

Quantum Chromodynamics is a Yang-Mills theory \cite{YM54} with gauge group SU(3) minimally coupled to quarks in the fundamental representation. In the presence of a single quark, its Lagrangian density $\ELL_{\text{QCD}}$ takes the form
\begin{equation}
\label{lagqcd}
\ELL_{\text{QCD}}=-\frac{1}{4}\,F_{\mu\nu}^{a}F^{a\,\mu\nu}+\overline{\psi}(i\gamma^{\mu}D_{\mu}-M)\psi\ .
\end{equation}
Here $\psi$ -- the quark field -- is a triplet of Dirac fields, $\overline{\psi}=\psi^{\dagger}\gamma^{0}$ is its Dirac conjugate, $M$ is its mass, the $\gamma^{\mu}$'s are matrices satisfying the Dirac algebra
\begin{equation}
\{\gamma_{\mu},\gamma_{\nu}\}=2\,\eta_{\mu\nu}\,\one\ ,
\end{equation}
where $\eta=\text{diag}(+1,-1,-1,-1)$ is the Minkowski metric, and the covariant derivative $D_{\mu}=D_{\mu}(A)$ acting on the fundamental representation is defined as
\begin{equation}
D_{\mu}=\partial_{\mu}-ig A_{\mu}^{a}T_{a}\ .
\end{equation}
In the above equation, $g$ is the strong coupling constant, the $T_{a}$'s ($a=1,\dots,8$) are the generators of the Lie algebra $\mf{su}(3)$ of SU(3) -- that is, they are $8$ linearly independent traceless $3\times3$ matrices --, chosen so as to satisfy the normalization condition
\begin{equation}
\Tr{T_{a}T_{b}}=\frac{1}{2}\,\delta_{ab}\ ,
\end{equation}
and $A_{\mu}^{a}$ -- the gluon field -- is an octet of vector fields also known as the gauge potential. The gluon field-strength tensor $F_{\mu\nu}^{a}=F_{\mu\nu}^{a}[A]$ is defined as
\begin{equation}
F_{\mu\nu}^{a}=\partial_{\mu}A_{\nu}^{a}-\partial_{\nu}A_{\mu}^{a}+gf^{a}_{bc}\,A_{\mu}^{b}A_{\nu}^{c}\ ,
\end{equation}
where the $f_{ab}^{c}$'s -- the structure constants -- define the commutation relations of $\mf{su}(3)$, 
\begin{equation}
[T_{a},T_{b}]=if_{ab}^{c}\,T_{c}\ ;
\end{equation}
$F_{\mu\nu}=F_{\mu\nu}^{a}T_{a}$ can equivalently be expressed in terms of the commutator of two covariant derivatives as
\begin{equation}
F_{\mu\nu}=\frac{i}{g}\,[D_{\mu},D_{\nu}]\ .
\end{equation}
Clearly, $f_{ab}^{c}=-f_{ba}^{c}$ and $F_{\mu\nu}^{a}=-F_{\nu\mu}^{a}$. The Jacobi identity $[X,[Y,Z]]+[Y,[Z,X]]+[Z,[X,Y]]=0$, valid for any triplet of square matrices $X,Y,Z$, translates into the relation
\begin{equation}\label{jacid}
f^{a}_{bc}f^{c}_{de}+f^{a}_{dc}f^{c}_{eb}+f^{a}_{ec}f^{c}_{bd}=0\ 
\end{equation}
for the structure constants.\\

In its full glory, the QCD Lagrangian can be expanded as
\begin{align}
\ELL_{\text{QCD}}&=-\frac{1}{2}\,\partial_{\mu}A_{\nu}^{a}\,\left(\partial^{\mu}A^{a\,\nu}-\partial^{\nu}A^{a\,\mu}\right)-gf^{a}_{bc}\,\partial_{\mu}A_{\nu}^{a}\,A^{b\,\mu}A^{c\,\nu}+\\
\notag &\quad-\frac{1}{4}\,g^{2}f^{a}_{bc}f^{a}_{de}\,A_{\mu}^{b}A_{\nu}^{c}A^{d\,\mu}A^{e\,\nu}+\psibar(i\gamma^{\mu}\partial_{\mu}-M)\psi+g\,\psibar\gamma^{\mu}T_{a}\psi A_{\mu}^{a}\ .
\end{align}
The classical field equations of QCD can be obtained by functionally differentiating the action $S_{\text{QCD}}$,
\begin{equation}
S_{\text{QCD}}=\int d^{4}x\ \ELL_{\text{QCD}}\ ,
\end{equation}
with respect to the gluon and quark fields. They read
\begin{align}
(i\gamma^{\mu}D_{\mu}-M)\psi&=0\ ,\label{fieldeq1}\\
D_{\mu}F^{a\,\mu\nu}+g\,\psibar\gamma^{\nu}T^{a}\psi&=0\ ,\label{fieldeq2}
\end{align}
where the covariant derivative $D_{\mu}$ acts on $F_{\mu\nu}$ -- and, more generally, on any object in the adjoint representation of SU(3) -- as
\begin{align}
D_{\mu}F^{\sigma\nu}=\partial_{\mu}F^{\sigma\nu}-ig[A_{\mu},F^{\sigma\nu}]\ ,
\end{align}
so that
\begin{align}
D_{\mu}F^{a\,\mu\nu}=\partial_{\mu}F^{a\,\mu\nu}+g\,f^{a}_{bc}\,A_{\mu}^{b}F^{c\,\mu\nu}\ .
\end{align}

QCD is a gauge theory in that it possesses a local symmetry -- specifically, a local SU(3) symmetry. In order to see this, consider the transformation defined by $A_{\mu}=A_{\mu}^{a}T_{a}\to A_{\mu}^{U}$, $\psi\to \psi^{U}$, where
\begin{equation}\label{transf}
A_{\mu}^{U}=U\left(A_{\mu}+\frac{i}{g}\,\partial_{\mu}\right)U^{\dagger}\ ,\qquad\qquad \psi^{U}=U\psi\ ,
\end{equation}
and $U=U(x)$ is a function from Minkowski space to the group of $3\times 3$ unitary matrices of determinant 1 -- i.e., to the group SU(3). Using the transformation laws
\begin{align}
\label{trasfaux}
D_{\mu}(A^{U})\psi^{U}=UD_{\mu}(A)\psi\ ,\qquad\qquad F_{\mu\nu}[A^{U}]=UF_{\mu\nu}[A]U^{\dagger}\ ,\\
\notag \overline{\psi^{U}}=\psibar U^{\dagger}\ ,\qquad\qquad F_{\mu\nu}^{a}F^{a\,\mu\nu}=2\Tr{F_{\mu\nu}F^{\mu\nu}}\ ,
\end{align}
it is easy to show that $\ELL_{\text{QCD}}$ remains invariant under the local SU(3) transformation in Eq.~\eqref{transf}. In particular, the QCD Lagrangian is constant over the gauge orbits, i.e. over the sets of field configurations which are related to one-another via a gauge transformation (the ``gauge-equivalent'' configurations): given a solution of the QCD field equations \eqref{fieldeq1} and \eqref{fieldeq2}, all of its gauge-equivalent configurations also solve the equations.

When globalized, the local SU(3) symmetry of the QCD action leads to the conservation of an octet $J^{\mu}_{a}$ of Noether currents,
\begin{equation}
J^{\mu}_{a}=\psibar\gamma^{\mu}T_{a}\psi+f_{abc}\,F^{b\,\mu\nu}A_{\nu}^{c}\ ,\qquad\qquad \partial_{\mu}J^{\mu}_{a}=0\ .
\end{equation}
Altogether, the $J^{\mu}_{a}$'s form the color current. The color current corresponds to invariance under the infinitesimal global transformation
\begin{equation}
\delta A_{\mu}^{a}=gf^{a}_{bc}\,A_{\mu}^{b}\chi^{c}\ ,\qquad\qquad \delta\psi=ig\chi^{a}T_{a}\psi\ ,
\end{equation}
obtained by first setting $U=e^{ig\chi}$ with $\chi=\chi^{a}T_{a}$ infinitesimal in Eq.~\eqref{transf}, so that
\begin{equation}\label{transfinf}
\delta A_{\mu}=D_{\mu}\chi=\partial_{\mu}\chi-ig[\chi,A_{\mu}]\ ,\qquad\qquad \delta\psi=ig\chi\psi\ ,
\end{equation}
and then taking the $\chi^{a}$'s to be constant. Observe that, in terms of the color current, the field equations for the gluon field read
\begin{equation}
\partial_{\mu}F^{a\,\mu\nu}=-gJ^{a\,\nu}\ .
\end{equation}\\

For future reference, we remark that the definitions presented in this section remain valid, with obvious modifications, for any Yang-Mills theory whose local symmetry group is a compact semi-simple group. In particular, they apply to arbitrary SU(N) Yang-Mills theories once we replace the quark triplet with an $N$-dimensional Dirac multiplet and the $T^{a}$'s with a set of $N_{A}=N^{2}-1$ linearly independent traceless $N\times N$ matrices ($a=1,\dots,N_{A}$). Moreover, the definitions can be extended to multiple quark fields by simply taking the QCD Lagrangian to be
\begin{equation}
\ELL_{\text{QCD}}=-\frac{1}{4}\,F_{\mu\nu}^{a}F^{a\,\mu\nu}+\sum_{f}\overline{\psi}_{f}(i\gamma^{\mu}D_{\mu}-M_{f})\psi_{f}\ ,
\end{equation}
where $f$ denotes the flavor of the quark and $M_{f}$ is the corresponding mass. The other equations must be changed accordingly. For example, in the presence of multiple quark fields, the color current becomes
\begin{equation}
J^{\mu}_{a}=\sum_{f}\psibar_{f}\gamma^{\mu}T_{a}\psi_{f}+f_{abc}\,F^{b\,\mu\nu}A_{\nu}^{c}\ .
\end{equation}
In order to keep the expressions simple, in what follows we will restrict ourselves to a single flavor of quark.

\subsection{Quantizing QCD: the Faddeev-Popov action}
\label{sec:fpquant}

Quantum Chromodynamics is usually quantized by making use of the functional formalism. In the functional formalism, given any quantum operator $\mc{O}$, its vacuum expectation value (VEV) $\avg{\mc{O}}$ is computed by averaging its classical counterpart -- which for simplicity we will also denote with $\mc{O}$ -- over the set of field configurations, using as the weighting factor the complex exponential $e^{iS}$ of the classical action. In other words,
\begin{equation}\label{avgdef}
\avg{\mc{O}}=\frac{\int \mc{D}\mathscr{F}\ e^{iS[\mathscr{F}]}\ \mc{O}[\mathscr{F}]}{\int \mc{D}\mathscr{F}\ e^{iS[\mathscr{F}]}}\ ,
\end{equation}
where $\mathscr{F}$ denotes a generic field and $\mc{D}\mathscr{F}$ is the measure over the complete set of fields. When working with gauge theories over a continuum spacetime, this definition has to be somewhat modified. The reason for this lies in the fact that, as remarked in the previous section, the action of any gauge theory is constant over the set of gauge-equivalent configurations. Since there are infinitely many such configurations, the integrand in the denominator of Eq.~\eqref{avgdef} is infinite. If, in addition, the operator $\mc{O}$ is gauge invariant, then the integrand in the numerator of Eq.~\eqref{avgdef} will be infinite as well. In order to cure these infinities, one resorts to a procedure first devised by L. D. Faddeev and V. Popov (FP) \cite{FP67}. In what follows, we will review the FP procedure in the context of QCD and of the covariant gauges.\\

Let us start from a generic QCD path integral $I$,
\begin{equation}
I=\int\mc{D}A\mc{D}\psibar\mc{D}\psi\ e^{iS_{\text{QCD}}[A,\psi,\psibar]}\ \mc{O}[A,\psi,\psibar]\ .
\end{equation}
If we assume $\mc{O}$ to be gauge invariant, then $I$ is infinite and thus ill-defined. Nonetheless, if we managed to factorize the infinity so that
\begin{equation}\label{piinffact}
I=\mc{C}_{\infty}\cdot I_{\text{finite}}\ ,
\end{equation}
where $\mc{C}_{\infty}$ is an infinite constant and $I_{\text{finite}}$ is finite, then the average of the operator $\mc{O}$,
\begin{equation}\label{bks693}
\avg{\mc{O}}=\frac{\mc{C}_{\infty}\cdot I_{\text{finite}}}{\int \mc{D}\mathscr{F}\ e^{iS[\mathscr{F}]}}\ ,
\end{equation}
would be well-defined provided that $\int \mc{D}\mathscr{F}\ e^{iS[\mathscr{F}]}$ contains the same infinite factor $\mc{C}_{\infty}$ that appears in the numerator of Eq.~\eqref{bks693}.

In order to show that this factorization can indeed be performed, let us re-write $I$ as
\begin{equation}\label{abc123}
I=\mc{N}^{-1}\int\mc{D}A\mc{D}\psibar\mc{D}\psi\mc{D}F\ e^{iS_{\text{QCD}}[A,\psi,\psibar]-i\int d^{4}x\ \frac{1}{2\xi}\,F^{a}F^{a}}\ \mc{O}[A,\psi,\psibar]\ ,
\end{equation}
where $F^{a}$ is a new set of integration fields, $\xi$ is a (non-negative) constant and
\begin{equation}
\mc{N}=\int\mc{D}F\ e^{-i\int d^{4}x\ \frac{1}{2\xi}\,F^{a}F^{a}}\ .
\end{equation}
Since the ill-definedness of $I$ is caused by gauge invariance, meaning that
\begin{align}\label{gaugeinv}
S_{\text{QCD}}[A^{U},\psi^{U},\overline{\psi^{U}}]&=S_{\text{QCD}}[A,\psi,\psibar]\ ,\qquad\qquad \mc{O}[A^{U},\psi^{U},\overline{\psi^{U}}]=\mc{O}[A,\psi,\psibar]\ ,\\
\notag &\qquad\qquad\mc{D}A^{U}\mc{D}\overline{\psi^{U}}\mc{D}\psi^{U}=\mc{D}A\mc{D}\psibar\mc{D}\psi\ ,
\end{align}
where we denoted with $A^{U}$ and $\psi^{U}$ the gauge-transformed version of $A$ and $\psi$ as in Eq.~\eqref{transf}, to make the integral well-defined we must manipulate the integrand in Eq.~\eqref{abc123} so as to introduce in the action terms which break gauge symmetry. This can be achieved by changing variables of integration from $F^{a}$ to $\lambda^{a}$, setting
\begin{equation}\label{gaugechoice}
F^{a}=F^{a}[\lambda]=\partial^{\mu}(A_{\mu}^{U(\lambda)})^{a}\ ,\qquad\qquad U(\lambda)=e^{ig\lambda^{a}T_{a}}\ .
\end{equation}
The resulting integral reads
\begin{equation}
I=\mc{N}^{-1}\int\mc{D}A\mc{D}\psibar\mc{D}\psi\mc{D}\lambda\ e^{iS_{\text{QCD}}[A,\psi,\psibar]-i\int d^{4}x\ \frac{1}{2\xi}\,(\partial\cdot A^{U(\lambda)})^{2}}\ \det\left(\frac{\delta (\partial\cdot A^{U(\lambda)})}{\delta \lambda}\right)\mc{O}[A,\psi,\psibar]\ ,
\end{equation}
where the determinant can be explicitly computed to be equal to\footnote{It is precisely at this point in the derivation that the existence of Gribov copies -- see the \hyperref[chpt:intro]{Introduction} and Sec.~\ref{sec:gzrev}~-- spoils the validity of the Faddeev-Popov procedure. If the operator $\partial^{\mu} D_{\mu}$ has zero modes, then the determinant vanishes and the change of variables $F^{a}\to \lambda^{a}$ cannot be performed consistently. As discussed in our introduction to the Screened Massive Expansion, we will disregard this issue as it is not relevant in the UV, nor potentially in the IR once dynamical mass generation is accounted for.}
\begin{equation}\label{fpdet}
\det\left(\frac{\delta (\partial\cdot A^{U(\lambda)})}{\delta \lambda}\right)=\det\left(\partial^{\mu}D_{\mu}(A^{U(\lambda)})\Lambda(\lambda)\right)\ ,
\end{equation}
and, for each value of the index $a$, $\Lambda_{a}(\lambda)$ is the matrix
\begin{equation}
\Lambda_{a}(\lambda)=-\frac{i}{g}\,\frac{\partial e^{ig\lambda}}{\partial \lambda^{a}}e^{-ig\lambda}\ .
\end{equation}
The latter lives in the adjoint representation; thus, its covariant derivative $D_{\mu}\Lambda$ reads
\begin{equation}
D_{\mu}(A)\Lambda_{a}=\partial_{\mu}\Lambda_{a}-ig[A_{\mu},\Lambda_{a}]\ .
\end{equation}

The crucial thing to notice about the determinant in Eq.~\eqref{fpdet} is that part of it decouples from the rest of the integral $I$ thanks to gauge invariance itself. Indeed, factorizing the determinant as
\begin{equation}
\det\left(\partial^{\mu}D_{\mu}(A^{U(\lambda)})\Lambda(\lambda)\right)=\det\left(\partial^{\mu}D_{\mu}(A^{U(\lambda)})\right)\det\left(\Lambda(\lambda)\right)
\end{equation}
and using the relations in Eq.~\eqref{gaugeinv}, we can rewrite the integral $I$ as
\begin{align}
I=&\mc{N}^{-1}\int\mc{D}A^{U}\mc{D}\overline{\psi^{U}}\mc{D}\psi^{U}\mc{D}\lambda\ e^{iS_{\text{QCD}}[A^{U},\psi^{U},\overline{\psi^{U}}]-i\int d^{4}x\ \frac{1}{2\xi}\,(\partial\cdot A^{U})^{2}}\times\\
\notag &\qquad\qquad\times\det\left(\partial^{\mu}D_{\mu}(A^{U})\right)\det\left(\Lambda(\lambda)\right)\mc{O}[A^{U},\psi^{U},\overline{\psi^{U}}]\ ,
\end{align}
where $U=U(\lambda)$. A simple renaming of variables $A^{U}\to A$, $\psi^{U}\to \psi$ now yields
\begin{align}\label{ialmost}
I=&\mc{N}^{-1}\left(\int\mc{D}\lambda\det\left(\Lambda(\lambda)\right)\right)\times\\
\notag &\times\left(\int\mc{D}A\mc{D}\psibar\mc{D}\psi\ e^{iS_{\text{QCD}}[A,\psi,\psibar]-i\int d^{4}x\ \frac{1}{2\xi}\,(\partial\cdot A)^{2}}\det\left(\partial^{\mu}D_{\mu}(A)\right)\mc{O}[A,\psi,\psibar]\right)\ ,
\end{align}
where the first line is a multiplicative constant which does not depend on the operator $\mc{O}$. On the second line, we see that the integrand is no longer gauge invariant: a transformation $A\to A^{U}$, $\psi\to \psi^{U}$, while still leaving the integration measure and the action $S_{\text{QCD}}$ invariant, changes the product $(\partial\cdot A)^{2}$ and the gauge potential inside the determinant. In particular, the second factor in brackets is finite. This is precisely what we were looking for: going back to Eq.~\eqref{piinffact}, we can identify the constant $\mc{C}_{\infty}$ with the $\mc{O}$-independent quantity
\begin{equation}
\mc{C}_{\infty}=\mc{N}^{-1}\left(\int\mc{D}\lambda\det\left(\Lambda(\lambda)\right)\right)\ .
\end{equation}

The operator $\partial^{\mu}D_{\mu}(A)$ in Eq.~\eqref{ialmost} and its determinant are respectively known as the Faddeev-Popov operator and the Faddeev-Popov determinant. The Faddeev-Popov determinant can be computed in terms of so-called \textit{ghost fields} by observing that, given a pair of Grassmann -- that is, anticommuting -- fields $c$ and $\overline{c}$ and an operator $\mc{M}$,
\begin{equation}
\int\mc{D}\cbar\mc{D}c\ \exp\left\{i\int d^{4}x\ \cbar\, \mc{M}c\right\}\propto\det(-\mc{M})\ .
\end{equation}
Therefore, by introducing two octets of Grassman fields $c^{a}$ and $\cbar^{a}$, we can rewrite the finite part of the integral $I$ in Eq.~\eqref{ialmost} as
\begin{equation}
I_{\text{finite}}=\int\mc{D}A\mc{D}\psibar\mc{D}\psi\mc{D}\cbar\mc{D}c\ e^{iS_{\text{FP}}[A,\psi,\psibar,c,\cbar]}\ \mc{O}[A,\psi,\psibar]\ ,
\end{equation}
where $S_{\text{FP}}$ is known as the Faddeev-Popov action and reads
\begin{equation}
S_{\text{FP}}=\int d^{4}x\ \ELL_{\text{FP}}=S_{\text{QCD}}+\int d^{4}x\ \left\{-\frac{1}{2\xi}\,(\partial\cdot A)^{2}+\partial^{\mu}\cbar^{a}D_{\mu}c^{a}\right\}\ .
\end{equation}
We remark that, in order to obtain the last term, we have performed a partial integration.

The Faddeev-Popov action $S_{\text{FP}}$ is not gauge invariant. Term by term, its Lagrangian is given by
\begin{align}\label{fplag}
\ELL_{\text{FP}}&=-\frac{1}{2}\,\partial_{\mu}A_{\nu}^{a}\,\left(\partial^{\mu}A^{a\,\nu}-\partial^{\nu}A^{a\,\mu}\right)-\frac{1}{2\xi}\,\partial^{\mu}A_{\mu}^{a}\partial^{\nu}A_{\nu}^{a}-gf^{a}_{bc}\,\partial_{\mu}A_{\nu}^{a}\,A^{b\,\mu}A^{c\,\nu}+\\
\notag &\quad-\frac{1}{4}\,g^{2}f^{a}_{bc}f^{a}_{de}\,A_{\mu}^{b}A_{\nu}^{c}A^{d\,\mu}A^{e\,\nu}+\psibar(i\gamma^{\mu}\partial_{\mu}-M)\psi+g\,\psibar\gamma^{\mu}T_{a}\psi A_{\mu}^{a}+\\
\notag &\quad+\partial^{\mu}\cbar^{a}\partial_{\mu}c^{a}+gf^{a}_{bc}\,\partial^{\mu}\cbar^{a}A_{\mu}^{b}c^{c}\ .
\end{align}
The fields $c^{a}$ and $\cbar^{a}$ are respectively known as Faddeev-Popov ghosts and antighosts. They are fermionic fields in that they anticommute with each other (and with the quark fields, which are also Grassmann/anticommuting fields); nonetheless, they have no spin structure, as can be seen from their kinetic term $\partial^{\mu}\cbar^{a}\partial_{\mu}c^{a}$, which is typical of a scalar field. Their role is to act as ``negative'' degrees of freedom by removing unphysical contributions to the computed physical quantities of the theory.

The constant $\xi$ is known as the \textit{gauge parameter} and it defines the \textit{covariant gauge} in which the calculation is carried out. The VEV of any gauge-invariant operator which can be expressed solely in terms the gluon and the quark fields is easily seen to be independent of $\xi$. Indeed, since for any such operator $\mc{O}$
\begin{equation}
\avg{\mc{O}}=\frac{\int\mc{D}A\mc{D}\psibar\mc{D}\psi\ e^{iS_{\text{QCD}}}\ \mc{O}}{\int\mc{D}A\mc{D}\psibar\mc{D}\psi\ e^{iS_{\text{QCD}}}}=\frac{\int\mc{D}A\mc{D}\psibar\mc{D}\psi\mc{D}\cbar\mc{D}c\ e^{iS_{\text{FP}}}\ \mc{O}}{\int\mc{D}A\mc{D}\psibar\mc{D}\psi\mc{D}\cbar\mc{D}c\ e^{iS_{\text{FP}}}}\ ,
\end{equation}
where strict equalities hold between all of the members, the right-hand side of the equality must be $\xi$-independent because the middle average also is. On the other hand, the averages of operators which are not gauge invariant, or those of operators which also involve the ghost and antighost fields, can -- and in general do -- depend on the gauge parameter. This is due to the fact that, since the FP procedure is only applicable to gauge-invariant operators, these averages are ill-defined with respect to the classical QCD action, and their calculation makes sense only a posteriori, by making use of the gauge-fixed, $\xi$-dependent Faddeev-Popov action.\\

In the quantum setting, the FP action replaces the classical QCD action. For future reference, let us write down its field equations. By functionally differentiating Eq.~\eqref{fplag} with respect to the gluon, quark, ghost and antighost fields, we obtain
\begin{align}\label{feqfp}
(i\gamma^{\mu}D_{\mu}-M)\psi=0\ ,\qquad\qquad \partial^{\mu}D_{\mu}c^{a}=0\ ,\qquad\qquad D_{\mu}\partial^{\mu}\cbar^{a}=0\ ,\\
\notag D_{\mu}F^{a\,\mu\nu}+\frac{1}{\xi}\,\partial^{\nu}\partial^{\mu}A_{\mu}^{a}+g\,\psibar\gamma^{\nu}T^{a}\psi-gf^{a}_{bc}\,\partial^{\nu}\cbar^{b}c^{c}=0\ .
\end{align}
Observe that the field equations of the ghost and antighost fields do not coincide: in general, $\partial^{\mu}D_{\mu}\neq D_{\mu}\partial^{\mu}$.\\

To end this section, we should mention that the covariant gauges are not the only class of gauges used to quantize the QCD action. Two popular alternatives for fixing the gauge in QCD are the axial gauges, defined by choosing the functional $F^{a}$ in Eq.~\eqref{gaugechoice} according to
\begin{equation}
F^{a}=n^{\mu}(A_{\mu}^{U})^{a},
\end{equation}
where $n^{\mu}$ is a constant vector, and the Coulomb gauge, defined by the choice
\begin{equation}
F^{a}=\vec{\nabla}\cdot (\vec{A^{U}})^{a}\ ,
\end{equation}
where only the spatial divergence of the (spatial component of the) gauge field is involved. Both of these have the disadvantage of making the calculations more difficult by breaking the Lorentz invariance of the action. A third alternative, the maximal abelian gauge, defined by
\begin{equation}
F^{A}=\partial^{\mu}A_{\mu}^{A}+gf^{A}_{IB}\,a^{I\,\mu}A^{B}_{\mu}\ ,\qquad\qquad F^{I}=\partial^{\mu}a_{\mu}^{I}\ ,
\end{equation}
where the $a^{I}$'s are the diagonal components of the gauge field $A_{\mu}=A^{a}_{\mu}T_{a}$, whereas the $A^{A}$'s are its off-diagonal components, breaks global SU(3) symmetry. We will go no further in discussing these alternatives.

\subsection{The partition function and the quantum effective action}
\label{sec:partfunct}

In this section we briefly review the definition of the partition function and of the quantum effective action and recall how these are used in quantum field theory. For the sake of definiteness, we will employ the Faddeev-Popov action of QCD for their formulation, although the same formalism applies to any field theory.\\

Consider the VEV of a time-ordered product $T\{A_{\mu_{1}}^{a_{1}}(x_{1})\cdots A_{\mu_{n}}^{a_{n}}(x_{n})\}$ of gluon field operators\footnote{It can be proved -- see e.g. \cite{PS95}, Chapter 9 -- that the path integral of a product of fields computed at different spacetime points is actually equal to the quantum average of the time-ordered product of the corresponding quantum operators, rather than to the average of their simple product.},
\begin{equation}
\Tavg{A_{\mu_{1}}^{a_{1}}(x_{1})\cdots A_{\mu_{n}}^{a_{n}}(x_{n})}=\frac{\int\mc{D}A\mc{D}\psibar\mc{D}\psi\mc{D}\cbar\mc{D}c\ e^{iS_{\text{FP}}}\ A_{\mu_{1}}^{a_{1}}(x_{1})\cdots A_{\mu_{n}}^{a_{n}}(x_{n})}{\int\mc{D}A\mc{D}\psibar\mc{D}\psi\mc{D}\cbar\mc{D}c\ e^{iS_{\text{FP}}}}\ .
\end{equation}
It is easy to see that, if we define a functional $Z[j]$ as
\begin{equation}
Z[j]=\int\mc{D}A\mc{D}\psibar\mc{D}\psi\mc{D}\cbar\mc{D}c\ e^{iS_{\text{FP}}+i\int d^{4}x\ A^{a}_{\mu}j^{\mu}_{a}}\ ,
\end{equation}
where the $j^{\mu}_{a}$'s are classical external currents, then
\begin{equation}
\Tavg{A_{\mu_{1}}^{a_{1}}(x_{1})\cdots A_{\mu_{n}}^{a_{n}}(x_{n})}=\left[(-i)^{n}Z^{-1}[j]\,\frac{\delta^{n}Z[j]}{\delta j^{\mu_{1}}_{a_{1}}(x_{1})\cdots \delta j^{\mu_{n}}_{a_{n}}(x_{n})}\right]_{j=0}\ ,
\end{equation}
where $\delta/\delta j^{\mu}_{a}(x)$ denotes a functional derivative with respect to $j^{\mu}_{a}$. More generally, we can define a \textit{partition function} $Z[j,j^{c},j^{\cbar},j^{\psi},j^{\psibar}]$ as
\begin{equation}
Z[j,j^{c},j^{\cbar},j^{\psi},j^{\psibar}]=\int\mc{D}A\mc{D}\psibar\mc{D}\psi\mc{D}\cbar\mc{D}c\ e^{iS_{\text{FP}}+i\int d^{4}x\ \left(A^{a}_{\mu}j^{\mu}_{a}+c^{a}j^{c}_{a}+\cbar^{a}j^{\cbar}_{a}+\psi^{A}j^{\psi}_{A}+\psibar^{A}j^{\psibar}_{A}\right)}\ ,
\end{equation}
where the external currents $j^{c,\cbar,\psi,\psibar}$ are Grassmann fields and the index $A$ is a multi-index that enumerates the Dirac and color components of the quark field. The time-ordered product of gluon, ghost and quark operators can be computed by differentiating the partition function with respect to the corresponding external currents\footnote{Care has to be taken when differentiating with respect to Grassmann variables such as the fermionic currents, since then the functional derivatives anticommute with the Grassmann fields. In what follows, we will distinguish between left and right derivatives as in \cite{Wei96}, denoting the first with a subscript $L$ and the second with a subscript $R$.}, multiplying the result by the inverse of the partition function and by an appropriate power of $i$, and finally setting the currents to zero. Due to this property, the partition function is also known as the generator of the Green functions of the theory.

If we define a quantity $W[j]$ as
\begin{equation}
W[j]=-i\ln Z[j]\ ,
\end{equation}
where with $j$ we have collectively denoted the external currents corresponding to the gluon, ghost and quark fields, we find that, for example,
\begin{align}
\frac{\delta^{2}W[j]}{\delta j^{\mu}_{a}(x)\delta j^{\nu}_{b}(y)}\bigg|_{j=0}&=i\left(\Tavg{A_{\mu}^{a}(x)A_{\nu}^{b}(y)}-\langle A_{\mu}^{a}(x)\rangle\langle A_{\nu}^{b}(y)\rangle\right)=\\
\notag&=i\Tavg{A_{\mu}^{a}(x)A_{\nu}^{b}(y)}_{\text{conn.}}\ .
\end{align}
In the general case, by differentiating the functional $W[j]$ with respect to the external currents, we obtain \textit{connected} Green functions. Therefore, $W[j]$ is known as the generator of the connected Green functions of the theory.\\

An interesting application of the generator of the connected Green function is the computation of the VEV of the elementary quantum fields of the theory. For the gluon field, we have
\begin{align}
\frac{\delta W[j]}{\delta j^{\mu}_{a}(x)}=\avg{A_{\mu}^{a}(x)}_{j}\ ,
\end{align}
where the subscript $j$ on the right-hand side denotes that the average is computed in the presence of the classical external currents. Similarly,
\begin{align}
\frac{\delta_{R} W[j]}{\delta j_{a}^{c}(x)}&=\avg{c^{a}(x)}_{j}\ ,\qquad\qquad\ \ \,\frac{\delta_{R} W[j]}{\delta j_{a}^{\cbar}(x)}=\avg{\cbar^{a}(x)}_{j}\ ,\\
\notag\frac{\delta_{R} W[j]}{\delta j_{A}^{\psi}(x)}&=\avg{\psi^{A}(x)}_{j}\ ,\qquad\qquad\,\frac{\delta_{R} W[j]}{\delta j_{A}^{\psibar}(x)}=\avg{\psibar^{A}(x)}_{j}\ .
\end{align}
By converse, we can define a set of currents $j_{\text{cl.}}[F_{\text{cl.}}]=j_{\text{cl.}}[A_{\text{cl.}},\psi_{\text{cl.}},\psibar_{\text{cl.}},c_{\text{cl.}},\cbar_{\text{cl.}}]$ such that
\begin{align}
\notag\frac{\delta_{R} W[j]}{\delta j^{a}_{\mu}(x)}\bigg|_{j=j_{\text{cl.}}}&=A_{\text{cl.}\,\mu}^{a}(x)\ ,\\
\frac{\delta_{R} W[j]}{\delta j_{a}^{c}(x)}\bigg|_{j=j_{\text{cl.}}}&=c_{\text{cl.}}^{a}(x)\ ,\qquad\qquad\ \frac{\delta_{R} W[j]}{\delta j_{a}^{\cbar}(x)}\bigg|_{j=j_{\text{cl.}}}=\cbar_{\text{cl.}}^{a}(x)\ ,\\
\notag\frac{\delta_{R} W[j]}{\delta j_{A}^{\psi}(x)}\bigg|_{j=j_{\text{cl.}}}&=\psi_{\text{cl.}}^{A}(x)\ ,\qquad\qquad\,\frac{\delta_{R} W[j]}{\delta j_{A}^{\psibar}(x)}\bigg|_{j=j_{\text{cl.}}}=\psibar_{\text{cl.}}^{A}(x)
\end{align}
for a given set of classical fields $A_{\text{cl.}\,\mu}^{a},c_{\text{cl.}}^{a},\cbar_{\text{cl.}}^{a},\psi_{\text{cl.}},\psibar_{\text{cl.}}$. If we differentiate the functional $\Gamma[F_{\text{cl.}}]$ defined as
\begin{align}
\Gamma[F_{\text{cl.}}]=W[j_{\text{cl.}}[F_{\text{cl.}}]]-\int d^{4}x\left\{A_{\text{cl.}\,\mu}^{a}\,j_{\text{cl.}\,a}^{\mu}+c^{a}_{\text{cl.}}\,j_{\text{cl.}\, a}^{c}+\cbar^{a}_{\text{cl.}}\,j_{\text{cl.}\, a}^{\cbar}+\psi^{A}_{\text{cl.}}\,j_{\text{cl.}\, A}^{\psi}+\psibar^{A}_{\text{cl.}}\,j_{\text{cl.}\, A}^{\psibar}\right\}
\end{align}
with respect to the fields $F_{\text{cl.}}$, we obtain
\begin{align}
\notag\frac{\delta_{L}\Gamma[F_{\text{cl.}}]}{\delta A_{\text{cl.}\,\mu}^{a}(x)}&=-j_{\text{cl.}\,a}^{\mu}(x)\ ,\\
\frac{\delta_{L}\Gamma[F_{\text{cl.}}]}{\delta c_{\text{cl.}}^{a}(x)}&=-j_{\text{cl.}\,a}^{c}(x)\ ,\qquad\qquad\, \frac{\delta_{L}\Gamma[F_{\text{cl.}}]}{\delta \cbar_{\text{cl.}}^{a}(x)}=-j_{\text{cl.}\,a}^{\cbar}(x)\ ,\\
\notag\frac{\delta_{L}\Gamma[F_{\text{cl.}}]}{\delta \psi_{\text{cl.}}^{A}(x)}&=-j_{\text{cl.}\,A}^{\psi}(x)\ ,\qquad\qquad \frac{\delta_{L}\Gamma[F_{\text{cl.}}]}{\delta \psibar_{\text{cl.}}^{A}(x)}=-j_{\text{cl.}\,A}^{\psibar}(x)\ ,
\end{align}
In particular, since $j_{\text{cl.}}=0$ if and only if $A_{\text{cl.}\,\mu}^{a}(x)=\avg{A_{\mu}^{a}(x)}_{j=0}$, $\psi_{\text{cl.}}(x)=\avg{\psi(x)}_{j=0}$, etc., we see that solving the equations
\begin{equation}
\frac{\delta_{L}\Gamma[F_{\text{cl.}}]}{\delta F_{\text{cl.}}(x)}=0
\end{equation}
for the fields $F_{\text{cl.}}(x)$ is equivalent to finding the vacuum expectation values $\avg{F(x)}$ in the absence of external currents. For this reason, $\Gamma[F_{\text{cl.}}]$ is known as the \textit{quantum effective action} -- or, in brief, the effective action -- of the theory. From now on, we will drop the ``cl.'' subscript from the arguments of $\Gamma$.

When differentiated three or more times, the quantum effective action can be shown to generate the 1-particle irreducible (1PI) Green functions of the theory \cite{PS95}. For this reason, $\Gamma[F]$ is also known as the generator of the 1PI Green functions. The second derivatives of the effective action, on the other hand, are equal, modulo factors of $i$, to the functional inverses of the propagators of the fields \cite{PS95}:
\begin{align}
\int d^{4}z\ \Tavg{F^{A}(x)F^{C}(z)}_{j_{\text{cl.}}[F]}\frac{\delta^{2}_{L}\Gamma[F]}{\delta F^{C}(z)\delta F^{B}(y)}=i\delta^{A}_{B}\,\delta(x-y)\ 
\end{align}
where the indices $A$, $B$, $C$ denote the different fields and their spacetime and color components.

The effective action can be used to study how the symmetries of a quantum field theory affect its Green functions. Indeed, it can be proved \cite{PS95,Wei96} that, for any transformation $\delta F^{A}$ of the fields,
\begin{equation}
\int d^{4}x\ \avg{\delta F^{A}(x)}_{j_{\text{cl.}}[F]}\frac{\delta_{L}\Gamma[F]}{\delta F^{A}(x)}=\avg{\delta S_{\text{FP}}}\ ,
\end{equation}
where $\delta S_{\text{FP}}$ is the transformation of the action corresponding to $\delta F^{A}$. If the action is invariant under $\delta F^{A}$ -- that is, if $\delta S_{\text{FP}}=0$ --, then the effective action will satisfy the invariance property
\begin{equation}\label{effactsym}
\int d^{4}x\ \avg{\delta F^{A}(x)}_{j_{\text{cl.}}[F]}\frac{\delta_{L}\Gamma[F]}{\delta F^{A}(x)}=0\ .
\end{equation}
Relations of this kind are used to prove the renormalizability of QCD -- and, more generally, of the Yang-Mills theories -- in the covariant gauges, by exploiting a symmetry of the action known as BRST symmetry. The latter is the subject of the next section.

\subsection{BRST symmetry, the Slavnov-Taylor identities and the Nielsen identities}
\label{sec:brst}

In Section \ref{sec:fpquant} we saw that in order to quantize QCD it is necessary to fix a gauge. By construction, the action that results from the Faddeev-Popov procedure is no longer gauge invariant; nonetheless, in 1975, C. Becchi, A. Rouet and R. Stora \cite{BRS75} and I. V. Tyutin \cite{Tyu75} showed that the FP action still possesses a symmetry which can be regarded as a remnant of the full gauge symmetry. This symmetry is today known as the BRST symmetry.\\

In order to define the BRST transformations, we first need to introduce a so-called Nakanishi-Lautrup (NL) field $B^{a}$ \cite{Lau66,Nak66} in the FP action. Observing that
\begin{equation}
e^{-\frac{i}{2\xi}\int d^{4}x\ (\partial\cdot A)^{2}}\propto\int\mc{D}B\ e^{i\int d^{4}x\ \left(\frac{\xi}{2}B^{a}B^{a}+B^{a}\partial\cdot A^{a}\right)}=\int\mc{D}B\ e^{i\int d^{4}x\ \left(\frac{\xi}{2}B^{a}B^{a}-\partial^{\mu}B^{a}A^{a}_{\mu}\right)}\ ,
\end{equation}
the FP Lagrangian can be rewritten as
\begin{equation}\label{lagfpb}
\ELL_{\text{FP}}=-\frac{1}{4}\,F_{\mu\nu}^{a}F^{a\,\mu\nu}+\psibar(i\gamma^{\mu}D_{\mu}-M)\psi+\frac{\xi}{2}\,B^{a}B^{a}-\partial^{\mu}B^{a}A^{a}_{\mu}+\partial^{\mu}\cbar^{a}D_{\mu}c^{a}\ .
\end{equation}
When using the above expression for $\ELL_{\text{FP}}$, it is understood that the path integrals are to be computed by also integrating over the configurations of the NL field.

Consider now the following set of transformations:
\begin{align}\label{brsttrans}
\notag \delta A_{\mu}^{a}&=\epsilon D_{\mu}c^{a}\ ,\\
\notag \delta\psi&=i\epsilon g\, c^{a}T_{a}\psi\ ,\\
\delta c^{a}&=-\frac{1}{2}\epsilon g f^{a}_{bc}\,c^{b}c^{c}\ ,\\
\notag \delta\cbar^{a}&=\epsilon B^{a}\ ,\\
\notag \delta B^{a}&=0\ ,
\end{align}
where $\epsilon$ is a constant Grassmann parameter. Going back to Eq.~\eqref{transfinf}, we see that the first two lines in Eq.~\eqref{brsttrans} are formally identical to an infinitesimal SU(3) gauge transformation of the gluon and quark fields with the transformation parameters $\chi^{a}(x)$ taken to be equal to $\epsilon c^{a}(x)$. It follows that the first two terms in Eq.~\eqref{lagfpb} -- that is, the terms which make up the classical QCD action -- are left invariant by Eqs.~\eqref{brsttrans}. As for the other terms, we have
\begin{align}
&\delta\left(\frac{\xi}{2}\,B^{a}B^{a}-\partial^{\mu}B^{a}A^{a}_{\mu}+\partial^{\mu}\cbar^{a}D_{\mu}c^{a}\right)=\\
\notag&=-\epsilon\,\partial^{\mu}B^{a}D_{\mu}c^{a}+\epsilon\,\partial^{\mu}B^{a}D_{\mu}c^{a}+\partial^{\mu}\overline{c}^{a}\delta(D_{\mu}c^{a})=\\
\notag&=\partial^{\mu}\overline{c}^{a}\delta(D_{\mu}c^{a})\ .
\end{align}
The transformation of the covariant derivative $D_{\mu}c^{a}$ can be shown to vanish as a consequence of the Jacobi identity -- Eq.~\eqref{jacid} --,
\begin{equation}
\delta(D_{\mu}c^{a})=0\ .
\end{equation}
Therefore, we find that the FP Lagrangian -- and the action $S_{\text{FP}}$ with it -- is invariant under the BRST transformations defined by Eqs.~\eqref{brsttrans},
\begin{equation}
\delta\ELL_{\text{FP}}=0\ .
\end{equation}

A second way to prove the invariance of the FP action under the BRST transformations is via the nilpotency of the latter. Let $s$ be the operator such that $\delta F=\epsilon s F$, where $F$ is one of the QCD fields. Explicitly,
\begin{align}\label{brsttranss}
\notag sA_{\mu}^{a}&= D_{\mu}c^{a}\ ,\\
\notag s\psi&=ig\, c^{a}T_{a}\psi\ ,\\
sc^{a}&=-\frac{g}{2} f^{a}_{bc}\,c^{b}c^{c}\ ,\\
\notag s\cbar^{a}&= B^{a}\ ,\\
\notag sB^{a}&=0\ .
\end{align}
A straigthforward calculation shows that $s^{2}F=0$ for every $F$. It follows that $s^{2}=0$, i.e., the BRST transformations are \textit{nilpotent}. In terms of the BRST operator $s$, the FP Lagrangian can be rewritten as
\begin{equation}\label{lagfpb2}
\ELL_{\text{FP}}=\ELL_{\text{QCD}}+s\left(\frac{\xi}{2}\,B^{a}\cbar^{a}-A^{a}_{\mu}\partial^{\mu}\cbar^{a}\right)\ .
\end{equation}
Since $s\ELL_{\text{QCD}}=0$ and $s^{2}=0$, the BRST invariance of the FP Lagrangian follows.\\

Being global symmetries of the FP action, the BRST transformations have an associated conserved current $j_{B}^{\mu}$, given by
\begin{align}
j_{B}^{\mu}=-F^{a\,\mu\nu}D_{\nu}c^{a}+B^{a}D^{\mu}c^{a}-g\,\psibar\gamma^{\mu}T_{a}\psi\, c^{a}+\frac{g}{2}\,f^{a}_{bc}\partial^{\mu}\cbar^{a}c^{b}c^{c}\ ,\quad\quad \partial_{\mu}j^{\mu}_{B}=0\ .
\end{align}
The corresponding charge $Q_{B}=\int d^{3}x\ j^{0}_{B}$, called the BRST charge, is nilpotent and self-adjoint as a quantum operator\footnote{In what follows, we will not delve into the fascinating subject of the canonical quantization of the FP action. The interested reader is referred to \cite{KO78a,KO78b,Oji78,KO79a,KO79b,KO79c} for a complete treatment of the topic, and to Appendix \ref{app:cqfp} for a short summary of the formalism.},
\begin{equation}
Q_{B}^{2}=0\ ,\qquad\qquad Q_{B}^{\dagger}=Q_{B}\ .
\end{equation}
$Q_{B}$ generates the BRST transformations, in the sense that
\begin{equation}\label{brstchargecomm}
[Q_{B},F]_{\mp}=- i sF\ ,
\end{equation}
where the commutator $[\cdot,\cdot]_{-}=[\cdot,\cdot]$ applies to bosonic operators, whereas the anticommutator $[\cdot,\cdot]_{+}=\{\cdot,\cdot\}$ applies to fermionic operators. Using Eq.~\eqref{brstchargecomm}, the gluon field equations in Eq.~\eqref{feqfp} -- which in the presence of the NL field read
\begin{align}
D_{\mu}F^{a\,\mu\nu}-\partial^{\nu}B^{a}+g\,\psibar\gamma^{\nu}T^{a}\psi-gf^{a}_{bc}\,\partial^{\nu}\cbar^{b}c^{c}=0\ ,\qquad\qquad \xi B^{a}+\partial\cdot A^{a}=0
\end{align}
-- can be rewritten as
\begin{equation}
\partial_{\mu}F^{a\,\mu\nu}+g J^{a\,\nu}-i\left\{Q_{B},D^{\nu}\cbar^{a}\right\}=0\ ,
\end{equation}
where $J^{\mu}_{a}$, the color current, now includes contributions from both from the ghost and NL fields,
\begin{equation}
J^{\mu}_{a}=\psibar\gamma^{\mu}T_{a}\psi+f_{abc}\left(F^{b\,\mu\nu}A_{\nu}^{c}+A^{b,\mu}B^{c}+\cbar^{b}D^{\mu}c^{c}-\partial^{\mu}\cbar^{b}c^{c}\right)\ .
\end{equation}
In particular, on the solutions of the field equations, the color charge $Q^{a}=\int d^{3}x\ j^{a\,0}$ can be evaluated as
\begin{align}
Q^{a}&=\frac{1}{g}\int d^{3}x\, \left(\partial_{i}F^{a\,0i}+i\left\{Q_{B},D_{0}\cbar^{a}\right\}\right)=\\
\notag&=\frac{1}{g}\oint d^{2}x_{i}\ F^{0i}+\frac{1}{g}\int d^{3}x\, \left\{Q_{B},D_{0}\cbar^{a}\right\}\ ,
\end{align}
where $i=1,\dots,3$ are spatial indices. In the literature, this expression for $Q^{a}$ has been used to tentatively link the confinement of color charge to the asymptotic behavior of the gluon field strength tensor $F_{\mu\nu}^{a}$ as $|\vec{r}|\to \infty$ \cite{Oji78,KO79a}.\\

BRST symmetry is a powerful tool for exploring the properties of QCD and, more generally, of the gauge theories. Modern proofs of the perturbative renormalizability of the Yang-Mills theories in the covariant gauges exploit the BRST invariance of the FP action to determine which kind of divergences can appear in their Green functions \cite{Wei96}. BRST symmetry is also used to prove the perturbative unitarity of the scattering matrix in the context of the gauge theories, once a gauge has been fixed \cite{KO79a}. This is done by classifying the states of the theory according to the \textit{cohomology} of the BRST charge $Q_{B}$: the physical states $\ket{\text{phys}}$ -- that is, the states which can be realized in the physical world -- are identified with the BRST-closed states of zero ghost charge $Q_{c}$,
\begin{equation}
Q_{B}\ket{\text{phys}}=0\ ,\qquad\qquad Q_{c}\ket{\text{phys}}=0\ ,
\end{equation}
where $Q_{c}$ generates the transformations $c^{a}\to e^{\lambda}c^{a}$, $\cbar^{a}\to e^{-\lambda}c^{a}$, easily seen to be an additional symmetry of the FP Lagrangian. The physical Hilbert space, defined as the quotient $(\text{Ker}\{Q_{B}\}\cap\text{Ker}\{Q_{c}\})/\text{Im}\{Q_{B}\}$, can then be shown to carry a positive-definite inner product, a property which is essential for interpreting mathematical quantities such as $\braket{\text{phys}^{\prime}|\text{phys}}$ as transition amplitudes from one physical state to another.\\

BRST symmetry imposes relations known as Slavnov-Taylor identities (STI) \cite{Tay71,Sla72} between the Green functions of the gauge theories. The STI are often derived from Eq.~\eqref{effactsym} by choosing as $\delta F$ the BRST transformations of the elementary fields $F$, $\delta F=sF$; nonetheless, they are easier to obtain using the operator formalism (see Appendix~\ref{app:cqfp}).

Quite generally, we might say that any identity of the form
\begin{equation}\label{stiop}
\bra{0}[Q_{B},\mc{O}]_{\mp}\ket{0}=0\ ,
\end{equation}
where $\ket{0}$ is the vacuum state of the theory, $\mc{O}$ is an arbitrary operator and the upper (resp. lower) sign applies to bosonic (resp. fermionic) operators, is a Slavnov-Taylor identity. That Eq.~\eqref{stiop} is indeed an identity can be seen by explicitly writing out the anti/commutator and observing that, since the vacuum must be a physical state -- i.e., $Q_{B}\ket{0}=0$ --,
\begin{equation}
\bra{0}[Q_{B},\mc{O}]_{\mp}\ket{0}=\bra{0}Q_{B}\mc{O}\pm\mc{O}Q_{B}\ket{0}=0\ .
\end{equation}
From Eqs.~\eqref{brstchargecomm} and \eqref{stiop}, it follows that the VEV of the BRST transformation of any operator $\mc{O}$ vanishes,
\begin{equation}
\avg{s\mc{O}}=0\ .
\end{equation}

As an example of a STI, consider the (time-ordered) BRST transformation of the product $B^{a}(x)\cbar^{b}(y)$. Since
\begin{align}
0=\Tavg{s\left(B^{a}(x)\cbar^{b}(y)\right)}=\Tavg{B^{a}(x)B^{b}(y)}\ ,
\end{align}
BRST invariance tells us that the two-point function of the NL field vanishes exactly. A second STI is given by
\begin{align}
0=\Tavg{s\left(A_{\mu}^{a}(x)\cbar^{b}(y)\right)}=\Tavg{D_{\mu}c^{a}(x)\cbar^{b}(y)+A_{\mu}^{a}(x)B^{b}(y)}\ , 
\end{align}
that is,
\begin{align}\label{stiab}
\Tavg{A_{\mu}^{a}(x)B^{b}(y)}=-\Tavg{D_{\mu}c^{a}(x)\cbar^{b}(y)}\ .
\end{align}
The content of the above identity can be unpacked by making use of the field equations and of the canonical anticommutation relations for the ghost fields. After taking the divergence of the right-hand side of the equation, we obtain
\begin{align}
\partial^{\mu}_{(x)}\Tavg{D_{\mu}c^{a}(x)\cbar^{b}(y)}=\avg{\left\{D_{0}c^{a}(x),\cbar^{b}(y)\right\}}\,\delta(x^{0}-y^{0})=-i\delta^{ab}\delta(x-y)\ ,
\end{align}
where the first delta distribution comes from the derivative of the time-ordering operator and we have used the operatorial identities
\begin{equation}
\partial^{\mu}D_{\mu}c^{a}=0\ ,\qquad\qquad \{\cbar^{a}(\vec{x},t),D_{0}c^{b}(\vec{y},t)\}=-i\delta^{ab}\delta^{(3)}(\vec{x}-\vec{y})
\end{equation}
(see Appendix~\ref{app:cqfp}). By Lorentz symmetry, it follows that
\begin{equation}
\Tavg{D_{\mu}c^{a}(x)\cbar^{b}(y)}=\int \frac{d^{4}p}{(2\pi)^{4}}\ e^{-ip\cdot (x-y)}\ \frac{p_{\mu}}{p^{2}}\,\delta^{ab}\ .
\end{equation}
Eq.~\eqref{stiab} thus tells us that the correlator between the $A$ and the $B$ field is given by
\begin{equation}
\Tavg{A_{\mu}^{a}(x)B^{b}(y)}=\int \frac{d^{4}p}{(2\pi)^{4}}\ e^{-ip\cdot (x-y)}\ \frac{-p_{\mu}}{p^{2}}\,\delta^{ab}\ .
\end{equation}

The last result can be used to derive a fundamental property of the gluon propagator. If in the last equation we replace the NL field with the gluon field by applying the field equation
\begin{equation}
B^{a}=-\frac{1}{\xi}\,\partial^{\mu}A_{\mu}^{a}\ ,
\end{equation}
we find that the following relation holds for the correlator of two gluon fields:
\begin{equation}
\Tavg{A_{\mu}^{a}(x)\partial^{\nu} A^{b}_{\nu}(y)}=\int \frac{d^{4}p}{(2\pi)^{4}}\ e^{-ip\cdot (x-y)}\ \xi\,\frac{p_{\mu}}{p^{2}}\,\delta^{ab}\ .
\end{equation}
Since the components of the gluon field commute with each other, the latter is equivalent to
\begin{equation}\label{def456}
\partial^{\nu}_{(y)}\Tavg{A_{\mu}^{a}(x)A^{b}_{\nu}(y)}=\int \frac{d^{4}p}{(2\pi)^{4}}\ e^{-ip\cdot (x-y)}\ \xi\,\frac{p_{\mu}}{p^{2}}\,\delta^{ab}\ .
\end{equation}
Let now $\Delta_{\mu\nu}^{ab}(p)$ be the Fourier-transform of the gluon propagator $\Tavg{A_{\mu}^{a}(x)A_{\nu}^{b}(y)}$,
\begin{equation}
\Tavg{A_{\mu}^{a}(x)A_{\nu}^{b}(y)}=\int \frac{d^{4}p}{(2\pi)^{4}}\ e^{-ip\cdot (x-y)}\ \Delta_{\mu\nu}^{ab}(p)\ .
\end{equation}
By Lorentz symmetry, $\Delta_{\mu\nu}^{ab}(p)$ can be expressed in terms of two scalar functions $\Delta_{T}^{ab}(p)$, $\Delta_{L}^{ab}(p)$, so that
\begin{equation}
\Delta_{\mu\nu}^{ab}(p)=\Delta_{T}^{ab}(p)\, t_{\mu\nu}(p)+\Delta_{L}^{ab}(p)\, \ell_{\mu\nu}(p)\ ,
\end{equation}
where $t_{\mu\nu}(p)$ and $\ell_{\mu\nu}(p)$ are, respectively, the transverse and longitudinal projectors
\begin{equation}
t_{\mu\nu}(p)=\eta_{\mu\nu}-\frac{p_{\mu}p_{\nu}}{p^{2}}\ ,\qquad\qquad\ell_{\mu\nu}(p)=\frac{p_{\mu}p_{\nu}}{p^{2}}\ .
\end{equation}
Eq.~\eqref{def456} then tells us that $\Delta_{L}(p)=-i\xi/p^{2}$, that is
\begin{equation}
\Delta_{\mu\nu}^{ab}(p)=\Delta_{T}^{ab}(p)\, t_{\mu\nu}(p)+\frac{-i\xi}{p^{2}}\,\delta^{ab}\, \ell_{\mu\nu}(p)\ .
\end{equation}
The longitudinal component of the gluon propagator is thus constrained by BRST symmetry to be proportional to the gauge parameter $\xi$ and to have a pole at $p^{2}=0$.

The STI obtained from applying the BRST operator to polynomials of higher degree in the fields enforce relations between the higher-order Green functions. For instance, the vanishing of $\langle T\{s(A_{\mu}^{a}(x)A_{\nu}^{b}(y)\cbar^{c}(z))\}\rangle$ implies that
\begin{equation}
\langle T\{A_{\mu}^{a}(x)A_{\nu}^{b}(y)B^{c}(z)\}\rangle+\langle T\{D_{\mu}c^{a}(x)A_{\nu}^{b}(y)\cbar^{c}(z)\}\rangle+\langle T\{A_{\mu}^{a}(x)D_{\nu}^{b}c(y)\cbar^{c}(z))\}\rangle=0\ ,
\end{equation}
whereas that of $\langle T\{s(\psi(x)\psibar(y)\cbar^{b}(z))\}\rangle$ yields
\begin{equation}
\langle T\{\psi(x)\psibar(y)B^{b}(z)\}\rangle=-ig\langle T\{T_{a}\psi(x)\psibar(y)\,c^{a}(x)\cbar^{b}(z)\}\rangle+ig\langle T\{\psi(x)\psibar(y)T_{a}\,c^{a}(y)\cbar^{b}(z)\}\rangle\ .
\end{equation}\\

In the context of the covariant gauges, BRST symmetry finds yet another application in the derivation of the so-called Nielsen identities (NI) \cite{Nie75,PS85,BLS95}. The NI describe how the Green functions of a gauge theory vary with the gauge parameter $\xi$. They are extremely useful when studying features such as the gauge dependence of the poles of the propagators, or more generally of any Green function.

In order to derive the general form of the Nielsen identities, we start by observing that, being $\frac{i\xi}{2}\int d^{4}x\ B^{a}B^{a}$ the only $\xi$-dependent term in the FP action, given an arbitrary operator $\mc{O}$,
\begin{equation}
\frac{\partial}{\partial \xi}\avg{\mc{O}}=\frac{i}{2}\int d^{4}x\ \Tavg{\mc{O}B^{a}(x)B^{a}(x)}\ ,
\end{equation}
where we have dropped a disconnected product of the form $\avg{\mc{O}}\avg{B^{a}B^{a}}$ since, as we saw earlier, $\Tavg{B^{a}(x)B^{b}(y)}=0$. Thanks to BRST symmetry, the last equation can be rewritten as
\begin{align}
\frac{\partial}{\partial \xi}\avg{\mc{O}}&=\frac{i}{2}\int d^{4}x\ \Tavg{\mc{O}\,s\left[B^{a}(x)\cbar^{a}(x)\right]}=\\
\notag&=\mp\frac{i}{2}\int d^{4}x\ \Tavg{\left(s\mc{O}\right)\,B^{a}(x)\cbar^{a}(x)}\ ,
\end{align}
where the upper (resp. lower) sign applies to bosonic (resp. fermionic) operators. The Nielsen identity for the Green function $\avg{\mc{O}}$ is then obtained by writing out explicitly the BRST variation of the operator $\mc{O}$ in the above equation. For instance, the NI for the gluon propagator reads
\begin{align}\label{nielsgl}
\frac{\partial}{\partial\xi}\Tavg{A_{\mu}^{a}(x)A_{\nu}^{b}(y)}&=-\frac{i}{2}\int d^{4}z\, \Big[\Tavg{D_{\mu}c^{a}(x)A_{\nu}^{b}(y)B^{c}(z)\cbar^{c}(z)}+\\
\notag&\qquad\qquad\quad\ +\Tavg{A_{\mu}^{a}(x)D_{\nu}c^{b}(y)B^{c}(z)\cbar^{c}(z)}\Big]\ ,
\end{align}
whereas the one for the quark propagator is given by
\begin{align}\label{nielsqk}
\frac{\partial}{\partial\xi}\Tavg{\psi(x)\psibar(y)}&=-\frac{1}{2}\int d^{4}z\, \Big[\Tavg{T_{a}\psi(x)\psibar(y)c^{a}(x)B^{b}(z)\cbar^{b}(z)}+\\
\notag&\qquad\qquad\quad\ -\Tavg{\psi(x)\psibar(y)T_{a}c^{a}(x)B^{b}(z)\cbar^{b}(z)}\Big]\ .
\end{align}

By taking the Fourier transform of the Nielsen identities, one is able to study how the poles of the corresponding Green function vary with the gauge. Let us show how this works for the case of the gluon propagator. Let $\mathcal{F}^{ab}_{\mu\nu}(x)$ be the function defined by
\begin{equation}
\mathcal{F}^{ab}_{\mu\nu}(x-y)=\frac{i}{2}\int d^{4}z\ \left\{\Tavg{D_{\mu}c^{a}(x)A_{\nu}^{b}(y)B^{c}(z)\cbar^{c}(z)}+(x\leftrightarrow y,\mu\leftrightarrow\nu,a\leftrightarrow b)\right\}\ .
\end{equation}
In terms of the Fourier transform $\mathcal{F}_{\mu\nu}^{ab}(p)$ of $\mathcal{F}^{ab}_{\mu\nu}(x)$, the NI in Eq.~\eqref{nielsgl} can be expressed in momentum space as
\begin{equation}
\frac{\partial}{\partial\xi}\,\Delta_{\mu\nu}^{ab}(p)=-\mathcal{F}_{\mu\nu}^{ab}(p)\ ,
\end{equation}
or, dropping the indices,
\begin{equation}
\frac{\partial}{\partial\xi}\,\Delta(p)=-\mathcal{F}(p)\ .
\end{equation}
A Nielsen identity for the inverse gluon propagator is then easily derived:
\begin{equation}
\frac{\partial}{\partial \xi}\,\Delta^{-1}(p)=\Delta^{-1}(p)\cdot \mathcal{F}(p)\cdot \Delta^{-1}(p)\ .
\end{equation}
Due to the orthogonality of the transverse and longitudinal projectors, $t(p)\cdot\ell(p)=0$, the transverse and the longitudinal components of the above equation decouple: if we set
\begin{equation}\label{nielsglinv}
\Delta^{-1}(p)=\Delta_{T}^{-1}(p)\ t(p)+\Delta^{-1}_{L}(p)\ \ell(p)\ ,\qquad \mc{F}(p)=\mc{F}_{T}(p)\ t(p)+\mc{F}_{L}(p)\ \ell(p)\ ,
\end{equation}
using the idempotency relations $t(p)\cdot t(p)=t(p)$, $\ell(p)\cdot \ell(p)=\ell(p)$ and dropping the color structure, from Eq.~\eqref{nielsglinv} we obtain the two equations
\begin{equation}
\frac{\partial}{\partial\xi}\,\Delta^{-1}_{T}(p)=\mc{F}_{T}(p)\,\Delta^{-2}_{T}(p)\ ,\qquad\qquad \frac{\partial}{\partial\xi}\,\Delta^{-1}_{L}(p)=\mc{F}_{L}(p)\,\Delta^{-2}_{L}(p)\ .
\end{equation}
It is quite simple to check that the longitudinal identity is indeed satisfied: to do this, it suffices to observe that\footnote{Here, like before, we use the ghost field equations and anticommutation relations.}
\begin{align}
&\partial^{\mu}_{(x)}\Tavg{D_{\mu}c^{a}(x)A_{\nu}^{b}(y)B^{c}(z)\cbar^{c}(z)}=-i\delta(x-z)\Tavg{A_{\nu}^{b}(y)B^{a}(x)}=\\
\notag&\qquad\qquad\qquad=-i\delta(x-z)\int\frac{d^{4}p}{(2\pi)^{4}}e^{-ip\cdot(x-y)}\ \frac{p_{\nu}}{p^{2}}\,\delta^{ab}
\end{align}
implies
\begin{equation}
\mc{F}_{L}(p)=\frac{i}{p^{2}}\ ,
\end{equation}
so that, with $\Delta_{L}^{-1}(p)=ip^{2}/\xi$, the longitudinal identity explicitly reads
\begin{equation}
\frac{\partial}{\partial\xi}\left(\frac{ip^{2}}{\xi}\right)=\frac{i}{p^{2}}\left(\frac{ip^{2}}{\xi}\right)^{2}=-\frac{ip^{2}}{\xi^{2}}\ .
\end{equation}

Of more interest is the transverse identity, since in this case the exact expression of the transverse gluon propagator is not known a priori. To see how information on the gauge dependence of the transverse pole can be obtained from the first of Eq.~\eqref{nielsglinv}, we start by noticing that, since the function $\mc{F}^{ab}_{\mu\nu}(x)$ can be rewritten as
\begin{equation}
\mathcal{F}^{ab}_{\mu\nu}(x-y)=-\frac{i}{2\xi}\int d^{4}z\ \left\{\Tavg{D_{\mu}c^{a}(x)A_{\nu}^{b}(y)\partial\cdot A^{c}(z)\cbar^{c}(z)}+(x\leftrightarrow y,\mu\leftrightarrow\nu,a\leftrightarrow b)\right\}\ ,
\end{equation}
due to the presence of two $A$ fields in its definition, the Fourier transform $\mc{F}^{ab}_{\mu\nu}(p)$ will contain the gluon propagator as a factor, that is
\begin{equation}
\mc{F}(p)=F_{T}(p)\,\Delta_{T}(p)\ t(p)+F_{L}(p)\,\Delta_{L}(p)\ \ell(p)
\end{equation}
for a pair of functions $F_{T}(p)$ and $F_{L}(p)$ whose zeros and poles, in general, are different from the poles of $\Delta_{T}(p)$ and $\Delta_{L}(p)$. We have already verified that this is the case for the longitudinal component, for which $F_{L}(p)=-1/\xi$. If we plug the last equation into the first of Eq.~\eqref{nielsglinv}, we now obtain
\begin{equation}
\frac{\partial}{\partial\xi}\,\Delta^{-1}_{T}(p,\xi)=F_{T}(p,\xi)\,\Delta^{-1}_{T}(p,\xi)\ ,
\end{equation}
where we have made the gauge dependence of the transverse propagator and of the function $F_{T}$ explicit.

Consider now what happens to the pole of $\Delta_{T}$ as $\xi$ is changed. The transverse pole $p_{0}(\xi)$ is defined as the solution of the equation
\begin{equation}
\Delta^{-1}_{T}(p_{0}(\xi),\xi)=0\ .
\end{equation}
By taking the total derivative of this equation with respect to the gauge parameter, we find that
\begin{align}
0=\frac{d}{d\xi}\,\Delta^{-1}_{T}(p_{0}(\xi),\xi)&=\frac{\partial}{\partial\xi}\,\Delta^{-1}_{T}(p_{0}(\xi),\xi)+\frac{\partial}{\partial p}\Delta^{-1}_{T}(p_{0}(\xi),\xi)\,\frac{dp_{0}}{d\xi}(\xi)=\\
\notag&=F_{T}(p_{0}(\xi),\xi)\,\Delta^{-1}_{T}(p_{0}(\xi),\xi)+\frac{\partial}{\partial p}\Delta^{-1}_{T}(p_{0}(\xi),\xi)\,\frac{dp_{0}}{d\xi}(\xi)=\\
\notag&=\frac{\partial}{\partial p}\Delta^{-1}_{T}(p_{0}(\xi),\xi)\,\frac{dp_{0}}{d\xi}(\xi)\ .
\end{align}
Since the momentum-derivative of $\Delta^{-1}_{T}$, in general, is different from zero\footnote{Actually, it can be shown that the derivative vanishes if the pole is found at $p=0$ for every value of the gauge parameter. However, if this is the case, the gauge-parameter independence of the pole holds trivially anyways, so this does not contradict our proof.}, the last equation implies that
\begin{equation}
\frac{dp_{0}}{d\xi}(\xi)=0\ ,
\end{equation}
that is, the position of the pole does not depend on the gauge parameter $\xi$. This exact property of the strong interactions will be extremely useful when formulating a possible modification of the QCD perturbative series in Chapter \ref{chpt:sme}.

\section{Ordinary perturbation theory, the strong coupling constant and the infrared breakdown of perturbative QCD}
\renewcommand{\rightmark}{\thesection\ \ \ Ordinary PT, $\alpha_{s}$ and the IR breakdown of pQCD}

Having reviewed the definition of Quantum Chromodynamics, we are now in a position to discuss the merits and failures of its standard perturbative formulation. When dealing with a quantum field theory, one rarely knows how to exactly solve the equations which describe its underlying physics. Approximate methods must thus be devised in order to turn abstract expressions into physical predictions. Perturbation theory is one such method. By assuming that the interactions only slightly affect the behavior of an otherwise free theory, it provides an expansion of the quantities of interest in powers of the coupling constant. If this assumption is accurate, the higher-order terms in the expansion will make a negligible contribution to the overall result. The approximation then consists in truncating the perturbative series at some predefined, finite order.

In the context of Quantum Chromodynamics, the results obtained by employing ordinary perturbation theory have proved to be very accurate in the high-energy regime. This is made possible by the asymptotic freedom which is typical of the non-abelian gauge theories: given a suitable definition for the energy dependence of the coupling constant, it can be shown that, as the energy increases, the coupling decreases like the inverse of a logarithm. Thus, at very high energies, the non-abelian gauge theories resemble free theories of massless gauge bosons and Dirac fermions.

If we reverse our perspective, however, we see that the perturbative method necessarily entails an increase of the coupling constant at low energies. And as the coupling increases, perturbation theory no longer can be trusted, since -- at the very least -- the higher-order terms in the perturbative series will become less and less negligible. The situation is only made worse by the fact that, at low energies, the coupling computed within the standard perturbative framework increases so fast that it becomes infinite at a finite energy, thus developing what is known in the literature as an infrared Landau pole. The existence of an IR Landau pole in the strong coupling constant puts the final word on the validity of ordinary perturbative QCD (pQCD) at low energies.\\

In what follows, we will start our discussion by reviewing the standard set-up of perturbation theory in the context of QCD.

\subsection{The standard perturbative expansion and Feynman rules of QCD}
\label{sec:opqcd}

Let $\mc{O}$ be an arbitrary operator and $\avg{\mc{O}}$ be its VEV. As we saw in Sec.~\ref{sec:fpquant}, $\avg{\mc{O}}$ can be computed in the functional formalism as
\begin{equation}\label{ghi789}
\avg{\mc{O}}=\frac{\int\mc{D}\mathscr{F}\ e^{iS_{\text{FP}}}\ \mc{O}}{\int\mc{D}\mathscr{F}\ e^{iS_{\text{FP}}}}\ .
\end{equation}
In the above equation, the Faddeev-Popov action can be naturally split into two terms,
\begin{align}
S_{\text{FP}}=S_{0}+S_{\text{int.}}\ ,
\end{align}
where the zero-order action $S_{0}$ and the interaction action $S_{\text{int.}}$ are defined as
\begin{equation}
S_{0}=\lim_{g\to 0}\ S_{\text{FP}}\ ,\qquad\qquad S_{\text{int.}}=S_{\text{FP}}-S_{0}\ .
\end{equation}\cleannlnp
Explicitly,
\begin{align}
\label{sqcd0}S_{0}&=\int d^{4}x\ \bigg\{-\frac{1}{2}\,\partial_{\mu}A_{\nu}^{a}(\partial^{\mu}A^{a\,\nu}-\partial^{\nu}A^{a\,\mu})-\frac{1}{2\xi}\,\partial^{\mu}A_{\mu}^{a}\partial^{\nu}A_{\nu}^{a}+\\
\notag&\qquad\qquad\qquad+\psibar(i\gamma^{\mu}\partial_{\mu}-M)\psi+\partial^{\mu}\cbar^{a}\partial_{\mu}c^{a}\bigg\}\ ,\\
\label{sqcdint0}S_{\text{int}}&=\int d^{4}x\ \bigg\{-gf^{a}_{bc}\,\partial_{\mu}A_{\nu}^{a}\,A^{b\,\mu}A^{c\,\nu}-\frac{1}{4}\,g^{2}f^{a}_{bc}f^{a}_{de}\,A_{\mu}^{b}A_{\nu}^{c}A^{d\,\mu}A^{e\,\nu}+\\
\notag&\qquad\qquad\qquad+g\,\psibar\gamma^{\mu}T_{a}\psi A_{\mu}^{a}+gf^{a}_{bc}\,\partial^{\mu}\cbar^{a}A_{\mu}^{b}c^{c}\bigg\}\ .
\end{align}

If we assume that the interaction terms contained in $S_{\text{int.}}$ contribute to $\avg{\mc{O}}$ with a small correction over the corresponding zero-order result, that is, $\avg{\mc{O}}\approx\avg{\mc{O}}_{0}$, where
\begin{equation}\label{def567}
\avg{\mc{O}}_{0}=\frac{\int\mc{D}\mathscr{F}\ e^{iS_{0}}\ \mc{O}}{\int\mc{D}\mathscr{F}\ e^{iS_{0}}}\ ,
\end{equation}
then it makes sense to seek an expansion of Eq.~\eqref{ghi789} in powers of $g$. Of course, such an hypothesis is sensible only provided that the coupling constant $g$ is not too large.

The power-expansion of $\avg{\mc{O}}$ can be obtained as follows. We first rewrite the exponentials in Eq.~\eqref{ghi789} as
\begin{align}\label{jkl234}
\avg{\mc{O}}&=\left(\frac{\int\mc{D}\mathscr{F}\ e^{iS_{0}}\ \mc{O}\,\sum_{n=0}^{+\infty}\ \frac{1}{n!}\,(iS_{\text{int.}})^{n}}{\int\mc{D}\mathscr{F}\ e^{iS_{0}}}\right)\left(\frac{\int\mc{D}\mathscr{F}\ e^{iS_{0}}\ \sum_{n=0}^{+\infty}\ \frac{1}{n!}\,(iS_{\text{int.}})^{n}}{\int\mc{D}\mathscr{F}\ e^{iS_{0}}}\right)^{-1}=\\
\notag&=\left(\sum_{n=0}^{+\infty}\ \frac{1}{n!}\,\Tavg{\mc{O}(iS_{\text{int.}})^{n}}_{0}\right)\left(\sum_{n=0}^{+\infty}\ \frac{1}{n!}\,\Tavg{(iS_{\text{int.}})^{n}}_{0}\right)^{-1}\ ,
\end{align}
where, as in Eq.~\eqref{def567}, the averages denoted with the subscript $0$ are to be computed with respect to the zero-order action. Then, in order to evaluate the averages, we observe that the free action $S_{0}$ is quadratic in the fields. More precisely, in momentum space,
\begin{align}
-iS_{0}=\int \frac{d^{4}p}{(2\pi)^{4}}\ \left\{\frac{1}{2}\,A_{\mu}^{a}(-p)\,[\Delta^{-1}_{0}(p)]^{\mu\nu}_{ab}\,A_{\nu}^{b}+\psibar(p)\,S_{M}^{-1}(p)\,\psi(p)+\cbar^{a}(p)\,[\mc{G}_{0}^{-1}(p)]_{ab}\,c^{b}(p)\right\}\ ,
\end{align}
where $\Delta_{0}(p)$, $S_{M}(p)$ and $\mc{G}_{0}(p)$ are, respectively, the zero-order gluon, quark and ghost propagator,
\begin{align}\label{qcdprop0}
\Delta_{0\,\mu\nu}^{ab}(p)&=\int d^{4}x\ e^{ip\cdot x}\ \Tavg{A_{\mu}^{a}(x)A_{\nu}^{b}(0)}_{0}=\frac{-i}{p^{2}}\,\delta^{ab}\,\left[t_{\mu\nu}(p)+\xi\ell_{\mu\nu}(p)\right]\ ,\\
S_{M}(p)&=\int d^{4}x\ e^{ip\cdot x}\ \Tavg{\psi(x)\psibar(0)}_{0}=\frac{i}{\slashed{p}-M}\,\one_{3\times 3}\ ,\\
\mc{G}_{0}^{ab}(p)&=\int d^{4}x\ e^{ip\cdot x}\ \Tavg{c^{a}(x)\cbar^{b}(0)}_{0}=\frac{i}{p^{2}}\, \delta^{ab}\ ,\label{qcdprop0f}
\end{align}
with $\slashed{p}=\gamma^{\mu}p_{\mu}$, and their inverses are given by
\begin{align}
[\Delta^{-1}_{0}(p)]^{\mu\nu}_{ab}(p)&=ip^{2}\,\delta_{ab}\,\left(t^{\mu\nu}(p)+\frac{1}{\xi}\ell^{\mu\nu}(p)\right)\ ,\\
S_{M}^{-1}(p)&=\int d^{4}x\ e^{ip\cdot x}\ \Tavg{\psi(x)\psibar(0)}_{0}=-i(\slashed{p}-M)\,\one_{3\times 3}\ ,\\
[\mc{G}_{0}^{-1}(p)]_{ab}&=\int d^{4}x\ e^{ip\cdot x}\ \Tavg{c^{a}(x)\cbar^{b}(0)}_{0}=-i p^{2}\, \delta_{ab}\ .
\end{align}
En passing, we note that the zero-order gluon and ghost propagators -- having a pole at $p^{2}=0$ -- are those of a massless particle, whereas that of the quark is massive (provided that $M\neq 0$).

Now, since the interaction action $S_{\text{int.}}$ is polynomial in the fields, every term of the form
\begin{equation}\label{ghi890}
\Tavg{\mc{O}(iS_{\text{int.}})^{n}}_{0}=\frac{\int\mc{D}\mathscr{F}\ e^{-(-iS_{0})}\ \mc{O}(iS_{\text{int.}})^{n}}{\int\mc{D}\mathscr{F}\ e^{-(-iS_{0}})}
\end{equation}
under the summation sign in Eq.~\eqref{jkl234} is made up of Gaussian integrals. The technique for computing such integrals is well-known. In the quantum-field-theoretical setting, they are usually evaluated by employing so-called Feynman rules, by virtue of which averages taken with respect to the zero-order action $S_{0}$ are expressed as sums of Feynman diagrams. To see where the Feynman rules come from, observe that the relation
\begin{equation}\label{jkl345}
\Tavg{F^{A_{1}}\cdots F^{A_{n}}}_{0}=\sum_{\sigma}\ (\pm1)^{s_{\sigma}}\Tavg{F^{A_{\sigma(1)}}F^{A_{\sigma(2)}}}_{0}\cdots \Tavg{F^{A_{\sigma(n-1)}}F^{A_{\sigma(n)}}}_{0}\ ,
\end{equation}
where $\sigma$ is a permutation of the indices and $s_{\sigma}=0,1$ depends on whether the permutation changes the order of the Grassmann variables, holds for the Gaussian averages. In particular, the left-hand side of Eq.~\eqref{jkl345} can be expressed as a sum of products of zero-order propagators, whose explicit form we have already reported in Eqs.~\eqref{qcdprop0}-\eqref{qcdprop0f}. In the context of Eq.~\eqref{ghi890}, in addition to the fields, each interaction term contained in $(iS_{\text{int.}})^{n}$ contributes to these products with a spacetime integral and with a multiplicative factor which characterizes the interaction\footnote{Interaction terms containing derivatives can be Fourier-transformed to momentum space, where the derivatives are replaced by factors of momentum times $-i$.}.

\begin{figure}[H]
\centering
\includegraphics[width=0.73\textwidth]{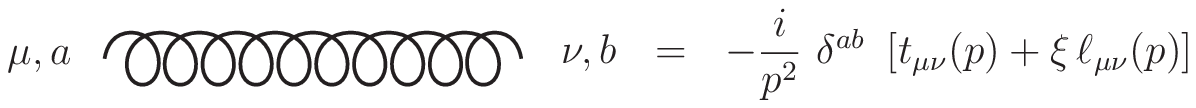}
\vspace{5pt}
\caption{Zero-order gluon propagator}\label{fig:gluprop0}
\end{figure}

\begin{figure}[H]
\centering
\includegraphics[width=0.48\textwidth]{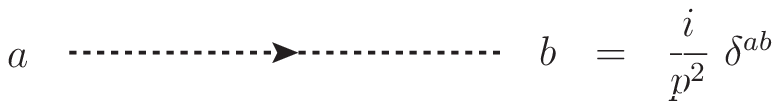}
\vspace{5pt}
\caption{Zero-order ghost propagator}\label{fig:ghprop}
\end{figure}

\begin{figure}[H]
\centering
\includegraphics[width=0.48\textwidth]{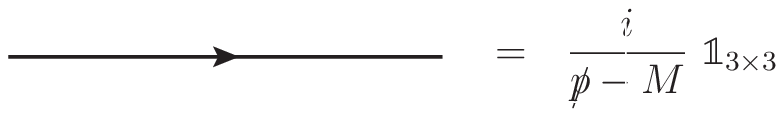}
\vspace{5pt}
\caption{Zero-order quark propagator}\label{fig:qkprop}
\end{figure}

\begin{figure}[H]
\centering
\includegraphics[width=0.88\textwidth]{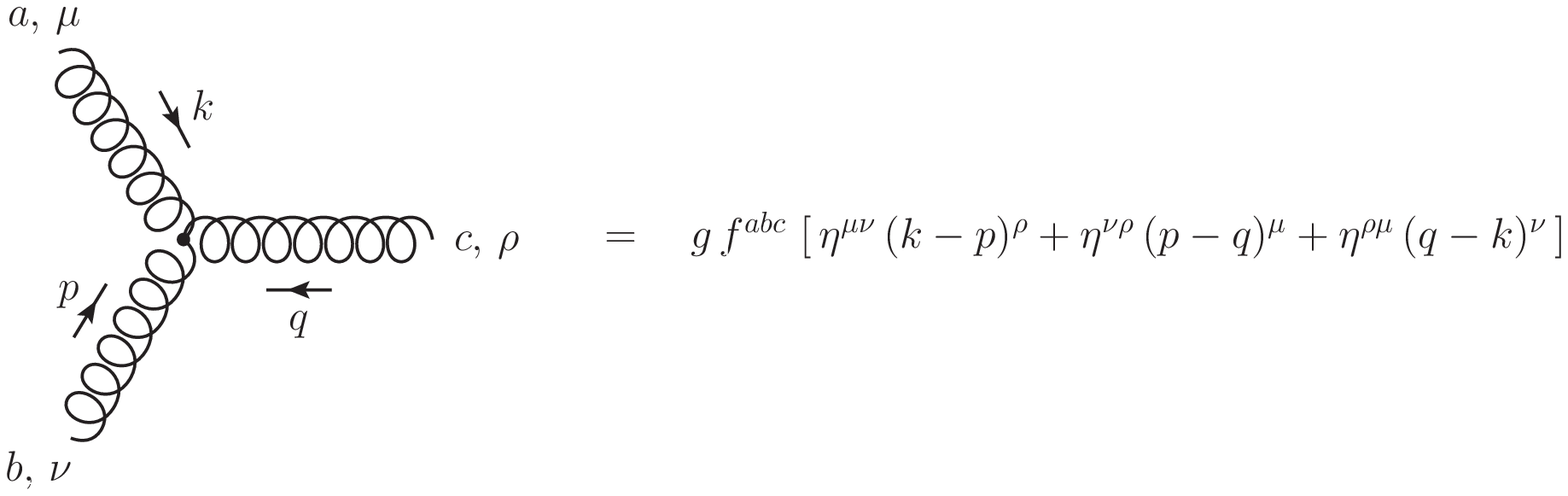}
\\
\caption{3-gluon vertex}\label{fig:3gl}
\end{figure}

\begin{figure}[H]
\centering
\includegraphics[width=0.81\textwidth]{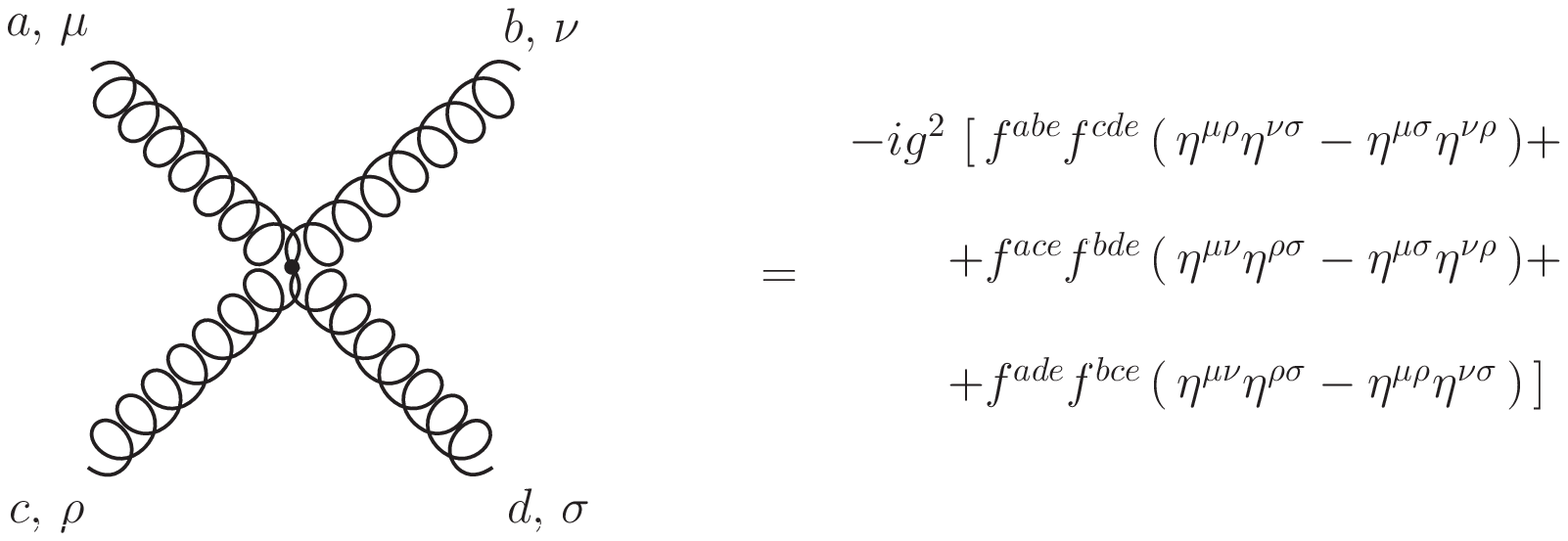}
\vspace{20pt}
\caption{4-gluon vertex}\label{fig:4gl}
\end{figure}

\begin{figure}[H]
\centering
\includegraphics[width=0.52\textwidth]{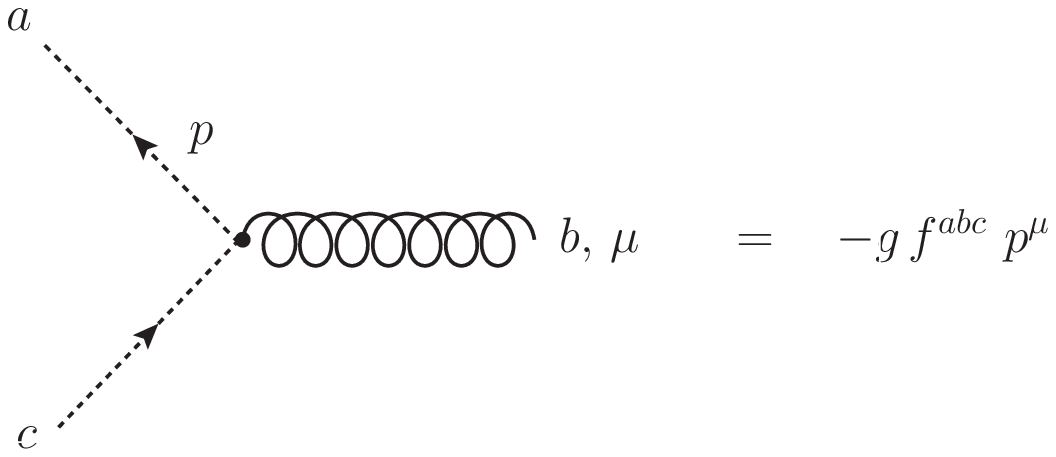}
\vspace{16pt}
\caption{Ghost-gluon vertex}\label{fig:ghgl}
\end{figure}

\begin{figure}[H]
\centering
\includegraphics[width=0.52\textwidth]{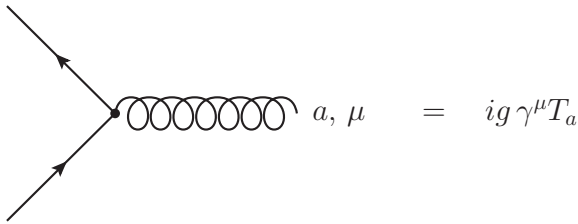}
\vspace{16pt}
\caption{Quark-gluon vertex}\label{fig:qkgl}
\end{figure}

The summands in Eq.~\eqref{jkl345} -- together with the interaction factors and integrals -- can be represented pictorially by drawing a vertex for every spacetime point that appears in the expression, and a line connecting one pair of vertices for every propagator. The correspondence between the graphical components of the resulting diagrams and the analytical expressions which must be used for the actual calculations is provided by the Feynman rules of the theory. Those of the standard perturbative expansion of QCD are displayed in Figs.~\ref{fig:gluprop0}-\ref{fig:qkgl}. Looking back at Eq.~\eqref{sqcdint0}, we see that the vertices can be read out straight from the interaction action, once the cubic and quartic gluon interactions have been symmetrized with respect to the gluon momenta and Minkowski and color indices.
\vspace{5mm}
\vspace{7mm}
\vspace{5mm}
\newpage
\topskip0pt
\vspace*{\fill}
\vspace{4mm}
\vspace{4mm}
\vspace{4mm}
\vspace*{\fill}
\newpage
\vspace*{5mm}
\vspace{5mm}
As the last step in our derivation, a combinatorial argument \cite{PS95} can be used to prove that
\begin{equation}
\sum_{n=0}^{+\infty}\ \frac{1}{n!}\,\Tavg{\mc{O}(iS_{\text{int.}})^{n}}_{0}=\left(\sum_{n=0}^{+\infty}\ \frac{1}{n!}\,\Tavg{\mc{O}(iS_{\text{int.}})^{n}}_{0,\text{conn.}}\right)\left(\sum_{n=0}^{+\infty}\ \frac{1}{n!}\,\Tavg{(iS_{\text{int.}})^{n}}_{0}\right)\ ,
\end{equation}
where $\Tavg{\mc{O}(iS_{\text{int.}})^{n}}_{0,\text{conn.}}$ is given by the subset of diagrams in which all the interaction vertices coming from $(iS_{\text{int.}})^{n}$ are connected to at least one vertex coming from the ``external'' operator $\mc{O}$. Therefore, going back to Eq.~\eqref{jkl234}, we obtain our final expression for $\avg{\mc{O}}$ in the form
\begin{equation}
\avg{\mc{O}}=\sum_{n=0}^{+\infty}\ \frac{1}{n!}\,\Tavg{\mc{O}(iS_{\text{int.}})^{n}}_{0,\text{conn.}}\ ,
\end{equation}
where we remark that the averages are to be computed diagrammatically using the Feynman rules of the theory. Clearly, with the exception of the $n=0$ term which is equal to $\avg{\mc{O}}_{0}$, the remaining terms under the sign of summation will be proportional to powers of the coupling constant $g$.\\

Under what conditions does the power-expansion of the Green functions fail? The first and most obvious answer is that it fails when the coupling constant is too large. If this is the case, higher-order terms will generally be larger than the lower-order ones, and only retaining the first few terms of the expansion will yield a bad approximation.

Ordinary perturbation theory is also unsuitable for computing Green functions which receive non-negligible contributions from quantities which are not analytical in the coupling constant. An example of these are non-perturbative mass scales of the form
\begin{equation}
\Lambda=\mu e^{-\frac{\kappa}{g^{2}}}\ ,
\end{equation}
with $\kappa$ a constant and $\mu$ a second mass scale, which cannot be expanded in powers of $g$.

A third situation in which the perturbative series is unable to capture the exact behavior of a Green function is when the latter has features that can only be described by resumming an infinite number of diagrams. This is what happens, for instance, when computing some four- and higher-point Green functions which contain information on bound states. The latter is lost when a purely perturbative truncation is performed.

Finally, perturbation theory can fail if the perturbative series does not converge. If this is the case, it may happen that adding higher-order terms to the perturbative series improves the approximation up until some finite order, after which the higher-order corrections actually start to make the approximation worse. The worst-kept secret in quantum field theory is that perturbation theory is believed, in fact, not to converge\footnote{And is actually \textit{known} not to converge in the framework of Quantum Electrodynamics \cite{Dys52}.}. Nonetheless, as far as Quantum Chromodynamics is concerned, going to higher perturbative orders has so far only improved the match with the experimental data.\\

To end this section, we make an observation that will be useful later on in Chapter~\ref{chpt:sme}. Going back to Eq.~\eqref{ghi890}, we see that the correspondence between (perturbative) quantum averages and Feynman diagrams exists simply because the zero-order action $S_{0}$, being quadratic in the fields, allows us to compute the former in terms of Gaussian integrals. Indeed, the derivation which brought us from the exact Green function $\avg{\mc{O}}$ to its perturbative expression only rests upon this basic property of $S_{0}$. As a consequence, if instead of the $S_{0}$ given by Eq.~\eqref{sqcd0} we had chosen any other quadratic action $S_{0}^{\prime}$, the procedure we followed in the previous pages would have yielded the same final result, with $S_{\text{int.}}$ replaced by $S_{\text{int.}}^{\prime}=S_{\text{FP}}-S_{0}^{\prime}$. Of course, such a result would no longer be an expansion of the Green function in powers of the coupling constant. In some situations, however, this may be precisely what we need in order to account for intrinsically non-perturbative effects.

\subsection{Regularization and renormalization}
\label{sec:regren}

When perturbation theory is used to compute the Green functions of a quantum field theory, it may happen that some of the integrals which appear in the expressions do not converge. The resulting divergences are called UV divergences if they arise from integrals $I$ which in the high-momentum limit of their domain of integration behave like
\begin{align}\label{uvdivdef}
I\sim \int\frac{d^{4}q}{(2\pi)^{4}}\ \frac{1}{q^{\kappa_{\text{UV}}}}
\end{align}
with $\kappa_{\text{UV}}\leq4$, or IR divergences if they arise from integrals which in the low-momentum limit of the integration domain behave like
\begin{align}
I\sim \int\frac{d^{4}q}{(2\pi)^{4}}\ \frac{1}{q^{\kappa_{\text{IR}}}}
\end{align}
with $\kappa_{\text{IR}}\geq4$. Since in local quantum field theories such as QCD momenta with a negative power can enter the perturbative series only via the propagators, these theories are free of IR divergences when they describe fields which are all massive: if this is the case, then the low-energy limit of the propagators $\lim_{q\to 0}\pm i(q^{2}-m^{2})^{-1}=\mp im^{-2}$ prevents the appearance of powers of momenta $\kappa_{\text{IR}}\geq 4$ as $q\to 0$. On the other hand, if the theory contains massless fields, then its Green functions can be IR-divergent.

As we saw in the last section, within the ordinary perturbative formulation of QCD the gluon and ghost fields are both treated as massless; IR divergences, thus, do indeed show up in the standard QCD perturbative series. The usual way to deal with these divergences is to limit oneself to the computation of quantities which are IR-finite\footnote{A comprehensive discussion of this topic in the context of standard perturbation theory can be found e.g. in \cite{Wei95}, Chapter 13.}. In what follows, we will not discuss further the IR-finiteness of QCD, given that the non-perturbative generation of a mass for the gluon is expected to cure the IR divergences. Instead, we will give a brief overview of the status of QCD for what concerns its finiteness in the UV.\newpage

By making use of dimensional arguments \cite{Wei95}, it can be shown that the Green functions of any quantum field theory whose coupling constants have non-positive mass dimension do contain UV divergences. QCD with its adimensional strong coupling $g$ is, of course, no exception. In order to remove the divergences and obtain meaningful physical predictions, the perturbative series of QCD must undergo a procedure known as \textit{renormalization}.

Renormalization is carried out in two steps. The first of these, called \textit{regularization}, consists in redefining the perturbative series so that the UV divergences only appear when a suitable limit is taken. As a consequence of regularization, the series is made finite, but also dependent on a \textit{regulator}, which needs to be removed at the end the calculation. The second step is the renormalization proper, and consists in absorbing the diverging terms into the free parameters of the theory. Once these terms disappear from the equations, the regulator can be removed safely, and the procedure as a whole will have left behind a finite expression.

The most commonly employed regularization scheme when dealing with gauge theories is \textit{dimensional regularization} (dimreg). If we go back to Eq.~\eqref{uvdivdef}, after generalizing from the $4$-dimensional spacetime to a $d$-dimensional one,
\begin{align}
I\sim \int\frac{d^{d}q}{(2\pi)^{d}}\ \frac{1}{q^{\kappa_{\text{UV}}}}\ ,
\end{align}
we see that, if $d$ strictly smaller than $\kappa_{\text{UV}}$, then the integral $I$ does actually converge. In particular, we could compute $I$ in an arbitrary dimension $d$ and, as long as the latter is left undetermined, the resulting expression would remain finite. This is what dimreg does. By redefining the theory so that it fields live in $d=4-\epsilon$ dimensions, dimreg provides finite results which depend on the regulator $\epsilon$. In the regularized expressions, the UV divergences appear as poles of the form $1/\epsilon^{\kappa}$ -- with $\kappa$ a positive integer -- which must be removed by renormalization before taking the $\epsilon\to0$ limit.

When dimensionally regularizing a theory, one must be careful in redefining any quantity in such a way that the corresponding $4$-dimensional limit has the correct dimensions. This inevitably leads to the introduction of a mass scale $\mu$ into the expressions. For instance, since in $d$ dimensions the strong coupling constant $g_{(d)}$ has mass dimension $(4-d)/2$, in going from $4$ to $d$ dimensions the coupling must be defined as $g_{(d)}=\mu^{\frac{4-d}{2}}\,g_{(4)}$, where the physical interpretation of the mass scale $\mu$ depends on the choice of renormalization scheme which is later made to remove the divergences.

Behind renormalization lies the observation that the parameters which are contained in the Lagrangian are not actually measured in the experiments; rather, what is measured are suitable combinations of Green functions. A similar argument can be made for the quantum fields. For instance, in the LSZ formula for the calculation of cross sections \cite{PS95}, the propagators -- and thus the corresponding fields -- appear in such a way that multiplying the latter by some constant does not change the result. Since neither the value of the parameters nor the normalization of the fields can be measured in the experiments, we are free to redefine both. The parameters and fields which were originally present in the Lagrangian are referred to as \textit{bare} quantities, whereas their redefined versions are called \textit{renormalized} quantities. Explicitly, in the context of QCD, one introduces a set of multiplicative constants $Z_{A}$, $Z_{c}$, $Z_{\psi}$, $Z_{g}$, $Z_{\xi}$ and $Z_{M}$ such that\footnote{It can be proved that the ghost and antighost fields, as well as the different color components of the same field, can be renormalized using the same multiplicative factor, see e.g. \cite{Wei96}.}
\newpage
\begin{align}\label{multrenqcd}
A_{B\,\mu}^{a}=Z_{A}^{1/2}\,A_{R\,\mu}^{a}\ ,&\quad\quad c^{a}_{B}=Z_{c}^{1/2}\,c^{a}_{R}\ ,\quad\quad \cbar^{a}_{B}=Z_{c}^{1/2}\,\cbar^{a}_{R}\ ,\quad\quad \psi_{B}=Z_{\psi}^{1/2}\,\psi_{R}\ ,\\
\notag &g_{B}=Z_{g}\,g_{R}\ ,\quad\quad \xi_{B}=Z_{\xi}\,\xi_{R}\ ,\quad\quad M_{B}=Z_{M}\,M_{B}\ ,
\end{align}
where the subscripts $B$ and $R$ denote, respectively, bare and renormalized quantities. The Green functions are similarly renormalized. For instance, for the propagators, we have
\begin{align}
\Tavg{A_{R\,\mu}^{a}(x)A_{R\,\nu}^{b}(y)}&=Z_{A}^{-1}\,\Tavg{A_{B\,\mu}^{a}(x)A_{B\,\nu}^{b}(y)}\ ,\\
\Tavg{c_{R}^{a}(x)\cbar_{R}^{b}(y)}&=Z_{c}^{-1}\,\Tavg{c_{B}^{a}(x)\cbar_{B}^{b}(y)}\ ,\\
\Tavg{\psi_{R}(x)\psibar_{R}(y)}&=Z_{\psi}^{-1}\,\Tavg{\psi_{B}(x)\psibar_{B}(y)}\ .
\end{align}

Consider what happens if the bare propagators in the above equations contain UV divergences. If such divergences appear as multiplicative factors, then they can be absorbed into the renormalization factors $Z_{A}$, $Z_{c}$ and $Z_{\psi}$, and the resulting renormalized propagators will be finite. In the process of factorizing the divergences, the bare couplings and masses may also need to be renormalized (and usually they do). The property of a theory by which all of its Green functions can be made UV-finite by a renormalization of the kind defined in Eq.~\eqref{multrenqcd} is called \textit{multiplicative renormalizability}. By making use of BRST symmetry, it can be shown that QCD is indeed a multiplicatively renormalizable theory \cite{Wei96}.

While the diverging terms contained in the $Z$-factors are fixed by the requirement that the renormalized Green functions be finite, no such constraint exists with respect to the choice of the finite part of the renormalization factors. This has two implications. First of all, renormalization can be carried out in a multitude of schemes, depending on which \textit{renormalization conditions} are chosen to fix the (finite terms of the) $Z$-factors. Second of all, the renormalized Green functions are renormalization-scheme-dependent. Thus, when comparing Green functions -- or, more generally, any renormalized quantity -- computed by different methods, one must be careful to take into account any difference in definition that might come from renormalization.

In QCD, there is no single universally adopted scheme for the renormalization of coupling, masses and fields. Nevertheless, three renormalization schemes are often used in the literature. These are the minimal-subtraction (MS), modified-minimal-subtraction ($\overline{\text{MS}}$) and momentum-subtraction (MOM) schemes.

The MS and $\overline{\text{MS}}$ schemes are defined specifically for dimensionally regularized quantities. In dimreg, it can be shown that, as $\epsilon=4-d\to 0$, the divergences take the general form
\begin{align}
\left(\frac{2}{\epsilon}+\ln 4\pi -\gamma_{E}\right)^{\kappa}
\end{align}
for some positive integer $\kappa$, where $\gamma_{E}\approx0.5772$ is the Euler-Mascheroni constant. In the MS scheme, the renormalization constants are chosen so that only the $\epsilon$-poles are removed from the equations. In the $\overline{\text{MS}}$ scheme, on the other hand, also the $\ln4\pi$ and $\gamma_{E}$ constants are removed. The highest-order perturbative results obtained so far in QCD are computed in the MS and $\overline{\text{MS}}$ schemes.

In the MOM scheme, the $Z$-factors are chosen so that the renormalized Green functions are equal to some specified quantity at fixed spacelike momentum $p^{2}=-\mu^{2}$ -- often, to their tree-level value. For instance, the MOM-scheme gluon and ghost renormalization factors $Z_{A}$ and $Z_{c}$ are defined by the requirement that the exact transverse gluon and\cleannlnp ghost propagators $\Delta_{T}(p^{2};\mu)$ and $\mc{G}(p^{2};\mu)$ renormalized at the scale $\mu$, when evaluated at $p^{2}=-\mu^{2}$, be equal to
\begin{equation}
\Delta_{T}(p^{2};\mu)\bigg|_{p^{2}=-\mu^{2}}=\frac{-i}{-\mu^{2}}\ ,\qquad\qquad\mc{G}(p^{2};\mu)\bigg|_{p^{2}=-\mu^{2}}=\frac{i}{-\mu^{2}}\ .
\end{equation}
The MOM scheme is especially suitable for situations in which finite-mass effects cannot be neglected when renormalizing the theory. We will explain precisely what this means in Chapter~\ref{chpt:sme}.

In the next section, we will see that renormalization causes the parameters of the theory to acquire a dependence on the energy scale. The analysis of the parameters' scale dependence is the subject of study of the Renormalization Group approach.

\subsection{The Renormalization Group, the running coupling constant and the Landau pole}
\label{sec:rgint}

As we saw both in the context of dimensional regularization and in defining the MOM scheme, when renormalizing a quantum field theory one is forced to introduce a momentum scale into the expressions. Let us explore in more detail how this works and what use can be made of this aspect of renormalization.\\

In dimreg, in order for the fields, couplings and masses to have the correct mass dimension in the $\epsilon=4-d\to 0$ limit, their corresponding renormalized quantities must be defined as
\begin{align}\label{multrenqcddimreg}
&A_{B\,\mu}^{a}=\mu^{-\frac{\epsilon}{2}}\,Z_{A}^{1/2}\,A_{R\,\mu}^{a}\ ,\quad\ c^{a}_{B}=\mu^{-\frac{\epsilon}{2}}Z_{c}^{1/2}\,c^{a}_{R}\ ,\quad\ \cbar^{a}_{B}=\mu^{-\frac{\epsilon}{2}}Z_{c}^{1/2}\,\cbar^{a}_{R}\ ,\quad\ \psi_{B}=\mu^{-\frac{\epsilon}{2}}Z_{\psi}^{1/2}\,\psi_{R}\ ,\\
\notag &\qquad\qquad\qquad g_{B}=\mu^{-\frac{\epsilon}{2}}Z_{g}\,g_{R}\ ,\qquad\qquad \xi_{B}=Z_{\xi}\,\xi_{R}\ ,\qquad\qquad M_{B}=Z_{M}\,M_{R}\ ,
\end{align}
where $\mu$ is an energy scale. Since the bare quantities do not depend on $\mu$, it follows that the $Z$-factors and the renormalized quantities, in general, are $\mu$-dependent.

In the MS and \MSbar\ scheme, by definition, the renormalization constants only contain $\epsilon$-poles (plus $\mu$-independent constants in the latter scheme); these poles, of course, are multiplied by factors of the coupling constant. Without loss of generality, we can assume that the coupling appears in the $Z$-factors in its renormalized form -- i.e., as $g_{R}$ --, for in any case $g_{B}$ can be expressed in terms of the former. When the MS and \MSbar\ renormalization constants are expressed as a series in $\epsilon^{-1}$ and $g_{R}$, they clearly depend on the scale $\mu$ only implicitly, through the coupling constant $g_{R}$. By looking at Eq.~\eqref{multrenqcddimreg}, then, we see that all of the renormalized quantities in the MS and \MSbar\ schemes must have an explicit dependence on the dimreg scale $\mu$. In particular, it makes sense to ask how they vary as functions of $\mu$.

Consider the renormalized gluon propagator: in dimreg,
\begin{equation}\label{lds934}
\Delta_{\mu\nu}^{ab}(x;\mu)=\Tavg{A_{R\,\mu}^{a}(x)A_{R\,\nu}^{b}(0)}=\mu^{\epsilon}\,Z_{A}^{-1}\,\Tavg{A_{B\,\mu}^{a}(x)A_{B\,\nu}^{b}(0)}\ .
\end{equation}
Since in the MS/\MSbar\ schemes, as we just saw, the renormalized fields depend explicitly on the dimreg scale $\mu$, the renormalized propagators do as well (hence the notation with $\mu$ as an argument of $\Delta_{\mu\nu}^{ab}$). By taking the total derivative of both members of Eq.~\eqref{lds934} with respect to $\mu$ and switching to momentum space, we obtain the following equation for the renormalized gluon propagator:
\begin{equation}\label{rgglu}
\left(\mu\frac{\partial}{\partial\mu}+\beta_{g}\,\frac{\partial}{\partial g_{R}}+\gamma_{M}\,M_{R}\,\frac{\partial}{\partial M_{R}}+\gamma_{\xi}\,\xi_{R}\,\frac{\partial}{\partial \xi_{R}}+\gamma_{A}\right)\,\Delta_{\mu\nu}^{ab}(p;\mu)=0\ ,
\end{equation}
where the \textit{beta} and \textit{gamma} functions -- the latter being also known as \textit{anomalous dimensions}~-- are defined as
\begin{align}
\beta_{g}=\mu\frac{dg_{R}}{d\mu}\ ,\quad\quad \gamma_{M}=\frac{\mu}{M_{R}}\frac{dM_{R}}{d\mu}\ ,\quad\quad \gamma_{\xi}=\frac{\mu}{\xi_{R}}\frac{d\xi_{R}}{d\mu}\ ,\quad\quad \gamma_{A}=\frac{\mu}{Z_{A}}\frac{dZ_{A}}{d\mu}\ .
\end{align}
From the second line of Eq.~\eqref{multrenqcddimreg}, we compute that
\begin{align}
\beta_{g}=\frac{\epsilon}{2}\,g_{R}-\frac{\mu}{Z_{g}}\frac{d Z_{g}}{d\mu}\,g_{R}\ ,\qquad\qquad\gamma_{M}=-\frac{\mu}{Z_{M}}\frac{dZ_{M}}{d\mu}\ ,\qquad\qquad \gamma_{\xi}=-\frac{\mu}{Z_{\xi}}\frac{dZ_{\xi}}{d\mu}\ .
\end{align}
Eq.~\eqref{rgglu} is called the Renormalization-Group (RG) equation for the gluon propagator (in the MS/\MSbar\ scheme). 

By proceeding as we did for the gluon propagator, one can derive RG equations for any of the Green functions of the theory. These equations are generally used for improving the convergence of the perturbative series, so that the perturbative approximation remains valid over a wide range of momenta. Roughly speaking, since at order $n$ the fixed-scale, renormalized perturbative expressions contain terms of the form
\begin{equation}
\left(g^{2}\,\ln\frac{-p^{2}}{\mu^{2}}\right)^{n}\ ,
\end{equation}
perturbation theory breaks down when $\left|g^{2}\,\ln\frac{-p^{2}}{\mu^{2}}\right|\sim 1$. This happens when the energy scale of the momenta is much larger or much smaller than that of the renormalization scale $\mu$. A way to tame these large logs, thus improving the approximation, is to make use of a \textit{sliding} renormalization scale. This is precisely what can be achieved by solving the RG equations.

As an example, let us go back to the gluon propagator. For simplicity, we will only consider its transverse component. The solution of the RG equation for the transverse gluon propagator, renormalized in the MS/\MSbar\ scheme at the scale $\mu$, can be expressed as
\begin{equation}
\Delta_{T}\left(p^{2};g_{R}(\mu),M_{R}(\mu),\xi_{R}(\mu),\mu\right)=e^{-\int_{\mu_{0}}^{\mu}\frac{d\mu^{\prime}}{\mu^{\prime}}\,\gamma_{A}(\mu^{\prime})}\,\Delta_{T}\left(p^{2};g_{R}(\mu_{0}),M_{R}(\mu_{0}),\xi_{R}(\mu_{0}),\mu_{0}\right)\ ,
\end{equation}
where $\mu_{0}$ is a second mass scale and we have written down explicitly all the parameters on which the propagator depends. If we take $\mu_{0}$ to be equal to $p=\sqrt{-p^{2}}$, then we can put the last equation in the form
\begin{equation}
\Delta_{T}\left(p^{2};g_{R}(\mu),M_{R}(\mu),\xi_{R}(\mu),\mu\right)=e^{\int_{\mu}^{p}\frac{d\mu^{\prime}}{\mu^{\prime}}\,\gamma_{A}(\mu^{\prime})}\,\Delta_{T}\left(p^{2};g_{R}(p),M_{R}(p),\xi_{R}(p),p\right)\ .
\end{equation}
Thus, by making use of the RG equations, the propagator evaluated at the momentum $p$ and renormalized at the scale $\mu$ can be expressed as a function of the parameters renormalized at the scale $p$, rather than on those renormalized at the scale $\mu$. Since at $p^{2}=-\mu^{2}$ the logs of the form $\log(-p^{2}/\mu^{2})$ vanish, the right-hand side of the equation does not contain large logs. As a result, the perturbative series will not break down when $\left|g^{2}\,\ln\frac{-p^{2}}{\mu^{2}}\right|\sim 1$, but rather when $g^{2}(p)\sim 1$.\\

In the context of the MOM scheme, the RG improvement of the Green functions works in a slightly different, but non-dissimilar way. Going back to Eq.~\eqref{multrenqcddimreg} and recalling that, in the MOM scheme, the renormalized quantities are defined by fixing their value at some specified momentum scale $\mu_{\text{MOM}}$ which bears no connection to the dimreg scale $\mu$, we see that the renormalized fields, couplings and masses, together with the renormalization factors, must necessarily depend on $\mu_{\text{MOM}}$. For instance, since the transverse gluon propagator's MOM renormalization condition reads
\begin{equation}
Z_{A}^{-1}D_{B,T}(p^{2}=-\mu_{\text{MOM}}^{2})=D_{T}(p^{2}=-\mu_{\text{MOM}}^{2};\mu_{\text{MOM}})=\frac{i}{\mu_{\text{MOM}}^{2}}\ ,
\end{equation}
where we have denoted with $D_{B,T}(p^{2})$ the bare transverse gluon propagator, it is clear that $Z_{A}$ \textit{explicitly} depends on $\mu_{\text{MOM}}$:
\begin{equation}
Z_{A}=-i\mu_{\text{MOM}}^{2}\,D_{B,T}(-\mu_{\text{MOM}}^{2})\ .
\end{equation}
This is at variance with what happens in the MS and \MSbar\ schemes with respect to the dimreg scale $\mu$. Then, since $A_{B\,\mu}^{a}=\mu^{-\frac{\epsilon}{2}}\,Z_{A}^{1/2}\,A_{R\,\mu}^{a}$, we see that the renormalized gluon field -- and the renormalized gluon propagator with it -- must explicitly depend on the MOM scale $\mu_{\text{MOM}}$ as well. The same applies to the other renormalized fields, parameters and Green functions. Thus, whereas in the MS and \MSbar\ schemes we were interested in investigating the dependence of the Green functions on the dimreg scale $\mu$, in the MOM scheme we study their dependence with respect to the MOM scale $\mu_{\text{MOM}}$.

The derivation of the RG equations in the MOM scheme follows the same steps undertaken to derive those of the MS and \MSbar\ schemes. The result is formally identical to its MS/\MSbar\ counterpart, with $\mu$ everywhere replaced by $\mu_{\text{MOM}}$. The only formal difference with respect to the MS/\MSbar\ case is that, since in the MOM scheme we do not differentiate with respect to the dimreg scale $\mu$, the MOM-scheme beta function loses an $\epsilon$-term:
\begin{equation}
\beta_{g}\big|_{\text{MOM}}=-\frac{\mu_{\text{MOM}}}{Z_{g}}\frac{d Z_{g}}{d\mu_{\text{MOM}}}\,g_{R}\ .
\end{equation}
Despite the similarities between the MS/\MSbar\ and MOM RG equations, we should notice that the beta functions and anomalous dimensions computed in these schemes are very different from each other. Indeed, as we remarked before, the MS/\MSbar\ $Z$-factors -- and the beta and gamma functions with them -- do not explicitly depend on the renormalization scale; the very opposite holds in the MOM scheme. Nonetheless, the solutions of the RG equations are formally the same in each of the three schemes. As a consequence, the removal of the large logs by the RG improvement of the perturbative series can be carried out in the MOM scheme as well.\\

From the perspective of the Renomalization Group, the strength of the interaction is measured in terms of the momentum-scale-dependent coupling\footnote{From now on we drop the $R$ subscript to denote renormalized quantities.} $g(\mu)$. The latter is known as the \textit{running coupling constant} and is defined as the solution of the beta-function equation
\begin{equation}\label{betaeq}
\mu\frac{dg}{d\mu}=\beta_{g}\ ,
\end{equation}
where the explicit form of $\beta_{g}$ depends on the renormalization scheme in which $g$ is defined. What does the beta function of QCD look like in the MS/\MSbar\ scheme? An explicit one-loop calculation \cite{PS95} carried out in ordinary perturbation theory shows that, for a Yang-Mills theory with gauge group SU(N) minimally coupled to $n_{f}$ Dirac fields in the fundamental representation,
\begin{equation}\label{beta1lsun}
\beta_{g}=-\frac{\beta_{0}g^{3}}{16\pi^{2}}\ ,
\end{equation}
where the beta-function coefficient $\beta_{0}$ is given by
\begin{equation}\label{beta01lsun}
\beta_{0}=\frac{11}{3}N-\frac{2}{3}n_{f}\ .
\end{equation}
Clearly, as long as $n_{f}<\frac{11}{2}N$, the one-loop beta function is negative. As a consequence, the corresponding running coupling constant will decrease with the momentum scale. This behavior is know as asymptotic freedom, and is typical of the non-abelian gauge theories. For QCD we have $N=3$ and $n_{f}$ at most\footnote{We say ``at most'' because, as it turns out, the number $n_{f}$ which must be plugged into the beta-function equation at the scale $\mu$ is actually equal to the number of fermions whose masses are smaller than $\mu$.} equal to $6$; therefore, $\beta_{0}\geq 7$: QCD is an asymptotically free theory.

Asymptotic freedom is the reason why standard pQCD works so well in the UV regime. If we assume that there exists an energy scale at which the perturbative truncation of the QCD Green functions constitutes a good approximation, then the Renormalization Group tells us that at larger scales, being the coupling smaller, the perturbative series will converge even better. Higher orders in perturbation theory will become more and more negligible, and the behavior described by Eq.~\eqref{beta1lsun} will become (asymptotically) exact. Of course, the assumption is proved right by comparing the theoretical predictions to the experimental results.

The negativity of the beta function, however, also implies that at lower energies the coupling becomes larger. To one loop, we can exactly solve the beta-function equation and obtain
\begin{equation}\label{tyx128}
g^{2}(\mu)=\frac{g^{2}(\mu_{0})}{1+\frac{\beta_{0}g^{2}(\mu_{0})}{16\pi^{2}}\,\ln(\mu^{2}/\mu_{0}^{2})}
\end{equation}
for the one-loop running coupling of standard perturbation theory. Whereas in the $\mu\to\infty$ limit $g^{2}(\mu)$ goes to zero like the inverse of a logarithm, in the IR the coupling increases faster and faster, until it develops a pole at a finite scale $\Lambda$ whose value is given by
\begin{equation}\label{lambdapole}
\Lambda=\mu_{0}\,\exp\left(-\frac{8\pi^{2}}{\beta_{0}g^{2}(\mu_{0})}\right)\ .
\end{equation}
Such a singularity is known as an IR Landau pole, and outright invalidates the one-loop standard perturbative approximation of QCD at low energies. The situation does not improve at all when higher-order corrections are included in the perturbative series: if we express the beta function's power expansion as
\begin{equation}
\beta_{g}=-\frac{g^{3}}{16\pi^{2}}\,\sum_{n=0}^{+\infty}\ \beta_{n}\,\left(\frac{g^{2}}{16\pi^{2}}\right)^{n}\ ,
\end{equation}
then we find that, for QCD, $\beta_{n}>0$ at least up to $n=4$, which is the current limit of the perturbative calculations. As a result, as the perturbative order is increased, the running coupling diverges at scales which are even larger than the $\Lambda$ defined by Eq.~\eqref{lambdapole}.

Does this means that QCD ceases to be a self-consistent theory at low energies? Of course not. Before hitting the Landau pole, the running coupling computed in standard perturbation theory has long become too large for the ordinary perturbative approximation to be trusted. The Landau pole is, in fact, an artifact of standard pQCD: as we will see starting from the next chapter, other approaches allow us to compute running coupling constants which not only are finite in the IR, but also remain relatively small at low energies. We will review some of these approaches in the following chapters.\\

In conclusion, while asymptotic freedom makes sure that ordinary perturbation theory constitutes a good approximation in the UV, the existence of an IR Landau pole in the pQCD running coupling constant signals the breakdown of the method in the low-energy regime. Due to the failure of pQCD at low energies, alternative approaches have to be devised for studying the IR dynamics of the strong interactions.

\chapter{Non-perturbative techniques and results in Quantum Chromodynamics}
\renewcommand{\leftmark}{\thechapter\ \ \ Non-perturbative techniques and results in QCD}
\renewcommand{\rightmark}{\thechapter\ \ \ Non-perturbative techniques and results in QCD}
\label{chpt:npmet}

In this chapter we will briefly review some of the techniques that are employed to study the non-perturbative regime of Quantum Chromodynamics and describe results upon which we will rely during the rest of this thesis. In Sec.~\ref{sec:lqcdrev} we will go over the definition of lattice QCD within the framework of pure Yang-Mills theory and report the results of \cite{DOS16} regarding the Landau-gauge gluon propagator and the Taylor-scheme running coupling. In Sec.~\ref{sec:operev} we will formulate the Operator Product Expansion in a general setting and show how non-vanishing vacuum condensates can contribute to the Green functions of QCD with terms that cannot be derived by ordinary perturbative methods. In Sec.~\ref{sec:gzrev} we will review the Gribov-Zwanziger approach to the existence of the Gribov copies and discuss the analytical properties of the Gribov-Zwanziger zero-order gluon propagator both in the absence and in the presence of condensates. Finally, in Sec.~\ref{sec:cfrev} we will describe the set-up of the Curci-Ferrari model and report the one- and two-loop results of \cite{GPRT19} concerning the RG-improved gluon propagator.

\section{Lattice QCD}
\renewcommand{\rightmark}{\thesection\ \ \ Lattice QCD}
\label{sec:lqcdrev}

\subsection{Set-up}

Recall that the quantum average $\avg{\mc{O}}$ of an arbitrary operator $\mc{O}$ can be computed as
\begin{equation}\label{vkq943}
\avg{\mc{O}}=\frac{\int\mc{D}\mathscr{F}\ e^{iS}\,\mc{O}}{\int\mc{D}\mathscr{F}\ e^{iS}}\ ,
\end{equation}
where $S$ is the classical action of the theory and $\mc{D}\mathscr{F}$ is the measure over the field configurations. Due to the oscillatory nature of the exponential $e^{iS}$, the integrals in Eq.~\eqref{vkq943} converge very poorly when evaluated numerically by making use of Monte Carlo techniques. This can however be fixed by going from Minkowski spacetime to Euclidean spacetime\footnote{The dynamical content of the theory is then recovered by performing an analytic continuation of the Euclidean results back to Minkowski space.}.

The ($4$-dimensional) Euclidean spacetime is defined by taking the fourth component $\tau=x^{4}$ of the Euclidean position $4$-vector to be equal to $it=ix^{0}$, where $t=x^{0}$ is the time component of the Minkowski position $4$-vector. Despite being a real quantity by definition, $\tau$ is referred to as the \textit{imaginary time}. In the action $S$ of the theory, the integration measure, the spacetime derivatives and the fields can be replaced by corresponding quantities defined in Euclidean spacetime:\newpage
\begin{align}
\notag dx^{0}dx^{1}dx^{2}dx^{3}&=-idx^{4}dx^{1}dx^{2}dx^{3}\ ,\\
\partial_{0}=\partial/\partial x^{0}&=i\partial/\partial x^{4}=i\partial_{4}\ ,\\
\notag V^{0}&=-iV^{4}\ ,\\
\notag V_{0}&=iV_{4}\ ,
\end{align}
where $V^{0},V_{0}$ and $V^{4},V_{4}$ are, respectively, the time components of a Minkowski vector or covector and the imaginary-time components of a Euclidean vector or covector. In terms of its Lagrangian $\mc{L}$, the action then reads
\begin{equation}
S=\int dx^{0}dx^{1}dx^{2}dx^{3}\ \mc{L}=-i\int dx^{4}dx^{1}dx^{2}dx^{3}\ \mc{L}\big|_{x^{0}=-ix^{4}}=iS_{E}\ ,
\end{equation}
where $S_{E}$, defined as
\begin{equation}
S_{E}=-\int dx^{4}\,dx^{1}dx^{2}dx^{3}\ \mc{L}\big|_{x^{0}=-ix^{4}} \,
\end{equation}
is the \textit{Euclidean action} of the theory. As long as the energy density is positive-definite, $S_{E}$ turns out to be a non-negative quantity. It follows that, in Euclidean space, the average
\begin{equation}\label{vkq944}
\avg{\mc{O}}=\frac{\int\mc{D}\mathscr{F}\ e^{-S_{E}}\,\mc{O}}{\int\mc{D}\mathscr{F}\ e^{-S_{E}}}
\end{equation}
has nice convergence properties thanks to the damped nature of the real exponential $e^{-S_{E}}$. In practice, when Monte Carlo techniques are employed to compute averages like $\avg{\mc{O}}$, $e^{-S_{E}}$ is used as the probability density for sampling the configuration space.

Dropping the label $E$, the Euclidean action $S_{\text{YM}}$ of pure Yang-Mills theory reads
\begin{equation}\label{fjm948}
S_{\text{YM}}=\int d^{4}x\ \frac{1}{4}\,F_{\mu\nu}^{a}F^{a\,\mu\nu}\ ,
\end{equation}
where $d^{4}x=dx^{4}dx^{1}dx^{2}dx^{3}$ and in terms of the component $A_{4}^{a}$ and of derivatives with respect to $x^{4}$ the gluon field-strength tensor $F^{a}_{\mu\nu}$ is still defined as $F_{\mu\nu}^{a}=\partial_{\mu}A_{\nu}^{a}-\partial_{\nu}A_{\mu}^{a}+gf^{a}_{bc}\,A^{b}_{\mu}A^{c}_{\nu}$. In Eq.~\eqref{fjm948}, the spacetime indices are raised and lowered using the Euclidean metric $\delta=\text{diag}(+1,+1,+1,+1)$ in place of the Minkowski metric. Clearly, $S_{\text{YM}}\geq0$.\\

The Euclidean Yang-Mills action is discretized on the lattice \cite{SO04} by defining group variables $U_{\mu}(x)\in$~SU(3) over the links that connect any pair of neighboring lattice sites, where the index $\mu$ refers to the direction of the link. $U_{\mu}(x)$ is then interpreted in terms of the (Euclidean) gauge field $A_{\mu}$ as
\begin{equation}
U_{\mu}(x)=e^{ig_{0}aA_{\mu}(x+a\hat{e}_{\mu}/2)}+O(a^{3})\ ,
\end{equation}
where $g_{0}$ is the bare lattice coupling, $a$ is the lattice spacing, $\hat{e}_{\mu}$ is a unit spacetime vector in the direction $\mu$ and $A_{\mu}(x+a\hat{e}_{\mu}/2)=A_{\mu}^{a}(x+a\hat{e}_{\mu}/2)\,T_{a}$ is the gauge field evaluated midpoint in the link. By converse, up to higher-order corrections in the lattice spacing,
\begin{equation}
A_{\mu}(x+a\hat{e}_{\mu}/2)=\frac{1}{2ig_{0}a}\left(U_{\mu}(x)-U^{\dagger}_{\mu}(x)\right)+O(a^{2})\ .
\end{equation}

The lattice action itself is defined in terms of so-called \textit{Wilson loops} \cite{Wil74} $U_{\mu\nu}(x)$,
\begin{equation}
U_{\mu\nu}(x)=U_{\mu}(x)U_{\nu}(x+a\hat{e}_{\mu})U^{\dagger}_{\mu}(x+a\hat{e}_{\nu})U^{\dagger}_{\nu}(x)\ .
\end{equation}
It can be shown \cite{Rot97} that, in the limit of vanishing lattice spacing $a$, the Wilson loops can be expanded as
\begin{equation}
U_{\mu\nu}=\one+ig_{0}a^{2}F_{\mu\nu}-\frac{g^{2}_{0}a^{4}}{2}F_{\mu\nu}F_{\mu\nu}+O(a^{6})\ ,
\end{equation}
where $F_{\mu\nu}=F^{a}_{\mu\nu}T_{a}$ and in the third term no sum is implied over the spacetime indices. In particular, by taking the trace of the above expression, we find that
\begin{equation}
\text{Tr}\left\{U_{\mu\nu}\right\}=3-\frac{g^{2}_{0}a^{4}}{4}\,\sum_{a}\,F_{\mu\nu}^{a}F_{\mu\nu}^{a}+O(a^{6})\ .
\end{equation}
The \textit{Wilson action} $S_{\text{W}}$, defined as
\begin{equation}
S_{\text{W}}=\frac{6}{g_{0}^{2}}\sum_{x,\mu<\nu}\ \left(1-\frac{1}{3}\,\text{Tr}\left\{U_{\mu\nu}(x)\right\}\right)\ ,
\end{equation}
is then easily seen to reduce to the Yang-Mills action in the $a\to 0$ limit. Within the lattice approach, $S_{\text{W}}$ is taken to be the defining action of pure Yang-Mills theory.

Since the number of integration variables -- that is, of link variables $U_{\mu}(x)$ -- is finite for a lattice of finite volume, the gauge invariance of the Wilson action\footnote{The gauge transformations act on the link variables as $U_{\mu}(x)\to g(x)U_{\mu}(x)g^{\dagger}(x+a\hat{e}_{\mu})$, where the $g(x)$'s are SU(3) matrices defined at each lattice site \cite{SO04}.} poses no issue of finiteness for averages computed by lattice methods. In particular, no analogue of the Faddeev-Popov procedure is strictly required to be performed when carrying out lattice calculations. Nonetheless, the gauge still needs to be fixed if one wants to evaluate the vacuum expectation values of gauge-dependent quantities such as the gluon or the ghost propagator. For the topics of covariant gauge fixing and of the effects of the Gribov copies within the lattice approach, we refer e.g. to \cite{SO04,DOS16}; for the evaluation of the ghost propagator in covariant gauges, we refer to \cite{BBCO15,CDMO18a}. A review of some of the most common techniques employed to discretize the quark fields on the lattice can be found in \cite{Rot97}.

\subsection{Results}
\label{sec:lqcdrevres}

\begin{figure}[H]
\centering
\includegraphics[width=0.70\textwidth]{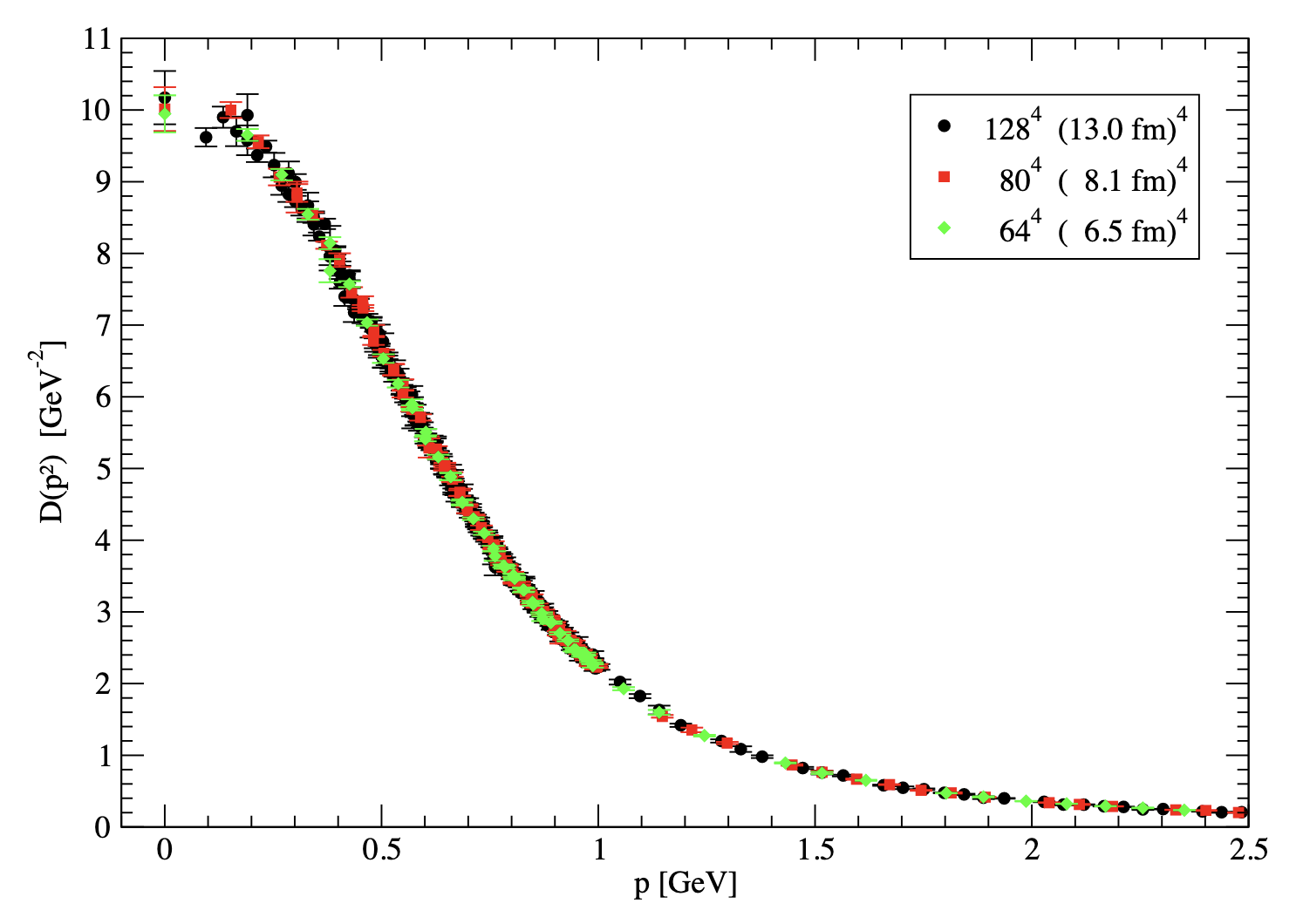}
\vspace{5pt}
\caption{Euclidean Landau-gauge transverse gluon propagator computed on the lattice for different lattice volumes. Figure from \cite{DOS16}.}\label{fig:latglupropdata}
\end{figure}

In the \hyperref[chpt:intro]{Introduction} we saw that the transverse component of the gluon propagator computed on the lattice is found to saturate to a finite, non-zero value in the limit of vanishing momenta, implying that in the infrared the gluons acquire a mass. In Fig.~\ref{fig:latglupropdata} we display an example of such behavior, provided to us by the Landau-gauge pure Yang-Mills results of \cite{DOS16}. In order to push the calculations below the GeV scale, these need to be performed on lattices of very large volumes. In the figure, the results obtained for three such volumes, of sides $6.5$, $8.1$ and $13.0$~fm at fixed lattice spacing $a=0.10$~fm, are shown. Crucially, neither the finiteness of the gluon propagator, nor its saturation value are found to be strongly dependent on the volume. This proves that the massiveness of the propagator is not an artifact of the finite-volume approximation. An analogous conclusion can be reached with respect to the dependence on the lattice spacing \cite{DOS16}, which however turns out to be larger than the volume dependence.

\begin{figure}[H]
\centering
\includegraphics[width=0.70\textwidth]{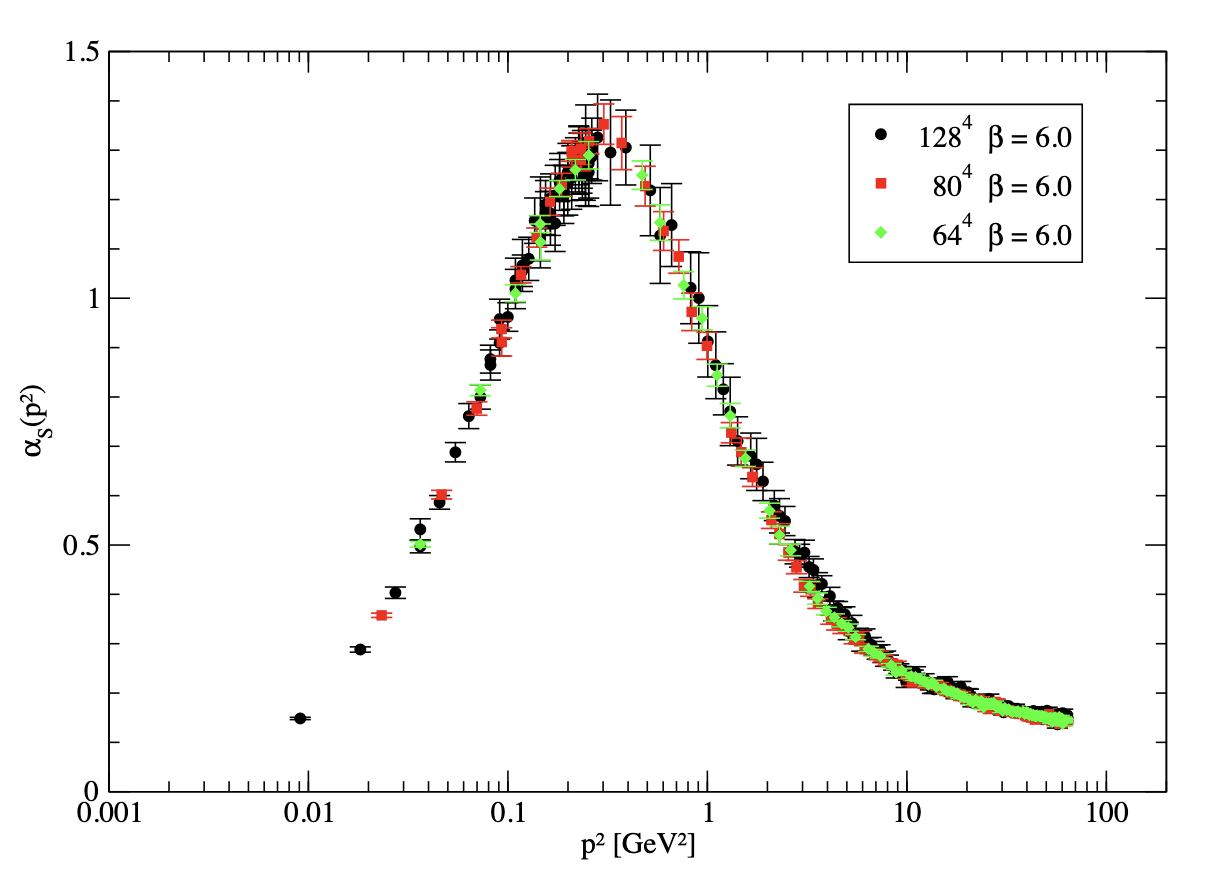}
\vspace{5pt}
\caption{Taylor running coupling computed on the lattice and in the Landau gauge for different lattice volumes. Figure from \cite{DOS16}.}\label{fig:latcoupdata}
\end{figure}

In Fig.~\ref{fig:latcoupdata} we display the lattice data for the running coupling of pure Yang-Mills theory reported in \cite{DOS16}. In the latter, as is customary for lattice calculations carried out in the Landau gauge, the coupling was computed from the gluon and the ghost propagators within the so-called \textit{Taylor scheme}\footnote{See Secs.~\ref{sec:smemomtay} and \ref{sec:dynmodrgimp}.}. As we can see, instead of diverging at a finite scale as in in ordinary pQCD, the Taylor coupling hits a maximum at $p\approx 0.6$~GeV and then decreases to zero in the limit of vanishing momenta. The fact that the coupling remains finite and moderately small at all scales proves on the one hand that the Landau pole predicted by ordinary pQCD is an artifact of the expansion, and on the other hand that a perturbation theory for infrared Quantum Chromodynamics may still be viable, provided that non-perturbative effects such as the dynamical generation of a gluon mass are accounted for.
\vspace{5mm}
\vspace{5mm}
\vspace{5mm}
The lattice results show that Quantum Chromodynamics is infrared-finite, displaying dynamical mass generation in the gluon sector. While in this section we have only reported data obtained in the framework of pure Yang-Mills theory, similar calculations carried out while including the quarks \cite{BHLP04,BHLP07,IMSS07,SO10,ABBC12,BBDM14,ZBDR19,CZBD20} essentially paint the same picture of the low-energy dynamics of full QCD.

\section{The Operator Product Expansion}
\renewcommand{\rightmark}{\thesection\ \ \ The Operator Product Expansion}
\label{sec:operev}

\subsection{Set-up}
The Operator Product Expansion (OPE) approach\footnote{The validity of the OPE in the context of asymptotically free theories was proved in 1970 by Zimmermann \cite{Zim70}.} assumes that the product of two local quantum operators $\mc{O}_{1}(x_{1})$ and $\mc{O}_{2}(x_{2})$ evaluated in the $x_{1},x_{2}\to x$ limit can be expressed as a linear combination of local operators $\mc{O}_{n}(x)$ evaluated at $x$,
\begin{equation}\label{fsl948}
\mc{O}_{1}(x_{1})\mc{O}_{2}(x_{2})\to\sum_{n}\ C_{12}^{n}(x_{1}-x_{2})\,\mc{O}_{n}(x)\ ,
\end{equation}
where the coefficient functions $C^{n}_{12}(x_{1}-x_{2})$ only depend on the separation $x_{1}-x_{2}\to 0$. By making use of Eq.~\eqref{fsl948}, one is able to compute the first non-perturbative contributions to the small-distance -- equivalently, to the high-energy -- behavior of the Green functions of QCD in terms of vacuum condensates. Let us see how this works.

For simplicity, let $x_{2}=0$ and rename $x_{1}$ as $x$. By taking the time-ordered VEV of Eq.~\eqref{fsl948}, one obtains the OPE of the Green function corresponding to the operators $\mc{O}_{1}$ and $\mc{O}_{2}$,
\begin{equation}\label{fsl949}
\Tavg{\mc{O}_{1}(x)\mc{O}_{2}(0)}\to\sum_{n}\ C_{12}^{n}(x)\,\avg{\mc{O}_{n}(0)}\qquad\quad(x\to0)\ .
\end{equation}
Given an arbitrary local operator $\mc{O}_{n}(x)\slashed{\propto} \one$, where $\one$ is the identity operator, the VEV $\avg{\mc{O}_{n}(x)}=\avg{\mc{O}_{n}(0)}$ is known as a vacuum condensate. Condensates are intrinsically non-perturbative objects, in that their value depends on the low-energy content of the theory. On the contrary, the coefficient functions which multiply the condensates in the OPE, being evaluated at small separations $x$, are entirely determined by the high-energy behavior of the operators. In asymptotically free theories such as QCD, these coefficients can be computed explicitly by resorting to ordinary perturbative techniques which we will review shortly. Once the OPE coefficients have been computed, Eq.~\eqref{fsl949} provides us with a high-energy expression for the Green function $\Tavg{\mc{O}_{1}(x)\mc{O}_{2}(0)}$ that includes non-perturbative contributions due to the condensates.

In the $x\to 0$ limit, the dominant contributions to the OPE are given by the condensates of lower mass dimension. This a consequence of the fact that the functions $C_{12}^{n}(x)$, having canonical mass dimension $\kappa_{n}=d_{1}+d_{2}-d_{n}$ -- where $d_{1,2,n}$ are the mass dimensions of the operators $\mc{O}_{1,2,n}$ --, scale like $C_{12}^{n}(\lambda x)\approx \lambda^{-\kappa_{n}}C_{12}^{n}(x)$, and thus tend to zero more rapidly as the dimension $d_{n}$ of the operator $\mc{O}_{n}$ increases. For this reason, only the lower-dimensional condensates are usually retained in the OPE.

The OPE coefficients $C_{12}^{n}(x)$ can be computed as follows. Starting from the coefficient $C_{12}^{\one}(x)$ associated to the identity operator,
\begin{equation}
\Tavg{\mc{O}_{1}(x)\mc{O}_{2}(0)}\to C_{12}^{\one}(x)+\sum_{\mc{O}_{n}\neq\one}\ C_{12}^{n}(x)\,\avg{\mc{O}_{n}(0)}\ ,
\end{equation}
we see that in the above expression $C_{12}^{\one}(x)$ is the only term not to be multiplied by a condensate. Therefore it must be equal to the perturbative value of the Green function,
\begin{equation}
C_{12}^{\one}(x)=\Tavg{\mc{O}_{1}(x)\mc{O}_{2}(0)}_{\text{pert.}}\ .
\end{equation}
As for the other coefficients, these can be computed by multiplying the OPE of $\mc{O}_{1}(x)\mc{O}_{2}(0)$ to a product of operators $\mc{O}_{3}(x_{3})\cdots \mc{O}_{N}(x_{N})$ and then taking the time-ordered VEV of the resulting expression,
\begin{equation}\label{fsl950}
\Tavg{\mc{O}_{1}(x)\mc{O}_{2}(0)\mc{O}_{3}(x_{3})\cdots \mc{O}_{N}(x_{N})}\to\sum_{n}\ C_{12}^{n}(x)\,\Tavg{\mc{O}_{n}(0)\mc{O}_{3}(x_{3})\cdots \mc{O}_{N}(x_{N})}\ .
\end{equation}
As long as the spacetime points $x_{3},\cdots, x_{N}$ are kept sufficiently far from the origin, no new condensates arise in Eq.~\eqref{fsl950}. In particular, we can evaluate both sides of the relation perturbatively in the limit of vanishing $x$ and of very large $x_{3},\cdots,x_{N}$, and then determine the functions $C_{12}^{n}(x)$ by matching the $x$-dependence of the right-hand side to that of the left-hand side. In practice, this is usually done in momentum space, where the momenta associated to the operators $\mc{O}_{3},\cdots,\mc{O}_{N}$ are kept small while the expression is expanded in powers of the momentum $p$ associated to the separation $x$ as $p\to \infty$.

\subsection{Results}
\label{sec:operevres}

By making use of the OPE formalism, one can show that, at high energies, a non-vanishing quark condensate $\avg{\psibar\psi}$ contributes to the quark propagator with a term that plays the role of a quark mass. To see this, assume that the quark is massless and consider the OPE of the product $\psi(x)\psibar(0)$,
\begin{equation}\label{onj951}
\psi(x)\psibar(0)\to C_{\one}(x)\,\one+C_{\psibar\psi}(x)\,\psibar\psi(0)+\cdots\ .
\end{equation}
In the above expression, the dots denote contributions coming both from higher-dimensional local operators and from operators which vanish as soon as the time-ordered VEV of $\psi(x)\psibar(0)$ is taken\footnote{Also, in Eq.~\eqref{onj951} we are disregarding the dimension-$2$ operators $A^{2}$ and $\overline{c}c$, since these do not contribute to the mass-like term under consideration. The former can be shown to correct the quark $Z$-function already to lowest order in perturbation theory -- see e.g. \cite{LS88}.}. The corresponding OPE for the quark propagator $S(x)$ reads
\begin{equation}
S(x)\to S_{\text{pert.}}(x)+C_{\psibar\psi}(x)\avg{\psibar\psi}\qquad\quad(x\to 0)\ ,
\end{equation}
or, in (Minkowski) momentum space,
\begin{equation}
S(p)\to\frac{i Z(p^{2})}{\slashed{p}}+c_{\psibar\psi}\,\frac{\avg{\psibar\psi}}{(p^{2})^{2}}\qquad\quad(p\to\infty)\ ,
\end{equation}
where $Z(p^{2})$ is the perturbative quark $Z$-function and by dimensional counting $c_{\psibar\psi}$ is a dimensionless coefficient. To lowest order in the coupling, since $Z(p^{2})=1+O(g^{2})$ and $c_{\psibar\psi}=O(g^{2})$, we can set $Z(p^{2})c_{\psibar\psi}\approx c_{\psibar\psi}$, so that
\begin{align}
S(p)\to\frac{iZ(p^{2})}{\slashed{p}}\left(1-\frac{ic_{\psibar\psi}}{\slashed{p}}\frac{\avg{\psibar\psi}}{p^{2}}\right)\approx\frac{iZ(p^{2})}{\slashed{p}\left(1+\frac{ic_{\psibar\psi}}{\slashed{p}}\frac{\avg{\psibar\psi}}{p^{2}}\right)}=\frac{iZ(p^{2})}{\slashed{p}-\mathcal{M}(p^{2})}\ ,
\end{align}
where the mass function $\mc{M}(p^{2})$ is given by
\begin{equation}
\mathcal{M}(p^{2})=-ic_{\psibar\psi}\,\frac{\avg{\psibar\psi}}{p^{2}}\ .
\end{equation}
The coefficient $c_{\psibar\psi}$ can be explicitly computed to lowest order in the coupling to yield \cite{LS88}
\begin{equation}\label{dkn495}
c_{\psibar\psi}=-i\frac{3N_{A}g^{2}}{8N^{2}}\qquad\Longrightarrow\qquad\mc{M}(p^{2})=-\frac{3 N_{A}g^{2}}{8N^{2}}\frac{\avg{\psibar\psi}}{p^{2}}\ 
\end{equation}
in the Landau gauge, where $N_{A}=N^{2}-1$ is the dimension of the gauge group SU(N). In particular, we see that, even when the quark is massless, a non-vanishing quark condensate $\avg{\psibar\psi}$ can provide its propagator with a mass function.

A word of caution is in order. As discussed in the last section, the OPE is only valid in the high-energy limit. That of being massive or massless, on the other hand, is a low-energy property of the fields. Therefore, the function $\mc{M}(p^{2})$ defined by Eq.~\eqref{dkn495} should not be interpreted as a fully-fledged non-perturbative mass function, but rather as the high-energy limit of the actual mass function, which cannot be derived within the framework of the OPE. What is of interest to us here is that -- at variance with ordinary perturbation theory~-- the OPE approach predicts that the mass of a massless quark can be different from zero as a non-perturbative consequence of the existence of condensates.\\

Like the quark propagator, the gluon propagator as well can be expanded by making use of a suitable OPE\footnote{To leading order in the dimension of the condensates -- that is, modulo condensates of dimension $3$ and higher -- no other terms are present in the gluon OPE as far as the propagator is concerned \cite{LS88}.},
\begin{equation}
A_{\mu}^{a}(x)A_{\nu}^{b}(0)\to C_{\one}(x)^{ab}_{\mu\nu}\,\one+C_{A^{2}}(x)^{ab}_{\mu\nu}\,A^{2}(0)+\cdots\ .
\end{equation}
The OPE for the (Minkowski) momentum-space gluon propagator $\Delta_{\mu\nu}^{ab}(p)=\Delta_{\mu\nu}(p)\delta^{ab}$ then reads
\begin{equation}
\Delta_{\mu\nu}(p)\to -\frac{i J(p^{2})}{p^{2}}\,t_{\mu\nu}(p)+c_{A^{2}}\,\frac{\avg{A^{2}}}{(p^{2})^{2}}\,t_{\mu\nu}(p)\approx-\frac{i J(p^{2})}{p^{2}\left(1-i c_{A^{2}}\frac{\avg{A^{2}}}{p^{2}}\right)}\,t_{\mu\nu}(p)\ ,
\end{equation}
where $J(p^{2})$ is the perturbative gluon dressing function, $c_{A^{2}}$ is a dimensionless coefficient, the calculations are carried out in the Landau gauge so that $\Delta_{\mu\nu}(p)\propto t_{\mu\nu}(p)$ and we have used $J(p^{2})c_{A^{2}}\approx c_{A^{2}}$ to $O(g^{2})$ in order to bring the condensate to the denominator. If we define a constant $m^{2}$ as
\begin{equation}
m^{2}=ic_{A^{2}}\avg{A^{2}}\ ,
\end{equation}
then the OPE tells us that in the high-energy limit
\begin{equation}
\Delta_{\mu\nu}(p)\to-\frac{iJ(p^{2})}{p^{2}-m^{2}}\,t_{\mu\nu}(p)\ ,
\end{equation}
where in the Landau gauge the coefficient $c_{A^{2}}$ is computed to be \cite{LS88}
\begin{equation}
c_{A^{2}}=-i\frac{Ng^{2}}{4N_{A}}\qquad\Longrightarrow\qquad m^{2}=\frac{N g^{2}}{4N_{A}}\,\avg{A^{2}}\ .
\end{equation}

Like in the quark sector, we see that the OPE predicts that a high-energy mass term, proportional to $\avg{A^{2}}$ \cite{CNZ99,GSZ01,GZ01,BLLM01}, is generated for the gluon propagator which cannot be accounted for by ordinary perturbation theory. Nonetheless, there is one crucial difference between the $\avg{A^{2}}$ and the $\avg{\psibar\psi}$ condensates: while $\psibar\psi$ is a gauge-invariant operator, the $A^{2}$ operator is not; it follows that the VEV $\avg{A^{2}}$ may be expected to be non-zero only if gauge invariance is broken by the vacuum\footnote{Incidentally, as we will discuss in Chpt.~\ref{chpt:smeapp}, in the limit of zero quark mass the condensate $\avg{\psibar\psi}$ should vanish as well because of a global symmetry known as chiral symmetry. Such a symmetry is indeed known to be broken by the vacuum.}. This would have disastrous consequences on the self-consistency of the theory. In Chapter~\ref{chpt:dynmod} we will show that, thanks to the existence of a (non-local) gauge-invariant generalization of the operator $A^{2}$, the condensate $\avg{A^{2}}$ can actually be different from zero in the Landau gauge without spoiling any fundamental symmetry of the strong interactions\footnote{We should remark, however, that condensates of non-local operators do not enter in the OPE of gauge-invariant operators.}.

\section{The Gribov-Zwanziger approach}
\renewcommand{\rightmark}{\thesection\ \ \ The Gribov-Zwanziger approach}
\label{sec:gzrev}

\subsection{Set-up and results}

In Sec.~\ref{sec:fpquant} we saw that the Faddeev-Popov procedure routinely employed to fix a gauge for the QCD action involves the introduction of the determinant of the FP operator $\partial^{\mu}D_{\mu}$ in the path integrals of the theory. If the FP operator has zero modes, meaning that there exist algebra fields $\chi^{a}$ such that $\partial^{\mu}D_{\mu}\chi^{a}=0$, then such a determinant vanishes; as a consequence, the FP procedure is invalidated\footnote{If the gauge field $A$ is sufficiently small, the FP operator is just a perturbation of the Laplacian operator. Since the latter has no zero modes that vanish at large distances, the FP procedure remains valid in the perturbative regime.}.

From the perspective of gauge invariance, the zero modes of the FP operator can be used to construct gauge transformations that relate different field configurations of the FP action to one another. Consider for instance the Landau gauge, defined by\footnote{The implication is a consequence of the fact that the exponential $e^{-\frac{i}{2\xi}\int d^{4}x\ (\partial\cdot A)^{2}}$ that appears in the QCD path integral reduces to a delta functional $\delta(\partial\cdot A)$ in the limit of vanishing $\xi$.}
\begin{equation}
\xi=0\qquad\Longrightarrow\qquad\partial^{\mu}A_{\mu}^{a}=0\ ,
\end{equation}
where $\xi$, like in Sec.~\ref{sec:fpquant}, is the gauge parameter. An infinitesimal gauge transformation of the gluon field $A_{\mu}^{a}$ with parameters $\chi^{a}$ yields
\begin{equation}
A_{\mu}^{a}\to\widetilde{A}_{\mu}^{a}=A_{\mu}^{a}+D_{\mu}\chi^{a}\ .
\end{equation}
If $\chi^{a}$ is a zero mode of $\partial^{\mu}D_{\mu}$, then $\partial^{\mu}A_{\mu}=0$ implies $\partial^{\mu}\widetilde{A}_{\mu}=0$, so that both $A_{\mu}^{a}$ and $\widetilde{A}_{\mu}^{a}$ belong to the configuration space of the Landau-gauge FP action. In other words, the FP procedure fails to fix a gauge for QCD. Two or more field configurations related to each other by gauge transformations constructed by making use of zero modes of the FP operator are called Gribov copies.

The existence of zero modes of the FP operator was proved by Gribov in 1978 \cite{GRI78}. In order to address the issue of the Gribov copies, he proposed that, in the Landau gauge, the Euclidean partition function of QCD be restricted to the so-called Gribov region, defined by the positive-definiteness of the FP operator $-\partial^{\mu}D_{\mu}$,
\begin{equation}
\int\mc{D}\mathscr{F}\ e^{-S_{\text{FP}}}\big|_{\xi=0}\to \int\mc{D}\mathscr{F}\ e^{-S_{\text{FP}}}\big|_{\xi=0}\ \Theta\left(-\partial^{\mu}D_{\mu}\right)\ ,
\end{equation}\cleannlnp
where
\begin{equation}
\Theta(x)=\begin{cases}1\quad x>0\\0\quad x<0\end{cases}
\end{equation}
is the Heaviside function. A local and renormalizable action $S_{\text{GZ}}$ that implements the Gribov condition was discovered by Zwanziger in 1989 \cite{ZWA89b}. In Euclidean space, the Gribov-Zwanziger (GZ) action $S_{\text{GZ}}$ reads
\begin{align}
S_{\text{GZ}}&=S_{\text{FP}}\big|_{\xi=0}+\int d^{4}x\ \left(\overline{\phi}^{ac}_{\mu}\,K^{ab}\phi^{bc\,\mu}-\overline{\omega}^{ac}_{\mu}K^{ab}\omega^{bc\,\mu}+\gamma^{2}gf_{abc}\,A^{a\,\mu}(\phi^{bc}_{\mu}+\overline{\phi}^{bc}_{\mu})\right)+S_{\text{vac.}}\ ,
\end{align}
where $K=-\partial^{\mu}D_{\mu}(A)$, the commuting $\phi^{ab}_{\mu},\overline{\phi}_{\mu}^{ab}$ and anticommuting $\omega^{ab}_{\mu},\overline{\omega}^{ab}_{\mu}$ are new auxiliary fields, $\gamma$ is a variational parameter with the dimensions of a mass known as the \textit{Gribov parameter} and the vacuum term $S_{\text{vac.}}$ is given by
\begin{equation}
S_{\text{vac.}}=-4\gamma^{2}\mc{V}_{4} N_{A}\ ,
\end{equation}
with $\mc{V}_{4}$ the volume of $4$-dimensional Euclidean spacetime and $N_{A}=N^{2}-1$ the dimension of the gauge group SU(N). The value of the Gribov parameter is fixed by the requirement that the effective action of QCD be minimal with respect to its variation. This can be easily shown to be equivalent to the \textit{horizon condition}
\begin{equation}\label{gkl614}
\avg{H[A]}_{\gamma}=4\mc{V}_{4}N_{A}\ ,
\end{equation}
where $H[A]$ is the so-called \textit{horizon functional},
\begin{equation}
H[A]=g^{2}\int d^{4}x\ f_{abc}f_{dec}\,A_{\mu}^{b}\, [K^{-1}(A)]^{ad} A^{e\,\mu}\ .
\end{equation}
The horizon functional appears in the GZ action as soon as the auxiliary fields are integrated out:
\begin{equation}
S_{\text{GZ}}\to S_{\text{FP}}\big|_{\xi=0}+\gamma^{4}\left(H[A]-4\mc{V}_{4}N_{A}\right)\ .
\end{equation}
Since the derivative of the Euclidean effective action with respect to $\gamma$ is equal to the derivative of $W=-\ln Z$, where $Z$ is the Euclidean partition function of the theory, the horizon condition easily follows.

On the shell of Eq.~\eqref{gkl614}, the extra terms in the GZ action can be shown \cite{ZWA89b} to enforce the Gribov condition $\det(-\partial^{\mu}D_{\mu})>0$ to every order in perturbation theory. At the perturbative level, the horizon functional modifies the zero-order gluon propagator due to the fact that the FP operator reduces to the Laplacian as $gA\to 0$,
\begin{equation}
K^{ad}\to-\partial^{2}\,\delta^{ad}\ .
\end{equation}
Indeed, by making use of the SU(N) relation $f_{abc}f_{dbc}=N\delta_{ad}$, we see that, when expanded in powers of $gA$, the horizon functional contains a quadratic gluon term,
\begin{equation}
H[A]=Ng^{2}\int d^{4}x\ A_{\mu}^{a}\left(-\frac{1}{\partial^{2}}\right)A^{a\,\mu}+O(g^{3}A^{3})\ ,
\end{equation}
that shifts the zero-order gluon propagator from $\Delta_{0}(p^{2})=1/p^{2}$ to a GZ propagator $\Delta_{\text{GZ}}(p^{2})$ whose inverse can be read out directly from the full action. Modulo color structure
\begin{equation}
\Delta^{-1}_{\text{GZ}}(p^{2})=p^{2}+\frac{2Ng^{2}\gamma^{4}}{p^{2}}\ ,
\end{equation}
leading to
\begin{equation}\label{rgm494}
\Delta_{\text{GZ}}(p^{2})=\frac{p^{2}}{p^{4}+2Ng^{2}\gamma^{4}}\ .
\end{equation}
We then see that $\Delta_{\text{GZ}}(p^{2})$ vanishes in the $p^{2}\to 0$ limit.

Gluon propagators with a vanishing zero-momentum limit have been derived in functional approaches such as the Dyson-Schwinger Equations (the so-called scaling solutions, \cite{VHA97,AB98}) and studied in relation to the phenomenon of confinement \cite{Oji78,KO79a}. Today we know that the deep-infrared behavior of the propagator is less suppressed than predicted within the original GZ approach. To account for the zero-momentum finiteness of the propagator, in \cite{DGSV08,DSVV08,DOV10,DSV11} it was shown that a condensate of the form $\langle\overline{\phi}^{ab}_{\mu}\,\phi^{ab\,\mu}-\overline{\omega}^{ab}_{\mu}\omega^{ab\,\mu}\rangle$ could provide the GZ action with an additional quadratic gluon term. The $p^{2}\to 0$ limit of the resulting Refined Gribov-Zwanziger (RGZ) zero-order propagator $\Delta_{\text{RGZ}}(p^{2})$,
\begin{equation}\label{rgm495}
\Delta_{\text{RGZ}}(p^{2})=\frac{p^{2}+M^{2}}{p^{4}+M^{2}p^{2}+2Ng^{2}\gamma^{4}}\ ,
\end{equation}
is controlled by a mass parameter $M^{2}$ whose value can be determined dynamically and was proved to be different from zero \cite{DSV11}. Once the quadratic gluon condensate $\avg{A^{2}}$ is included in the RGZ formalism, a modified propagator of the form
\begin{equation}\label{rgm496}
\Delta_{\text{RGZ}}^{(\avg{A^{2}})}(p^{2})=\frac{p^{2}+M^{2}}{p^{4}+(M^{2}+m^{2})p^{2}+2Ng^{2}\gamma^{4}+M^{2}m^{2}}
\end{equation}
with $m^{2}\propto\avg{A^{2}}$ can be derived that is found to be in very good agreement with the lattice data up to momenta $\approx 1$-$1.5$~GeV \cite{DOV10}.

The GZ and RGZ propagators reported in Eqs.~\eqref{rgm494}-\eqref{rgm496} have the interesting property of possessing poles which, depending on the values of the parameters, can be complex conjugate. By looking at Eq.~\eqref{rgm496}, we see that this happens when
\begin{equation}
(M^{2}+m^{2})^{2}-4M^{2}m^{2}-8Ng^{2}\gamma^{4}=(M^{2}-m^{2})^{2}-8Ng^{2}\gamma^{4}<0\ .
\end{equation}
The GZ propagator clearly satisfies the above inequality, given that $M^{2}=m^{2}=0$ in that case, yielding purely imaginary poles at $p^{2}=\pm i\sqrt{2N}g\gamma^{2}$. As for the RGZ propagator, using the parameters
\begin{equation}
M^{2}=2.15\pm0.13\text{ GeV}^{2}\ ,\quad m^{2}=-1.81\pm0.14\text{ GeV}^{2}\ ,\quad 2Ng^{2}\gamma^{4}=4.16\pm0.38\text{ GeV}^{4}
\end{equation}
obtained in \cite{DOV10} by fitting the lattice data, one finds that the poles are complex conjugate for the central values of the fit, although the upper and lower bounds of the parameters do not exclude that the propagator may have a pair of real poles.

To end this section, we mention that proposals have been recently made that aim to extend the GZ and RGZ formalisms to covariant gauges other than the Landau gauge in a way that complies with BRST invariance \cite{CDFG15,CDFG16a,CDFG16b,CDPF17,CFPS17,CDGP18,MPPS19,DFPR19}. Such proposals reformulate the horizon condition -- Eq.~\ref{gkl614} -- in terms of a gauge-invariant gluon field $A^{h}$ which will be the subject of Chapter~\ref{chpt:dynmod}.
\newpage

\section{The Curci-Ferrari model}
\renewcommand{\rightmark}{\thesection\ \ \ The Curci-Ferrari model}
\label{sec:cfrev}

\subsection{Set-up}
\label{sec:cfrevsu}

The Curci-Ferrari (CF) model is defined by the Euclidean action $S_{\text{CF}}$ \cite{TW10,TW11}
\begin{equation}\label{fmw958}
S_{\text{CF}}=S_{\text{YM}}+S_{\text{g.f.}}+S_{m^{2}}\ ,
\end{equation}
where
\begin{equation}
S_{\text{g.f.}}=\int d^{4}x\, \left(iB^{a}\partial\cdot A^{a}+\cbar^{a}\partial^{\mu}D_{\mu}c^{a}\right)
\end{equation}
is the Landau-gauge FP gauge-fixing term\footnote{Note that the replacements $c^{a}\to -c^{a}$, $B^{a}\to iB^{a}$ and partial integrations with respect to Eq.~\eqref{lagfpb} -- which are customary in Euclidean space -- do not change the outcome of gauge fixing.}, whereas
\begin{equation}
S_{m^{2}}=\int d^{4}x\ \frac{1}{2}\,m^{2}A_{\mu}^{a}A^{a\mu}
\end{equation}
is a mass term for the gluons. Under an infinitesimal gauge transformation with parameters $\chi^{a}$, the variation of the mass operator $A_{\mu}^{a}A^{a\,\mu}$ is given by $\delta(A_{\mu}^{a}A^{a\,\mu})=2\partial_{\mu}\chi^{a}A^{a\,\mu}\neq 0$. It follows that the gluon mass term must be interpreted as being added to the action \textit{after} gauge fixing has been performed, for otherwise the lack of gauge invariance of $S_{\text{YM}}+S_{m^{2}}$ would forbid the FP procedure to be carried out in the first place.

Although the gluon mass term $S_{m^{2}}$ breaks the BRST invariance of the FP action,
\begin{equation}
sS_{m^{2}}=-m^{2}\int d^{4}x\ \partial_{\mu}c^{a}A^{a\,\mu}\neq 0\ ,
\end{equation}
where $s$ is the ordinary BRST operator,
\begin{equation}
sA_{\mu}^{a}=-D_{\mu}c^{a}\ ,\qquad sc^{a}=\frac{g}{2}f^{a}_{bc}\,c^{b}c^{c}\ ,\qquad s\cbar^{a}=iB^{a}\ ,\qquad sB^{a}=0\ ,
\end{equation}
the overall CF action still possesses a global symmetry which reduces to BRST symmetry in the limit of vanishing mass. The corresponding operator $s_{m^{2}}$ reads \cite{PRST21a}
\begin{equation}\label{fkm294}
s_{m^{2}}A_{\mu}^{a}=-D_{\mu}c^{a}\ ,\quad\ s_{m^{2}}c^{a}=\frac{g}{2}f^{a}_{bc}\,c^{b}c^{c}\ ,\quad\ s_{m^{2}}\cbar^{a}=iB^{a}\ ,\quad\ s_{m^{2}}B^{a}=im^{2}c^{a}\ ,
\end{equation}
and acts on the sum $S_{\text{g.f.}}+S_{m^{2}}$ as
\begin{equation}
s_{m^{2}}(S_{\text{g.f.}}+S_{m^{2}})=0\ .
\end{equation}
Clearly, $\lim_{m^{2}\to 0}s_{m^{2}}=s$. For $m^{2}\neq 0$, since
\begin{equation}
s_{m^{2}}^{2}\cbar^{a},\ s_{m^{2}}^{2}B^{a}\propto m^{2}\ ,
\end{equation}
the extended BRST operator $s_{m^{2}}$ is not nilpotent. While in general the lack of nilpotency of the BRST operator poses obstructions to the unitarity of a gauge theory, it has been argued in \cite{PRST21a} that, if the CF model displays color confinement, then it may still be possible to define a physical subspace with a positive-definite inner product within the CF Hilbert space as the kernel of $s_{m^{2}}^{2}$.

The invariance of $S_{\text{CF}}$ under the extended BRST transformations in Eq.~\eqref{fkm294} can be exploited to prove that the CF model is perturbatively renormalizable. The divergent parts of the CF renormalization factors can be shown to satisfy the constraints \cite{DVS03,TW11}
\begin{equation}\label{dkw495}
Z_{g}Z_{A}^{1/2}Z_{c}=1\ ,\qquad\qquad Z_{m^{2}}Z_{A}Z_{c}=1\ ,
\end{equation}
where $Z_{m^{2}}$ is the renormalization factor of the gluon mass parameter,
\begin{equation}
m_{B}^{2}=Z_{m^{2}}m^{2}\ .
\end{equation}

The two-point sector of the Curci-Ferrari model is usually renormalized within the so-called Infrared-Safe (IRS) scheme \cite{TW11}, defined by extending the relations in Eq.~\eqref{dkw495} to the finite terms of the renormalization factor and by imposing the conditions
\begin{equation}\label{sld435}
\Delta(\mu^{2};\mu^{2})=\frac{1}{\mu^{2}+m^{2}(\mu^{2})}\ ,\qquad \mc{G}(\mu^{2};\mu^{2})=\frac{1}{\mu^{2}}\ ,
\end{equation}
on the (transverse) gluon and ghost propagators $\Delta(p^{2};\mu^{2})$ and $\mc{G}(p^{2};\mu^{2})$ renormalized at the scale $\mu$. The mass term in the first of Eqs.~\eqref{sld435} is introduced in order to account for the presence of an analogous tree-level term in the fixed-scale CF gluon propagator,
\begin{equation}
\Delta(p^{2})=\frac{1}{Z_{A}\,p^{2}+Z_{A}Z_{m^{2}}\,m^{2}+\Pi^{(\text{CF})}_{T}(p^{2})}\ ,
\end{equation}
where $\Pi^{(\text{CF})}_{T}(p^{2})$ is the transverse component of the CF gluon polarization.

\subsection{Results}
\label{sec:cfrevres}

Within the IRS scheme, the running coupling $\alpha_{s}(\mu^{2})$ of the CF model is found not to develop Landau poles provided that the value of $\alpha_{s}(\mu_{0}^{2})$ at the initial renormalization scale $\mu_{0}$ is not too large \cite{GPRT19,PRST21a}. For such values of the coupling, the model is finite and self consistent in the infrared, admitting a perturbative expansion which can be improved order by order.

\begin{figure}[H]
\centering
\includegraphics[width=0.70\textwidth]{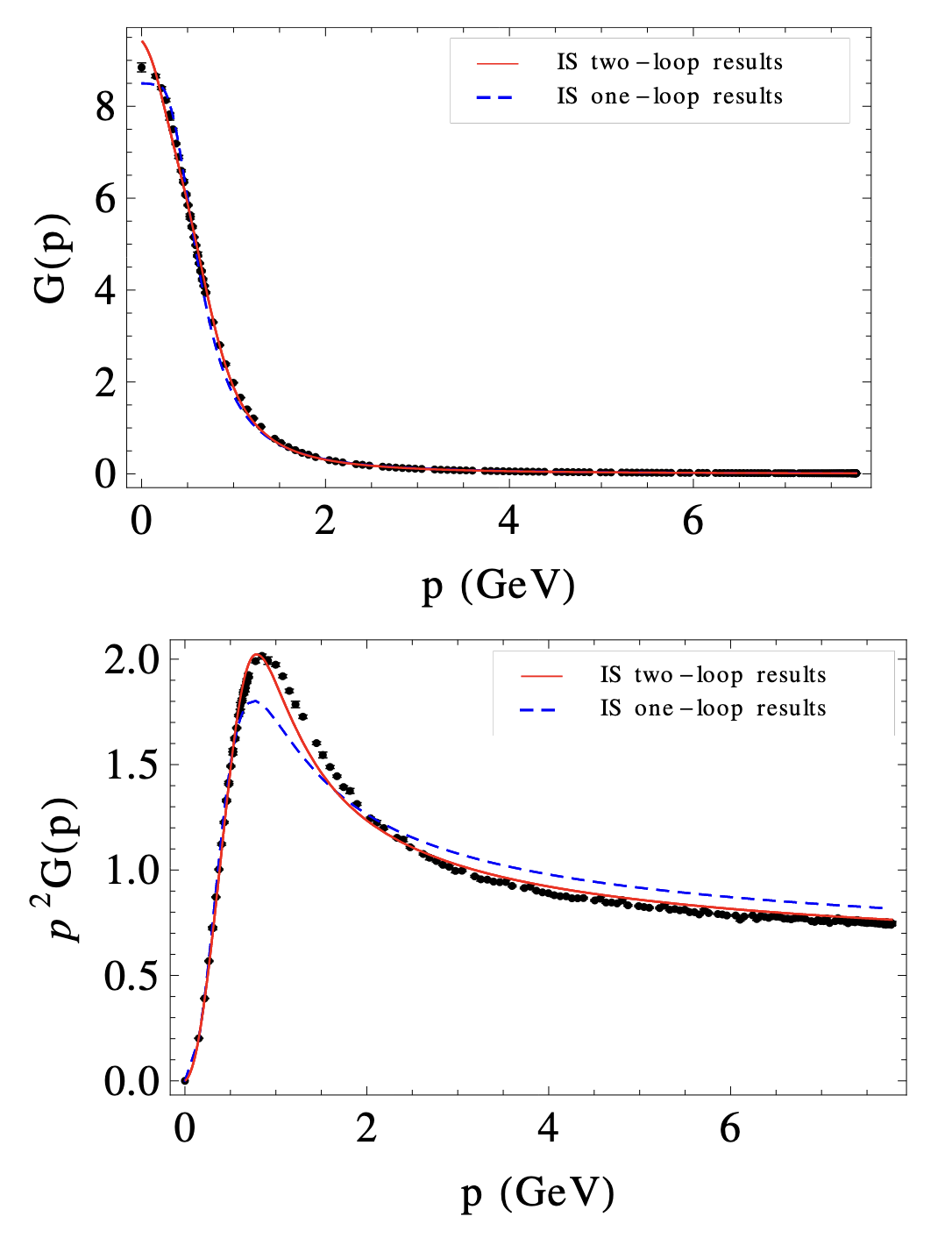}
\vspace{5pt}
\caption{Euclidean Landau-gauge transverse gluon propagator (top) and dressing function (bottom) computed to one and two loops within the Curci-Ferrari model, together with the lattice data from \cite{CMM08}. Figure from \cite{GPRT19}.}\label{fig:cfprops}
\end{figure}

In Fig.~\ref{fig:cfprops} we display the one- and two-loop RG-improved Euclidean gluon propagator and dressing function computed within the Curci-Ferrari model and renormalized in the IRS scheme. The figure is reported from \cite{GPRT19} and features lattice data from \cite{CMM08}. As we can see, the one-loop CF dressing function falls slightly above the lattice results in the UV and slightly below them at intermediate energies\footnote{For future reference, we note that a simple multiplicative normalization of the CF one-loop dressing function can brings its UV tail to match the lattice data at the price of an over-suppressed zero-momentum limit for the propagator.\label{note:cfir}}. Nonetheless, already at one loop the CF results accurately capture the behavior of the lattice propagator, displaying dynamical mass generation in the gluon sector and a non-zero saturation value for the propagator at vanishing momenta. Including the two-loop corrections sensibly improves the quantitative match with the lattice, bringing both the UV tail and the $p\approx 1$-$2$~GeV segment of the curve closer to the lattice results.\\

The lesson that can be learned from the achievements of the Curci-Ferrari model is that treating the gluons as massive at tree level yields a perturbative description of the infrared regime of QCD which agrees both qualitatively and quantitatively with the non-perturbative picture painted by the lattice calculations. The Screened Massive Expansion and the Dynamical Model, to be discussed in the next chapters, build on this notion to provide perturbative frameworks for low-energy QCD that accomplish the same objective without changing the overall Faddeev-Popov action.
\newpage
\vspace*{5mm}
\vspace{5mm}

\part{Massive perturbative models for infrared QCD}

\chapter{The Screened Massive Expansion}
\renewcommand{\leftmark}{\thechapter\ \ \ The Screened Massive Expansion}
\renewcommand{\rightmark}{\thechapter\ \ \ The Screened Massive Expansion}
\label{chpt:sme}

In Part I of this thesis we saw that the ordinary perturbation theory of Quantum Chromodynamics breaks down at low energies due to the presence of a Landau pole in its running coupling constant. In order to investigate the infrared regime of the strong interactions, one has to resort to alternative, non-perturbative computational methods, such as lattice QCD or the Dyson-Schwinger Equations. Both of these approaches offer a solid perspective on the low-energy behavior of the transverse gluon propagator. They clearly show that, in the infrared, instead of diverging as would be typical for a massless field, the propagator changes concavity and saturates to a finite value \cite{LSWP98a,LSWP98b,BBLW00,BBLW01,AN04,AP06,ABP08,AP08,HV13}. In other words, the gluons acquire a dynamically generated mass.

The occurrence of dynamical mass generation (DMG) in the gluon sector of QCD is a feature of primary interest both from a phenomenological and from a theoretical point of view. Phenomenologically, since in the IR a massive gluon behaves very differently from a massless one, we should expect the low-energy dynamics of the strong interactions to be deeply affected by DMG. For instance, as discussed in some early studies \cite{PP80,CF94,LW96,CF97,MN00,Fie02,LMM05}, a gluon mass would bring about both phase-space effects and explicit modifications to the scattering amplitudes in the evaluation of the QCD cross sections. Some (non-conclusive) experimental evidence for the IR-massiveness of the gluon has been presented in the literature by now -- see the \hyperref[chpt:intro]{Introduction} for a brief review.

From a purely theoretical perspective, on the other hand, DMG poses new challenges to our understanding and ability to make analytical predictions in the context of QCD. In this respect, it is useful to take the Higgs mechanism of the electroweak interactions as a benchmark. In the electroweak sector of the Standard Model, the quarks and the charged leptons acquire a mass as a consequence of their being coupled to the Higgs field. The Higgs mechanism is triggered by the non-vanishing VEV of the latter, which is a direct consequence of the Higgs potential having a minimum for some non-zero value of the Higgs field. In particular, the magnitude of the quarks' and charged leptons' masses is in direct relation to the parameters which are present in the Higgs Lagrangian; moreover, the Higgs mechanism is mainly classical in nature, as it already occurs at the classical level, without the aid of quantum corrections. DMG for the quarks and charged leptons in the electroweak sector can be easily accommodated into the standard formalism (and studied with the ordinary computational methods) of the gauge theories, resulting in a dynamical symmetry breaking of the global gauge group which we know how to deal with analytically.

Conversely, the dynamical generation of a mass for the gluon is a genuinely quantum phenomenon. It is not caused by the interaction of the gluon field with the VEV of some other field, but rather it happens through self-interaction. The scale that sets the units for the gluon mass cannot be read off directly from the QCD Lagrangian, but is itself generated by the strong interactions. In addition to there being no evidence that it leads to dynamical symmetry breaking, DMG for the gluons does not play well with the standard analytical method used in QCD -- namely, with ordinary perturbation theory. Indeed, as we discussed in the \hyperref[chpt:intro]{Introduction}, a mass for the gluons cannot be generated at any finite order in standard perturbation theory due to the constraints imposed by gauge invariance on the form of the radiative corrections which appear in the QCD perturbative series. In other words, it is an intrinsically non-perturbative effect of the strong interactions. We remark that this is a limitation of pQCD which is not directly related to its infrared breakdown: one could imagine that ordinary pQCD remained valid at arbitrarily low energies, and DMG in the gluon sector would still be perturbatively forbidden by gauge invariance.\\

In the early 2010s, Tissier and Wschebor \cite{TW10,TW11} made a groundbreaking discovery: by adding a mass term for the gluons in the Landau-gauge Faddeev-Popov Lagrangian, they were able to derive propagators and vertices which accurately reproduced the lattice data in the deep-IR region of QCD (see e.g. Sec.~\ref{sec:cfrevres}). The strong running coupling computed in their so-called Curci-Ferrari (CF) model was furthermore shown to remain finite and moderately small down to arbitrarily small momenta -- a feature that had already been anticipated by the results of the lattice calculations. What is perhaps most surprising about the achievements of the CF model is that, at the most basic level, the latter uses nothing more than standard perturbative techniques. Of course, since the inclusion of a gluon mass term constitutes a modification of the FP Lagrangian, the CF model must be interpreted as an effective theory; indeed, it is not clear whether the model can be derived from first-principles QCD. Nonetheless, the success of the Curci-Ferrari model strongly suggests that, once DMG for the gluons is taken into account by the formalism, the IR regime of the strong interactions might actually be accessible by simple perturbative calculations. This raises a crucial question: may it be that standard perturbation theory fails at low energies because it implicitly assumes that the gluons remain massless down to arbitrarily small scales? If this is so, can a change of the expansion point of the QCD perturbative series fix its IR behavior?

An answer to these questions was provided in 2015 with the formulation of the Screened Massive Expansion of QCD \cite{SIR15a,SIR15b,SIR16b}. The SME is a simple modification of ordinary perturbation theory that consists in treating the transverse gluons as massive already at tree level, while leaving the Faddeev-Popov action unchanged. By expanding around massive, rather than massless transverse gluons, the SME implements DMG in the gluon sector from the get-go; the gluon mass itself can then be shown to screen the IR regime of QCD from the development of Landau poles, thus making the low-energy limit accessible to ordinary perturbative techniques.\\

The main objective of this chapter is to present the general features and main results of the Screened Massive Expansion with regard to the gauge sector of QCD. Since dynamical mass generation for the gluons occurs even in the absence of quarks, in what follows we will focus exclusively on pure Yang-Mills theory (YMT) -- that is, on QCD with no quarks. Working within the framework of YMT allows us to investigate the overall dynamics of the gluons without having to worry about the values of the quark masses. Moreover, it provides for better benchmarks on the infrared regime of the strong interactions, given that the highest-quality lattice calculations which are currently available for the gluon and ghost propagators are carried out in YMT rather than in full QCD.

This chapter is organized as follows. In Sec.~\ref{sec:smedef} we define the Screened Massive Expansion of YMT, discuss some of its features and present the first results for the one-loop gluon and ghost propagators. In Sec.~\ref{sec:smeopt} we show how the SME can be optimized by making use of arguments based on gauge invariance. This will allow us to reduce the number of free parameters in the expressions and make predictions from first principles. In Sec.~\ref{sec:smerg} we perform a Renormalization Group analysis and improvement of the Screened Massive Expansion. The strong running coupling constant $\alpha_{s}(p)$ computed in the SME will be shown to be free of Landau poles, provided that the value of the coupling at the initial renormalization scale is not too large. The absence of IR Landau poles in $\alpha_{s}(p)$ proves that the SME is indeed a self-consistent method, valid down to arbitrarily small energy scales.

Most of the contents of this chapter were originally presented in various papers published between 2016 and 2020. Specifically, the main sources for the three sections that make up the chapter are:
\begin{itemize}
\item Sec.~\ref{sec:smedef}: \cite{SIR16b,CS18,SC18,SIR19b},
\item Sec.~\ref{sec:smeopt}: \cite{SC18,SC22b,SIR19a,SIR19b},
\item Sec.~\ref{sec:smerg}: \cite{CS20}.
\end{itemize}
Refs. \cite{CS18}, \cite{SC18}, \cite{CS20} and \cite{SC22b} are attached as an insert to this thesis, to be found in Appendix~\ref{app:published}.

\section{Motivation, definition and first results}
\renewcommand{\rightmark}{\thesection\ \ \ Motivation, definition and first results}
\label{sec:smedef}

\subsection{Dynamical mass generation and perturbation theory: the set-up of the Screened Massive Expansion}
\label{sec:smesu}

In our review of the set-up of ordinary perturbation theory (Sec.~\ref{sec:opqcd}), we saw that the standard perturbative expansion of QCD is obtained by splitting the Faddeev-Popov action $S_{\text{FP}}$ as
\begin{equation}
S_{\text{FP}}=S_{0}+S_{\text{int.}}\ ,
\end{equation}
where, in the absence of quarks, the zero-order action $S_{0}$ can be expressed in momentum space as
\begin{align}\label{dks849}
S_{0}=i\int \frac{d^{4}p}{(2\pi)^{4}}\ \left\{\frac{1}{2}\,A_{\mu}^{a}(-p)\,[\Delta^{-1}_{0}(p)]^{\mu\nu}_{ab}\,A_{\nu}^{b}(p)+\cbar^{a}(p)\,[\mc{G}_{0}^{-1}(p)]_{ab}\,c^{b}(p)\right\}\ ,
\end{align}
with the zero-order gluon and ghost propagators $\Delta_{0}(p)$ and $\mc{G}_{0}(p)$ given by
\begin{equation}
\Delta_{0\,\mu\nu}^{ab}(p)=\frac{-i}{p^{2}}\,\delta^{ab}\,\left(t_{\mu\nu}(p)+\xi\,\ell_{\mu\nu}(p)\right)\ ,\qquad\qquad\mc{G}_{0}^{ab}(p)=\frac{i}{p^{2}}\,\delta^{ab}\ ,
\end{equation}
whereas the interaction terms in $S_{\text{int.}}$ explicitly read
\begin{align}
S_{\text{int}}&=\int d^{4}x\ \bigg\{-gf^{a}_{bc}\,\partial_{\mu}A_{\nu}^{a}\,A^{b\,\mu}A^{c\,\nu}-\frac{1}{4}\,g^{2}f^{a}_{bc}f^{a}_{de}\,A_{\mu}^{b}A_{\nu}^{c}A^{d\,\mu}A^{e\,\nu}+gf^{a}_{bc}\,\partial^{\mu}\cbar^{a}A_{\mu}^{b}c^{c}\bigg\}\ .
\end{align}
Since $\Delta_{0}(p)$ and $\mc{G}_{0}(p)$ have a pole at $p^{2}=0$, Eq.~\eqref{dks849} implies that at the tree level of standard perturbation theory the ghosts and the gluons are massless.

As we know by now, the strong interactions generate an infrared mass for the transverse gluons. However, in the framework of ordinary pQCD, gauge invariance prevents the quantum corrections from shifting the position of the transverse gluon pole to a non-zero value of $p^{2}$. This makes the standard perturbative expansion of QCD especially unsuitable to describe the low-energy limit of the strong interactions. In order to obtain a better approximation of the exact gluon propagator, we could try to reorganize the perturbative series in such a way that the transverse gluons are treated as massive already at tree level. In other words, we could look for a modification of perturbation theory capable of yielding a zero-order gluon propagator of the form
\begin{equation}\label{glupropm}
\Delta_{m\,\mu\nu}^{ab}(p)=\delta^{ab}\,\left(\frac{-it_{\mu\nu}(p)}{p^{2}-m^{2}}+\xi\,\frac{-i\ell_{\mu\nu}(p)}{p^{2}}\right)\ ,
\end{equation}
where the pole of the longitudinal component of $\Delta_{m}(p)$ is left unshifted from $p^{2}=0$, since we know that $\Delta_{L}^{ab}(p)=\ell^{\mu\nu}(p)\,\Delta_{\mu\nu}^{ab}(p)=-i\frac{\xi}{p^{2}}\,\delta^{ab}$ is an exact identity for the dressed gluon propagator -- see Sec.~\ref{sec:brst}.

At the end of the derivation of the Feynman rules for ordinary pQCD in Sec.~\ref{sec:opqcd}, we observed that the perturbative expansion of QCD can be formally generalized to an arbitrary zero-order action, provided that the latter is chosen to be quadratic in the fields. This comes in very handy, since the $\Delta_{m}(p)$ in Eq.~\eqref{glupropm} can be obtained as the zero-order gluon propagator associated to the quadratic action $S_{m}$ given by
\begin{align}\label{sfpm}
S_{m}=i\int \frac{d^{4}p}{(2\pi)^{4}}\ \left\{\frac{1}{2}\,A_{\mu}^{a}(-p)\,[\Delta^{-1}_{m}(p)]^{\mu\nu}_{ab}\,A_{\nu}^{b}(p)+\cbar^{a}(p)\,[\mc{G}_{0}^{-1}(p)]_{ab}\,c^{b}(p)\right\}\ .
\end{align}
In particular, a perturbative expansion of pure Yang-Mills theory around massive transverse gluons can be achieved by splitting the Faddeev-Popov action $S_{\text{FP}}$ as\footnote{Similar ideas can be found in the literature, see e.g. \cite{JP97,KPP97}.}
\begin{equation}\label{msplit}
S_{\text{FP}}=S_{m}+S_{\text{int.}}^{\prime}\ ,
\end{equation}
where $S_{m}$ is given by Eq.~\eqref{sfpm}, whereas by definition
\begin{equation}
S_{\text{int.}}^{\prime}=S_{\text{FP}}-S_{m}\ .
\end{equation}
Explicitly, since $S_{m}=S_{0}+\delta S$ with
\begin{equation}
\delta S=i\int \frac{d^{4}p}{(2\pi)^{4}}\ \frac{1}{2}\,A_{\mu}^{a}(-p)\,[\Delta^{-1}_{m}(p)-\Delta^{-1}_{0}(p)]^{\mu\nu}_{ab}\,A_{\nu}^{b}(p)\ ,
\end{equation}
the modified interaction action $S_{\text{int.}}^{\prime}$ is equal to $S_{\text{int.}}-\delta S$, where
\begin{equation}
-\delta S=-i\int \frac{d^{4}p}{(2\pi)^{4}}\ \frac{1}{2}\,A_{\mu}^{a}(-p)\,\Gamma^{\mu\nu}_{ab}(p)\,A_{\nu}^{b}(p)
\end{equation}
and the two-point gluon vertex $\Gamma_{ab}^{\mu\nu}(p)$ contained in the interaction term $-\delta S$ reads
\begin{equation}\label{glumct}
\Gamma^{\mu\nu}_{ab}(p)=-im^{2}\,t^{\mu\nu}(p)\,\delta_{ab}\ .
\end{equation}

\begin{figure}[H]
\centering
\includegraphics[width=0.73\textwidth]{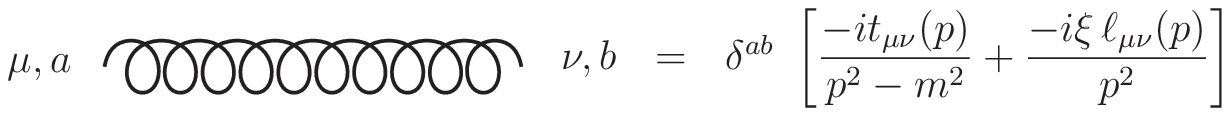}
\vspace{5pt}
\caption{Zero-order massive gluon propagator}\label{fig:glupropm}
\end{figure}

\begin{figure}[H]
\centering
\includegraphics[width=0.61\textwidth]{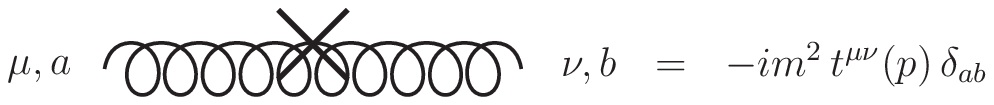}
\vspace{5pt}
\caption{Gluon mass counterterm}\label{fig:glumct}
\end{figure}

The Feynman rules associated to the split defined by Eq.~\eqref{msplit} can be read off directly from the action terms $S_{m}$ and $S_{\text{int.}}^{\prime}$. By construction, the zero-order gluon propagator is given by Eq.~\eqref{glupropm} -- see Fig.~\ref{fig:glupropm} --, whereas the zero-order ghost propagator and the 3-gluon, 4-gluon and ghost-gluon vertices are the same as those of standard perturbation theory -- Figs.~\ref{fig:ghprop}, \ref{fig:3gl}, \ref{fig:4gl} and \ref{fig:ghgl}, respectively. In addition to the ordinary interaction vertices, a new 2-gluon vertex -- depicted by a cross in Fig.~\ref{fig:glumct} -- must be included in the calculations; this is the vertex known as the \textit{gluon mass counterterm}, and corresponds to the quantity $\Gamma^{\mu\nu}_{ab}(p)$ in Eq.~\eqref{glumct}. In what follows, we will refer to Feynman diagrams containing one or more gluon mass counterterms as \textit{crossed diagrams}.

The gluon mass counterterm $\Gamma^{\mu\nu}_{ab}(p)$ arises from the shift of the zero-order action $S_{0}\to S_{m}=S_{0}+\delta S$, which must be balanced by an opposite shift in the interaction action $S_{\text{int.}}\to S_{\text{int.}}^{\prime}=S_{\text{int.}}-\delta S$ in order for the total Faddeev-Popov action to be left unchanged. $\Gamma^{\mu\nu}_{ab}(p)$ is a non-perturbative vertex, in that it is not proportional to any power of the strong coupling constant $g$; conversely, it is equal -- modulo a factor of $-i$ and the transverse, color-diagonal tensor structure $t^{\mu\nu}(p)\,\delta_{ab}$ -- to the \textit{gluon mass parameter} $m^{2}$, which is the very same parameter that appears in the zero-order massive gluon propagator.

Due to the presence of the gluon mass counterterm, the massive expansion defined by Eq.~\eqref{msplit} is intrinsically non-perturbative in nature. More precisely, the perturbative series obtained by expanding the Green functions in powers of $S_{\text{int.}}^{\prime}$ is a power series in both the coupling constant and the gluon mass parameter $m^{2}$, with the caveat that the latter also appears inside the loop integrals via the zero-order gluon propagator. For this reason, when computing a quantity to some fixed order in the coupling constant, one must be careful in choosing which crossed diagrams are to be included in the calculation. Indeed, it is easy to convince oneself that the two-point nature of the gluon mass counterterm implies that there are infinitely many crossed diagrams at any finite order in the coupling constant. We will have more to say about this in the next section.
\vspace{5mm}
\vspace{5mm}
\vspace{5mm}
How do we know that the perturbative series defined by the massive split in Eq.~\eqref{msplit} yields a better approximation than ordinary perturbation theory? Ultimately, only calculations carried out a posteriori can tell. Nonetheless, a preliminary indication can already be obtained by making use of a variational tool known as the Gaussian Effective Potential.

Roughly speaking, the Gaussian Effective Potential (GEP) is a first-order approximation of the vacuum energy density $\mc{E}$ of a quantum field theory computed under the hypothesis that the vacuum state can be described as a Gaussian functional of the fields. Since this hypothesis holds true for the (quadratic) perturbative vacua which are used to set up perturbation theory, the GEP can be exploited to study which kind of perturbative expansion is better suited to approximate the exact theory.

In the language of perturbation theory, provided that the elementary fields $F$ do not acquire a VEV, the GEP $V_{G}$ can be defined as
\begin{equation}\label{spe039}
V_{G}=\frac{i}{\mc{V}_{4}}\ln \int \mc{D}F\ e^{iI_{0}}-\frac{i}{\mc{V}_{4}}\,\avg{I_{\text{int.}}}_{0}\ .
\end{equation}
Here $\mc{V}_{4}$ is the four-dimensional volume of spacetime, $I_{0}$ is the zero-order action of the theory, $I_{\text{int.}}$ is the interaction action, the full action is given by the sum $I_{0}+I_{\text{int.}}$ and the subscript $0$ denotes that the VEV of $I_{\text{int.}}$ is to be computed with respect to the zero-order integration measure $\mc{D}F\,e^{iI_{0}}$. The crucial property of the GEP is that $V_{G}$ can be shown to be greater than or equal to the exact energy density $\mc{E}$ of the theory,
\begin{equation}
e^{-i\mc{E}\mc{V}_{4}}=\int\mc{D}F\ e^{i(I_{0}+I_{\text{int.}})}\ .
\end{equation}
This is a consequence of the so-called Jensen-Feynman inequality \cite{Fey98}, which holds for the Gaussian integrals if the elementary fields $F$ are not Grassmann fields.

Since the zero-order and interaction terms $I_{0}$ and $I_{\text{int.}}$ in Eq.~\eqref{spe039} are only restricted by the requirements that 1. they sum to the total action of the theory, and 2. that $I_{0}$ be quadratic in the fields, we are free to choose any quadratic zero-order action $I_{0}$, define $I_{\text{int.}}$ as the difference between the full action and $I_{0}$, and the inequality $V_{G}\geq\mc{E}$ will be fulfilled. In particular, in the context of pure Yang-Mills theory, we can take $I_{0}$ to be equal to $S_{m}$, $I_{\text{int.}}=S_{\text{int.}}^{\prime}$, and the resulting GEP will be a function of the mass parameter $m^{2}$ which appears in the zero-order gluon propagator $\Delta_{m}$. The best approximation to the vacuum energy density of Yang-Mills theory will then be provided by the value of the mass parameter which, by minimizing the GEP $V_{G}(m^{2})$, pushes the value of the potential closer to the exact result $\mc{E}$ \footnote{Earlier we stated that the Jensen-Feynman inequality holds if the functional averages do not involve Grassmann fields. Of course, in the path integrals of pure Yang-Mills theory we do integrate over Grassmann fields -- namely, the ghost and antighost fields. Nonetheless, it can be shown that in the framework of pure Yang-Mills theory the Jensen-Feynman inequality is saturated by \textit{maximizing} the contribution due to the ghosts. This is equivalent to requiring that the latter do not acquire a mass in the infrared; see \cite{CS18} for a detailed treatment of this technical aspect of the GEP.}.

An explicit calculation shows that, in an arbitrary covariant gauge \cite{Com19}, the mass-dependent GEP for a pure Yang-Mills theory with gauge group SU(N) is given by \cite{CS18}
\begin{equation}\label{aqd}
V_{G}(m^{2})=\frac{3N_{A}m^{4}}{128\pi^{2}}\ \left(\alpha\,\ln^{2}\frac{m^{2}}{m_{0}^{2}}+2\, \ln\frac{m^{2}}{m_{0}^{2}}-1\right)\ ,
\end{equation}
where $N_{A}$ is the dimension of the gauge group and $\alpha$ is a rescaled coupling defined by
\begin{equation}
\alpha=\frac{9N\alpha_{s}}{8\pi}\ ,\qquad\qquad \alpha_{s}=\frac{g^{2}}{4\pi}\ ,
\end{equation}
and $m_{0}^{2}$ is an arbitrary non-zero mass scale, generated by the renormalization of the potential, whose explicit value cannot be computed from first principles since pure Yang-Mills theory is scale-free at the classical level. We stress that in Eq.~\eqref{aqd} there is nothing special about the scale $m_{0}^{2}$, other than it being different from zero: we could redefine $m_{0}^{2}$ by multiplying it by some arbitrary constant, and the GEP would just acquire an additional term proportional to $m^{4}$. Also, we observe that the GEP is gauge-independent, in the sense that $V_{G}(m^{2})$ does not depend on the gauge parameter $\xi$ despite having been computed in a general covariant gauge.

\begin{figure}[H]
\centering
\includegraphics[width=0.47\textwidth,angle=270]{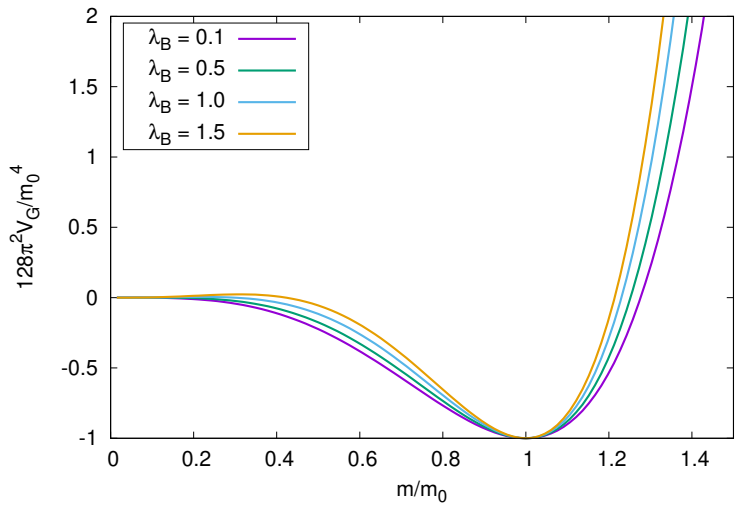}
\vspace{5pt}
\caption{The Gaussian Effective Potential of pure Yang-Mills theory, normalized by a factor of $\frac{128\pi^{2}}{3N_{A}m_{0}^{4}}$ and computed for different values of $\lambda_{B}=16\pi^{2}\alpha$. Figure from \cite{CS18}.}\label{fig:gep0}
\end{figure}

By differentiating $V_{G}$ with respect to $m^{2}$, we find that -- regardless of the value of the coupling constant -- the GEP has a local minimum at $m^{2}=m_{0}^{2}\neq 0$, where
\begin{equation}
V_{G}(m^{2}=m_{0}^{2})=-\frac{3N_{A}m^{4}_{0}}{128\pi^{2}} < 0\ .
\end{equation}
By contrast, at $m^{2}=0$ -- which is also a local minimum for $V_{G}$ -- the GEP vanishes, thus attaining a value which is greater than $V_{G}(m^{2}=m_{0}^{2})$. This can be clearly seen in Fig.~\ref{fig:gep0}, where the GEP is plotted as a function of the ratio $m/m_{0}$ for different values of the coupling constant.\newpage

The fact that the GEP is globally minimized by a non-zero value of the gluon mass parameter signals that a perturbative expansion that treats the transverse gluons as massive provides a better approximation to the vacuum energy density of pure Yang-Mills theory in comparison to one in which the gluons are massless. In other words, the true vacuum of pure Yang-Mills theory better resembles the vacuum of free massive transverse gluons, rather than that of free massless ones. In particular, we may expect the massive expansion defined by the split in Eq.~\eqref{msplit} to reproduce more faithfully the infrared regime of the strong interactions at finite order in its perturbative series, when compared to the ordinary massless perturbation theory. As we will see in the following sections, this is indeed the case: the Screened Massive Expansion allows us to incorporate the phenomenon of dynamical mass generation non-trivially into the formalism of QCD, and to analytically compute gluon and ghost propagators which are in excellent agreement with the results of the lattice calculations down to the deep IR.
\vspace{5mm}
\vspace{5mm}

\subsection{General properties of the Screened Massive Expansion}
\label{sec:genpropsme}

In the process of replacing the zero-order massless gluon propagator $\Delta_{0}$ with a massive one, the Screened Massive Expansion introduces a gluon mass parameter $m^{2}$ in the perturbative series of QCD. As it stands, $m^{2}$ is a new free parameter whose value will need to be fixed in order to be able to make physical predictions. Since pure Yang-Mills theory is scale-free at the classical level, $m^{2}$ -- together with the renormalization scale -- is the only dimensionful parameter of the theory. Therefore, choosing a value for the gluon mass parameter is equivalent to setting the energy units of the theory\footnote{This is not the case for full QCD with massive quarks, as of course the quark masses are also dimensionful parameters. Nonetheless, the corrections to the gluon mass due to the presence of quarks turn out to be negligible as far as the overall scale of $m$ is concerned \cite{Sir16a}. Therefore, even in the context of full QCD, when discussing the deep IR behavior of the gluon propagator one can safely assume that $m$ is the only relevant energy scale. The same holds true for the ghost propagator, since the corrections to the latter due to the quarks are higher-order in perturbation theory.}. Of course, this cannot be done from first principles, but must instead be done a posteriori by making comparisons with the experiments, or with the lattice data.

We should mention that $m$ must not be interpreted as the actual mass of the gluon. Indeed, whereas at tree level the pole of the transverse gluon propagator is found at $p^{2}=m^{2}$, the quantum corrections radically change the analytic structure of the dressed propagator already at the one-loop order. In particular, we shall see that the transverse gluon propagator computed in the SME possesses a pair of complex-conjugate poles. Within the Screened Massive Expansion, $m$ must simply be regarded as a mass scale which determines the infrared behavior of the gluon propagator. 

The presence of the new free parameter $m^{2}$ in the equations of the SME, when taken at face value, may seem to reduce the predictive power of the approximation. While this is certainly true if we limit ourselves to using the plain SME, we anticipate that optimization methods can be devised to fix the value of some of the free parameters of the expansion from first principles, thus restoring the predictivity of the technique. Interestingly, these methods are ultimately based on the delicate balance that exists between the massiveness of the gluon and the need to preserve gauge/BRST invariance. We will come back to this topic in both Sec.~\ref{sec:smeopt} and Sec.~\ref{sec:smerg}.\\

As a consequence of the massiveness of the zero-order gluon propagator $\Delta_{m}$, some of the loop diagrams of the SME contain \textit{mass divergences} -- that is, diverging terms proportional to the gluon mass parameter $m^{2}$. These divergences cannot be absorbed by the ordinary renormalization counterterms of QCD, since the total Faddeev-Popov action -- which is left unchanged by the SME -- contains no gluon mass term, and cannot therefore accommodate a gluon mass renormalization counterterm.

Nonetheless, the mass divergences can be shown to disappear as soon as a sufficient number of crossed diagrams is included in the equations. Heuristically, this can be explained by arguing that since $\Delta_{m}\to \Delta_{0}$ in the high-energy ($|p^{2}|\gg m^{2}$) limit, the Screened Massive Expansion cannot modify the UV behavior of QCD; in particular, it will not modify the structure of the UV-diverging terms in its perturbative series. If any of the diagrams is found to contain a mass divergence, then other diagrams must exist that contain opposite mass divergences to cancel it. This must be possible order by order in the coupling constant.

A more rigorous argument requires us to analyze the structure of the crossed diagrams which appear at a fixed order in $g$ in the perturbative series of the SME. Let $\mc{D}_{0}$ be a diagram which does not contain gluon mass counterterms, and denote with $\mc{D}_{1}$ the sum of all of the diagrams which can be obtained from $\mc{D}_{0}$ by inserting a single gluon mass counterterm in one of its internal gluon lines. Clearly, $\mc{D}_{1}$ is of the same order in $g$ as $\mc{D}_{0}$. Now, observe that the insertion of a single gluon counterterm $\Gamma(p)$ into a gluon line $\Delta_{m}(p)$ can be achieved by replacing $\Delta_{m}(p)\to \Delta_{m}(p)\cdot \Gamma(p)\cdot \Delta_{m}(p)$ under the sign of integral; moreover,
\begin{align}
\Delta_{m}(p)\cdot \Gamma(p)\cdot \Delta_{m}(p)&=\frac{im^{2}}{(p^{2}-m^{2})^{2}}\,t(p)=-m^{2}\frac{\partial}{\partial m^{2}}\,\Delta_{m}(p)\ .
\end{align}
It is then easy to see that
\begin{equation}
\mc{D}_{1}=-m^{2}\frac{\partial}{\partial m^{2}}\,\mc{D}_{0}\ .
\end{equation}
If $\mc{D}_{0}$ contains a mass divergence, then the sum $\mc{D}_{0}+\mc{D}_{1}$, which thanks to the above equation can be expressed as
\begin{equation}
\mc{D}_{0}+\mc{D}_{1}=\left(1-m^{2}\frac{\partial}{\partial m^{2}}\right)\,\mc{D}_{0}\ ,
\end{equation}\cleannlnp
will not contain any, given that the operator $1-m^{2}\frac{\partial}{\partial m^{2}}$ kills any term linear in $m^{2}$ inside $\mc{D}_{0}$. What about the other crossed diagrams, $\mc{D}_{2}$, $\mc{D}_{3}$, $\dots$, $\mc{D}_{n}$, which contain 2, 3, $\dots$, $n$ gluon mass counterterms? Just like $\mc{D}_{1}$, they are given by
\begin{equation}\label{bwl620}
\mc{D}_{n}=\frac{(-m^{2})^{n}}{n!}\,\frac{\partial^{n}}{\partial(m^{2})^{n}}\,\mc{D}_{0}\ .
\end{equation}
In particular, since every $\partial/\partial m^{2}$ derivative increases the number of propagators under the sign of integral, the diagrams in $\mc{D}_{2}$ -- which contain at least three gluon propagators -- are superficially convergent, as well as any $\mc{D}_{n}$ with $n\geq 3$. Therefore, once the mass divergences are cancelled in subdiagrams by summing the latter to their crossed counterparts, the rest of the perturbative series is also convergent, as far as the divergences proportional to $m^{2}$ are concerned.

Eq.~\eqref{bwl620} helps us understand the precise relation that exists between the SME and ordinary perturbation theory. Indeed, if we start from a single diagram $\mc{D}_{0}(m^{2})$ -- where the dependence on $m^{2}$ is made explicit for reasons that will become clear in a moment~-- and sum to it all of the diagrams $\mc{D}_{n}(m^{2})$ that can be obtained by inserting $n$ gluon mass counterterms in its internal gluon lines, we find that
\begin{align}\label{gnd840}
\sum_{n=0}^{+\infty}\,\mc{D}_{n}(m^{2})&=\left(\sum_{n=0}^{\infty}\frac{(-m^{2})^{n}}{n!}\,\frac{\partial^{n}}{\partial(m^{2})^{n}}\right)\,\mc{D}_{0}(m^{2})=\\
\notag&=e^{-\lambda\frac{\partial}{\partial m^{2}}}\Big|_{\lambda=m^{2}}\,\mc{D}_{0}(m^{2})=\\
\notag&=\mc{D}_{0}(m^{2}-\lambda)\Big|_{\lambda=m^{2}}=\mc{D}_{0}(0)\ .
\end{align}
In other words, by summing an uncrossed diagram $\mc{D}_{0}(m^{2})$ to all of its crossed counterparts $\mc{D}_{n}(m^{2})$ ($n\geq1$), we obtain the former computed in ordinary pQCD, $\mc{D}_{0}(0)$. This can be seen explicitly at the level of the single gluon line, for which an identity analogous to Eq.~\eqref{gnd840} reads
\begin{equation}
\sum_{n=0}^{+\infty}\ \Delta_{m}(p)\cdot[\Gamma(p)\cdot\Delta_{m}(p)]^{n}=\Delta_{m}(p)\cdot \frac{1}{\one-\Gamma(p)\cdot\Delta_{m}(p)}=\Delta_{0}(p)\ ,
\end{equation}
where we have used
\begin{equation}
\frac{1}{\one-\Gamma(p)\cdot\Delta_{m}(p)}=\frac{1}{1+\frac{m^{2}}{p^{2}-m^{2}}}\,t(p)+\ell(p)=\frac{p^{2}-m^{2}}{p^{2}}\,t(p)+\ell(p)\ .
\end{equation}
It is then clear that in the Screened Massive Expansion we are not interested in resumming all of the crossed diagrams: if we did so, we would be back to ordinary massless perturbation theory. On the contrary, at any finite order in the coupling constant $g$, we must choose a finite number of crossed diagrams to include in the calculation.\\

Unfortunately, to date, no first-principles argument has been found to constrain the maximum number of crossed diagrams that are to be retained in a fixed-loop-order calculation. Nonetheless, the principle of renormalizability, together with a principle of minimality, are still able to provide us with a useful criterion for the truncation of the SME perturbative series. At a fixed loop order $\ell$, in order to obtain a renormalizable result, one must at the very least include all of the crossed diagrams which are needed to remove the mass divergences arising from the uncrossed diagrams. Among these crossed and uncrossed diagrams, there will be a subset of diagrams with a maximal number of vertices $N$. In the spirit of perturbation theory, once some diagrams with $\ell^{\prime}\leq\ell$ loops and $N^{\prime}\leq N$ vertices are included, then \textit{all} of the diagrams with the same properties should be included as well. Since there is no need to add other diagrams, by a principle of minimality we shall not do so. When following these criteria, the SME can be interpreted as a \textit{double} expansion in both the number $\ell$ of loops and the number $N$ of vertices.

In Secs.~\ref{sec:ghprsme} and \ref{sec:glprsme}, the double expansion will be used to derive explicit expressions for the one-loop ghost and gluon propagators. There we will see that the uncrossed loops that contribute to the gluon propagator contain mass divergences which are removed by corresponding crossed loops with a maximum number of vertices $N=3$. As a consequence, our calculations will include every diagram with at most $1$ loop and at most $3$ vertices.

\subsection{The ghost propagator}
\label{sec:ghprsme}

In the present and in the following section we start reviewing some of the results which have been obtained to one-loop by making use of the Screened Massive Expansion of pure Yang-Mills theory. For the sake of conciseness, we will not go through the explicit derivation of the analytic expressions; the details of the calculations can be found in \cite{SIR16b,SC18,SIR19b} (see Appendix~\ref{app:published} for \cite{SC18}). Let us begin from the dressed ghost propagator.\\

The dressed ghost propagator $\mc{G}^{ab}(p)$, defined as the Fourier transform
\begin{equation}
\mc{G}^{ab}(p)=\int d^{4}x\ e^{ip\cdot x}\ \Tavg{c^{a}(x)\cbar^{b}(0)}\ ,
\end{equation}
can be expressed in terms of a single scalar function $\mc{G}(p^{2})$ as
\begin{equation}
\mc{G}^{ab}(p)=\mc{G}(p^{2})\,\delta^{ab}\ .
\end{equation}
The function $\mc{G}(p^{2})$, in turn, can be computed by resumming all the insertions of the one-particle-irreducible (1PI) ghost self-energy $\Sigma(p^{2})$,
\begin{equation}
\mc{G}(p^{2})=\frac{i}{Z_{c}\,p^{2}-\Sigma(p^{2})}\ ,
\end{equation}
where $Z_{c}$, the ghost field renormalization factor, is required for removing the diverging terms contained in $\Sigma(p^{2})$.
\vspace{5mm}
\begin{figure}[H]
\centering
\includegraphics[width=0.68\textwidth]{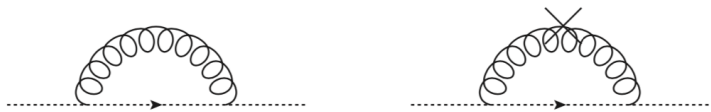}
\vspace{5pt}
\caption{Diagrams for the one-loop SME ghost self-energy}\label{fig:ghose}
\end{figure}
\vspace{5mm}
To one loop, the diagrams which contribute to the ghost self-energy are those displayed in Fig.~\ref{fig:ghose} \cite{SIR16b,SIR19b}. The crossed loop on the right of the figure is included in the calculation in order to be consistent with the choice of diagrams made for the computation of the gluon propagator. More precisely, as anticipated in the last section, to obtain the one-loop gluon and ghost SME propagators we shall include every one-loop diagram which has a maximum of three interaction vertices. An explicit calculation carried out in a generic covariant gauge within the framework of the Screened Massive Expansion of pure Yang-Mills SU(N) theory yields \cite{SIR16b,SIR19b}
\begin{equation}\label{siggh}
\Sigma(p^{2})=\frac{\alpha}{4}\,p^{2}\,\left(1-\frac{\xi}{3}\right)\left(\frac{2}{\epsilon}-\ln\frac{m^{2}}{\overline{\mu}^{2}}\right)-\alpha\,p^{2}\,\left(G(s)-\frac{2}{3}-\frac{\xi}{12}\ln s\right)
\end{equation}
for the one-loop, 1PI dimensionally regularized ghost self-energy. In the above equation, $\mubar=\sqrt{4\pi}\mu e^{-\gamma_{E}/2}$ is the energy scale generated by dimreg, $\alpha$ is a rescaled coupling constant defined as
\begin{equation}
\alpha=\frac{3N\alpha_{s}}{4\pi}\ ,\qquad\qquad \alpha_{s}=\frac{g^{2}}{4\pi}\ ,
\end{equation}
$s=-p^{2}/m^{2}$, and $G(s)$ is the function given by
\begin{equation}\label{tya450}
G(s)=\frac{1}{12}\left[\frac{(1+s)^{2}(2s-1)}{s^{2}}\,\ln(1+s)-2s\ln s+\frac{1}{s}+2\right]\ .
\end{equation}

In order to remove the divergence in Eq.~\eqref{siggh}, we choose the ghost field renormalization factor $Z_{c}$ to be equal to
\begin{equation}\label{gth620}
Z_{c}=1+\frac{\alpha}{4}\,\left[\left(1-\frac{\xi}{3}\right)\left(\frac{2}{\epsilon}-\ln\frac{m^{2}}{\overline{\mu}^{2}}\right)+\frac{8}{3}+4g_{0}\right]\ ,
\end{equation}
where $g_{0}$ is an adimensional constant which selects the renormalization scheme for the ghost propagator. In the above equation, the $\epsilon$-pole coincides with the one contained in the corresponding MS renormalization factor \cite{IZ06}. This was to be expected, since -- as discussed in Sec.~\ref{sec:genpropsme} -- the Screened Massive Expansion does not modify the UV behavior of the theory, and in particular its UV divergences.

To one loop, the renormalized ghost propagator, which reads
\begin{equation}\label{vls049}
\mc{G}(p^{2})=\frac{i}{p^{2}[1+\alpha (G(s)-\xi\ln s/12+g_{0})]}\ ,
\end{equation}
can be rewritten in such a way that the coupling constant formally disappears from the equations. In order to do so, we define two adimensional constants $G_{0}$ and $Z_{\mc{G}}$ as
\begin{equation}\label{pwj920}
G_{0}=\frac{1}{\alpha}+g_{0}\ ,\qquad\qquad Z_{\mc{G}}=\frac{1}{\alpha}\ ,
\end{equation}
in terms of which $\mc{G}(p^{2})$ can be expressed as
\begin{equation}\label{ghpropsme}
\mc{G}(p^{2})=\frac{iZ_{\mc{G}}}{p^{2}(G(s)-\xi\ln s/12 +G_{0})}\ .
\end{equation}
While in the context of the SME the multiplicative factor $Z_{\mc{G}}$ is given by Eq.~\eqref{pwj920}, when comparing $\mc{G}(p^{2})$ with ghost propagators computed by other methods we are free to interpret $Z_{\mc{G}}$ as an independent variable. This is a consequence of the arbitrariness in the choice of the normalization for the quantum fields, which makes the corresponding propagators comparable only modulo multiplicative factors. On the other hand, the additive constant $G_{0}$ contains information on both the value of the strong coupling constant and the renormalization scheme in which the one-loop ghost propagator is defined.\\

As the first step in our analysis of the SME ghost propagator, let us explore its asymptotic behavior\footnote{\label{note:log}We should remark that the true high-energy limit of the propagators can only be studied by improving their momentum behavior via the use of the Renormalization Group equations (Sec.~\ref{sec:rgint}). Thus, when in the context of a fixed-scale calculation we say that a propagator has a certain high-energy limit, we interpret this to be a formal, rather than an actual, property of the propagator.}. At large and small momenta -- corresponding, respectively, to $s\to \infty$ and $s\to 0$ --, the function $G(s)$ which appears in the denominator of the propagator $\mc{G}(p^{2})$ has the following limits:
\begin{equation}\label{fvj242}
\lim_{s\to \infty}\,G(s)=\frac{1}{4}\,\ln s+\frac{1}{3}\ ,\qquad\qquad \lim_{s\to 0}\, G(s)=\frac{5}{24}\ .
\end{equation}
By plugging the first of these into Eq.~\eqref{siggh}, we find that at high energies the ghost self-energy reduces to
\begin{equation}\label{nfe629}
\Sigma(p^{2})\to\frac{\alpha}{4}\,p^{2}\,\left(1-\frac{\xi}{3}\right)\left(\frac{2}{\epsilon}-\ln\frac{-p^{2}}{\overline{\mu}^{2}}\right)\ .
\end{equation}
It should come as no surprise that this is the ghost self-energy computed in ordinary perturbation theory \cite{IZ06}: again, as we said, the SME does not modify the high-energy limit of the theory. Because of Eq.~\eqref{nfe629}, at large momenta the ghost propagator goes to zero like $1/p^{2}\,\ln(-p^{2})$.

At the other end of the spectrum, as $p^{2}\to 0$, the function $G(s)$ approaches a constant. It follows that the ghost propagator diverges in the infrared:
\begin{equation}
\lim_{p^{2}\to 0}\,\mc{G}(p^{2})=\lim_{p^{2}\to 0}\,\frac{iZ_{\mc{G}}}{p^{2}(5/24-\xi\ln (-p^{2}/m^{2})/12 +G_{0})}=\infty\ .
\end{equation}
In particular, in the framework of the SME, the ghosts remain massless. It is worth noticing that, whereas the masslessness of the ghost is independent of the gauge, the way in which $\mc{G}(p^{2})$ tends to infinity as $p^{2}\to 0$ very much is: while in the zero-momentum limit $p^{2}\mc{G}(p^{2})$ remains finite in the Landau gauge ($\xi=0$), the same is not true in any other gauge, for then $p^{2}\mc{G}(p^{2})$ itself goes to zero like $1/\ln(-p^{2})$ \footnote{The infrared-suppression of the ghost dressing function in a general covariant gauge $\xi\neq 0$ compared to its Landau-gauge counterpart was observed in a recent lattice calculation \cite{CDMO18a}.}.\\

\begin{figure}[H]
\centering
\includegraphics[width=0.46\textwidth,angle=270]{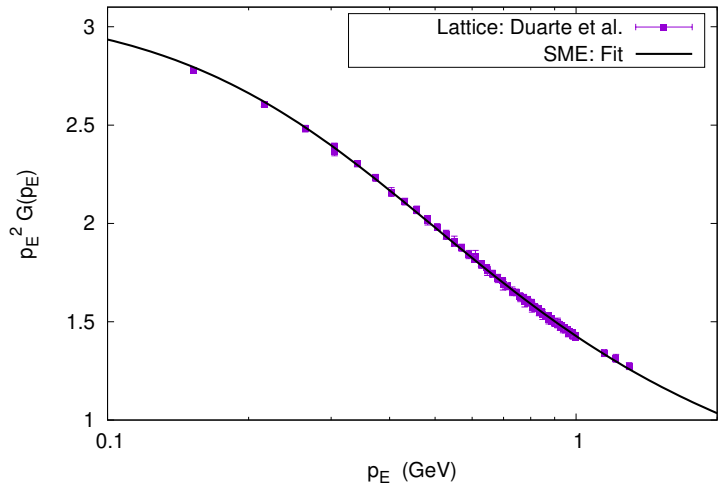}
\vspace{5pt}
\caption{Euclidean ghost dressing function in the Landau gauge ($\xi=0$). Solid curve: one-loop SME result with the parameters of Tab.~\ref{tab:fitdatagh}. Squares: lattice data from \cite{DOS16}.}\label{fig:ghproplandfit}
\end{figure}

How does the ghost propagator computed in the Screened Massive Expansion compare with the lattice data? The answer is shown in Fig.~\ref{fig:ghproplandfit}, where we display a fit of the Euclidean ghost dressing function $p^{2}_{E}\mc{G}_{E}(p^{2}_{E})$ evaluated in the Landau gauge ($\xi=0$) together with the lattice data of \cite{DOS16}. We recall that $\mc{G}_{E}(p_{E}^{2})$ is defined as
\begin{equation}
\mc{G}_{E}(p_{E}^{2})=i\mc{G}(-p_{E}^{2})\ ,
\end{equation}
with the Euclidean momentum $p_{E}^{2}\geq 0$.

The fit in Fig.~\ref{fig:ghproplandfit} was performed by using as free parameters the multiplicative and additive renormalization constants $Z_{\mc{G}}$ and $G_{0}$ defined in Eq.~\eqref{ghpropsme}, while fixing the gluon mass parameter $m$ to 0.654 GeV. The latter was obtained by fitting the SME transverse gluon propagator, as described in the next section. Moreover, in order to avoid (at this stage) the use of the Renormalization Group, the data was cut at a momentum of 2 GeV. Finally, we should mention that the normalization of the ghost propagator is the one provided by the lattice data, that is, we did not perform a further re-normalization of the propagator. A summary of the values obtained from the fit is reported in Tab.~\ref{tab:fitdatagh}.

As we can see, in the deep IR region the Screened Massive Expansion of the ghost propagator manages to accurately reproduce the Landau-gauge lattice results. At low momenta, the Euclidean ghost dressing function $p^{2}_{E}\mc{G}_{E}(p^{2}_{E})$ first changes concavity and then saturates to a finite value, implying that as $p^{2}\to 0$ the ghost propagator diverges to infinity like $1/p^{2}$, as previously anticipated. This behavior is consistent with that of the decoupling solution obtained by solving the Dyson-Schwinger Equations for pure Yang-Mills theory \cite{AN04,AP06,ABP08,AP08,HV13}.
\vspace{5mm}
\vspace{5mm}
\begin{table}[H]
\def\arraystretch{1.2}
\centering
\begin{tabular}{cc|c}
\hline
$G_{0}$&$Z_{\mc{G}}$&$m$ (GeV)\\
\hline
\hline
$0.1464$&$1.0994$&$0.654$\\
\hline
\end{tabular}
\vspace{3mm}
\caption{Parameters obtained from fitting the lattice data of \cite{DOS16} for the Landau-gauge Euclidean ghost propagator in the range $0$-$2$ GeV at fixed $m=0.654$ GeV.}\label{tab:fitdatagh}
\end{table}
\vspace{5mm}

\subsection{The gluon propagator}
\label{sec:glprsme}

The dressed gluon propagator $\Delta_{\mu\nu}^{ab}(p)$ is defined as the Fourier transform
\begin{equation}
\Delta_{\mu\nu}^{ab}(p)=\int d^{4}x\ e^{ip\cdot x}\ \Tavg{A_{\mu}^{a}(x)A_{\nu}^{b}(0)}\ .
\end{equation}
Since -- as we saw in Sec.~\ref{sec:brst} -- the longitudinal projection of $\Delta_{\mu\nu}^{ab}(p)$ is exactly equal to $-i\xi/p^{2}$, the gluon propagator can be expressed in terms of a single unknown scalar function $\Delta(p^{2})$:
\begin{equation}
\Delta_{\mu\nu}^{ab}(p)=\left(\Delta(p^{2})\,t_{\mu\nu}(p)+\frac{-i\xi}{p^{2}}\,\ell_{\mu\nu}(p)\right)\delta^{ab}\ .
\end{equation}
Modulo the color structure, $\Delta(p^{2})$ is the transverse component of the gluon propagator.

$\Delta(p^{2})$ can be computed by resumming all the insertions of the 1PI gluon polarization\footnote{At variance with ordinary pQCD, the SME gluon polarization can develop a longitudinal component $\Pi_{L}(p^{2})\neq0$ when computed to finite order in the gluon mass counterterm. Nonetheless, we have seen that by resumming all the crossed diagrams associated to an uncrossed SME diagram the ordinary pQCD result is recovered. Therefore, we can assume that such a resummation has been performed and set $\Pi_{L}(p^{2})=0$.}, whose transverse component we denote by $\Pi(p^{2})$:
\begin{equation}\label{gnr649}
\Delta(p^{2})=\frac{-i}{Z_{A}\,p^{2}-m^{2}-\Pi(p^{2})}\ .
\end{equation}
In the above equation, the gluon field renormalization factor $Z_{A}$ is needed to absorb the diverging terms in $\Pi(p^{2})$, whereas the presence of the mass term $m^{2}$ is a consequence of the massiveness of the zero-order SME propagator.

\begin{figure}[H]
\centering
\includegraphics[width=0.25\textwidth]{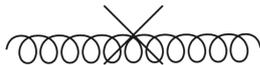}
\caption{Single-counterterm diagram}\label{fig:gluctdiag}
\end{figure}

\begin{figure}[H]
\centering
\includegraphics[width=\textwidth]{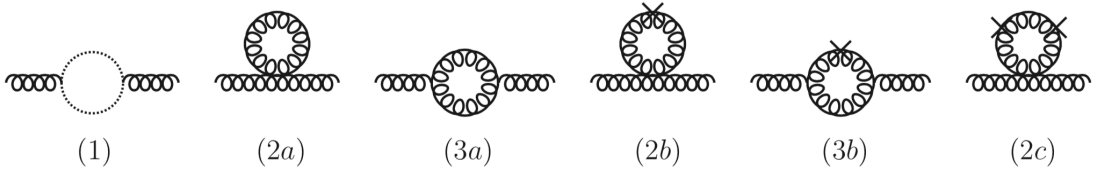}
\caption{Loop diagrams for the one-loop SME gluon polarization}\label{fig:glupoldiag}
\end{figure}

To one loop, the 1PI gluon polarization receives contributions from the diagrams shown in Figs.~\ref{fig:gluctdiag} and \ref{fig:glupoldiag}. The single-counterterm diagram in Fig.~\ref{fig:gluctdiag} is easily computed to be equal to $-m^{2}$. It follows that the 1PI gluon polarization can be expressed as
\begin{equation}
\Pi(p^{2})=-m^{2}+\Pi_{\text{loop}}(p^{2})\ ,
\end{equation}
where the function $\Pi_{\text{loop}}(p^{2})$ collects the contributions to $\Pi(p^{2})$ due to the loops of the perturbative series. By plugging the last equation into Eq.~\eqref{gnr649}, we see that the transverse gluon propagator can be written as
\begin{equation}\label{kgq692}
\Delta(p^{2})=\frac{-i}{Z_{A}\,p^{2}-\Pi_{\text{loop}}(p^{2})}\ .
\end{equation}
The expression \eqref{kgq692} for $\Delta(p^{2})$ is especially important, since it proves that, in the framework of the Screened Massive Expansion, the generation of a gluon mass is not the trivial consequence of having forced a mass term into the zero-order gluon propagator. Indeed, if $\Pi_{\text{loop}}(p^{2})$ vanished in the zero-momentum limit, then the gluon propagator -- having a pole at $p^{2}=0$ -- would remain massless. Instead, as we shall see in a moment, $\Pi_{\text{loop}}(p^{2})$ actually turns out to be proportional to the gluon mass parameter $m^{2}$ when computed at $p^{2}=0$. As a result, the propagator is finite in the zero-momentum limit and a mass is generated for the gluon. Coming from the loops of the expansion, such a mass is a truly dynamical effect of the interactions.
\vspace{5mm}
\vspace{5mm}
The diagrams which make up $\Pi_{\text{loop}}(p^{2})$ at the one-loop order are depicted in Fig.~\ref{fig:glupoldiag}. Those denoted by (1), (2a) and (3a) are just the ordinary one-loop diagrams of standard perturbation theory, although we should keep in mind that -- at variance with pQCD -- diagrams (2a) and (3a) are computed by making use of the massive gluon propagator $\Delta_{m}$ \footnote{As such, diagrams (2a) and (3a) can also be found in the gluon polarization of the Curci-Ferrari model.}. Due to the mass which runs inside the loops, both of these diagrams can be shown to contain mass divergences. As discussed in Sec.~\ref{sec:genpropsme}, these divergences cannot be renormalized by making use of the gluon field renormalization factor $Z_{A}$ \footnote{That $Z_{A}$ is only able to absorb divergences proportional to the momentum squared $p^{2}$ is clear from Eq.~\eqref{kgq692}.}. In order to eliminate the mass divergences, the crossed diagrams (2b) and (3b) must also be included in the calculation. This is sufficient because, as we saw in Sec.~\ref{sec:genpropsme},
\begin{equation}
\Pi_{(2b/3b)}(p^{2})=-m^{2}\frac{\partial}{\partial m^{2}}\,\Pi_{(2a/3a)}(p^{2})\ ,
\end{equation}
so that if the polarization term $\Pi_{(2a/3a)}(p^{2})$ associated to diagram (2a/3a) contains a divergence proportional to $m^{2}$, then the sum
\begin{equation}
\Pi_{(2a/3a)}(p^{2})+\Pi_{(2b/3b)}(p^{2})=\left(1-m^{2}\frac{\partial}{\partial m^{2}}\right)\Pi_{(2a/3a)}(p^{2})
\end{equation}
does not. Finally, since diagram (3b) is a one-loop, three-vertex diagram, we also include the only other diagram which has these properties -- namely, diagram (2c).
\vspace{5mm}
\vspace{5mm}
An explicit calculation carried out in a general covariant gauge within the framework of the Screened Massive Expansion of pure Yang-Mills SU(N) theory shows that, to one loop and in dimensional regularization \cite{SIR16b,SC18},
\begin{equation}
\Pi_{\text{loop}}(p^{2})=\frac{\alpha}{3}\, \left(\frac{13}{6}-\frac{\xi}{2}\right)\, p^{2}\left(\frac{2}{\epsilon}-\ln\frac{m^{2}}{\overline{\mu}^{2}}\right)-\alpha\, p^{2}\left(F(s)+\xi F_{\xi}(s)+\mc{C}\right)\ .
\end{equation}
In the above equation, like in the last section, $\mubar=\sqrt{4\pi}\mu e^{-\gamma_{E}/2}$ is the energy scale generated by dimreg, $\alpha=\frac{3N\alpha_{s}}{4\pi}$ is a rescaled coupling constant and $s=-p^{2}/m^{2}$. Moreover, $\mc{C}$ is an unessential constant which disappears after renormalization, and $F(s)$ and $F_{\xi}(s)$ are the functions defined as
\begin{align}
\label{ffunct}F(s)&=\frac{5}{8s}+\frac{1}{72}\ \left[L_{a}(s)+L_{b}(s)+L_{c}(s)+R_{a}(s)+R_{b}(s)+R_{c}(s)\right]\ ,\\
F_{\xi}(s)&=\frac{1}{4s}-\frac{1}{12}\left[2s\ln s-\frac{2(1-s)(1-s^{3})}{s^{3}}\ln(1+s)+\frac{3s^{2}-3s+2}{s^{2}}\right]\ ,
\end{align}
where the logarithmic functions $L_{a}(s)$, $L_{b}(s)$ and $L_{c}(s)$ and the rational functions $R_{a}(s)$, $R_{b}(s)$ and $R_{c}(s)$ are given by
\begin{align}
\notag L_{a}(s)&=\frac{3s^{3}-34s^{2}-28s-24}{s}\ \sqrt{\frac{4+s}{s}}\ \ln\left(\frac{\sqrt{4+s}-\sqrt{s}}{\sqrt{4+s}+\sqrt{s}}\right)\ ,\\
L_{b}(s)&=\frac{2(1+s)^{2}}{s^{3}}\ (3s^{3}-20s^{2}+11s-2)\ \ln(1+s)\ ,\\
\notag L_{c}(s)&=(2-3s^{2})\ \ln s\ ,
\end{align}
\begin{align}
\notag R_{a}(s)&=-\frac{4+s}{s}\ (s^{2}-20s+12)\ ,\\
\label{rfunct}R_{b}(s)&=\frac{2(1+s)^{2}}{s^{2}}\ (s^{2}-10s+1)\ ,\\
\notag R_{c}(s)&=\frac{2}{s^{2}}+2-s^{2}\ .
\end{align}

In order to remove the divergence in $\Pi_{\text{loop}}(p^{2})$, the gluon field renormalization constant $Z_{A}$ must be chosen according to
\begin{equation}\label{gth621}
Z_{A}=1+\frac{\alpha}{3}\, \left(\frac{13}{6}-\frac{\xi}{2}\right)\left(\frac{2}{\epsilon}-\ln\frac{m^{2}}{\overline{\mu}^{2}}\right)+\alpha(f_{0}-\mc{C})\ ,
\end{equation}
where $f_{0}$ is an adimensional constant that selects the renormalization scheme for the gluon propagator. Just like in the ghost sector, the $\epsilon$-pole in $Z_{A}$ is the one that appears in the MS gluon field renormalization constant \cite{IZ06}, yet again confirming that the SME does not modify the UV behavior of the theory.

As we did for the ghost propagator, we can rewrite the transverse gluon propagator,
\begin{equation}\label{vms472}
\Delta(p^{2})=\frac{-i}{p^{2}[1+\alpha (F(s)+\xi\,F_{\xi}(s)+f_{0})]}\ ,
\end{equation}
in such a way that the coupling constant disappears from the equation. To do so, we define two adimensional constants $F_{0}$ and $Z_{\Delta}$ as
\begin{equation}\label{snt638}
F_{0}=\frac{1}{\alpha}+f_{0}\ ,\qquad\qquad Z_{\Delta}=\frac{1}{\alpha}\ ,
\end{equation}
in terms of which
\begin{equation}\label{ksb805}
\Delta(p^{2})=\frac{-iZ_{\Delta}}{p^{2}(F(s)+\xi\,F_{\xi}(s)+F_{0})}\ .
\end{equation}
Again, $Z_{\Delta}$ can be taken to be a free constant when the SME gluon propagator is compared to one computed by different methods (see Sec.~\ref{sec:glprsme}), whereas $F_{0}$ contains information on both the coupling constant and the renormalization scheme in which $\Delta(p^{2})$ is defined.\\

In the high- and low-momentum limits $s\to \infty$ and $s\to 0$, the functions $F(s)$ and $F_{\xi}(s)$ have the asymptotic behavior
\begin{equation}\label{fvj243}
\lim_{s\to \infty}\ F(s)=\frac{17}{18}+\frac{13}{18}\,\ln s\ ,\qquad\qquad \lim_{s\to 0}\ F(s)=\frac{5}{8s}\ ,
\end{equation}
\begin{equation}
\lim_{s\to\infty}\ F_{\xi}(s)=-\frac{1}{6}\, \ln s -\frac{1}{12}\ ,\qquad\qquad \lim_{s\to0}\ F_{\xi}(s)=\frac{1}{4s}\ .
\end{equation}
As a consequence, in the UV, the SME gluon polarization $\Pi(p^{2})$ reduces to
\begin{equation}
\Pi(p^{2})\to\frac{\alpha}{3}\, \left(\frac{13}{6}-\frac{\xi}{2}\right)\, p^{2}\left(\frac{2}{\epsilon}-\ln\frac{-p^{2}}{\overline{\mu}^{2}}\right)\ ,
\end{equation}
which is the same expression that is obtained in standard perturbation theory \cite{IZ06}. In particular, just like the ghost propagator, the transverse gluon propagator tends to zero as $1/p^{2}\ln(-p^{2})$ in the high-energy limit (see Note~\ref{note:log} in Sec.~\ref{sec:glprsme}). On the other hand, at vanishing momentum,
\begin{equation}
\Delta(0)=\frac{-iZ_{\Delta}}{-M_{\xi}^{2}}\ ,
\end{equation}
where $M_{\xi}^{2}$ is the $\xi$-dependent mass scale given by
\begin{equation}
M_{\xi}^{2}=\frac{5m^{2}}{8}\left(1+\frac{2\xi}{5}\right)\ .
\end{equation}
Since $\Delta(0)$ is finite, the gluon propagator is evidently massive.\\

In order to investigate the origin of the gluon mass in the framework of the SME, it is interesting to analyze what kind of contributions are made to the scale $M_{\xi}^{2}$ -- equivalently, to the polarization term $\Pi_{\text{loop}}(0)$ -- by the loop diagrams in Fig.~\ref{fig:glupoldiag}. As we stated earlier, diagram (1) is just an ordinary pQCD diagram. For this reason, we do not expect it to generate a mass for the gluon, and indeed we find that $\Pi_{(1)}(p^{2}=0)=0$. On the other hand, diagrams (2a), (2b), (3a) and (3b) are all computed by making use of the massive propagator $\Delta_{m}$, and all of them contain mass divergences. As a consequence, their $p^{2}\to 0$ limits are UV-divergent, and to quantify their contribution to $M_{\xi}^{2}$ it only makes sense to consider the sums (2a+2b) and (3a+3b). This being said, we find that the mass scale $M_{\xi}^{2}$ is distributed as follows between the one-loop diagrams (1) to (3c):
\begin{equation}
\Pi_{(1)}(0)=0\ ,
\end{equation}
\begin{align}
\notag\Pi_{(2a)}(0)+\Pi_{(2b)}(0)&=-\frac{3\alpha}{4}\ m^{2}\ ,\\
\Pi_{(2c)}(0)&=\frac{3\alpha}{8}\ m^{2}\ ,\\
\notag\Pi_{(3a)}(0)+\Pi_{(3b)}(0)&=\alpha\left(1+\frac{\xi}{4}\right) m^{2}\ .
\end{align}
All of the one-loop diagrams, with the exception of the ordinary ghost loop, contribute to the zero-momentum finiteness of the gluon propagator. Notably, the diagrams which do not vanish in the $p^{2}\to 0$ limit all involve the self-interaction of gluons, be it mediated by 3-gluon or by 4-gluon vertices. The breakdown of the contributions to $\Pi_{\text{loop}}(0)$ clearly illustrates that the generation of a dynamical mass for the gluons in QCD -- as described within the framework of the SME -- is the consequence of the non-abelian nature of the interaction between the gauge fields.

\begin{figure}[H]
\centering
\includegraphics[width=0.45\textwidth]{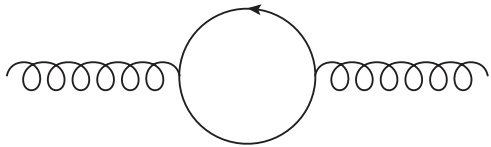}
\caption{Quark loop (lowest-order full-QCD gluon polarization)}\label{fig:glqkloop}
\end{figure}

\begin{figure}[H]
\centering
\includegraphics[width=0.46\textwidth,angle=270]{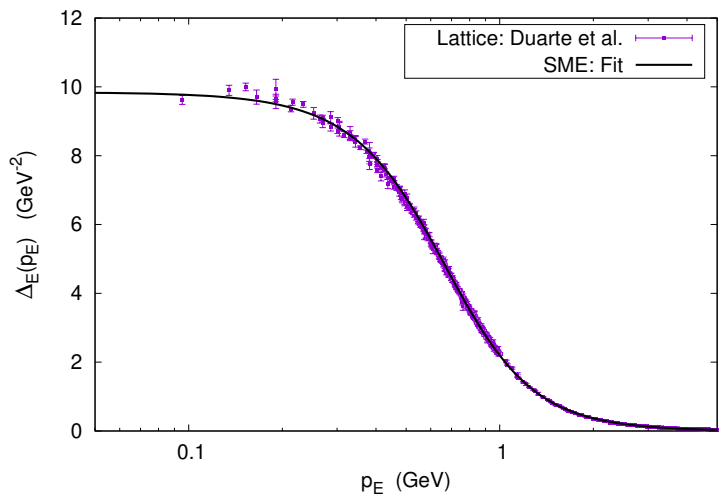}
\vspace{5pt}
\caption{Euclidean transverse gluon propagator in the Landau gauge ($\xi=0$). Solid line: one-loop SME result with the parameters of Tab.~\ref{tab:fitdata}. Squares: lattice data from \cite{DOS16}.}\label{fig:gluproplandfit}
\end{figure}

We remark that this would still hold true had we included the quarks in the calculation. Indeed, the quark loop which in full QCD contributes to the gluon polarization to lowest order in the coupling (Fig.~\ref{fig:glqkloop}), not containing internal gluon lines, has the same expression both in the SME and in ordinary perturbation theory; hence it vanishes in the limit $p^{2}\to 0$, just like the ghost loop $\Pi_{(1)}(p^{2})$. As an aside, we note that the vanishing of the quark loop at $p^{2}=0$, combined with the lack of the necessary gauge self-interaction vertices, also implies that the SME would not predict the occurrence of DMG for the photons if it were applied to Quantum Electrodynamics. This is further confirmation of the non-triviality of DMG in the framework of the Screened Massive Expansion of QCD.
\vspace{\fill}
\vspace{\fill}
\newpage
Having discussed the main features of the SME gluon propagator, it is now time to compare our results with the lattice data. In Fig.~\ref{fig:gluproplandfit} we show the Landau-gauge ($\xi=0$) Euclidean transverse gluon propagator $\Delta_{E}(p_{E}^{2})$, defined as\footnote{We should mention that the analytic continuation of results obtained in Minkowski space to the Euclidean space and vice-versa is far from being as trivial as Eq.~\eqref{tkm495} suggests \cite{SC22a}. This is especially true for the gluon propagator, which -- as we will see in Sec.~\ref{sec:smeopt} -- possesses a pair of complex-conjugate poles. In what follows we will disregard this issue and take Eq.~\eqref{tkm495} as a definition.}
\begin{equation}\label{tkm495}
\Delta_{E}(p_{E}^{2})=-i\Delta(-p_{E}^{2})\qquad\quad (p_{E}^{2}\geq 0)\ ,
\end{equation}
together with the lattice data of \cite{DOS16}. The plot was obtained by fitting the free parameters $F_{0}$, $Z_{\Delta}$ and $m$ in Eq.~\eqref{ksb805} over the Euclidean momentum range $p_{E}\in[0,4]$~GeV; this is at variance with the ghost dressing function, for which we had cut the data at $p_{E}=2$~GeV. Like in the previous section, we did not perform a further re-normalization of the propagator, using instead the normalization provided by the lattice. The outcome of the fit is reported in Tab.~\ref{tab:fitdata}. For future reference, we note that the fitted value of the parameter $F_{0}$ is $-0.8872$.

As we can see from the figure, the SME gluon propagator accurately reproduces the lattice data down to very small momenta. As the momentum decreases, the propagator changes concavity and saturates to a finite value $\approx10$~GeV$^{-2}$, corresponding to $\Delta^{-1/2}(0)\approx 0.3$~GeV. The energy units are set by the gluon mass parameter, whose fitted value is found to be $m\approx 0.654$~GeV, in general agreement with values derived by other methods -- see e.g. \cite{Fie02}.
\vspace{\fill}
\vspace{\fill}
\begin{table}[H]
\def\arraystretch{1.2}
\centering
\begin{tabular}{ccc}
\hline
$F_{0}$&$m$ (GeV)&$Z_{\Delta}$\\
\hline
\hline
$-0.8872$&$0.6541$&$2.6308$\\
\hline
\end{tabular}
\\
\caption{Parameters obtained by fitting the lattice data of \cite{DOS16} for the Landau-gauge Euclidean transverse gluon propagator in the range $0$-$4$ GeV.}\label{tab:fitdata}
\end{table}
\vspace{\fill}
\newpage
The remarkable precision with which the infrared behavior of the gluon propagator is predicted by the SME substantiates the hypothesis that a shift of the expansion point of QCD is able both to incorporate dynamical mass generation and to avoid the low-energy breakdown of perturbation theory, yielding results which remain valid down into the deep IR. Nonetheless, up to this point, all of our predictions required the fitting of free parameters which do not exist in standard perturbation theory. In the next section, we will discuss how principles of gauge invariance can be exploited to fix some of their values a priori, thus restoring the predictive power of the Screened Massive Expansion.

\section{Optimization of the Screened Massive Expansion}
\renewcommand{\rightmark}{\thesection\ \ \ Optimization of the Screened Massive Expansion}
\label{sec:smeopt}

In order to compare the one-loop ghost and gluon propagators computed in the Screened Massive Expansion with the results of the Landau-gauge lattice calculations, in Secs.~\ref{sec:ghprsme} and \ref{sec:glprsme} we performed fits which made use of a number of free parameters. For the gluon propagator, these were the gluon mass parameter $m^{2}$, the multiplicative renormalization constant $Z_{\Delta}$ and the additive renormalization constant $F_{0}$; for the gluon dressing function, they were the multiplicative renormalization constant $Z_{\mc{G}}$ and the additive renormalization constant $G_{0}$. In total, these add up to one dimensionful parameter ($m^{2}$) plus \textit{four} adimensional renormalization constants. By comparison, the ordinary perturbative expressions contain far fewer free parameters. Indeed, going back to Eqs.~\eqref{vls049} and \eqref{vms472}, we see that, in the absence of the gluon mass parameter $m^{2}$, the one-loop analytical expressions for the ghost and gluon propagators would be determined in terms of the value of the coupling constant $\alpha$ (equivalently, $\alpha_{s}$) and of \textit{two} additive renormalization constants, one for each propagator. The latter are usually fixed by choosing appropriate renormalization conditions.

Where do the additional SME free parameters come from? The gluon mass parameter, of course, arises from the shift $\Delta_{0}\to \Delta_{m}$ that defines the Screened Massive Expansion. The additive renormalization constants $F_{0}$ and $G_{0}$, on the other hand, were introduced with the double purpose of keeping the renormalization of the propagators fully general and of removing the coupling constant from the expressions\footnote{We remark that, formally, the removal of $\alpha$ is only possible at the one-loop order, since higher-order corrections to the ghost self-energy/gluon polarization -- being proportional to powers $\alpha^{k}$ with $k\geq 2$ -- would make it impossible to redefine the additive renormalization constants like we did in Eqs.~\eqref{pwj920} and \eqref{snt638}, while yielding $\alpha$-independent propagators.}, thus reducing the number of parameters. As a by-product of this reparametrization, new multiplicative renormalization constants $Z_{\Delta}$ and $Z_{\mc{G}}$ appeared as factors in $\Delta(p^{2})$ and $\mc{G}(p^{2})$. When comparing the propagators with the results obtained by other methods (such as lattice QCD), the arbitrariness in the choice of the ghost and gluon fields' normalization, together with our decision not to fix the renormalization conditions for the SME expressions, allows us to interpret $Z_{\Delta}$ and $Z_{\mc{G}}$ as new free parameters.

As we saw in Secs.~\ref{sec:ghprsme} and \ref{sec:glprsme}, when expressed in terms of $m^{2}$, $F_{0}$, $G_{0}$, $Z_{\Delta}$ and $Z_{\mc{G}}$, the SME ghost and gluon propagators turn out to accurately reproduce the infrared lattice data. So accurately, in fact, that we may say that, at low energies, the parametrization introduced in the last section effectively incorporates most of the corrections coming from the higher orders in the perturbative series. Nonetheless, the number of free parameters in the expressions is far too large to interpret our results as first-principles analytical predictions. If we could somehow fix the values of the constants $F_{0}$ and $G_{0}$, then the predictive power of the expansion would be restored, leaving $Z_{\Delta}$ and $Z_{\mc{G}}$ as the only renormalization parameters and the gluon's $m^{2}$ as the scale that sets the physical units of the expansion. In this scenario, $m^{2}$ would play a role similar to that of the scale\footnote{That is, either the renormalization scale $\mu$ itself or the value of the QCD scale $\Lambda_{\text{QCD}}$, $\Lambda_{\text{QCD}}=\mu\exp(-8\pi^{2}/\beta_{0}g^{2}(\mu))$ -- see Sec.~\ref{sec:rgint}.} introduced in pQCD by the coupling constant $\alpha_{s}(\mu^{2})$.

The objective of this section is to show that the number of parameters of the SME can indeed be reduced by resorting to principles of gauge invariance \cite{SC18}. More precisely, we will exploit the gauge-parameter independence of the position of the poles of the gluon propagator in a general covariant gauge -- see the discussion on the Nielsen identities in Sec.~\ref{sec:brst} -- to fix the free gluon constant $F_{0}$. The latter can then be used to determine the ghost constant $G_{0}$ based on principles of minimal sensitivity. In what follows, we will not go into the details of the determination of $G_{0}$, limiting ourselves to present the main results of \cite{SIR19a,SIR19b} in Sec.~\ref{sec:glghopt}.

The \textit{optimized} one-loop expressions obtained by enforcing the gauge-parameter independence of the analytical structure of the gluon propagator will be shown to reproduce the infrared lattice data just as well as the fits of Secs.~\ref{sec:ghprsme} and \ref{sec:glprsme}. However, at variance with the fits, this will be achieved from first principles, with no external inputs other than a required energy scale in the form of the gluon mass parameter $m^{2}$.

\subsection{The gluon poles and the Nielsen identities in the context of the Screened Massive Expansion}
\label{sec:glupls}

By making use of the Nielsen identities, in Sec.~\ref{sec:brst} we showed that the position $p_{0}^{2}$ of the poles of the transverse gluon propagator, defined as the solution to the equation
\begin{equation}
\Delta^{-1}(p_{0}^{2},\xi)=0\ ,
\end{equation}
does not depend on the gauge parameter $\xi$. While this is a trivial property for massless propagators, since in that case $p_{0}^{2}=0$ for any $\xi$, if the gluon develops a mass its poles are shifted from the origin and, in principle, can be found at any non-zero value of the complex-$p^{2}$ plane; therefore, the gauge-parameter independence of the poles acquires a new significance in the presence of dynamical mass generation. Gauge-independent gluon poles are likely to play an important role in the evaluation of gauge-invariant physical quantities, which can be directly measured in the experiments\footnote{We should remark, however, that the Nielsen identities do not guarantee that the poles of the propagator do not depend on the gauge; technically speaking, they only prove that their position does not depend on the gauge parameter \textit{in a general covariant gauge}.}.

From an analytical perspective, the only constraint that the gluon poles must satisfy is that they either be real, or appear in complex-conjugate (c.c.) pairs. This is a direct consequence of the identity
\begin{equation}\label{vle830}
\Delta(\overline{p^{2}})=\overline{\Delta(p^{2})}\ ,
\end{equation}
where the overline denotes complex-conjugation, which holds for $\Delta(p^{2})$ since the latter is real for $p^{2}\in\Bbb{R}$, $p^{2}\leq 0$ and analytic away from its singularities. From Eq.~\eqref{vle830} it follows that if $\Delta^{-1}(p_{0}^{2})=0$, then also $\Delta^{-1}(\overline{p_{0}^{2}})=0$, implying that $p_{0}^{2}$ and $\overline{p_{0}^{2}}$ are both poles of the propagator.

The existence of c.c. poles in the propagators of interacting quantum fields has been linked in the literature to the phenomenon of confinement. This is due to the fact that, in the presence of c.c. poles, the K\"{a}ll\'{e}n-Lehman spectral representation which is derived by ordinary quantum-field-theoretical methods for the propagators of physical particles is invalidated \cite{SIR17a}: first of all, the spectral representation acquires an anomalous rational part which describes the contribution coming from the c.c. poles \cite{SIR17a,SC18}; additionally, the spectral function associated to the propagator is negative at both low and high momenta \cite{SC18,HK19}, in violation of the positivity constraints which hold for physical particles. From a dynamical perspective, the imaginary part of the c.c. poles causes an exponential damping of the coordinate-space Minkowski propagator at large times $|t|\to \infty$, a symptom that the degrees of freedom described by the propagator are removed from the asymptotic states of the theory \cite{SC22a}.\\

What does the Screened Massive Expansion tell us about the analytic structure of the transverse gluon propagator? If we take the parameters in Tab.~\ref{tab:fitdata} as an example, we find that -- as predicted by the Gribov-Zwanziger and Refined Gribov-Zwanziger approaches (Sec.~\ref{sec:gzrev}) -- the Landau-gauge ($\xi=0$) gluon propagator computed in the SME indeed possesses a pair of complex conjugate poles at
\begin{equation}
p_{0}^{2}=(0.4487\pm1.0209\, i)\,m^{2}\ ,
\end{equation}
which, with $m=0.654$~GeV, correspond to
\begin{equation}
p_{0}=(\pm 0.5784\pm0.3776\,i)\,\text{GeV}\ ,
\end{equation}
the four $\pm$ signs being independent from each other. More generally, since
\begin{equation}
\Delta^{-1}(p^{2},\xi)=iZ_{\Delta}^{-1}\,p^{2}\,J^{-1}(-p^{2}/m^{2},\xi)\ ,
\end{equation}
where the inverse dressing function $J^{-1}(s,\xi)$ reads
\begin{equation}
J^{-1}(s,\xi)=F(s)+\xi\,F_{\xi}(s)+F_{0}
\end{equation}
and is singular at $s=0$, the gluon poles are given by the zeros of $J^{-1}(-p^{2}/m^{2},\xi)$,
\begin{equation}
J^{-1}(-p^{2}_{0}/m^{2},\xi)=0\ ,
\end{equation}
the latter being equivalent to the pair of coupled equations
\begin{equation}\label{jsd728}
\text{Re}\left\{F(-p^{2}_{0}/m^{2})+\xi\,F_{\xi}(-p^{2}_{0}/m^{2})\right\}+F_{0}=0\ ,\qquad \text{Im}\left\{F(-p^{2}_{0}/m^{2})+\xi\,F_{\xi}(-p^{2}_{0}/m^{2})\right\}=0\ .
\end{equation}
Observe that the imaginary part of $J^{-1}(s,\xi)$ does not depend on the additive renormalization constant $F_{0}$, since the latter is a real quantity; moreover, $\text{Im}\{J^{-1}(s,\xi)\}$ does not vanish for arbitrary $s$, despite $J(s,\xi)$ being a real function for $s\in\Bbb{R}$, as the domain of $p^{2}$ -- equivalently, of $s$ -- is the whole complex plane, $p^{2}\in\Bbb{C}$.

In order to solve Eqs.~\eqref{jsd728}, it is understood that one must first fix a gauge $\xi$; the solution $p_{0}^{2}$ will then be a function of the parameters $m^{2}$, $F_{0}$ and $\xi$:
\begin{equation}
p_{0}^{2}=p_{0}^{2}(m^{2},F_{0},\xi)\ .
\end{equation}
As $\xi$ varies, we are actually allowed to change the values of both $F_{0}$ and $m^{2}$. Indeed, $F_{0}$ is a renormalization constant for the gluon propagator; since the latter is gauge dependent, $F_{0}$ also generally depends on $\xi$. $m^{2}$, on the other hand, is a mass parameter which is introduced in the Faddeev-Popov action \textit{after} fixing the gauge of the QCD Lagrangian. As such, there is no reason to force it to be a gauge-invariant parameter. With $F_{0}=F_{0}(\xi)$ and $m^{2}=m^{2}(\xi)$, the solutions of Eqs.~\eqref{jsd728} take the form
\begin{equation}\label{ghu452}
p_{0}^{2}(\xi)=p_{0}^{2}(m^{2}(\xi),F_{0}(\xi),\xi)\ .
\end{equation}

Are there solutions of Eqs.~\eqref{jsd728} such that $dp_{0}^{2}/d\xi\neq 0$? The answer is yes. For general values of the functions $m^{2}(\xi)$ and $F_{0}(\xi)$, the gluon propagator computed in the SME has poles which depend on the gauge parameter, in violation of the Nielsen identities. This happens because the massive shift that defines the SME breaks BRST symmetry at any finite order in perturbation theory, unless all the gluon mass counterterms are resummed. As a result, the Nielsen identities are not automatically satisfied \cite{SC22b}, and the position of the gluon poles can depend on $\xi$.

Nonetheless, since the shift does not change the total Faddeev-Popov Lagrangian, we should expect the gauge-parameter independence of the gluon poles to be recovered also in the framework of the SME. And indeed, the freedom in the choice of the functions $F_{0}(\xi)$ and $m^{2}(\xi)$ in Eqs.~\eqref{jsd728} and \eqref{ghu452} enables us to \textit{enforce} the $\xi$-independence of the poles. This can be done as follows. Suppose that we know that, in a gauge $\xi_{1}$, the exact gluon propagator has a pole at $p^{2}_{0}=p_{0}^{2}(\xi_{1})$. The Nielsen identities then tell us that in a gauge $\xi_{2}$ the propagator will have a pole at the same position $p_{0}^{2}(\xi_{2})=p_{0}^{2}(\xi_{1})$. In order for this to be true at one loop in the SME, the values of the parameters $m^{2}(\xi_{2})$ and $F_{0}(\xi_{2})$ need to be chosen so that
\begin{equation}
p_{0}^{2}(m^{2}(\xi_{2}),F_{0}(\xi_{2}),\xi_{2})=p_{0}^{2}(m^{2}(\xi_{1}),F_{0}(\xi_{1}),\xi_{1})\ ,
\end{equation}
where $m^{2}(\xi_{1})$ and $F_{0}(\xi_{1})$ are the parameters by which $p_{0}^{2}=p_{0}^{2}(\xi_{1})$ in the gauge $\xi_{1}$. In terms of Eq.~\eqref{jsd728}, this means that the value of the gluon mass parameter $m^{2}(\xi_{2})$ must be such that
\begin{equation}\label{fgr438}
\text{Im}\left\{F(-p^{2}_{0}/m^{2}(\xi_{2}))+\xi_{2}\,F_{\xi}(-p^{2}_{0}/m^{2}(\xi_{2}))\right\}=0\ .
\end{equation}
If this equation has a solution, then -- again by Eq.~\eqref{jsd728} -- $F_{0}(\xi_{2})$ will be given by
\begin{equation}\label{grt348}
F_{0}(\xi_{2})=-\text{Re}\left\{F(-p^{2}_{0}/m^{2}(\xi_{2}))+\xi_{2}\,F_{\xi}(-p^{2}_{0}/m^{2}(\xi_{2}))\right\}\ .
\end{equation}

The requirement that the gluon poles be gauge-parameter independent allows us to fix the value of $F_{0}(\xi)$ and $m^{2}(\xi)$ in any covariant gauge $\xi$ starting from their values $F_{0}(\xi_{0})$ and $m^{2}(\xi_{0})$ in an initial gauge $\xi_{0}$. To do so, one first computes the position of the pole $p_{0}^{2}=p_{0}^{2}(\xi_{0})$ in the gauge $\xi_{0}$ by solving the equation $J^{-1}(-p_{0}^{2}/m^{2}(\xi_{0}),\xi_{0})=0$ for $p_{0}^{2}$, and then uses the steps described above to obtain $F_{0}(\xi)$ and $m^{2}(\xi)$.

\begin{figure}[H]
\centering
\includegraphics[width=0.34\textwidth,angle=270]{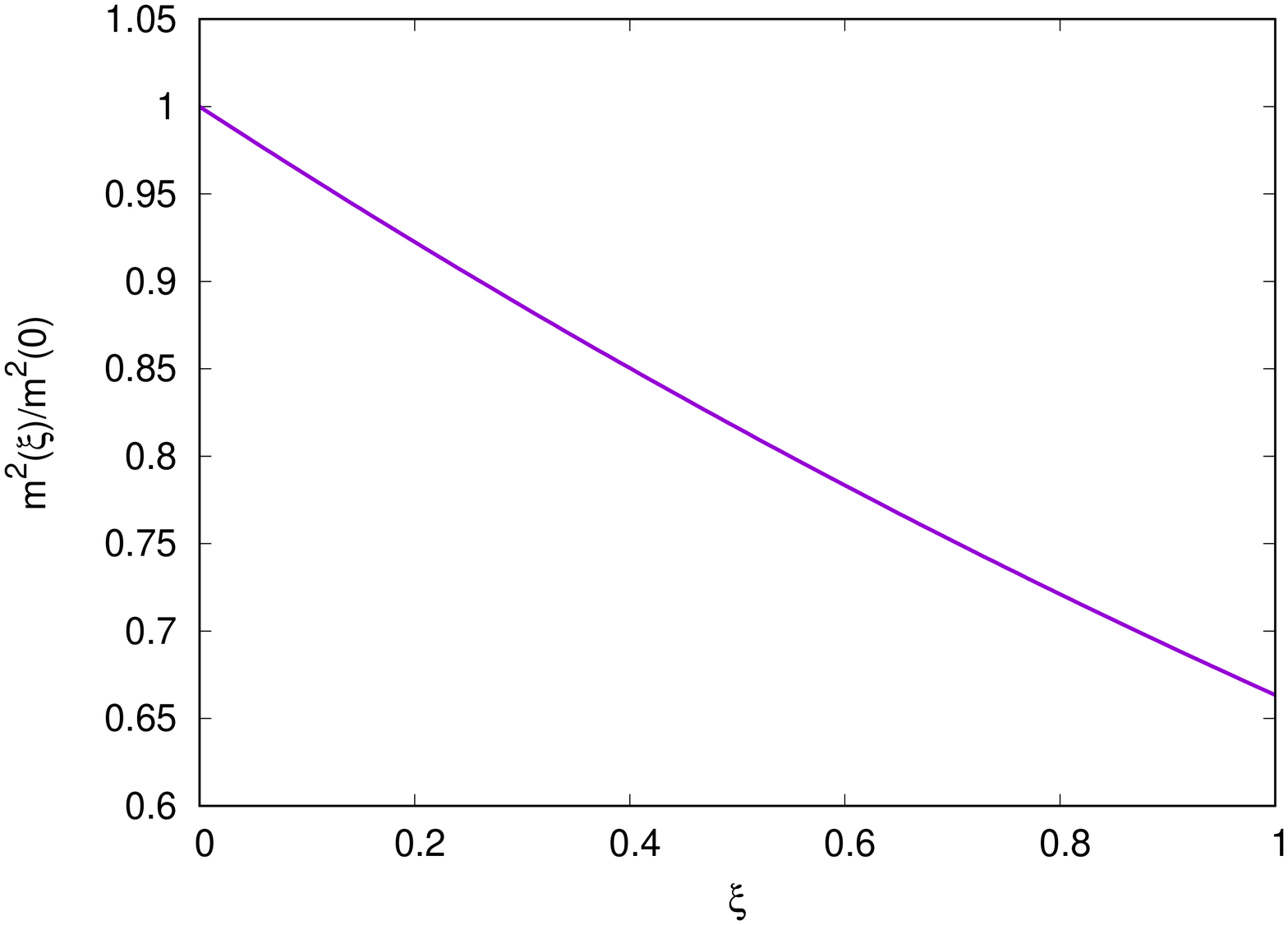}\hspace{2mm}\includegraphics[width=0.34\textwidth,angle=270]{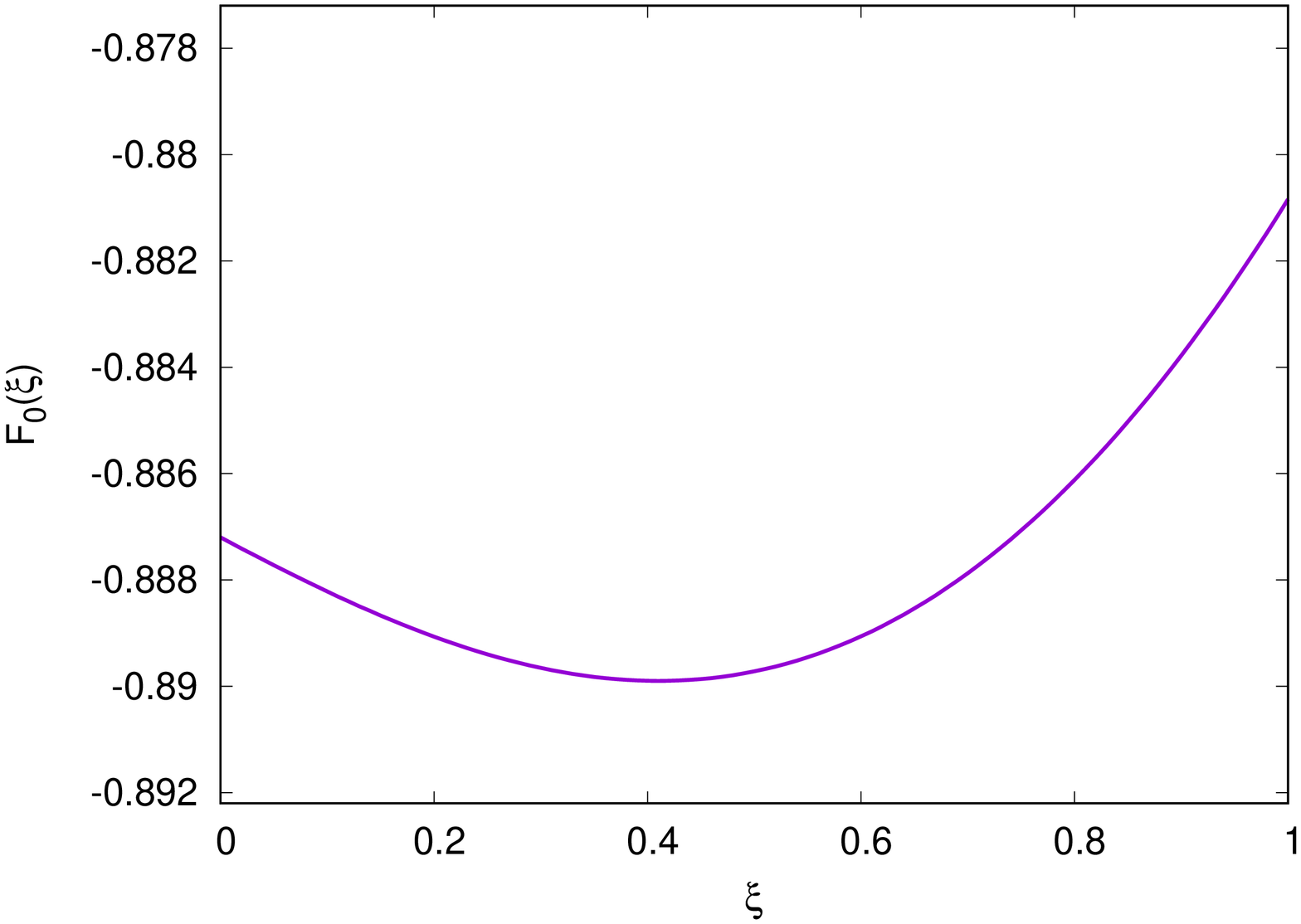}
\vspace{5pt}
\caption{Ratio $m^{2}(\xi)/m^{2}(0)$ (left) and function $F_{0}(\xi)$ (right) computed from the Landau-gauge ($\xi=0$) lattice fit values of Tab.~\ref{tab:fitdata} using Eqs.~\eqref{fgr438} and \eqref{grt348}.}\label{fig:fitdataxigauge}
\end{figure}

As an example of the application of this method, in Fig.~\ref{fig:fitdataxigauge} we show the ratio $m^{2}(\xi)/m^{2}(0)$ and the function $F_{0}(\xi)$ computed from the Landau-gauge ($\xi=0$) lattice fit values of Tab.~\ref{tab:fitdata}, Sec.~\ref{sec:glprsme} -- namely, $F_{0}(0)=-0.8872$ and $m(0)=0.6541$~GeV. As we can see, the gluon mass parameter decreases with the gauge, whereas the additive renormalization constant $F_{0}(\xi)$ is a non-monotonic function of $\xi$. So long as $m^{2}(\xi)$ and $F_{0}(\xi)$ are chosen like in the figure, the gluon poles computed in an arbitrary covariant gauge remain fixed at their $\xi=0$ value, $p_{0}=(\pm 0.5784\pm0.3776\,i)\,\text{GeV}$. We remark that Eqs.~\eqref{fgr438} and \eqref{grt348} are automatically satisfied by $\overline{p_{0}^{2}}$ if they are by $p_{0}^{2}$, explaining how invariance is achieved for both the c.c. gluon poles by a single choice of the functions $m^{2}(\xi)$ and $F_{0}(\xi)$.

Interestingly, the method we just described for determining $m^{2}(\xi)$ and $F_{0}(\xi)$ only works if the poles of the gluon propagator have a non-zero imaginary part (and thus appear in c.c. pairs), for otherwise Eq.~\eqref{fgr438} would be trivially solved by an arbitrary function $m^{2}(\xi)$ and we would be left with an infinite number of corresponding $F_{0}(\xi)$'s by Eq.~\eqref{grt348}. On the other hand, if the propagator has more than one pair of c.c. poles, then it may be impossible to enforce the gauge-parameter independence of \textit{all} the poles by making use of unique functions $m^{2}(\xi)$ and $F_{0}(\xi)$. It is remarkable that the single pair of c.c. poles found by fitting the lattice data prevents these issues from arising.\\

Once the gluon mass parameter and additive renormalization constant are fixed in an arbitrary covariant gauge by enforcing the gauge-parameter independence of the position of the poles, the SME expression for the gluon propagator is left to depend -- modulo multiplicative renormalization -- on just two real numbers: $m^{2}(\xi_{0})$ and $F_{0}(\xi_{0})$, both evaluated at an initial gauge $\xi_{0}$. By pushing forward with requirements of gauge invariance, it can be shown that the constant $F_{0}(\xi_{0})$ also can be fixed from first principles, thus completing the reduction of the number of free parameters needed to restore the predictivity of the SME in the gluon sector. This will be the subject of the next section.
\vspace*{5mm}
\vspace{5mm}

\subsection{The phases of the residues of the gluon propagator: restoring the predictivity of the Screened Massive Expansion in the gluon sector}
\label{sec:phaseopt}

While investigating the properties of the SME gluon propagator evaluated on the fit parameters of Tab.~\ref{tab:fitdata}, an unexpected discovery was made regarding the behavior of the residues at its poles: if the propagator is extended to an arbitrary covariant gauge by enforcing the gauge-parameter independence of the poles as described in Sec.~\ref{sec:glupls}, then the \textit{phases} of the its residues depend on the gauge parameter $\xi$ so slightly that they may also be regarded to be invariant.

\begin{figure}[H]
\centering
\includegraphics[width=0.46\textwidth,angle=270]{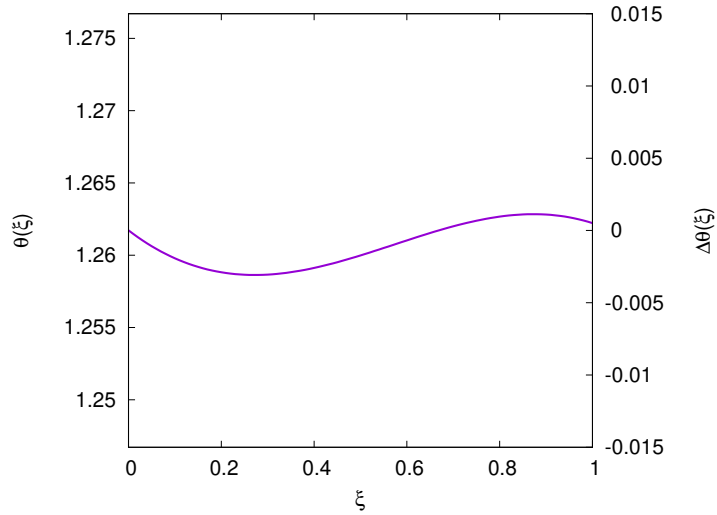}
\vspace{5pt}
\caption{Phase of the residue $\theta(\xi)$ (left axis) and phase difference $\Delta\theta(\xi)=\theta(\xi)-\theta(0)$ (right axis) as functions of the gauge $\xi$ computed by enforcing the gauge-parameter independence of the gluon poles obtained from the fitted Landau-gauge parameters of Tab.~\ref{tab:fitdata}. $\theta(0)\approx 1.2617$.}\label{fig:fitxitheta}
\end{figure}

In more detail, the functions $m^{2}(\xi)$ and $F_{0}(\xi)$ were derived by making use of Eqs.~\eqref{fgr438} and \eqref{grt348} with the Landau-gauge parameters of Tab.~\ref{tab:fitdata} as an input (see Fig.~\ref{fig:fitdataxigauge}). The residue $R(\xi)$ of the gluon propagator at $p_{0}^{2}=(0.4487+1.0209\, i)\,m^{2}(0)$, defined in terms of the Euclidean propagator as
\begin{equation}\label{resdef}
R(\xi)=\lim_{p^{2}_{E}\to -p_{0}^{2}}\ (p^{2}_{E}+p_{0}^{2})\Delta_{E}(p^{2}_{E},\xi)=|R(\xi)|e^{i\theta(\xi)}\ ,
\end{equation}
was then calculated as a function of the gauge parameter $\xi$. Over the range $\xi\in[0,1]$, the phase $\theta(\xi)$ was found to attain values
\begin{equation}
\theta(\xi)=1.262^{+0.09\%}_{-0.24\%}\ ,
\end{equation}
where the quoted value is the Landau-gauge phase, $\theta(0)=1.262$, whereas the maximum and minimum values were found at $\xi\approx 0.9$ and $\xi\approx 0.3$, respectively. In modulus, these are equal to $\theta(0)$ to within less than 2.5 parts in 1000. For reference, the function $\theta(\xi)$ is shown on an enlarged scale in Fig.~\ref{fig:fitxitheta}. Due to the analytical properties of the gluon propagator, these results also hold -- with obvious modifications -- for the conjugated pole $\overline{p_{0}^{2}}=(0.4487-1.0209\, i)\,m^{2}(0)$, which has residue $\overline{R(\xi)}$ and phase $-\theta(\xi)$.

If the gauge-parameter independence of the phases of gluon residues were an \textit{exact} property of the covariant gauges, this finding could hint to some degree of physical significance for the angle $\theta$. Such a notion is strengthened by the observation that $\theta$ does not depend on the renormalization of the propagator. Indeed, while the residue $R(\xi)$ can always be redefined by multiplying the propagator $\Delta(p^{2},\xi)$ by a (generally $\xi$-dependent) renormalization factor $Z(\xi)$, so that the absolute value $|R(\xi)|$ is both gauge and renormalization dependent, the same is not true for the phase $\theta(\xi)$: given that $Z(\xi)$ must be chosen real, the renormalization of the propagator cannot change the phase $\theta$, nor its gauge dependence.

\vspace{5mm}
\vspace{5mm}

Interestingly, in the context of the Screened Massive Expansion, the assumption that $\theta(\xi)$ be exactly gauge-parameter independent provides us with a criterion for fixing the value of the renormalization constant $F_{0}(\xi_{0})$ at some initial gauge $\xi_{0}$ \cite{SC18}. The idea that underlies this kind of optimization is that, if the phases of the residues of the exact gluon propagator do not depend on $\xi$, then the best approximation of the propagator is the one for which $\theta(\xi)$ varies the least with the gauge parameter. That this can be achieved by an appropriate choice of $F_{0}(\xi_{0})$ can be shown as follows. Going back to the definition of the residue -- Eq.~\eqref{resdef}~--, we see that
\begin{align}\label{vfn436}
R(\xi)&=\lim_{p_{E}^{2}\to -p_{0}^{2}}\,(p_{E}^{2}+p_{0}^{2})\Delta_{E}(p_{E}^{2},\xi)=\\
\notag&=\lim_{p_{E}^{2}\to -p_{0}^{2}}\,\frac{p_{E}^{2}+p_{0}^{2}}{\Delta_{E}^{-1}(p_{E}^{2},\xi)-\Delta_{E}^{-1}(-p_{0}^{2},\xi)}=\left(\frac{\partial\Delta^{-1}_{E}}{\partial p_{E}^{2}}\bigg|_{p_{E}^{2}=-p_{0}^{2}}\right)^{-1}\ .
\end{align}
In terms of the inverse dressing function $J^{-1}(s)$,
\begin{equation}
\Delta_{E}^{-1}(p_{E}^{2},\xi)=Z_{\Delta}^{-1}\,p_{E}^{2}\,J^{-1}(p_{E}^{2}/m^{2}(\xi))\ ,\qquad J^{-1}(-p_{0}^{2}/m^{2}(\xi))=0\ ,
\end{equation}
Eq.~\eqref{vfn436} reads
\begin{equation}
R(\xi)=\frac{Z_{\Delta}\,m^{2}(\xi)}{-p_{0}^{2}}\left(\frac{\partial J^{-1}}{\partial s}\bigg|_{s=-\frac{p_{0}^{2}}{m^{2}(\xi)}}\right)^{-1}.
\end{equation}
By denoting the adimensional ratios $m^{2}(\xi)/m^{2}(\xi_{0})$ and $-p_{0}^{2}/m^{2}(\xi_{0})$ respectively with $a(\xi)$ and $z_{0}$, the residue can be put in the form
\begin{equation}\label{dyt596}
R(\xi)=\frac{Z_{\Delta}\,a(\xi)}{z_{0}}\left(\frac{\partial J^{-1}}{\partial s}\bigg|_{s=\frac{z_{0}}{a(\xi)}}\right)^{-1}.
\end{equation}
In particular, $R(\xi)$ depends on $Z_{\Delta}$, $z_{0}$ and $a(\xi)$, and also on $F_{0}(\xi)$ and $\xi$ through $J^{-1}(s)$.

Now, at $\xi=\xi_{0}$, as detailed in Sec.~\ref{sec:glupls}, the position of the poles is found by solving the equation
\begin{equation}\label{otg560}
J^{-1}(-p_{0}^{2}/m^{2}(\xi_{0}))=0\qquad\Longleftrightarrow\qquad J^{-1}(z_{0})=0\ .
\end{equation}
Therefore, one does not need the mass parameter $m^{2}(\xi_{0})$ to compute $z_{0}$: the adimensional pole $z_{0}$ only depends on $F_{0}(\xi_{0})$ and on the gauge $\xi_{0}$ itself,
\begin{equation}
z_{0}=z_{0}(F_{0}(\xi_{0}),\xi_{0})\ .
\end{equation}
Moreover, in a general gauge $\xi\neq \xi_{0}$, to find $m^{2}(\xi)$ and $F_{0}(\xi)$ one needs to solve the equation
\begin{equation}
J^{-1}(-p_{0}^{2}/m^{2}(\xi))=0\qquad\Longleftrightarrow\qquad J^{-1}(z_{0}/a(\xi))=0\ ,
\end{equation}
i.e.
\begin{equation}\label{reg234}
F(z_{0}/a(\xi))+\xi\,F_{\xi}(z_{0}/a(\xi))+F_{0}(\xi)=0\ .
\end{equation}
The imaginary part of Eq.~\eqref{reg234} allows us to determine $a(\xi)$,
\begin{equation}
\text{Im}\left\{F(z_{0}/a(\xi))+\xi\,F_{\xi}(z_{0}/a(\xi))\right\}=0\ ,
\end{equation}
which therefore only depends on the gauge $\xi$ and on the parameters $F_{0}(\xi_{0})$, $\xi_{0}$ through $z_{0}$,
\begin{equation}
a(\xi)=a(\xi;F_{0}(\xi_{0}),\xi_{0})\ .
\end{equation}
The real part of Eq.~\eqref{reg234}, on the other hand, allows us to determine $F_{0}(\xi)$,
\begin{equation}
F_{0}(\xi)=-\text{Re}\left\{F(z_{0}/a(\xi))+\xi\,F_{\xi}(z_{0}/a(\xi))\right\}\ ,
\end{equation}
which therefore also only depends on $F_{0}(\xi_{0})$ and $\xi_{0}$:
\begin{equation}
F_{0}(\xi)=F_{0}(\xi;F_{0}(\xi_{0}),\xi_{0})\ .
\end{equation}
In other words, $a(\xi)$ and $F_{0}(\xi)$, as functions of $\xi$, are determined by the single constant $F_{0}(\xi_{0})$, and not by the mass scale $m^{2}(\xi_{0})$. This could have been anticipated by observing that $a(\xi)$ and $F_{0}(\xi)$ are adimensional parameters, while $m^{2}(\xi_{0})$ is a dimensionful scale.

\begin{figure}[H]
\centering
\includegraphics[width=0.47\textwidth,angle=270]{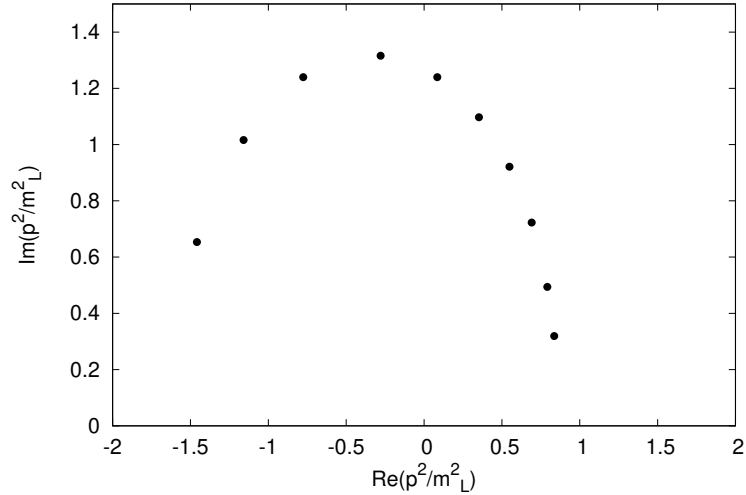}
\vspace{5pt}
\caption{Adimensional poles of the transverse gluon propagator as a function of $F_{0}(0)\in[-2.00,-0.10]\,$. Left to right: $F_{0}(0)=-2.00$, $-1.90$, $-1.75$, $-1.50$, $-1.25$, $-1.00$, $-0.75$, $-0.50$, $-0.25$, $-0.10$. For each pole $p_{0}^{2}$, its complex conjugate $\overline{p_{0}^{2}}$ is also a pole of the propagator. $m_{L}=m(0)$.}\label{fig:pospoles}
\end{figure}

Since $z_{0}$, $a(\xi)$ and $F_{0}(\xi)$ are all functions of $F_{0}(\xi_{0})$ and $\xi_{0}$ alone, then the residue $R(\xi)$ also is:
\begin{equation}
R(\xi)=R(\xi;F_{0}(\xi_{0}),\xi_{0})
\end{equation}
(here we have suppressed the dependence of $R(\xi)$ on $Z_{\Delta}$, since the latter does not affect the value of the phase $\theta(\xi)$). In particular, if the gauge-parameter independence of the position of the gluon poles is enforced by the method of Sec.~\ref{sec:glupls}, then phases of the residues in any covariant gauge are completely determined by the value of the additive renormalization constant $F_{0}(\xi_{0})$ at the gauge $\xi_{0}$. In what follows, we will take $\xi_{0}$ to be equal to the Landau gauge, $\xi_{0}=0$.
\vspace{5mm}
\begin{figure}[H]
\centering
\includegraphics[width=0.52\textwidth,angle=0]{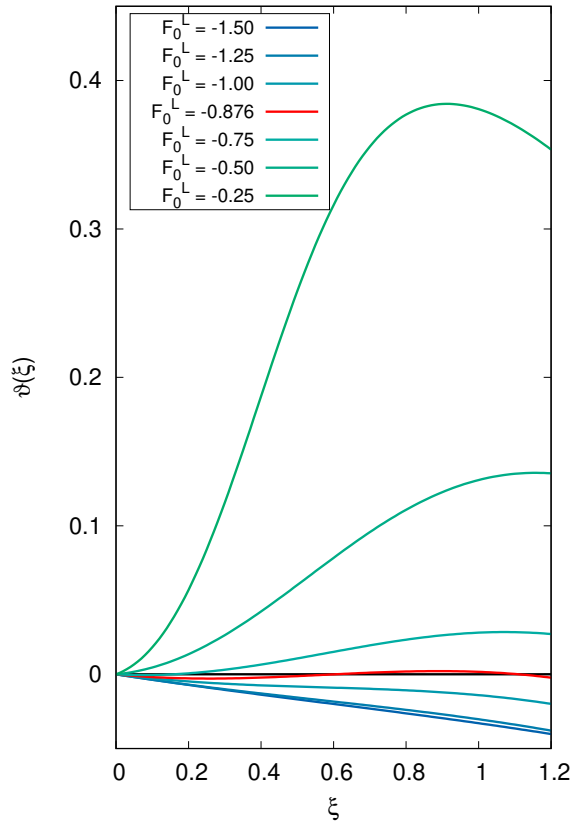}
\vspace{5pt}
\caption{Phase difference $\Delta\theta(\xi)=\theta(\xi)-\theta(0)$ as a function of $\xi$ for different values of $F_{0}(0)$ in the range $[-1.5,-0.25]$. $F_{0}^{L}=F_{0}(0)$.}\label{fig:phasefig}
\end{figure}
\vspace{5mm}
In Fig.~\ref{fig:phasefig} we show the phase differences $\Delta\theta(\xi)=\theta(\xi)-\theta(0)$ of the residues at the gluon pole $p_{0}^{2}$ as functions of the gauge parameter $\xi$ for different values of the additive renormalization constant $F_{0}(0)$ at $\xi=0$. We remark that, as $F_{0}(0)$ varies, the position of the pole $p_{0}^{2}$ also changes due to Eq.~\eqref{otg560}; in Fig.~\ref{fig:pospoles} we display the adimensional poles $p_{0}^{2}/m^{2}(0)$ for different values of $F_{0}(0)$, with the understanding that the corresponding $\overline{p_{0}^{2}}$'s are also poles of the transverse gluon propagator.

As we can see from Fig.~\ref{fig:phasefig}, the deviation of the phase $\theta(\xi)$ from its Landau-gauge value $\theta(0)$ is approximately zero only in a neighborhood of the fitted parameter $F_{0}(0)\approx-0.9$. This indicates that, analytically, the near gauge-parameter independence of the phase is a non-trivial feature of the one-loop SME approximation which is realized only when the SME propagator accurately reproduces the lattice result. It thus makes sense to seek a value of $F_{0}(0)$ that minimizes the dependence of $\theta(\xi)$ on the gauge parameter.

\begin{figure}[H]
\centering
\includegraphics[width=0.45\textwidth,angle=270]{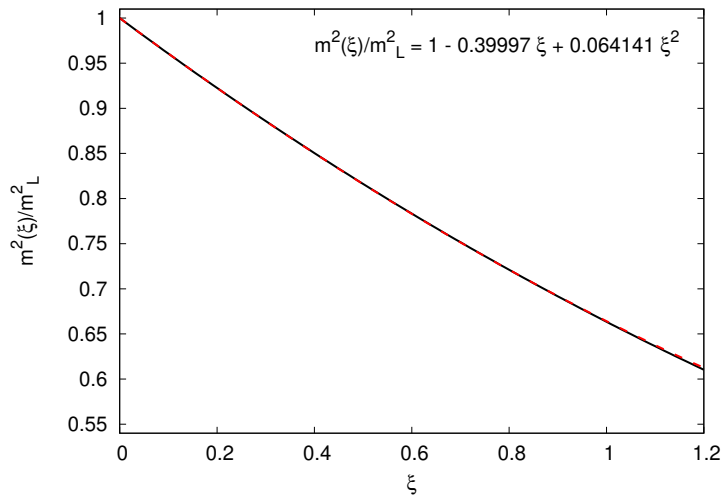}
\vspace{5pt}
\caption{Optimal mass parameter $m^{2}(\xi)$ as a function of $\xi$ computed at $F_{0}(0)=-0.876$. Dashed red line: polynomial approximation given by Eq.~\eqref{polymass}. $m_{L}=m(0)$.}\label{fig:massfitfig}
\end{figure}

\begin{figure}[H]
\centering
\includegraphics[width=0.46\textwidth,angle=270]{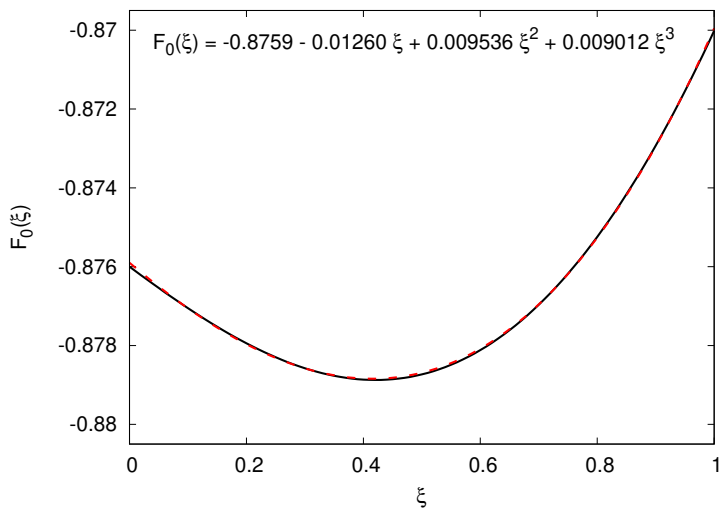}
\vspace{5pt}
\caption{Optimal additive normalization constant $F_{0}(\xi)$ as a function of $\xi$ computed at $F_{0}(0)=-0.876$. Dashed red line: polynomial approximation given by Eq.~\eqref{polyf0}.}\label{fig:f0fitfig}
\end{figure}

In order to determine the optimal value for the additive renormalization constant $F_{0}(0)$, in \cite{SC18} the maximum value of the difference $|\Delta\theta(\xi)|$ was minimized over the range $\xi\in[0,1.2]$. The resulting $F_{0}(0)$ was found to be equal to
\begin{equation}
F_{0}(0)=-0.876\ ,
\end{equation}
yielding the adimensional poles
\begin{equation}
p_{0}^{2}=(0.4575\pm1.0130\ i)\, m^{2}(0)\ .
\end{equation}
The functions $m^{2}(\xi)$ and $F_{0}(\xi)$ which enforce the gauge-parameter independence of the poles at the given value of $F_{0}(0)$ can be approximated by the polynomials
\BE\label{polymass}
m^{2}(\xi)\approx \left(1-0.39997\,\xi+0.064141\,\xi^{2}\right)\,m^{2}(0)
\EE
and
\BE\label{polyf0}
F_{0}(\xi)\approx -0.8759-0.01260\,\xi+0.009536\,\xi^{2}+0.009012\,\xi^{3}\ ,
\EE
shown in Figs.~\ref{fig:massfitfig} and \ref{fig:f0fitfig} together with the corresponding optimized curves. These approximations are quite accurate up to and beyond the Feynman gauge ($\xi=1$).

The phase $\theta(\xi)$ corresponding to $F_{0}(0)=-0.876$ -- highlighted in red in Fig.~\ref{fig:phasefig} -- was found to be
\begin{equation}
\theta(\xi)=1.262^{+0.22\%}_{-0.22\%}\ ,
\end{equation}
the quoted value being the Landau-gauge phase, $\theta(0)=1.262$, whereas the maximum and minimum deviations $\Delta\theta(\xi)=\pm 0.22\%\,\theta(0)$ were found at $\xi\approx 0.9$ and $\xi\approx0.2$, respectively, with $\text{max}_{\xi\in[0,1.2]}\left\{|\Delta\theta(\xi)|\right\}$ being less than $2.5$ parts in 1000. A summary of the optimized results is reported in Tab.~\ref{tab:paramtab}.
\vspace{\fill}
\vspace{\fill}
\newpage
\topskip0pt
\vspace*{\fill}
\vspace{5mm}
\vspace{5mm}
\begin{table}[H]
\def\arraystretch{0.4}
\centering
\begin{tabular}{|c|ccc|}
\hline
\hline
&&&\\
$F_{0}(0)=-0.876$&$\dfrac{p_{0}^{2}}{m^{2}(0)}=0.4575\pm1.0130\ i$&$\dfrac{p_{0}}{m(0)}=\pm0.8857\pm0.5718\ i$&$\theta=\pm\,1.262$\\
&&&\\
\hline
\hline
\end{tabular}
\vspace{5pt}
\caption{Position and phases of the residues of the gluon poles for $F_{0}(0)=-0.876$.}\label{tab:paramtab}
\end{table}
\vspace*{\fill}
\newpage
It is apparent that the optimization procedure described in this section yields parameters which lie very close to those obtained by fitting the lattice data: the relative difference between the optimized value of $F_{0}(0)$ and the one provided by the fit is around $1\%$, whereas the phases of the residues provided by the two approaches are equal up to the third decimal digit. As a result, as we will explicitly see in the next section, the optimized Euclidean Landau-gauge gluon propagator turns out to be undistinguishable from that which is obtained by fitting $F_{0}(0)$. This gives us confidence that the assumption that $d\theta/d\xi=0$ yields sensible results regardless of the specific criterion which is chosen to minimize the dependence of $\theta(\xi)$ on the gauge parameter $\xi$.

\subsection{The optimized gluon and ghost propagators}
\label{sec:glghopt}

Within the Screened Massive Expansion, by enforcing the gauge-parameter independence of the poles and phases of the residues as laid out in the previous sections, the gluon propagator can be computed from first principles in any covariant gauge. The optimized SME propagator,
\begin{equation}
\Delta_{E}(p_{E}^{2},\xi)=\frac{Z_{\Delta}(\xi)}{p_{E}^{2}\left[F(s/a(\xi))+\xi\,F_{\xi}(s/a(\xi))+F_{0}(\xi)\right]}\ ,
\end{equation}
where $s=p_{E}^{2}/m^{2}(0)$ and $a(\xi)=m^{2}(\xi)/m^{2}(0)$, is expressed in terms of the known functions $a(\xi)$ and $F_{0}(\xi)$ -- approximated by Eqs.~\eqref{polymass} and \eqref{polyf0} in Sec.~\ref{sec:phaseopt} --, of a multiplicative renormalization constant $Z_{\Delta}(\xi)$ and of the Landau-gauge gluon mass parameter $m^{2}(0)$.

Since pure Yang-Mills theory is scale free, the value of the gluon mass parameter $m^{2}(0)$ cannot be predicted from first principles, and must instead be fixed by a comparison with the experiments\footnote{Or with the lattice data, given that pure Yang-Mills theory is not realized in the physical world.}. In this respect, its status is similar to that of the QCD scale $\Lambda_{\text{QCD}}$ of ordinary perturbation theory (Sec.~\ref{sec:rgint}): it is $m^{2}(0)$ that sets the energy units of the approximation. The value of $Z_{\Delta}(\xi)$, on the other hand, is determined by the normalization conditions which are chosen for the propagator.

In what follows we will compare our optimized results with the lattice data. Unfortunately, most of the lattice calculations of the QCD propagators are carried out in the Landau gauge, due to numerical difficulties which arise on the lattice when trying to enforce the gauge-fixing conditions for $\xi\neq 0$. Therefore, we shall limit ourselves with making our main comparisons with the Landau-gauge lattice data, using Ref.~\cite{BBCO15} as the only available benchmark for the gluon propagator in covariant gauges $\xi\neq 0$.\\

\begin{figure}[H]
\centering
\includegraphics[width=0.48\textwidth,angle=270]{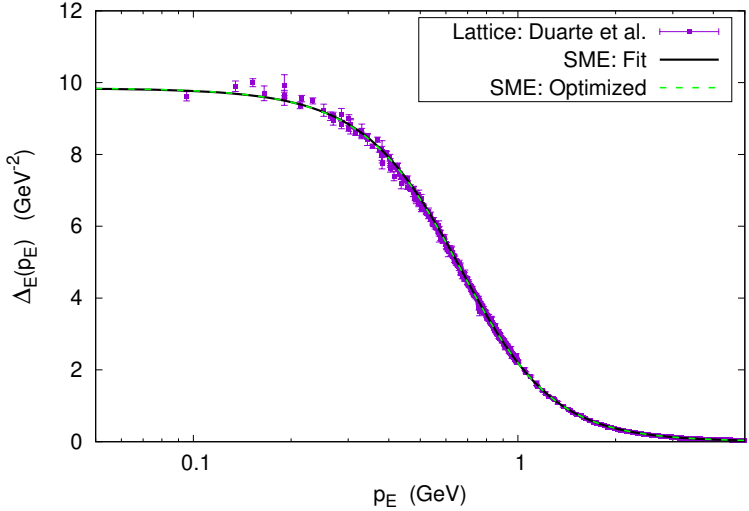}
\vspace{5pt}
\caption{Euclidean transverse gluon propagator in the Landau gauge ($\xi=0$). Solid black curve: SME with $F_{0}$ fitted from the lattice data (Tab.~\ref{tab:fitdata}). Dashed green curve: SME with $F_{0}$ optimized by gauge invariance ($F_{0}=-0.876$, Tab.~\ref{tab:optdata}). Squares: lattice data from \cite{DOS16}.}\label{fig:proplandopt}
\end{figure}

In Fig.~\ref{fig:proplandopt} the Landau-gauge Euclidean transverse gluon propagators computed in the SME are shown together with the lattice data of \cite{DOS16}. The solid black curve is the full fit already presented in Sec.~\ref{sec:glprsme} (see Tab.~\ref{tab:fitdata} for the fit parameters). The dashed green curve, on the other hand, is obtained by first fixing $F_{0}=-0.876$, as determined via the optimization procedure discussed in Secs.~\ref{sec:glupls} and \ref{sec:phaseopt}, and then fitting the gluon mass parameter $m^{2}(0)$ and the multiplicative renormalization constant $Z_{\Delta}(0)$. The outcome of the fit is reported in Tab.~\ref{tab:optdata}.

As we can see from the figure, the optimized and the fully-fitted propagators cannot be distinguished to the naked eye. The value of the gluon mass parameter obtained for the optimized curve, $m=0.6557$~GeV, is equal to the one obtained via the full fit ($m=0.6541$~GeV) to within less than 1\%; the same holds true for the multiplicative constants $Z_{\Delta}$. This happens because, as we noted in the last section, the optimized and the fitted Landau-gauge additive renormalization constants $F_{0}$ are extremely close to each other. In\cleannlnp Tab.~\ref{tab:paramtabfit} we report the dimensionful position of the gluon poles, computed from Tab.~\ref{tab:paramtab} by making use of the fitted value of $m^{2}(0)$ at $F_{0}(0)=-0.876$.

\begin{figure}[H]
\centering
\includegraphics[width=0.48\textwidth,angle=270]{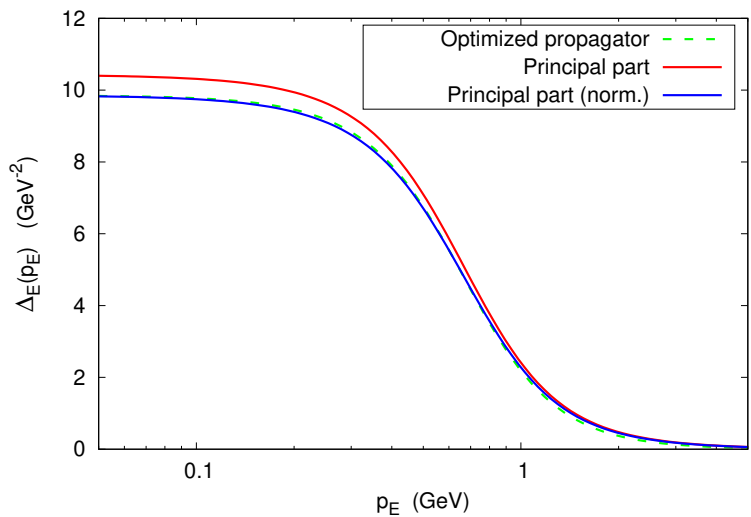}
\vspace{5pt}
\caption{Euclidean transverse gluon propagator in the Landau gauge ($\xi=0$). Dashed green curve: optimized propagator, with parameters given by Tab.~\ref{tab:optdata}. Solid red curve: principal part of the optimized propagator, with poles and residues given by Tabs.~\ref{tab:paramtab} and \ref{tab:paramtabfit}, $|R|=0.947\,Z_{\Delta}$. Solid blue curve: principal part of the optimized propagator, multiplied by a factor of $0.945$. $m=0.6557$~GeV.}\label{fig:glupropprinc}
\end{figure}

\begin{figure}[H]
\centering
\includegraphics[width=0.48\textwidth,angle=270]{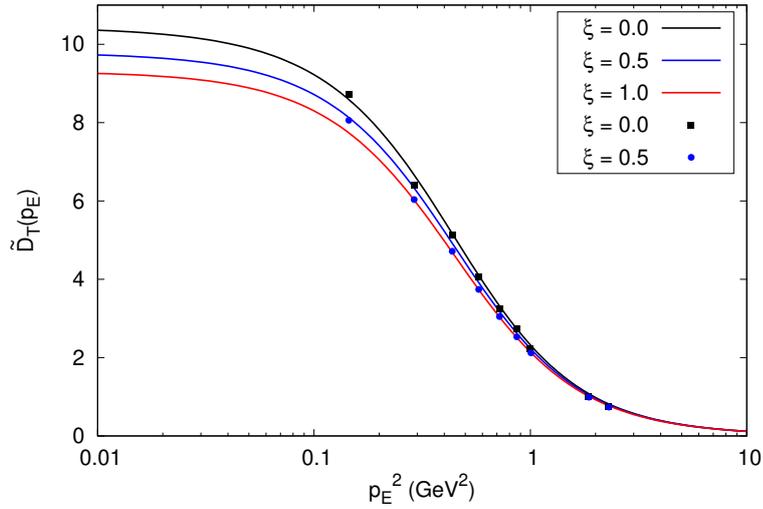}
\vspace{5pt}
\caption{Optimized SME Euclidean transverse gluon propagator computed in different gauges, renormalized at $\mu=4.317$ GeV. Lattice data from \cite{BBCO15}. $m(0)=0.6557$~GeV.}\label{fig:propgauge}
\end{figure}

The precision with which the optimized propagator reproduces the infrared lattice data substantiates the hypothesis that an accurate description of the low-energy dynamics of the gluon can be obtained from first principles by enforcing the gauge-parameter independence of the poles and of the phases of the residues.
\vspace{\fill}
\vspace{5mm}
\begin{table}[H]
\def\arraystretch{1.2}
\centering
\begin{tabular}{c|cc}
\hline
$F_{0}$&$m$ (GeV)&$Z_{\Delta}$\\
\hline
\hline
$-0.876$&$0.6557$&$2.6481$\\
\hline
\end{tabular}
\vspace{5pt}
\caption{Parameters obtained by fitting the lattice data of \cite{DOS16} for the Landau-gauge Euclidean transverse gluon propagator in the range $0$-$4$~GeV at fixed $F_{0}=-0.876$.}\label{tab:optdata}
\end{table}
\vspace{5mm}
\begin{table}[H]
\def\arraystretch{1.4}
\centering
\begin{tabular}{|c|cc|}
\hline
\hline
$F_{0}^{L}=-0.876$&$p_{0}^{2}=(0.1969\pm0.4359\ i)$ GeV$^{2}$&$p_{0}=(\pm0.5810\pm0.3751\ i)$ GeV\\
\hline
\hline
\end{tabular}
\vspace{5pt}
\caption{Dimensionful position of the gluon poles for $F_{0}(0)=-0.876$ and $m(0)=0.6557$~GeV. The latter is obtained by fitting the optimized gluon propagator to the lattice data of Ref.~\cite{DOS16}.}\label{tab:paramtabfit}
\end{table}
\vspace{\fill}
\newpage
For future reference, we mention that in the Landau gauge the gluon propagator is well approximated by multiplying the principal part $\Delta_{E}^{(\text{PP})}(p_{E}^{2})$ of the optimized propagator by a constant $\approx1$. $\Delta_{E}^{(\text{PP})}(p_{E}^{2},\xi)$ itself, in a general covariant gauge, is defined as
\begin{equation}
\Delta_{E}^{(\text{PP})}(p_{E}^{2},\xi)=\frac{R(\xi)}{p_{E}^{2}+p_{0}^{2}}+\frac{\overline{R(\xi)}}{p_{E}^{2}+\overline{p_{0}^{2}}}\ .
\end{equation}
In the Landau gauge, with $Z_{\Delta}=2.6481$ obtained by the fit, $|R|$ is found to be equal to
\begin{equation}
|R|=0.947\,Z_{\Delta}=2.508\ .
\end{equation}
The principal part $\Delta_{E}^{(\text{PP})}(p_{E}^{2})$ of the propagator (solid red curve) is displayed together with the Landau-gauge optimized propagator (dashed green curve) in Fig.~\ref{fig:glupropprinc}. As is clear from the figure, $\Delta_{E}^{(\text{PP})}(p_{E}^{2})$ makes up for the majority of the optimized propagator $\Delta_{E}(p_{E}^{2})$. Moreover, when normalized by a factor of $0.945$ (solid blue curve), the principal part provides a quite accurate approximation of the full optimized propagator up to momenta $\approx 4$~GeV.
\vspace{5mm}
\vspace{5mm}
Outside of the Landau gauge, the optimized gluon propagator behaves as displayed in Fig.~\ref{fig:propgauge}, where the lattice data for $\xi=0,0.5$ were extracted from Ref.~\cite{BBCO15}. In order to compare the propagator with the data, the former was renormalized in the MOM scheme,
\begin{equation}
\Delta_{E}(\mu^{2},\xi)=\frac{1}{\mu^{2}}
\end{equation}
at $\mu=4.317$~GeV, as reported in \cite{BBCO15}. The mass parameter $m(0)=0.6557$~GeV fitted from the lattice data of \cite{DOS16} was used to obtain the curves shown in the figure.\newpage
\vspace*{5mm}
\vspace{5mm}
\begin{figure}[H]
\centering
\includegraphics[width=0.48\textwidth,angle=270]{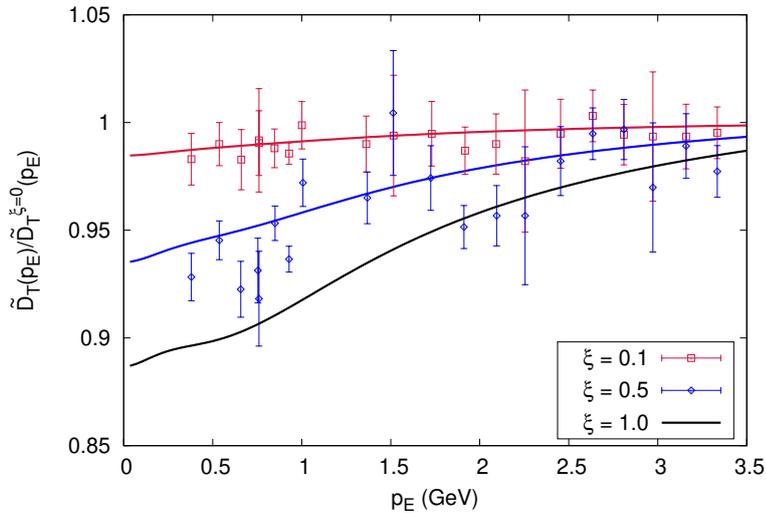}
\vspace{5pt}
\caption{Ratio between the optimized SME Euclidean gluon propagators computed in a general covariant gauge and in the Landau gauge ($\xi=0$), renormalized at $\mu=4.317$ GeV. Lattice data from \cite{BBCO15}.}\label{fig:propgaugerat}
\end{figure}
\vspace{5mm}
As the gauge increases, the lattice propagator remains massive and slightly decreases in value at fixed momentum. The optimized SME propagator reproduces this behavior, showing a good qualitative agreement with the lattice. When the ratios between the $\xi\neq 0$ propagators and their Landau-gauge counterpart are compared (Fig.~\ref{fig:propgaugerat}), no definite conclusion can be reached regarding the quantitative agreement between the analytical predictions and the lattice results due to the errors which affect the lattice data.

In the introduction to Sec.~\ref{sec:smeopt}, we anticipated that the ghost additive renormalization constant $G_{0}$ that appears in the expression
\begin{equation}
\mc{G}(p^{2})=\frac{iZ_{\mc{G}}}{p^{2}\left(G(s)-\xi\,\log(s)/12+G_{0}\right)}
\end{equation}
for the SME ghost propagator can be determined from the knowledge of the gluon constant $F_{0}$ by making use of principles of minimal sensitivity. This is possible because $F_{0}$ and $G_{0}$ are not actually fully independent parameters, being defined as
\begin{align}
F_{0}&=\frac{1}{\alpha}+f_{0}\ ,\\
\notag G_{0}&=\frac{1}{\alpha}+g_{0}
\end{align}
(see Secs.~\ref{sec:ghprsme} and \ref{sec:glprsme}), where $f_{0}$ and $g_{0}$ are adimensional constants which are fixed once renormalization conditions are imposed on the gluon and ghost propagator, respectively. As long as the renormalization conditions are left arbitrary, as we did in this chapter, the values of $F_{0}$ and $G_{0}$ can be chosen independent of each other. If, on the other hand, the renormalization conditions are specified, then the knowledge of $F_{0}$, together with that of $f_{0}$ and $g_{0}$, are sufficient to compute $G_{0}$.

A straightforward calculation allows us to make the dependence of $G_{0}$ and $F_{0}$ on the renormalization conditions and on $\alpha$ explicit. From the definition of the one-loop Euclidean gluon and ghost SME propagators $\Delta_{E}(p_{E}^{2})$ and $\mc{G}_{E}(p_{E}^{2})$, we find that\footnote{For simplicity, in the following expressions we omit the explicit dependence of $G_{0}$, $F_{0}$, $m^{2}$ and the propagators on the gauge parameter $\xi$.}
\begin{align}
G_{0}&=\frac{1}{\alpha(\mu)}\big(\mu^{2}\,\mc{G}_{E}(\mu^{2})\big)^{-1}-G(\mu^{2}/m^{2})+\frac{\xi}{12}\,\ln(\mu^{2}/m^{2})\ ,\label{g0mu}\\
F_{0}&=\frac{1}{\alpha(\mu)}\big(\mu^{2}\,\Delta_{E}(\mu^{2})\big)^{-1}-F(\mu^{2}/m^{2})-\xi\,F_{\xi}(\mu^{2}/m^{2})\ ,\label{f0mu}
\end{align}
where $\mu$ is the renormalization scale for the propagators in a MOM-like renormalization scheme. In particular, since by Eq.~\eqref{f0mu} the coupling is given in terms of the gluon functions as
\begin{equation}
\frac{1}{\alpha(\mu)}=\mu^{2}\Delta_{E}(\mu^{2})\left(F(\mu^{2}/m^{2})+\xi\,F_{\xi}(\mu^{2}/m^{2})+F_{0}\right)\ ,
\end{equation}
the ghost additive renormalization constant can be computed as
\begin{equation}
G_{0}=\frac{\Delta_{E}(\mu^{2})}{\mc{G}_{E}(\mu^{2})}\left(F(\mu^{2}/m^{2})+\xi\,F_{\xi}(\mu^{2}/m^{2})+F_{0}\right)-G(\mu^{2}/m^{2})+\frac{\xi}{12}\,\ln(\mu^{2}/m^{2})\ .
\end{equation}
In the above expression, the parameters $m^{2}=m^{2}(\xi)$ and $F_{0}=F_{0}(\xi)$ are known thanks to the optimization procedure presented in the previous sections. Therefore, the only unknowns are the gluon and the ghost propagators $\Delta_{E}(\mu^{2})$ and $\mc{G}_{E}(\mu^{2})$ at the scale $\mu$, and the scale $\mu$ itself. These are determined by the renormalization conditions.

In Refs.~\cite{SIR19a,SIR19b}, the propagators were defined in the so-called screened MOM (SMOM) scheme, by which\footnote{The renormalization conditions reported in Eq.~\eqref{sid485} are the same as those of the Curci-Ferrari's IRS scheme -- Sec.~\ref{sec:cfrevsu} --, the only difference being the independence of the gluon mass parameter $m^{2}$ from the renormalization scale.}
\begin{equation}\label{sid485}
\Delta_{E}(\mu^{2};\mu^{2})=\frac{1}{\mu^{2}+m^{2}}\ ,\qquad\qquad\mc{G}_{E}(\mu^{2};\mu^{2})=\frac{1}{\mu^{2}}\ .
\end{equation}
In the SMOM scheme, the ghost constant $G_{0}$ is explicitly given by
\begin{equation}
G_{0}=\left(1+\frac{m^{2}}{\mu^{2}}\right)^{-1}\left(F(\mu^{2}/m^{2})+\xi\,F_{\xi}(\mu^{2}/m^{2})+F_{0}\right)-G(\mu^{2}/m^{2})+\frac{\xi}{12}\,\ln(\mu^{2}/m^{2})\ .
\end{equation}
The optimal renormalization scale $\mu$ was then determined by requiring that $G_{0}$ be scale independent,
\begin{equation}\label{g0derc}
\frac{\partial G_{0}}{\partial \mu}=0\ .
\end{equation}
Such a requirement is made necessary by the observation, made in \cite{SIR19a}, that a $\mu$-dependent $G_{0}$ would spoil the multiplicative renormalizability of the theory if $F_{0}$ is taken to be scale independent, as is implied by the optimization procedure of Secs.~\ref{sec:glupls} and \ref{sec:phaseopt}. The principle of minimal sensitivity on the scale $\mu$ applied to $G_{0}$ provides us with a single optimal value of $G_{0}$. In the Landau gauge, this was found to be equal to
\begin{equation}
G_{0}=0.1452\ ,
\end{equation}
corresponding to $\mu=\mu^{\star}=1.004\,m(0)$ in Eq.~\eqref{g0derc}. Like for the gluon, the additive renormalization constant of the optimized ghost propagator is equal to the corresponding fitted value ($G_{0}=0.1464$, Tab.~\ref{tab:fitdataghsmom}) to within less than 1\%.
\vspace{5mm}
\begin{table}[H]
\def\arraystretch{1.2}
\centering
\begin{tabular}{cc|c}
\hline
$m$ (GeV)&$G_{0}$&$Z_{\mc{G}}$\\
\hline
\hline
$0.6557$&$0.1452$&$1.0959$\\
\hline
\end{tabular}
\vspace{5pt}
\caption{Parameter $Z_{\mc{G}}$ obtained by fitting the lattice data of \cite{DOS16} for the Landau-gauge Euclidean ghost propagator in the range $0$-$2$ GeV at fixed $m=0.6557$ GeV and $G_{0}=0.1452$.}\label{tab:fitdataghsmom}
\end{table}
\vspace{5mm}
\begin{figure}[H]
\centering
\includegraphics[width=0.48\textwidth,angle=270]{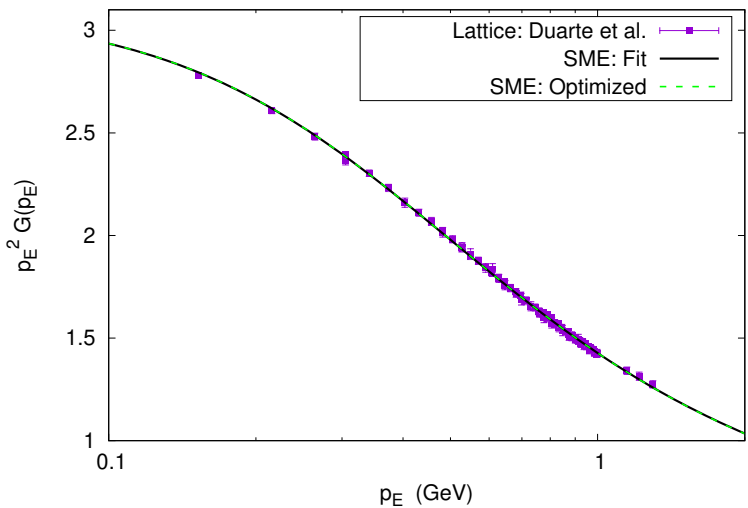}
\vspace{5pt}
\caption{Euclidean ghost dressing function in the Landau gauge ($\xi=0$). Solid black curve: SME with $G_{0}$ fitted from the lattice data (Tab.~\ref{tab:fitdatagh}). Dashed green curve: SME with $G_{0}$ optimized by gauge invariance and minimal sensitivity on the renormalization scale ($G_{0}=0.1452$, Tab.~\ref{tab:fitdataghsmom}). Squares: lattice data from \cite{DOS16}. $m=0.6557$~GeV.}\label{fig:ghproplandopt}
\end{figure}

In Fig.~\ref{fig:ghproplandopt} we show the Landau-gauge Euclidean ghost dressing functions computed in the SME together with the lattice data of \cite{DOS16}. The solid black curve is the fully-fitted dressing function already presented in Sec.~\ref{sec:ghprsme} (parameters in Tab.~\ref{tab:fitdatagh}). The dashed green curve, on the other hand, was obtained by fitting the multiplicative renormalization constant $Z_{\mc{G}}$ alone, as reported in Tab.~\ref{tab:fitdataghsmom}, using $m=0.6557$~GeV and $G_{0}=0.1452$.

As a consequence of the constants $G_{0}$ obtained by the two approaches being very close to each other, the optimized Landau-gauge ghost dressing function quantitatively matches the infrared lattice data just as well as its fully-fitted counterpart. This result validates the optimization procedure laid out for the ghost propagator, based on principles of gauge invariance and of minimal sensitivity. A thorough discussion of the gauge dependence of the optimized ghost propagator can be found in \cite{SIR19b} together with a comparison with the lattice data of \cite{CDMO18a}.\\

In the present section we have described optimization procedures that enable us to compute the gluon and the ghost propagators in the framework of the Screened Massive Expansion from first principles, using the gluon mass parameter $m^{2}(0)$ and two multiplicative renormalization constants -- one for each propagator -- as the only external inputs for the calculations. We have shown that enforcing the gauge-parameter independence of the poles and of the phases of the residues of the gluon propagator is sufficient to reduce the number of free parameters in the expressions, fixing the value of the gluon additive renormalization constant $F_{0}(\xi)$ and of the adimensional ratio $m^{2}(\xi)/m^{2}(0)$ in any gauge. Starting from the optimized $F_{0}$, we were able to determine an optimal value for the ghost additive renormalization constant $G_{0}$ by minimal sensitivity, which in turn provided us with an optimized expression for the ghost propagator. When compared with the lattice data in the Landau gauge, the optimized gluon and ghost propagators were found to accurately reproduce the infrared dynamics of pure Yang-Mills theory.

In order to test the validity of the Screened Massive Expansion over the full dynamical range of momenta, encompassing both the IR and the UV regimes, Renormalization Group methods (Sec.~\ref{sec:rgint}) need to be employed. The Renormalization Group analysis of the Screened Massive Expansion is the subject of the next section. Studying the dependence of the strong coupling constant $\alpha_{s}$ and of the propagators on the renormalization scale will allow us to address the topic of the breakdown of perturbation theory in the low-energy limit of QCD.

\section{Renormalization Group analysis of the Screened Massive Expansion in the Landau gauge}
\renewcommand{\rightmark}{\thesection\ \ \ RG analysis of the Screened Massive Expansion in the Landau gauge}
\label{sec:smerg}

As discussed in Sec.~\ref{sec:rgint}, the presence of large logs in the perturbative series of the Green functions of a quantum field theory spoils the validity of the fixed-scale approximations at energies much different from the renormalization scale. The Screened Massive Expansion is not immune from this issue, as is clearly illustrated by the fact that, in Secs.~\ref{sec:ghprsme}, \ref{sec:glprsme} and \ref{sec:glghopt}, we were able to compare the SME results with the lattice data only up to momenta of $2$~GeV (for the ghost dressing function) or $4$~GeV (for the gluon propagator).

While the renormalization scale is not explicitly present in the expressions for the SME propagators, we note that -- either by fitting the lattice data or by optimization -- our determination of the free parameters of the expansion was essentially made at low energies, effectively giving them the status of infrared parameters. Even the optimization of $F_{0}$ and of $G_{0}$ was more or less implicitly carried out in the low-energy regime: $F_{0}$ was determined by enforcing the gauge-parameter independence of the gluon poles, whose position is found at a scale set by the gluon mass parameter $m^{2}(0)$, whereas $G_{0}$ was fixed by minimizing its value with respect to a renormalization scale $\mu$, yielding $\mu=\mu^{\star}=1.004\,m(0)$ in the Landau gauge. From an analytical perspective, the SME propagators contain logarithms of the form $\ln(-p^{2}/m^{2})$ \footnote{And others such as $\ln(\sqrt{-p^{2}/m^{2}+4}\pm\sqrt{-p^{2}/m^{2}})$ and $\ln(-p^{2}/m^{2}+1)$.}, implying that, as far as the large logs are concerned, the SME is affected by the same problems as ordinary perturbation theory when the momenta are not of the order of the gluon mass parameter $m^{2}$.

In this section, we wish to extend the validity of the SME propagators to scales larger than the gluon mass parameter. Achieving this goal requires us to make use of Renormalization Group methods (Sec.~\ref{sec:rgint}), by which the coupling constant and the propagators are treated as functions of the renormalization scale. In order to perform the RG analysis of the SME, we will need to go back to the expressions \eqref{vls049} and \eqref{vms472} for the ghost and gluon propagators, in which the coupling constant appears explicitly. Moreover, we will need to fix appropriate renormalization conditions for the propagators, instead of leaving them arbitrary as we did up to this point. The analysis will be carried out in the Landau gauge, where the calculations are especially simple.

By integrating the RG equations, we will obtain expressions that -- at least in principle~-- remain valid over a wide range of momenta. In particular, we will see that the strong running coupling $\alpha_{s}(\mu^{2})$ computed in the Screened Massive Expansion does not develop Landau poles, provided that the value of the coupling $\alpha_{s}(\mu^{2}_{0})$ at the initial renormalization scale $\mu_{0}$ is sufficiently small. This is a crucial feature of the SME: on the one hand, it confirms that the massive shift that defines the expansion yields a self-consistent perturbative series; on the other hand, it proves that the infrared Landau pole that affects the running coupling of ordinary pQCD is just an artifact of the choice of a massless expansion point.

The one-loop RG-improved ghost and gluon propagators will be shown to be in good agreement with the lattice data from momenta of the order of 10~GeV down to $p\approx m$. Below this threshold, the coupling becomes quite too large for the one-loop approximation to provide sufficiently accurate results: in the deep infrared regime, the fixed-scale results of Sec.~\ref{sec:glghopt} still represent our best estimate for the behavior of the propagators. The overlap between the RG-improved and the fixed-scale propagators at momenta $p\approx m$ will be exploited in Sec.~\ref{sec:smergopt} to compute an optimal value of the coupling constant $\alpha_{s}(\mu^{2}_{0})$ at the initial renormalization scale $\mu_{0}$ as a function of the gluon mass parameter $m^{2}$. This will again leave us with expressions whose only free parameter is $m^{2}$, playing the same role as the QCD scale $\Lambda_{\text{QCD}}$ in ordinary perturbation theory.

\subsection{MOM-Taylor-scheme renormalization of the Screened Massive Expansion}
\label{sec:smemomtay}

In Sec.~\ref{sec:smedef} we saw that, in the Landau gauge ($\xi=0$), the one-loop Euclidean transverse gluon and ghost propagators $\Delta(p^{2})$ and $\mc{G}(p^{2})$ computed in the Screened Massive Expansion can be expressed as\footnote{For conciseness of notation, in the rest of this chapter we will drop the subscript $E$ from the quantities defined in Euclidean space. Thus the $p^{2}$ in Eq.~\eqref{shy589} is actually a Euclidean momentum squared.}
\begin{equation}\label{shy589}
\Delta(p^{2})=\frac{1}{p^{2}\left[1+\alpha\left(F(s)+f_{0}\right)\right]}\ ,\qquad\qquad\mc{G}(p^{2})=\frac{1}{p^{2}\left[1+\alpha\left(G(s)+g_{0}\right)\right]}\ ,
\end{equation}
where $s=p^{2}/m^{2}$, $\alpha$ is a rescaled coupling constant,
\begin{equation}
\alpha=\frac{3N\alpha_{s}}{4\pi}=\frac{3Ng^{2}}{16\pi^{2}}\ ,
\end{equation}
the functions $G(s)$ and $F(s)$ were defined in Eqs.~\eqref{tya450} and \eqref{ffunct}, and $f_{0}$ and $g_{0}$ are additive renormalization constants. The latter are fixed as soon as the renormalization conditions for the propagators are chosen. In \cite{CS20}, the RG analysis of the Screened Massive Expansion was carried out in two renormalization schemes: the MOM scheme, described in Secs.~\ref{sec:regren} and \ref{sec:rgint}, and the SMOM scheme, defined in Sec.~\ref{sec:glghopt}. During the rest of this chapter, we will focus only on the former.

In the MOM scheme, the gluon and ghost propagators are renormalized by fixing their value at a given renormalization scale $\mu$ so that
\begin{equation}\label{gjn379}
\Delta(\mu^{2})=\mc{G}(\mu^{2})=\frac{1}{\mu^{2}}\ .
\end{equation}
By looking at Eq.~\eqref{shy589}, we see that, to one loop, these renormalization conditions are equivalent to choosing
\begin{equation}\label{yht346}
f_{0}=-F(\mu^{2}/m^{2})\ ,\qquad\qquad g_{0}=-G(\mu^{2}/m^{2})\ .
\end{equation}
Going back to the explicit one-loop expressions for the ghost and gluon field renormalization factors $Z_{c}$ and $Z_{A}$ in terms of the constants $g_{0}$ and $f_{0}$ -- Eqs.~\eqref{gth620} and \eqref{gth621}~--,
\begin{align}
Z_{c}&=1+\frac{\alpha}{4}\,\left[\left(1-\frac{\xi}{3}\right)\left(\frac{2}{\epsilon}-\ln\frac{m^{2}}{\overline{\mu}^{2}}\right)+\frac{8}{3}\right]+\alpha g_{0}\ , \\
Z_{A}&=1+\frac{\alpha}{3}\, \left(\frac{13}{6}-\frac{\xi}{2}\right)\left(\frac{2}{\epsilon}-\ln\frac{m^{2}}{\overline{\mu}^{2}}\right)+\alpha(f_{0}-\mc{C})\ ,
\end{align}
where $\mubar=\sqrt{4\pi}\mu_{\text{d.r.}} e^{-\gamma_{E}/2}$, $\mu_{\text{d.r.}}$ being the scale introduced by dimensional regularization, Eq.~\eqref{yht346} yields the following MOM gluon and ghost field renormalization constants:
\begin{align}
Z_{A}&=1+\frac{\alpha}{3}\, \left(\frac{13}{6}-\frac{\xi}{2}\right)\left(\frac{2}{\epsilon}-\ln\frac{m^{2}}{\overline{\mu}^{2}}\right)-\alpha(F(\mu^{2}/m^{2})+\mc{C})\ ,\label{grs625}\\
Z_{c}&=1+\frac{\alpha}{4}\,\left[\left(1-\frac{\xi}{3}\right)\left(\frac{2}{\epsilon}-\ln\frac{m^{2}}{\overline{\mu}^{2}}\right)+\frac{8}{3}\right]-\alpha\,G(\mu^{2}/m^{2})\ .\label{grs626}
\end{align}

The gluon and ghost anomalous dimensions $\gamma_{A}$ and $\gamma_{c}$, defined as (see Sec.~\ref{sec:regren})
\begin{equation}
\gamma_{A}=\frac{\mu}{Z_{A}}\frac{dZ_{A}}{d\mu}\ ,\qquad\qquad\gamma_{c}=\frac{\mu}{Z_{c}}\frac{dZ_{c}}{d\mu}\ ,
\end{equation}
can be explicitly computed in the MOM scheme thanks to Eqs.~\eqref{grs625} and \eqref{grs626}. To lowest order in the coupling constant, they read
\begin{equation}\label{vjz582}
\gamma_{A}=-2\alpha\,\frac{\mu^{2}}{m^{2}}\,F^{\prime}(\mu^{2}/m^{2})\ ,\qquad\qquad \gamma_{c}=-2\alpha\,\frac{\mu^{2}}{m^{2}}\,G^{\prime}(\mu^{2}/m^{2})\ ,
\end{equation}
where
\begin{equation}
F^{\prime}(s)=\frac{dF}{ds}(s)\ ,\qquad\qquad G^{\prime}(s)=\frac{dG}{ds}(s)\ .
\end{equation}
Observe that in Eqs.~\eqref{grs625} and \eqref{grs626}, the terms which depend on the renormalization scale $\mu$ only implicitly through the coupling constant $\alpha$ do not contribute to the MOM anomalous dimensions to $O(\alpha_{s})$ since the derivative of $\alpha$ with respect to $\mu$, given by the beta function,
\begin{equation}\label{tsr453}
\mu\frac{d\alpha}{d\mu}=\mu\frac{d}{d\mu}\left(\frac{3Ng^{2}}{16\pi^{2}}\right)=2\,\frac{\alpha}{g}\,\beta_{g}\qquad\qquad\left(\mu\frac{dg}{d\mu}=\beta_{g}\right)\ ,
\end{equation}
is higher-order: being $\beta_{g}$ of order $g^{3}$, $\mu\, d\alpha/d\mu$ is of order $\alpha^{2}_{s}$. Similarly, if the gluon mass parameter were taken to depend on the renormalization scale, then $\mu\, dm^{2}/d\mu$ would be at least of order $\alpha_{s}$, so that the product $\alpha\,\mu\,dm^{2}/d\mu$ would be at least $O(\alpha_{s}^{2})$ and would not contribute to the one-loop anomalous dimensions.

In what follows, we will assume that the gluon mass parameter $m^{2}$ is independent from the renormalization scale. This assumption is motivated by an in-depth order-by-order analysis of the RG equations for the Screened Massive Expansion in the MOM scheme, which shows that, to one-loop, the derivative $\mu\,dm^{2}/d\mu$ drops out of the equations just like it does from the anomalous dimensions, implying that the $\mu$-dependence of the gluon mass parameter is arbitrary to $O(\alpha_{s})$ \cite{CS20}. The simplest solutions to the MOM RG equations thus have $m^{2}$ being independent from $\mu$.\\

With $Z_{A}$ and $Z_{c}$ defined in the MOM scheme, the renormalization of the SME is completed by specifying the renormalization conditions for the coupling constant. In the Landau gauge, the latter is conveniently defined in the so-called Taylor scheme, which we have already encountered in Chpt.~\ref{chpt:npmet}. Recalling that the renormalized coupling $g$ is defined as
\begin{equation}\label{eck544}
g=Z_{g}^{-1}\,g_{B},
\end{equation}
where $g_{B}$ is the bare coupling constant and $Z_{g}$ is the coupling renormalization factor, a theorem by Taylor \cite{Tay71} shows that, in the Landau gauge, the \textit{diverging} terms\footnote{That is, the terms which would make up the renormalization factors in the MS scheme.} of the renormalization factors $Z_{A}$, $Z_{c}$, and $Z_{g}$ satisfy the relation
\begin{equation}\label{hfi529}
\left(Z_{g}\,Z_{A}^{1/2}\,Z_{c}\right)_{\text{div.}}=1\ .
\end{equation}
The choice
\begin{equation}\label{hfi530}
Z_{g}=Z_{A}^{-1/2}Z_{c}^{-1}\ ,
\end{equation}
which extends the Taylor relation \eqref{hfi529} to the full renormalization factors, is thus consistent with the divergences of the theory, and its adoption yields a renormalized coupling constant known as the Taylor coupling.

In the Taylor scheme, the beta function $\beta_{g}$ takes on an especially simple form. Since by Eq.~\eqref{eck544}
\begin{equation}
\beta_{g}=\mu\frac{dg}{d\mu}=-\frac{\mu}{Z_{g}}\frac{dZ_{g}}{d\mu}\ ,
\end{equation}
because of the Taylor condition -- Eq.~\eqref{hfi530} -- we find that
\begin{equation}
\beta_{g}=\frac{g}{2}\,(\gamma_{A}+2\gamma_{c})\ .
\end{equation}
Thus the knowledge of the gluon and ghost anomalous dimensions is sufficient to compute the beta function associated to the Taylor coupling.

In the context of the Screened Massive Expansion, we can define a beta function $\beta_{\alpha}$ associated to the rescaled coupling $\alpha$ \footnote{Note the $\mu^{2}$ in $\beta_{\alpha}$.}:
\begin{equation}
\beta_{\alpha}=\frac{d\alpha}{d\ln\mu^{2}}\ .
\end{equation}
Using Eq.~\eqref{tsr453}, it is easy to verify that
\begin{equation}
\beta_{\alpha}=\frac{\alpha}{g}\,\beta_{g}\ .
\end{equation}
In particular, in the MOM-Taylor scheme, we find that
\begin{equation}\label{gte127}
\beta_{\alpha}=\frac{\alpha}{2}(\gamma_{A}+2\gamma_{c})=-\alpha^{2}\,\frac{\mu^{2}}{m^{2}}\,H^{\prime}(\mu^{2}/m^{2})\ ,
\end{equation}
where the function $H(s)$ is defined as
\begin{equation}
H(s)=F(s)+2G(s)
\end{equation}
and $H^{\prime}(s)=dH(s)/ds$.\\

As the next step in our RG analysis of the SME, in the following section we will analytically solve the RG equation for the running coupling constant and numerically integrate the RG-improved propagators defined in the MOM-Taylor scheme.

\subsection{The SME strong running coupling and RG-improved propagators}
\label{sec:rasrgimp}

To one loop, the Renormalization Group equation for the SME coupling constant $\alpha$ reads
\begin{equation}\label{rtg422}
\mu^{2}\frac{d\alpha}{d\mu^{2}}=\beta_{\alpha}=-\alpha^{2}\,\frac{\mu^{2}}{m^{2}}\,H^{\prime}(\mu^{2}/m^{2})\ .
\end{equation}
Since $m^{2}$ does not depend on $\mu$, the latter can be rewritten in terms of the adimensional variable $s$ as
\begin{equation}\label{rtg423}
\frac{d\alpha}{d\ln s}=-\alpha^{2}\,H^{\prime}(s)\ .
\end{equation}
The solution of Eq.~\eqref{rtg423} is easily found to be
\begin{equation}\label{hgt659}
\alpha(s)=\frac{\alpha(s_{0})}{1+\alpha(s_{0})\,[H(s)-H(s_{0})]}\ ,
\end{equation}
where $s_{0}=\mu_{0}^{2}/m^{2}$ is the adimensionalized initial renormalization scale and $\alpha(s_{0})$ is the rescaled coupling constant defined at the same scale.

The one-loop MOM-Taylor strong running coupling $\alpha_{s}(\mu^{2})$ is obtained from the last equation using $\alpha=3N\alpha_{s}/4\pi$. Explicitly,
\begin{align}\label{bgn432}
\alpha_{s}(\mu^{2})=\frac{\alpha_{s}(\mu_{0}^{2})}{1+\dfrac{3N\alpha_{s}(\mu_{0}^{2})}{4\pi}\,[H(\mu^{2}/m^{2})-H(\mu_{0}^{2}/m^{2})]}\ .
\end{align}
In the high-energy limit $\mu^{2},\mu_{0}^{2}\gg m^{2}$ ($s\gg1$), by Eqs.~\eqref{fvj242} and \eqref{fvj243},
\begin{equation}
F(s)\to \frac{13}{18}\,\ln s+\frac{17}{18}\ ,\qquad\qquad G(s)\to\frac{1}{4}\,\ln s+\frac{1}{3}\ ,
\end{equation}
the function $H(s)$ and the difference $H(\mu^{2}/m^{2})-H(\mu^{2}_{0}/m^{2})$ that appears in the denominator of $\alpha_{s}(\mu^{2})$ approach
\begin{equation}
H(s)\to\frac{11}{9}\,\ln s+\frac{29}{18}\ ,\qquad\quad H(\mu^{2}/m^{2})-H(\mu^{2}_{0}/m^{2})\to\frac{11}{9}\,\ln\left(\mu^{2}/\mu_{0}^{2}\right)\ .
\end{equation}
In particular, in this limit the gluon mass parameter disappears from $\alpha_{s}(\mu^{2})$, while the strong running coupling itself reduces to
\begin{align}
\alpha_{s}(\mu^{2})\to\frac{\alpha_{s}(\mu_{0}^{2})}{1+\dfrac{11N}{3}\dfrac{\alpha_{s}(\mu_{0}^{2})}{4\pi}\,\ln\left(\mu^{2}/\mu_{0}^{2}\right)}\ .
\end{align}
This is the result found in ordinary perturbation theory, Eq.~\eqref{tyx128}, with $n_{f}=0$ in the beta function coefficient $\beta_{0}$. In the low-energy limit, on the other hand, again by Eqs.~\eqref{fvj242} and \eqref{fvj243},
\begin{equation}
F(s)\to \frac{5}{8s}\ ,\quad G(s)\to \frac{5}{24}\qquad\Longrightarrow\qquad H(s)\to\frac{5}{8s}\ .
\end{equation}
Therefore, as $\mu^{2}\to 0$,
\begin{align}
\alpha_{s}(\mu^{2})\to\frac{32\pi}{15N}\frac{\mu^{2}}{m^{2}}\to 0\ :
\end{align}
the running coupling vanishes like $\mu^{2}$, as predicted by the lattice calculations -- see Sec.~\ref{sec:lqcdrevres}. This is not accidental: the lattice coupling, as well as ours, is defined in the Taylor scheme, starting from the behavior of the MOM gluon and ghost propagators. It will become clear by the end of this section that the vanishing of the MOM-Taylor coupling at zero momentum is the consequence of the gluon propagator becoming massive and of the gluon dressing function remaining finite at $p=0$. To one loop, the speed at which $\alpha_{s}(\mu^{2})$ approaches zero does not depend on the initial conditions of the RG flow (i.e. on $\alpha_{s}(\mu_{0}^{2})$ and $\mu_{0}^{2}$), but only on the value of the gluon mass parameter $m^{2}$.
\vspace{5mm}
\begin{figure}[H]
\centering
\includegraphics[width=0.48\textwidth,angle=270]{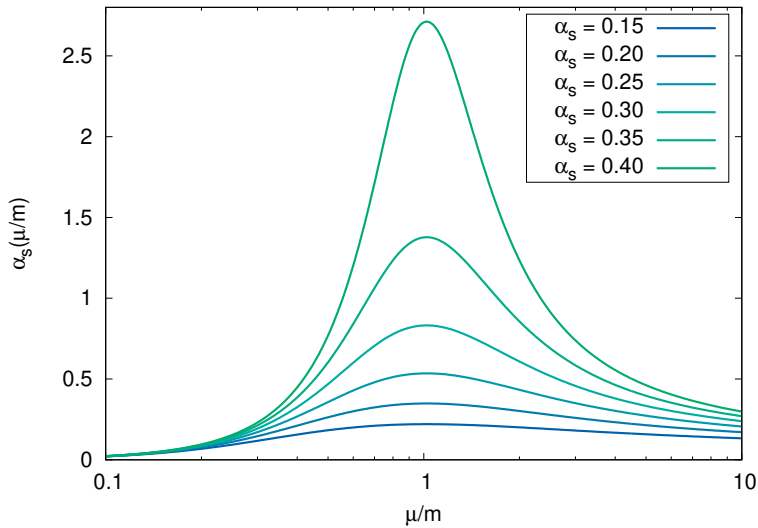}
\vspace{5pt}
\caption{One-loop MOM-Taylor SME running coupling $\alpha_{s}(\mu^{2})$ as a function of the adimensional scale $\mu/m$ for different initial values of the coupling $\alpha_{s}(\mu^{2}_{0})$ at the scale $\mu_{0}/m$ = 6.098. With $m=0.656$~GeV, this corresponds to $\mu_{0}=4$~GeV.}\label{fig:alphasvar}
\end{figure}
\vspace{5mm}
The one-loop MOM-Taylor SME running coupling $\alpha_{s}(\mu^{2})$ is shown in Fig.~\ref{fig:alphasvar} for $N=3$ as a function of $\sqrt{s}=\mu/m$ for different initial values of the coupling constant $\alpha_{s}(\mu^{2}_{0})$. The adimensional initial renormalization scale $\mu_{0}/m$ was chosen equal to $6.098$ so that, with our previous determination of the gluon mass parameter ($m=0.656$~GeV), $\mu_{0}=4$~GeV.

As we can see, for sufficiently small initial values of the coupling constant, the strong running coupling $\alpha_{s}(\mu^{2})$ computed in the SME has no Landau poles. Instead, it first increases in value as $\mu$ decreases, then hits a maximum at a fixed scale $\mu_{\star}=1.022\, m$, and finally it decreases to zero as $\mu\to 0$. That the position of the maximum is independent\cleannlnp from the value of the initial coupling $\alpha_{s}(\mu_{0}^{2})$ is a consequence of the one-loop beta function having a zero at the renormalization scale $\mu_{\star}$ such that
\begin{equation}
H^{\prime}(\mu^{2}_{\star}/m^{2})=0\ .
\end{equation}
The solution of this equation, $\mu_{\star}=1.022\, m$, can be obtained numerically and is clearly independent from $\alpha_{s}(\mu_{0}^{2})$.

For $\alpha_{s}(\mu_{0})\geq 0.469$, on the other hand, the running coupling develops an infrared Landau pole. To see this, suppose that $\alpha_{s}(\mu^{2})$ becomes infinite at the scale $\mu_{\text{pole}}^{2}$. This can only happen if
\begin{equation}
1+\frac{3N\alpha_{s}(\mu_{0}^{2})}{4\pi}\,[H(\mu^{2}_{\text{pole}}/m^{2})-H(\mu_{0}^{2}/m^{2})]=0\ ,
\end{equation}
implying that the initial value of the coupling is related to $\mu_{\text{pole}}^{2}$ via the equation
\begin{equation}
\alpha_{s}(\mu_{0}^{2})=\frac{4\pi}{3N[H(\mu_{0}^{2}/m^{2})-H(\mu^{2}_{\text{pole}}/m^{2})]}\ .
\end{equation}
Since $\mu_{\star}=1.022\, m$ is a minimum for $H(\mu^{2}/m^{2})$, we have $H(\mu^{2}/m^{2})\geq H(\mu^{2}_{\star}/m^{2})$ for any value of the renormalization scale $\mu$. It follows that
\begin{equation}
\alpha_{s}(\mu_{0}^{2})\geq\frac{4\pi}{3N[H(\mu_{0}^{2}/m^{2})-H(\mu^{2}_{\star}/m^{2})]}=\alpha_{s}^{(\text{thr.})}(\mu_{0}^{2})=0.469
\end{equation}
at $N=3$ and $\mu_{0}=6.098\,m$. We remark that the threshold value $\alpha_{s}^{(\text{thr.})}(\mu_{0})$ defined by the last equation depends on the initial renormalization scale $\mu_{0}$, as well as on the number of colors $N$.\\

Recall from Sec.~\ref{sec:rgint} that the RG-improved gluon and ghost propagators renormalized at the scale $\mu_{0}$ -- $\Delta(p^{2};\mu^{2}_{0})$ and $\mc{G}(p^{2};\mu^{2}_{0})$, respectively --, can be expressed in terms of their anomalous dimensions $\gamma_{A}$ and $\gamma_{c}$ as\footnote{The derivation of Eq.~\eqref{erb295} is identical in all respects to the one carried out in Sec.~\ref{sec:rgint} to obtain Eq.~\eqref{erb294}, so we will not repeat it in this section.}
\begin{align}
\Delta\left(p^{2};\mu^{2}_{0}\right)=e^{\int^{p}_{\mu_{0}}\frac{d\mu}{\mu}\,\gamma_{A}(\mu)}\,\Delta\left(p^{2};p^{2}\right)\ ,\label{erb294}\\
\mc{G}\left(p^{2};\mu^{2}_{0}\right)=e^{\int^{p}_{\mu_{0}}\frac{d\mu}{\mu}\,\gamma_{c}(\mu)}\,\mc{G}\left(p^{2};p^{2}\right)\ .\label{erb295}
\end{align}
In the MOM scheme, the values $\Delta\left(p^{2};p^{2}\right)$ and $\mc{G}\left(p^{2};p^{2}\right)$ of the propagators at a momentum equal to their renormalization scale are given by the renormalization conditions themselves, Eq.~\eqref{gjn379}, which we can rewrite as
\begin{equation}
\Delta\left(p^{2};p^{2}\right)=\mc{G}\left(p^{2};p^{2}\right)=\frac{1}{p^{2}}\ .
\end{equation}
By making use of the explicit expressions for the one-loop MOM gamma functions, Eq.~\eqref{vjz582}, we then find that the RG-improved one-loop SME propagators read
\begin{align}
\Delta\left(p^{2};\mu^{2}_{0}\right)=\frac{1}{p^{2}}\exp\left(-\int^{p^{2}/m^{2}}_{\mu_{0}^{2}/m^{2}}ds\ \alpha(s)\,F^{\prime}(s)\right)\ ,\label{rez934}\\
\mc{G}\left(p^{2};\mu^{2}_{0}\right)=\frac{1}{p^{2}}\exp\left(-\int^{p^{2}/m^{2}}_{\mu_{0}^{2}/m^{2}}ds\ \alpha(s)\,G^{\prime}(s)\right)\ .\label{rez935}
\end{align}
\newpage
\vspace*{5mm}
\begin{figure}[H]
\centering
\includegraphics[width=0.48\textwidth,angle=270]{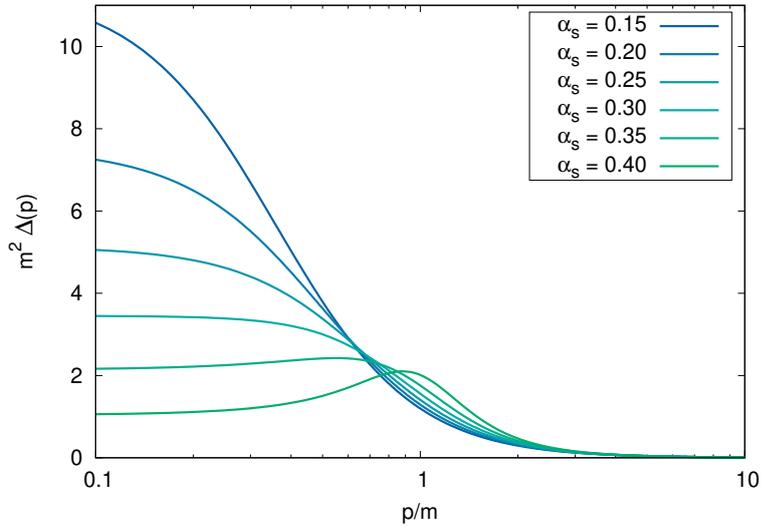}
\vspace{5pt}
\caption{One-loop Landau-gauge RG-improved Euclidean transverse gluon propagator computed in the MOM-Taylor scheme as a function of the adimensional momentum $p/m$ for different initial values of the coupling $\alpha_{s}(\mu^{2}_{0})$ at the scale $\mu_{0}/m$ = 6.098. Adimensionalized by $m^{2}$.}\label{fig:gluproprgvar}
\end{figure}
\vspace{5mm}
\begin{figure}[H]
\centering
\includegraphics[width=0.48\textwidth,angle=270]{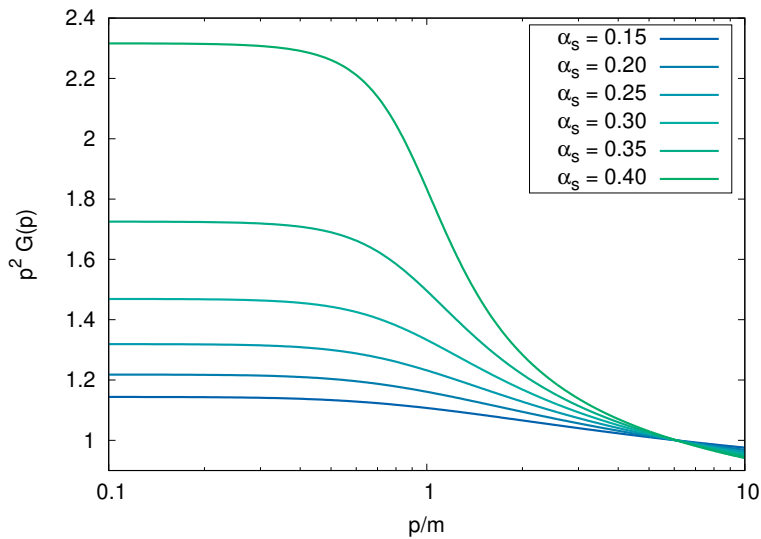}
\vspace{5pt}
\caption{One-loop Landau-gauge RG-improved Euclidean ghost dressing function computed in the MOM-Taylor scheme as a function of the adimensional momentum $p/m$ for different initial values of the coupling $\alpha_{s}(\mu^{2}_{0})$ at the scale $\mu_{0}/m$ = 6.098.}\label{fig:ghoproprgvar}
\end{figure}
\vspace{5mm}
The integrals in Eqs.~\eqref{rez934} and \eqref{rez935} cannot be evaluated analytically. In order to compute the RG-improved propagators, one thus has to resort to numerical integration. Nonetheless, as shown in \cite{CS20}, the asymptotic limits of $\Delta\left(p^{2};\mu^{2}_{0}\right)$ and $\mc{G}\left(p^{2};\mu^{2}_{0}\right)$ can still be put in closed form by making use of the $s\to 0$ and $s\to \infty$ expressions for $F(s)$, $G(s)$ and $\alpha(s)$. One then finds that in the UV, provided that $\mu_{0}^{2}\gg m^{2}$ as well as $p^{2}\gg m^{2}$,
\begin{equation}\label{dkz583}
\Delta\left(p^{2};\mu^{2}_{0}\right)\to\frac{1}{p^{2}}\left[\frac{\alpha_{s}(p^{2})}{\alpha_{s}(\mu_{0}^{2})}\right]^{\frac{13}{22}}\ ,\qquad\qquad \mc{G}\left(p^{2};\mu^{2}_{0}\right)\to\frac{1}{p^{2}}\left[\frac{\alpha_{s}(p^{2})}{\alpha_{s}(\mu_{0}^{2})}\right]^{\frac{9}{44}}\ ,
\end{equation}
which is just the ordinary pQCD result for $n_{f}=0$, while in the deep IR
\begin{equation}\label{bne495}
\Delta\left(p^{2};\mu^{2}_{0}\right)\to\frac{\kappa}{m^{2}}\ ,\qquad\qquad \mc{G}\left(p^{2};\mu^{2}_{0}\right)\to\frac{\kappa^{\prime}}{p^{2}}\ ,
\end{equation}
where $\kappa$ and $\kappa^{\prime}$ are constants. By Eq.~\eqref{bne495} we see that, at vanishing momenta, the RG-improved SME gluon and ghost propagators are massive and massless, respectively, just like their fixed-scale counterparts.

In Figs.~\ref{fig:gluproprgvar} and \ref{fig:ghoproprgvar} we show the RG-improved gluon propagator and ghost dressing function computed for different values of the initial coupling constant $\alpha_{s}(\mu_{0}^{2})<\alpha_{s}^{(\text{thr.})}(\mu_{0}^{2})$, again at the initial scale $\mu_{0}=6.098\,m$. Observe that, in the figures, both $\Delta(p^{2};\mu_{0}^{2})$ and the momenta are adimensionalized by appropriate factors of the gluon mass parameter $m$. The propagators clearly display the asymptotic behavior we just described. Notably, the gluon propagators attains a maximum for any value of the initial coupling\footnote{Though this is less clear from the figure for lower values of $\alpha_{s}(\mu_{0}^{2})$, the existence of a maximum for every initial value of the coupling can be proved analytically, see \cite{CS20}.}, whereas the ghost dressing function quickly saturates to a constant at momenta of the order of the gluon mass parameter.\\

For future reference, we make an observation on the relation between the Taylor coupling and the RG-improved propagators. Going back to Eqs.~\eqref{erb294} and \eqref{erb295}, it is interesting to compute the product between the gluon propagator and the square of the ghost propagator. A simple calculation shows that
\begin{align}
\Delta\left(p^{2};\mu^{2}_{0}\right)\,[\mc{G}\left(p^{2};\mu^{2}_{0}\right)]^{2}&=e^{\int^{p}_{\mu_{0}}\frac{d\mu}{\mu}\,[\gamma_{A}(\mu)+2\gamma_{c}(\mu)]}\ \Delta\left(p^{2};p^{2}\right)\,[\mc{G}\left(p^{2};p^{2}\right)]^{2}=\\
\notag&=e^{2\int^{p}_{\mu_{0}}\frac{d\mu}{\mu}\,\frac{\beta_{\alpha}}{\alpha}}\ \Delta\left(p^{2};p^{2}\right)\,[\mc{G}\left(p^{2};p^{2}\right)]^{2}=\\
\notag&=e^{\int^{p}_{\mu_{0}}\frac{d\mu}{\alpha}\,\frac{d\alpha}{d\mu}}\ \Delta\left(p^{2};p^{2}\right)\,[\mc{G}\left(p^{2};p^{2}\right)]^{2}=\\
\notag&=\frac{\alpha(p^{2}/m^{2})}{\alpha(\mu_{0}^{2}/m^{2})}\ \Delta\left(p^{2};p^{2}\right)\,[\mc{G}\left(p^{2};p^{2}\right)]^{2}\ ,
\end{align}
where we have used the Taylor condition in the form $\beta_{\alpha}=\frac{\alpha}{2}(\gamma_{A}+2\gamma_{c})$. It follows that the Taylor coupling can be expressed in terms of the propagators as
\begin{equation}\label{dvf341}
\alpha_{s}(\mu^{2})=\alpha_{s}(\mu_{0}^{2})\ \left[\frac{\Delta\left(\mu^{2};\mu^{2}_{0}\right)}{\Delta\left(\mu^{2};\mu^{2}\right)}\right]\left[\frac{\mc{G}\left(\mu^{2};\mu^{2}_{0}\right)}{\mc{G}\left(\mu^{2};\mu^{2}\right)}\right]^{2}\ .
\end{equation}
In particular, when $\Delta(p^{2})$ and $\mc{G}(p^{2})$ are renormalized in the MOM scheme, the latter reads
\begin{equation}\label{dvf342}
\alpha_{s}(\mu^{2})=\alpha_{s}(\mu_{0}^{2})\ \left[\mu^{2}\,\Delta\left(\mu^{2};\mu^{2}_{0}\right)\right]\left[\mu^{2}\,\mc{G}\left(\mu^{2};\mu^{2}_{0}\right)\right]^{2}\ .
\end{equation}
An explicit example of this relation is given by the UV limit of the SME propagators (equivalently, by the results of ordinary perturbation theory) reported in Eq.~\eqref{dkz583}. From Eq.~\eqref{dvf342}, it is clear that, in the MOM scheme, the finiteness of $\Delta(p^{2})$ and $p^{2}\mc{G}(p^{2})$ in the limit $p^{2}\to 0$ always implies that the Taylor coupling vanishes at zero momentum.\\

In this section we have explored the analytical properties of the MOM-Taylor scheme SME strong running coupling and presented results regarding the corresponding RG-improved gluon and ghost propagators. The absence of Landau poles from the SME running coupling $\alpha_{s}(\mu^{2})$ for sufficiently small initial values of the coupling constant is an essential check on the validity of the method. Indeed, if the running coupling computed in the SME had IR Landau poles regardless of the initial conditions of the RG flow (as it happens in ordinary pQCD), then the results obtained in Secs.~\ref{sec:smedef} and \ref{sec:smeopt} would bear no connection to the well-established UV behavior of the theory, having been obtained at a scale where the theory would essentially be undefined. On the contrary, the finiteness of $\alpha_{s}(\mu^{2})$ confirms both that the Screened Massive Expansion is self-consistent as a computational method, and that the dynamics of QCD can in principle be studied over an unlimited range of momenta by making use of the Renormalization Group.

We observe that, in the Screened Massive Expansion, the finiteness of the running coupling is made possible by the presence of a non-perturbative mass scale -- namely, the gluon mass parameter -- in the equations. To one loop, it is the vanishing of the adimensional function $H^{\prime}(s)$ in the MOM-scheme beta function -- see Eq.~\eqref{rtg422} -- that allows $\alpha_{s}(\mu^{2})$ to attain a maximum at a fixed renormalization scale $\mu=\mu_{\star}$, instead of diverging like in ordinary pQCD. In the absence of the gluon mass parameter, or in scale-independent renormalization schemes, $s\,H^{\prime}(s)$ would be replaced by a constant in $\beta_{\alpha}$, forcing the latter to be negative at every renormalization scale, as is the case in pQCD. Instead, around $\mu=\mu_{\star}=1.022\,m$, the running of the coupling slows down, stops and then changes sign as the renormalization scale decreases.

Of course, the fact that $\alpha_{s}(\mu^{2})$ diverges for large values of $\alpha_{s}(\mu_{0}^{2})$ calls for further investigations on the behavior of the RG-improved propagators. In particular, we need to make sure that the exact propagators, as given for instance by the lattice data, can be reproduced by making use of values of $\alpha_{s}(\mu_{0}^{2})$ smaller than the threshold at which the Landau pole appears, $\alpha_{s}^{(\text{thr.})}(\mu_{0}^{2})=0.469$ at $\mu_{0}/m=6.098$. In addition to this, the RG-improved SME propagators depend on two free parameters, $\alpha_{s}(\mu_{0}^{2})$ and $m^{2}$, at variance with their ordinary pQCD analogues -- which clearly only depend on the initial value of the coupling constant. Thus we find ourself in the same situation in which we were at the end of Sec.~\ref{sec:smedef}, with the propagators needing optimization in order for predictions to be made from first principles. The determination of an optimal value of $\alpha_{s}(\mu_{0}^{2})$ is the subject of the next section.

\subsection{Intermediate-energy matching with the fixed-scale optimized results and comparison with the lattice data}
\label{sec:smergopt}

While in principle the validity of the RG-improved results presented in the last section extends to arbitrary energy scales, in practice the one-loop approximation by which they were obtained is accurate only provided that the value of the running coupling is sufficiently small. By Fig.~\ref{fig:alphasvar} we see that, depending on the value of the initial coupling, the maximum $\alpha_{s}(\mu=1.022\,m)$ can become quite large for the perturbative standards. Therefore, we may expect the one-loop RG-improved propagators to deviate from the exact results at the corresponding momenta $p\approx m$, when integrated starting from the UV region.

The accuracy of the one-loop approximation can be tested by making use of the optimized fixed-scale (OFS) expressions derived in Sec.~\ref{sec:smeopt}. Since these are valid up to scales of the order of the gluon mass parameter, we expect the RG-improved and OFS results to overlap at least in the intermediate-energy region where $p\approx m$. Such an overlap, of course, will occur only for those values of the initial coupling $\alpha_{s}(\mu_{0}^{2})$ which yield a good approximation of the exact propagators. Matching the high-energy RG-improved results to their low-energy fixed-scale counterpart thus provides us with a way to fix an optimal value of $\alpha_{s}(\mu_{0}^{2})$ from first principles.

In \cite{CS20}, the intermediate-energy matching of the RG-improved and optimized fixed-scale results was carried out by comparing the respective running couplings as functions of the renormalization scale. In the context of the OFS approach, a Taylor coupling $\alpha_{s}^{(\text{OFS})}(\mu^{2})$ can be defined by making use of Eqs.~\eqref{dvf341} and \eqref{dvf342}\footnote{We remark that Eq.~\eqref{dvf341} can be derived from general principles of renormalizability which do not require the propagators to be RG improved, see \cite{CS20}. For this reason, it can also be employed in the context of non-improved approaches, like the fixed-scale approximation of Secs.~\ref{sec:smedef} and \ref{sec:smeopt}.}. Assuming that the fixed-scale propagators are renormalized in the MOM scheme, Eq.~\eqref{dvf342} can be rewritten as
\begin{equation}\label{ytt438}
\alpha_{s}^{(\text{OFS})}(\mu^{2})=\kappa\, \left[F(\mu^{2}/m^{2})+F_{0}\right]^{-1}\left[G(\mu^{2}/m^{2})+G_{0}\right]^{-2}\ ,
\end{equation}
where $F_{0}$ and $G_{0}$ take on the values obtained by optimization, whereas $\kappa$ is a constant that absorbs both $\alpha_{s}^{(\text{OFS})}(\mu_{0}^{2})$ and the multiplicative renormalization constants $Z_{\Delta}$ and $Z_{\mc{G}}$ contained in the OFS gluon and ghost propagators. Since $\alpha_{s}^{(\text{OFS})}(\mu_{0}^{2})$ is undefined in the fixed-scale approach\footnote{Recall that the fixed-scale expressions of Secs.~\ref{sec:smedef} and \ref{sec:smeopt} do not explicitly contain the coupling constant.}, in the above expression $\kappa$ is a free parameter.

The normalization of $\alpha_{s}^{(\text{OFS})}(\mu^{2})$ -- i.e., the value of $\kappa$ -- can be fixed by requiring that the former be equal to its RG counterpart $\alpha_{s}(\mu^{2})$ at an intermediate-energy renormalization scale $\mu_{1}$, yielding
\begin{equation}\label{dfj247}
\alpha_{s}^{(\text{OFS})}(\mu^{2})=\alpha_{s}(\mu_{1}^{2})\,\left[\frac{F(\mu_{1}^{2}/m^{2})+F_{0}}{F(\mu^{2}/m^{2})+F_{0}}\right]\left[\frac{G(\mu_{1}^{2}/m^{2})+G_{0}}{G(\mu^{2}/m^{2})+G_{0}}\right]^{2}\ .
\end{equation}
In \cite{CS20}, the scale $\mu_{1}$ was chosen equal to $1.372\,m$, corresponding to $0.9$~GeV for $m=0.656$~GeV. We note that, if the running couplings $\alpha_{s}^{(\text{OFS})}(\mu^{2})$ and $\alpha_{s}(\mu^{2})$ do match at intermediate energies, then the specific choice of the scale $\mu_{1}$ is irrelevant to the normalization of $\alpha_{s}^{(\text{OFS})}(\mu^{2})$, as long as $\mu_{1}\approx m$.

\begin{figure}[H]
\centering
\includegraphics[width=0.48\textwidth,angle=270]{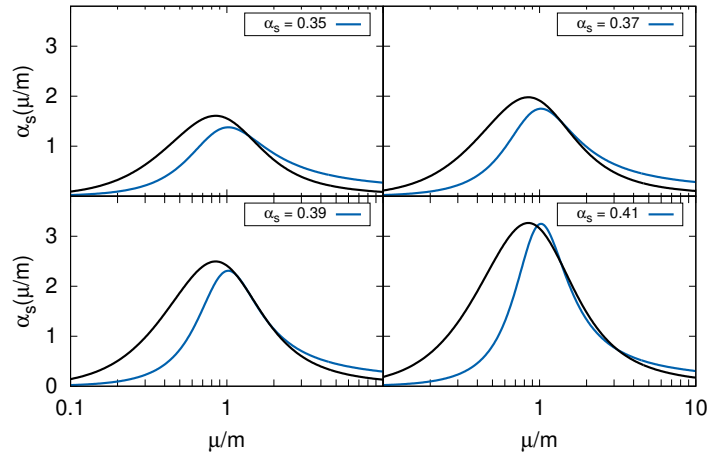}
\vspace{5pt}
\caption{$N=3$ optimized fixed-scale (black) and RG (blue) strong running couplings as a function of the adimensional renormalization scale $\mu/m$ for different values of the initial RG coupling $\alpha_{s}(\mu_{0}^{2})$ ($\mu_{0}=6.098\,m$). The matching scale is set to $\mu_{1}=1.372\,m$.}\label{fig:alphasmatchvar}
\end{figure}

In Fig.~\ref{fig:alphasmatchvar} we compare the OFS strong running coupling $\alpha_{s}^{(\text{OFS})}(\mu^{2})$ -- Eq.~\eqref{dfj247}~-- and its RG counterpart $\alpha_{s}(\mu^{2})$ for different values of the initial RG coupling constant $\alpha_{s}(\mu_{0}^{2})$. $\alpha_{s}^{(\text{OFS})}(\mu^{2})$ clearly shows the same behavior as $\alpha_{s}(\mu^{2})$. Nonetheless, despite having chosen $\kappa$ in such a way that the two functions coincide at $\mu=\mu_{1}$, the two running couplings are far from being equal for arbitrary values of $\alpha_{s}(\mu_{0}^{2})$; this is true in the low- and high-energy regimes, as well as at intermediate energies. There exists, however, an interval of values, centered around $\alpha_{s}(\mu_{0}^{2})\approx 0.39$, for which the OFS and RG running couplings overlap at renormalization scales $\mu\approx\mu_{1}$.

\begin{figure}[H]
\centering
\includegraphics[width=0.48\textwidth,angle=270]{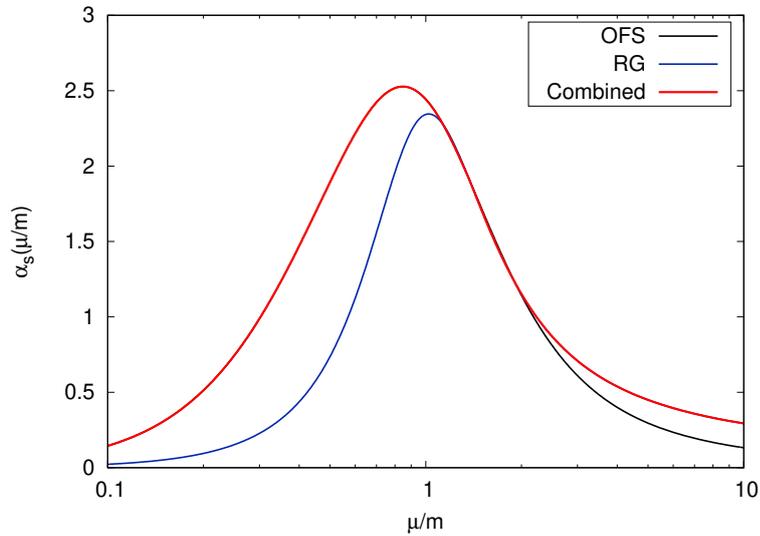}
\vspace{5pt}
\caption{$N=3$ optimized fixed-scale (black) and RG (blue) strong running couplings as a function of the adimensional renormalization scale $\mu/m$ for $\alpha_{s}(\mu_{0}^{2})=0.391$ ($\mu_{0}=6.098\,m$). Matching scale set to $\mu_{1}=1.372\,m$. The red curve is obtained by combining the low-energy OFS coupling and the high-energy RG coupling.}\label{fig:alphasmatchvaropt}
\end{figure}

In \cite{CS20} it was found that the relative difference between the two couplings is less than 1\% over the widest possible interval of momenta -- ranging from $\mu\approx1.1\,m$ to $\mu\approx2\,m$~-- when $\alpha_{s}(\mu_{0}^{2})$ is chosen equal to $0.391$. The running couplings corresponding to such a value are displayed in Fig.~\ref{fig:alphasmatchvaropt}. For $\alpha_{s}(\mu_{0}^{2})=0.391$, the maximum of $\alpha_{s}(\mu^{2})$ is found to be $\alpha_{s}(\mu_{\star}^{2})\approx2.34$, suggesting -- as previously anticipated -- that the one-loop RG-improved propagators may diverge from the exact results for $p\lesssim m$.

With $\alpha_{s}(\mu_{0}^{2})$ determined by the intermediate-scale matching with the OFS results, the gluon mass parameter $m^{2}$ is left as the only free parameter of the RG-improved propagators. Observe that the value of $m^{2}$ sets not only the energy scale of the propagators, but also that of the initial renormalization scale $\mu_{0}$ and of the matching scale $\mu_{1}$.
\newpage
\vspace*{5mm}
\vspace{5mm}
\vspace{5mm}

\begin{figure}[H]
\centering
\includegraphics[width=0.48\textwidth,angle=270]{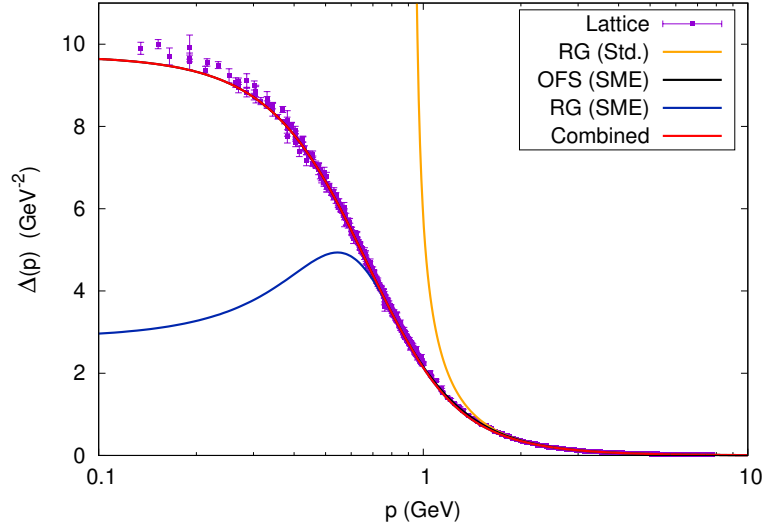}
\vspace{5pt}
\caption{Landau-gauge Euclidean transverse gluon propagator renormalized at the scale $\mu_{0}=4$~GeV. Black curve: optimized fixed-scale SME renormalized by matching at $\mu_{1}=0.9$~GeV. Blue curve: RG-improved SME. Red curve: combined OFS and RG-improved SME matched at $\mu_{1}$. Orange curve: ordinary pQCD. Squares: lattice data of \cite{DOS16}. $m=0.656$~GeV, $\alpha_{s}(\mu_{0})=0.391$.}\label{fig:gluproplandmatch}
\end{figure}

\begin{figure}[H]
\centering
\includegraphics[width=0.48\textwidth,angle=270]{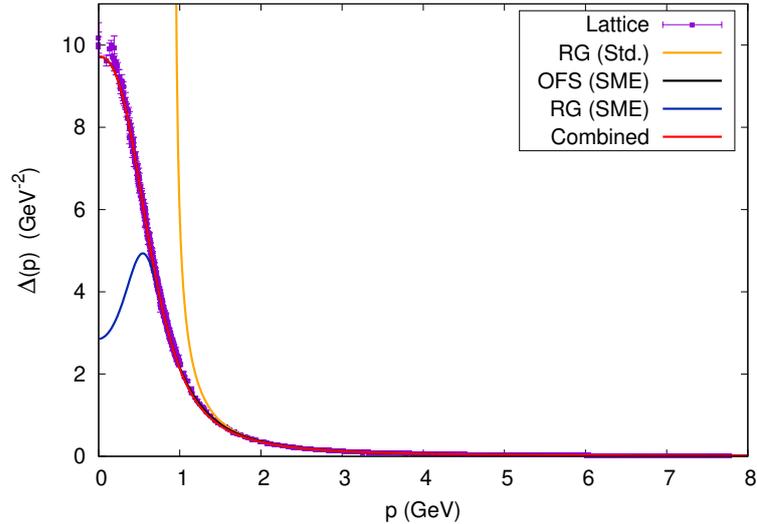}
\vspace{5pt}
\caption{Landau-gauge Euclidean transverse gluon propagator renormalized at the scale $\mu_{0}=4$~GeV with the lattice data of \cite{DOS16}. Linear $x$ axis. Curves as in Fig.~\ref{fig:gluproplandmatch}.}\label{fig:gluproplandmatchnolog}
\end{figure}

\begin{figure}[H]
\centering
\includegraphics[width=0.48\textwidth,angle=270]{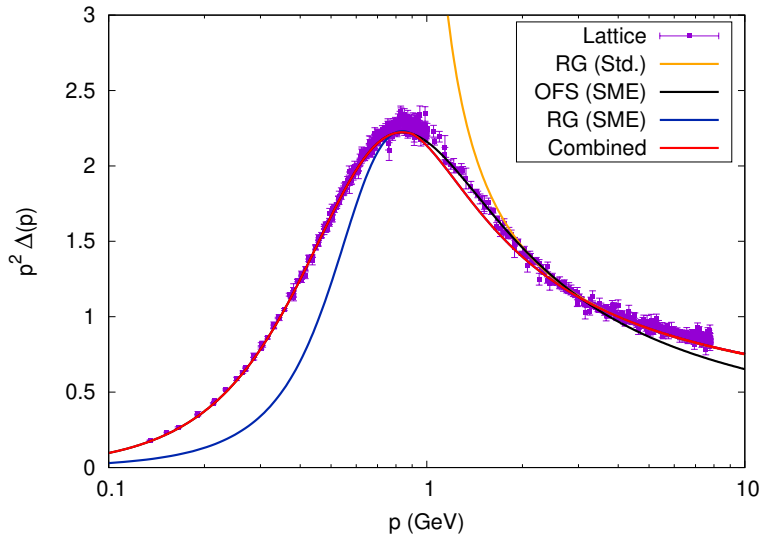}
\vspace{5pt}
\caption{Landau-gauge Euclidean transverse gluon dressing function renormalized at the scale $\mu_{0}=4$~GeV with the lattice data of \cite{DOS16}. Curves as in Fig.~\ref{fig:gluproplandmatch}.}\label{fig:gludresslandmatch}
\end{figure}

In Figs.~\ref{fig:gluproplandmatch}, \ref{fig:gluproplandmatchnolog} and \ref{fig:gludresslandmatch} we display the Euclidean transverse gluon propagator and dressing function computed by different methods within the Screened Massive Expansion, together with the lattice data of \cite{DOS16}. For obtaining each of the curves, the gluon mass parameter was set to $m=0.656$~GeV. As a result, as previously reported, the initial renormalization scale is equal to $\mu_{0}=4$~GeV, while the matching scale is $\mu_{1}=0.9$~GeV. We remark that the lattice data of \cite{DOS16} were originally renormalized at $4$~GeV, explaining our choice of $\mu_{0}$.
\vspace{\fill}
\vspace{5mm}
\vspace{\fill}
\newpage
\vspace*{5mm}
\vspace{5mm}
The optimized fixed-scale propagator and dressing function are shown in black in the figures. At variance with the previous sections, in Figs.~\ref{fig:gluproplandmatch}, \ref{fig:gluproplandmatchnolog} and \ref{fig:gludresslandmatch} these are normalized by matching their value with that of their RG analogues at $p=\mu_{1}$, rather than by fitting the multiplicative renormalization constant $Z_{\Delta}$.

The RG-improved gluon propagator and dressing function computed for the optimal value $\alpha_{s}(\mu_{0}^{2})=0.391$, on the other hand, are shown in blue. As expected, the propagator/dressing function deviates from the lattice data at $p\approx m$: for smaller values of the momentum, the OFS results (black curve, hidden behind the red curve in the IR, see ahead) still yield a better approximation of the lattice data. In particular, compared to the exact results, the RG-improved propagator appears strongly suppressed at low energies. This behavior is often observed in one-loop gluon propagators computed by massive perturbative methods\footnote{The Dynamical Model of Chapter~\ref{chpt:dynmod} explicitly shares this feature. As for the IRS Curci-Ferrari one-loop propagator reported in \cite{GPRT19}, the latter either matches the IR lattice data and misses the UV tail, or it reproduces well the UV tail and displays a suppressed IR limit when rescaled -- see Note~\ref{note:cfir}, Sec.~\ref{sec:cfrevres}.}; the two-loop results obtained within the Curci-Ferrari model \cite{GPRT19} suggest that taking into account the higher-order corrections to the propagator will enhance its IR limit.

At high energies, especially for $p>3$-$4$~GeV, the RG-improved results (hidden below the red curve in the figures, see ahead) generally show a better agreement with the lattice data in comparison to their OFS counterpart. Despite the former falling somewhat below the data in the range $\approx 1$-$3$~GeV, the RG-improved propagator still reproduces the lattice UV tail quite accurately up to $p\approx8$~GeV. At such large scales, the SME results are indistinguishable from those computed by ordinary perturbation theory, displayed in orange in the figures\footnote{For integrating the ordinary pQCD propagator, the value $\alpha_{s}(\mu_{0}^{2})=0.391$ was used.}.
The red curves in Figs.~\ref{fig:gluproplandmatch}, \ref{fig:gluproplandmatchnolog} and \ref{fig:gludresslandmatch} are obtained by combining the low-energy OFS gluon propagator with its high-energy RG-improved counterpart at the matching scale $\mu_{1}=0.9$~GeV \footnote{In particular, the combined propagator is superimposed to the OFS propagator (black curve) in the IR, and to the RG-improved propagator (blue curve) in the UV.}. The combined propagator/dressing function provides the best overall approximation of the lattice data over the whole momentum range $p\in[0,8]$~GeV.
\vspace{5mm}
\begin{figure}[H]
\centering
\includegraphics[width=0.48\textwidth,angle=270]{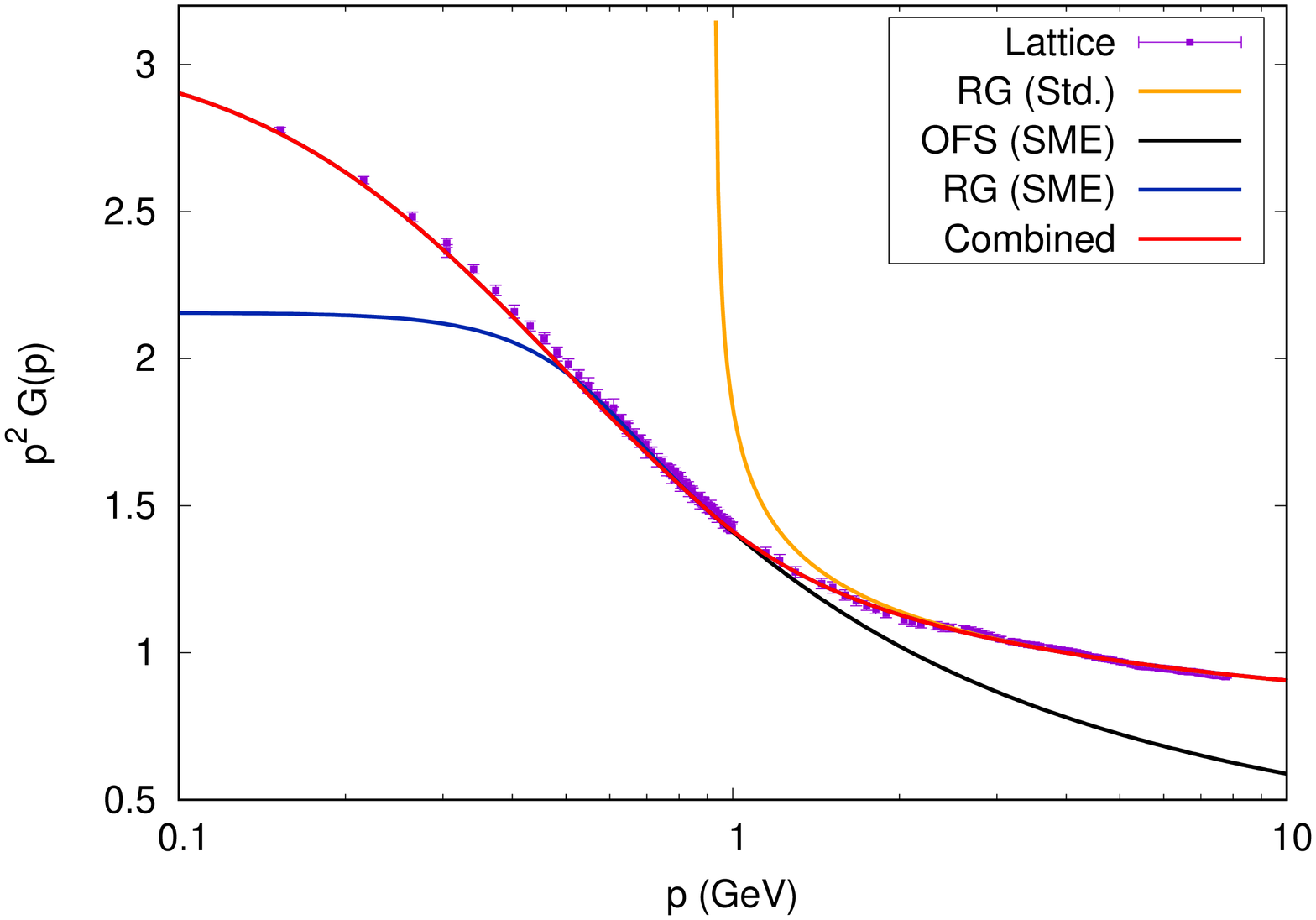}
\vspace{5pt}
\caption{Landau-gauge Euclidean ghost dressing function renormalized at the scale $\mu_{0}=4$~GeV with the lattice data of \cite{DOS16}. Curves as in Fig.~\ref{fig:gluproplandmatch}.}\label{fig:ghoproplandmatch}
\end{figure}
\vspace{5mm}
\begin{figure}[H]
\centering
\includegraphics[width=0.48\textwidth,angle=270]{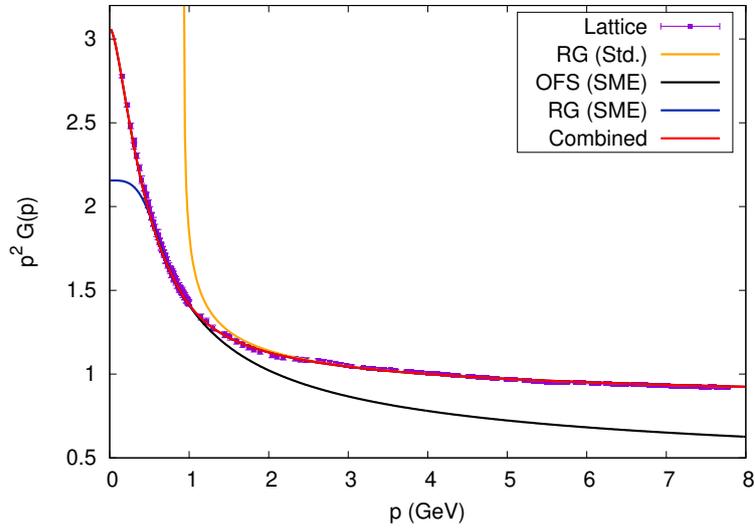}
\vspace{5pt}
\caption{Landau-gauge Euclidean ghost dressing function renormalized at the scale $\mu_{0}=4$~GeV with the lattice data of \cite{DOS16}. Linear $x$ axis. Curves as in Fig.~\ref{fig:gluproplandmatch}.}\label{fig:ghoproplandmatchnolog}
\end{figure}
\vspace{5mm}
In Figs.~\ref{fig:ghoproplandmatch} and \ref{fig:ghoproplandmatchnolog} we show the Landau-gauge Euclidean ghost dressing function with the same color coding and parameters $m^{2}$, $\alpha_{s}(\mu_{0}^{2})$, $\mu_{0}$ and $\mu_{1}$ as in the previous figures. The improvement brought about in the UV by the Renormalization Group is remarkable. While the OFS dressing function fails to reproduce the lattice data already at momenta $p\approx2$~GeV, its RG-improved counterpart is in excellent agreement with them up to $p\approx8$~GeV. At low momenta $p\lesssim m$, on the other hand, the RG-improved dressing function saturates too quickly, and the OFS results provide a better approximation, just like in the gluon sector.

As reported before, the propagators shown in Figs.~\ref{fig:gluproplandmatch} to \ref{fig:ghoproplandmatchnolog} use the gluon mass parameter $m^{2}$ as their only free parameter, the initial coupling constant $\alpha_{s}(\mu_{0}^{2})$ at $\mu_{0}=4$~GeV and normalization of the optimized fixed-scale results having been obtained by matching the OFS and RG running coupling at $\mu_{1}=0.9$~GeV. For $m^{2}$, we chose the value fitted in Sec.~\ref{sec:smeopt} -- namely, $m=0.656$~GeV -- starting from the lattice gluon propagator and the optimized-fixed scale expressions.

An alternative determination of $m^{2}$ can be obtained by making use of the combined low-energy OFS and high-energy RG-improved propagators as follows. First, the RG-improved propagators renormalized at the scale $\mu_{0}$ are defined as functions of $\alpha_{s}(\mu_{0}^{2})$ and $m^{2}$, with the dimensionful value of $\mu_{0}$ set to $4$~GeV following the lattice data of \cite{DOS16}. Then $\alpha_{s}(\mu_{0}^{2})$ is expressed as a function of the gluon mass parameter by making use of the running of the strong coupling, with the initial renormalization scale set to $\mu=6.098\,m$, where we know by optimization that $\alpha_{s}(\mu^{2})=\alpha_{s}^{\text{opt.}}=0.391$. Explicitly,
\begin{equation}
\alpha_{s}(\mu_{0}^{2})=\frac{\alpha_{s}^{\text{opt.}}}{1+\dfrac{3N\alpha_{s}^{\text{opt.}}}{4\pi}\,[H(\mu_{0}^{2}/m^{2})-H(6.098^{2})]}\ .
\end{equation}
Since $\alpha_{s}(\mu_{0}^{2})$ is completely determined by the value of $m^{2}$, the same will hold true for the RG-improved propagators. Finally, the latter are combined with the OFS propagators at the matching scale $p=\mu_{1}=1.372\,m$, retaining the RG-improved propagators for $p\geq \mu_{1}$ and the OFS propagators for $p\leq \mu_{1}$. The resulting combined propagators are functions of the gluon mass parameter alone. In particular, the value of $m^{2}$ can be determined by fitting the combined gluon propagator with the lattice data of \cite{DOS16} over the widest possible range of momenta.

\begin{figure}[H]
\centering
\includegraphics[width=0.34\textwidth,angle=270]{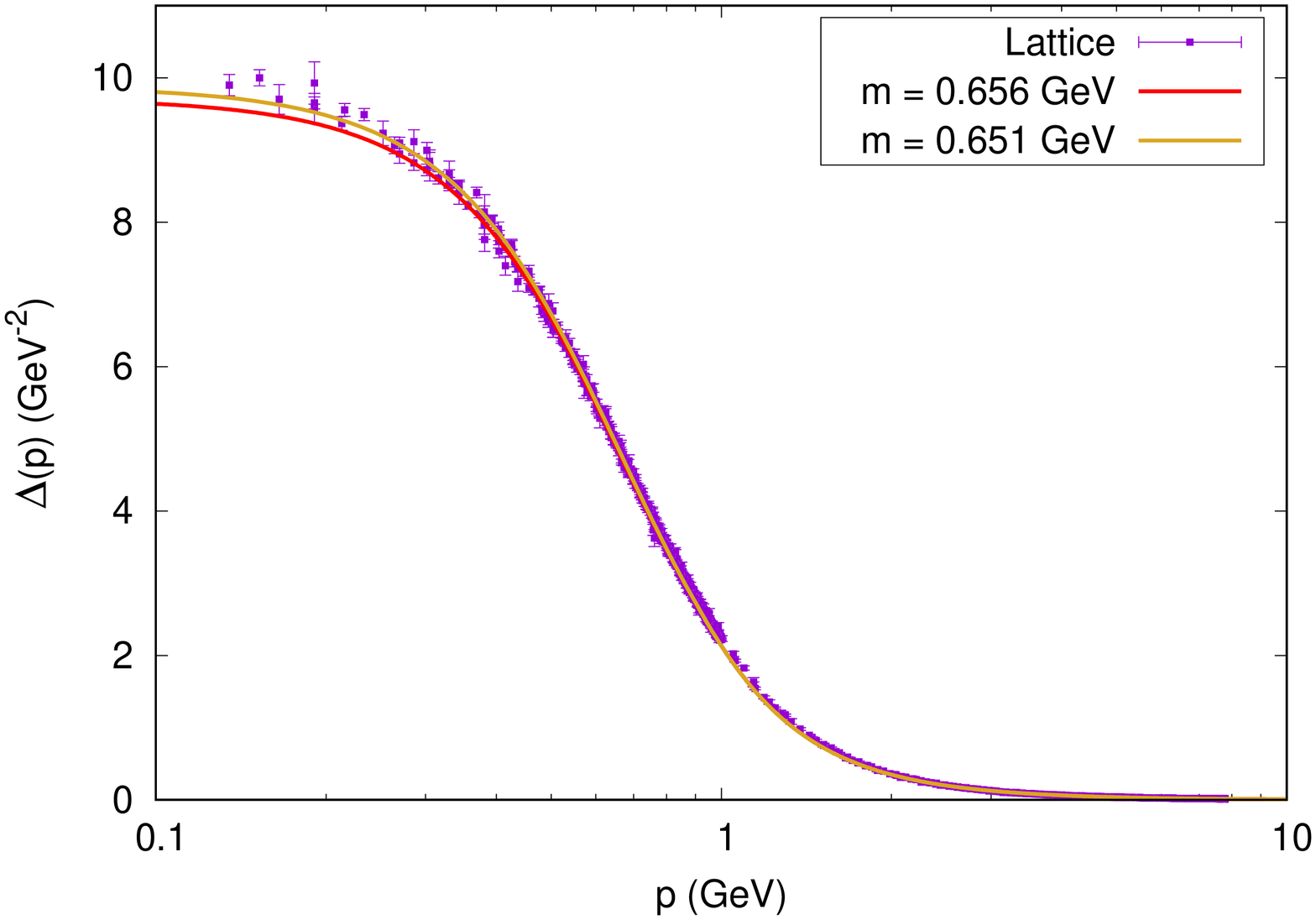}\hspace{2mm}\includegraphics[width=0.34\textwidth,angle=270]{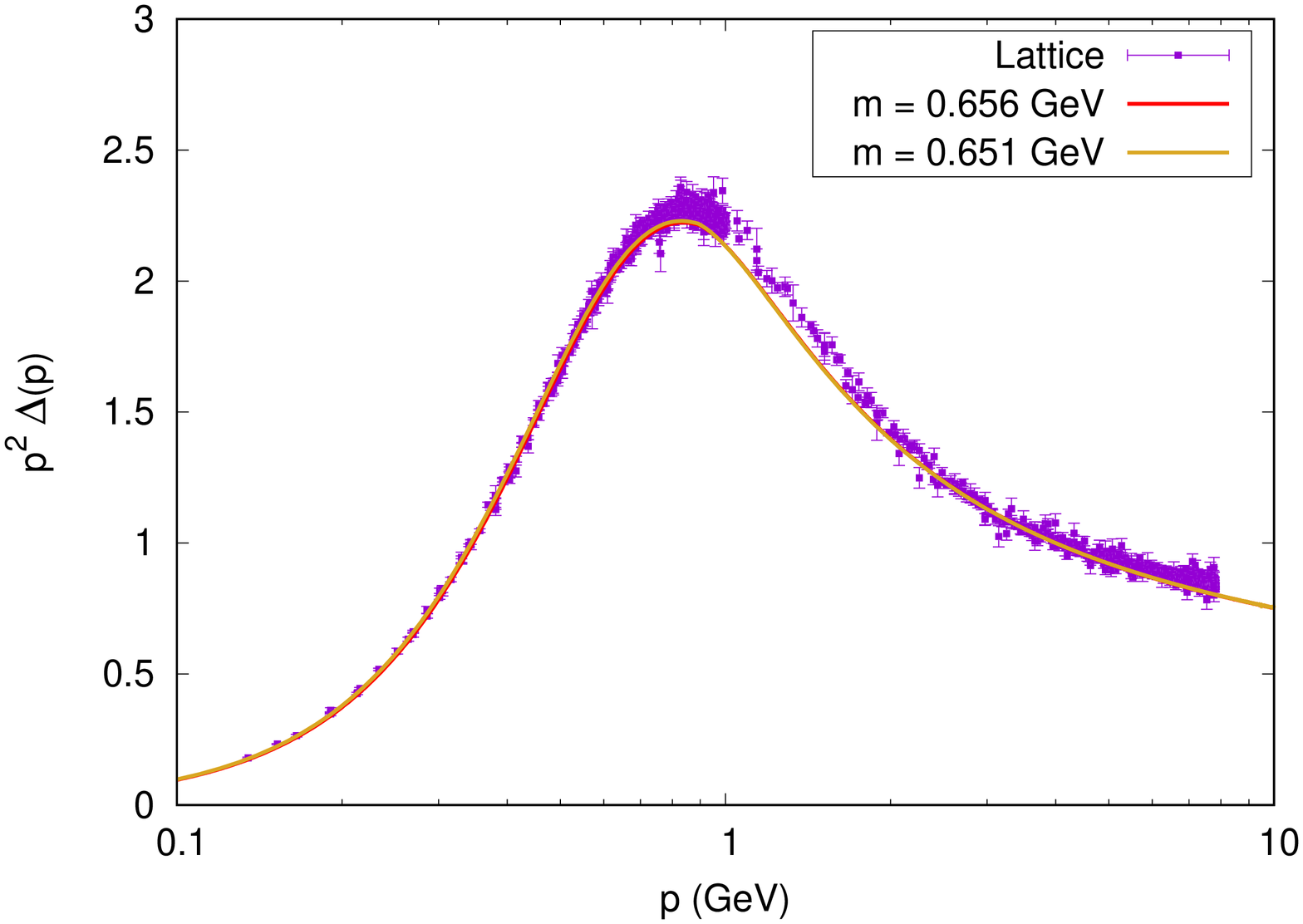}
\vspace{5pt}
\caption{Landau-gauge RG-improved Euclidean transverse gluon propagator (left) and dressing function (right) renormalized at the scale $\mu_{0}=4$~GeV, obtained by combining the low-energy OFS results with the high-energy RG-improved results. Lattice data from \cite{DOS16}.}\label{fig:gluprop0651}
\end{figure}

The outcome of the fit is shown in Fig.~\ref{fig:gluprop0651}. It was found that the value that best fits the lattice data of \cite{DOS16} is $m=0.651$~GeV, very close to our previous determination (also shown in the figure). The difference between the $m=0.651$~GeV and $m=0.656$~GeV combined propagators is minimal, being visible only in the deep IR, where the former is closer to the lattice data when compared to the latter. Analogous plots for the ghost dressing function can be found in \cite{CS20}; in the ghost sector, the $m=0.651$~GeV and $m=0.656$~GeV dressing functions are indistinguishable to the naked eye.\\

In this section we have discussed the optimization of the RG-improved propagators computed in the Screened Massive Expansion. While formally the RG-improved propagators depend both on the value of the coupling constant $\alpha_{s}(\mu_{0}^{2})$ at the initial renormalization scale $\mu_{0}$ and on the gluon mass parameter $m^{2}$, the requirement that the former match with their optimized fixed-scale counterpart at intermediate energy scales ($p\approx m$) allowed us to determine the value of $\alpha_{s}(\mu_{0}^{2})$ as a function of $m^{2}$. The resulting propagators show a good agreement with the lattice data for $p\gtrsim m$ up to $8$~GeV, but deviate from the exact results at low energies. This is due to the fact that, in the IR, the SME running coupling computed in the MOM-Taylor scheme attains a maximum $\alpha_{s}^{\text{max}}\approx 2.34$ which is quite large for the perturbative standards, thus preventing the one-loop approximation from fully capturing the low-energy behavior of the propagators.

For $p\lesssim m$, the optimized fixed-scale results of Sec.~\ref{sec:smeopt} still provide the best description of the lattice data. When the low-energy OFS propagators are combined with their high-energy RG-improved counterpart at the matching scale $\mu_{1}=1.372\,m$, the resulting combined functions succeed in reproducing the lattice data over the whole momentum range $p\in[0,8]$~GeV. By making use of these, one is able to obtain an alternative determination of the gluon mass parameter, $m=0.651$~GeV, which is very close to the value found in Sec.~\ref{sec:smeopt} by fitting the OFS gluon propagator ($m=0.656$~GeV).
\vspace{5mm}
\vspace{5mm}

\section{Conclusions}
\renewcommand{\rightmark}{\thesection\ \ \ Conclusions}
\label{sec:smeconc}

The results presented in this chapter show that the Screened Massive Expansion is capable of describing the dynamics of the gluons and ghosts from first principles over a wide range of momenta. After optimization, the gluon mass parameter $m^{2}$ is left as the only free parameter of the expansion both at low energies, where the optimized fixed-scale propagators of Sec.~\ref{sec:smeopt} provide the best approximation of the lattice data, and at high energies, where the RG-improved propagators derived in Sec.~\ref{sec:smergopt} manage to reproduce the lattice UV tails up to $p\approx 8$~GeV.

The absence of Landau poles from the SME running coupling, made possible by the explicit scale dependence of the beta function, confirms that the method is self-consistent and well-defined in the infrared. Moreover, it proves that the diverging of the ordinary pQCD coupling at low energies is just an artifact of the expansion, ultimately due to the failure of pQCD in accounting for the dynamical generation of an IR gluon mass.

The precision with which the Screened Massive Expansion is able to reproduce the lattice results already to one loop seems to indicate that most of the non-perturbative effects which shape the IR dynamics of the gauge sector of QCD can be incorporated in a perturbative series that treats the transverse gluons as massive at tree level.

\chapter{Applications of the Screened Massive Expansion}
\renewcommand{\leftmark}{\thechapter\ \ \ Applications of the Screened Massive Expansion}
\renewcommand{\rightmark}{\thechapter\ \ \ Applications of the Screened Massive Expansion}
\label{chpt:smeapp}

As an application of the Screened Massive Expansion presented in Chpt.~\ref{chpt:sme}, in this chapter we summarize the main findings of \cite{CS18}, \cite{CRBS21} and \cite{SC21} regarding the thermal behavior of the gluons in pure Yang-Mills theory and the dynamical generation of a mass for the quarks in full QCD. In Sec.~\ref{sec:smeft} we will extend the GEP analysis of Sec.~\ref{sec:smesu} to finite temperature \cite{CS18} and study the gluon propagator and its poles as functions of the temperature \cite{SC21}. In Sec.~\ref{sec:smeqk} we will make use of the techniques presented in the last chapter to show that a shift of the quark action similar to the one employed in the gluon sector is able to account for dynamical mass generation in the quark sector, as is expected in full QCD as a consequence of chiral symmetry breaking \cite{CRBS21}.

Refs. \cite{CS18}, \cite{CRBS21} and \cite{SC21} are attached as an insert to this thesis, to be found in Appendix~\ref{app:published}.

\section{The Screened Massive Expansion at finite temperature}
\renewcommand{\rightmark}{\thesection\ \ \ The Screened Massive Expansion at finite temperature}
\label{sec:smeft}

The Screened Massive Expansion can be extended to finite temperatures $T>0$ by making use of the formalism of thermal field theory (TFT). In the framework of TFT \cite{KG06}, the partition function and the quantum fields are defined in Euclidean space, with the imaginary-time variable $\tau=it$ taken to lie in the interval $\tau\in[0,\beta]$. The dimensionful quantity $\beta=1/T$ -- that is, the inverse temperature of the system -- tends to infinity as $T\to 0$; thanks to the boundary conditions imposed on the finite-temperature fields, the results of ordinary zero-temperature quantum field theory are recovered in such a limit.

The set-up of the Screened Massive Expansion at $T>0$ is discussed in depth in \cite{CS18} and \cite{SC21}. The shift of the expansion point of the perturbative series that defines the SME is replicated in the framework of TFT to obtain expressions that depend on the gluon mass parameter $m^{2}$ as well as on the temperature. Since $m^{2}$ is a mass scale whose value depends on the problem at hand, in the context of the SME the former can itself be taken to be a function of the temperature, $m^{2}=m^{2}(T)$.

\subsection{The Gaussian Effective Potential at finite temperature and the deconfinement phase transition}
\label{sec:gepft}

In Sec.~\ref{sec:smesu} we argued that treating the gluons as massive at tree level is expected to provide a better approximation of the low-energy dynamics of QCD compared to ordinary massless perturbation theory because such a choice minimizes the Gaussian Effective Potential of pure Yang-Mills theory. The GEP analysis carried out in Sec.~\ref{sec:smesu} can be repeated at finite temperature in order to check that this remains true for $T>0$ \cite{CS18}. At non-zero temperatures, the GEP can be interpreted as a first-order approximation of the temperature-dependent free energy density $\mc{F}(T)$ of the theory \cite{CS18}. Therefore, in what follows, we will denote the former by $\mc{F}_{G}(T,m)$.
\vspace{5mm}
\begin{figure}[H]
\centering
\includegraphics[width=0.55\textwidth,angle=270]{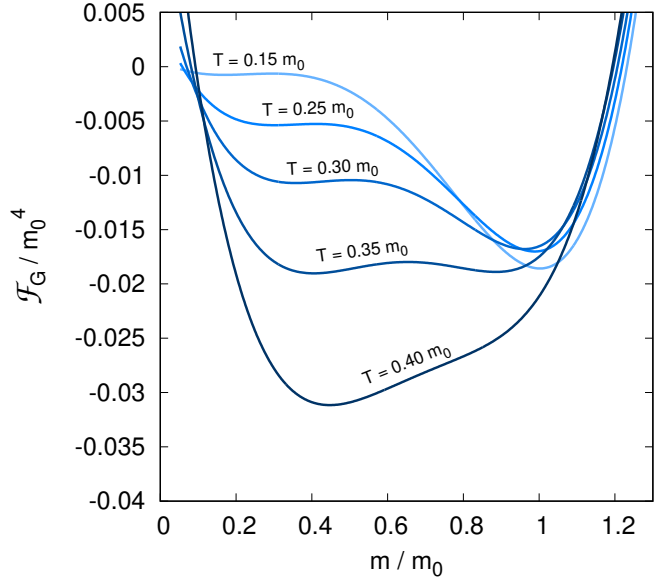}
\vspace{5pt}
\caption{Gaussian Effective Potential of pure Yang-Mills theory as a function of the gluon mass parameter for different values of the temperature. All dimensionful quantities are adimensionalized by factors of $m_{0}=m(T=0)$.}\label{fig:GEPT}
\end{figure}
\vspace{5mm}
\begin{figure}[H]
\centering
\includegraphics[width=0.48\textwidth,angle=270]{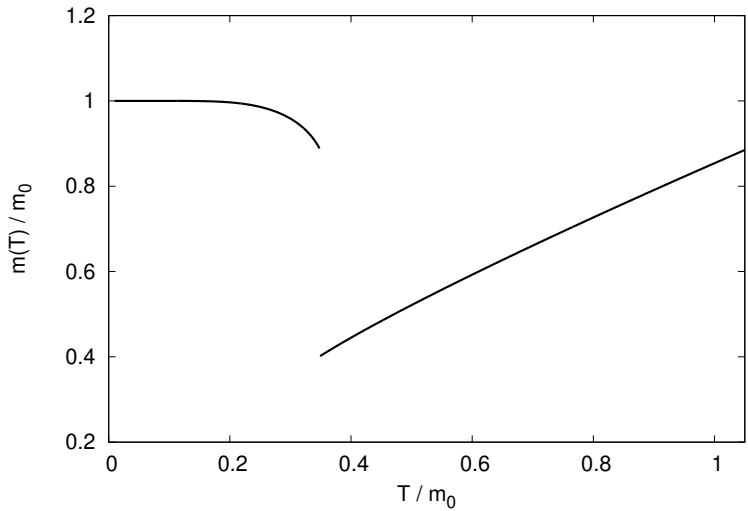}
\vspace{5pt}
\caption{Global minimum of the Gaussian Effective Potential of pure Yang-Mills theory as a function of the temperature. The discontinuity is found at $T=T_{c}\approx 0.35\,m_{0}$.}\label{fig:massT}
\end{figure}
\vspace{5mm}
The Gaussian Effective Potential $\mc{F}_{G}(T,m)$ of Yang-Mills theory is displayed in Fig.~\ref{fig:GEPT} as a function of the gluon mass parameter $m$ for different values of the temperature. In the figure, all the quantities are adimensionalized by the value of $m$ which minimizes the GEP at zero temperature, $m_{0}=m(T=0)$.

We recall from Sec.~\ref{sec:smesu} that for $T=0$ the GEP has a global minimum at $m=m_{0}\neq0$ and a local minimum at $m=0$. As we can see from Fig.~\ref{fig:GEPT}, as the temperature increases, the position of the $m=0$ minimum shifts towards larger values of the mass and the corresponding value of the GEP grows more negative, until the two minima align at $T= T_{c}\approx0.35\,m_{0}$. For $T>T_{c}$, the lower-mass minimum becomes the global minimum of the GEP, and its position starts to increase roughly linearly with the temperature.

The position of the global minimum of the GEP is shown in Fig.~\ref{fig:massT} as a function of temperature. At low temperatures, the value of $m$ which minimizes the GEP, denoted by $m(T)$, remains roughly equal to its $T=0$ value until it starts to decrease at $T\approx0.2\,m_{0}$. After hitting the critical temperature $T_{c}$, the high-temperature behavior of $m(T)$ becomes similar to that of the thermal masses computed for the massless particles in the framework of TFT \cite{KG06}, $m(T)\propto T$. For this reason, we may interpret $T=T_{c}$ as the temperature at which the gluon ceases to be massive and starts to behave as a massless particle.
\vspace{5mm}
\begin{figure}[H]
\centering
\hspace{-24mm}\includegraphics[width=0.45\textwidth,angle=270]{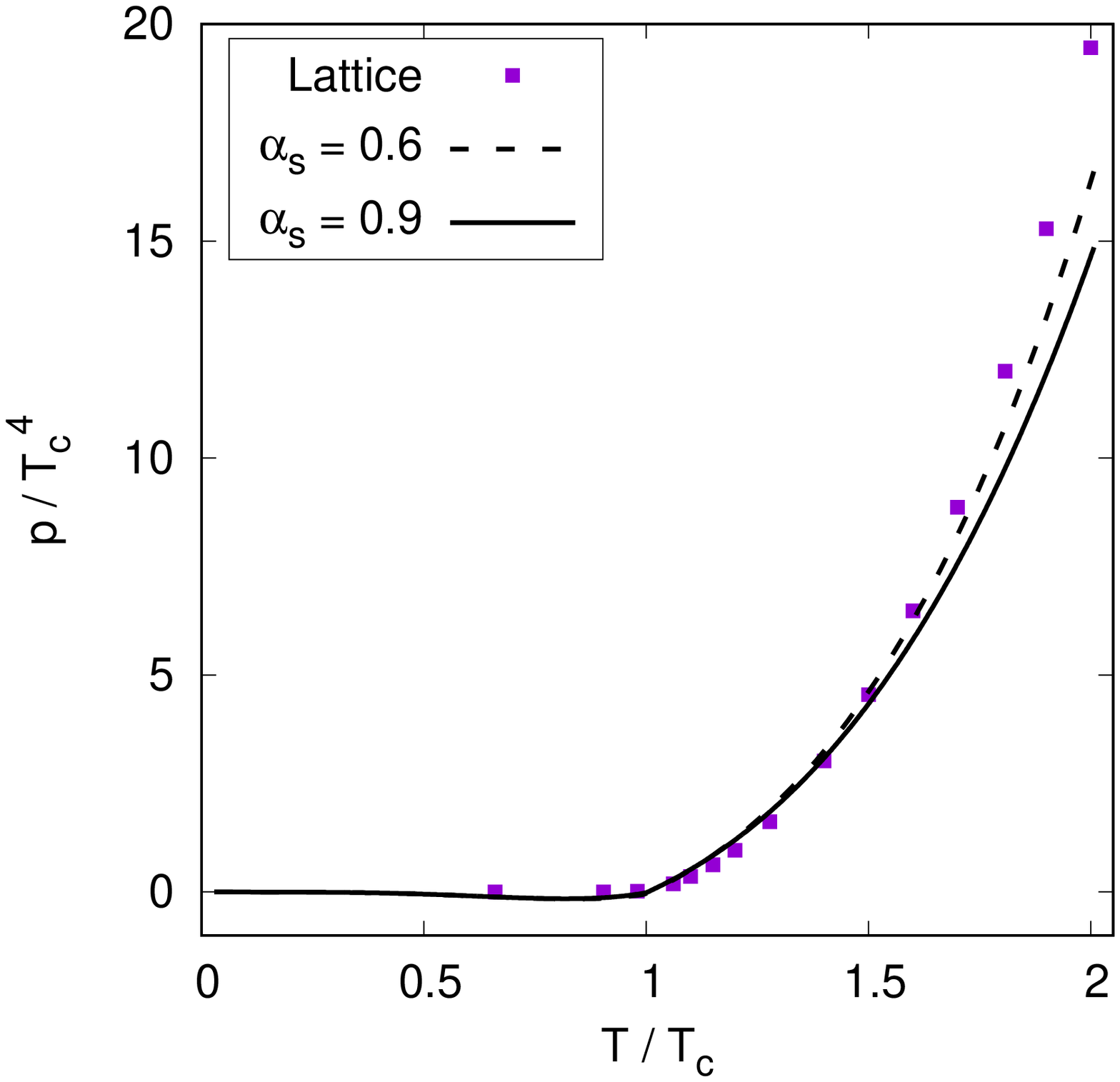}\hspace{-22mm}\includegraphics[width=0.45\textwidth,angle=270]{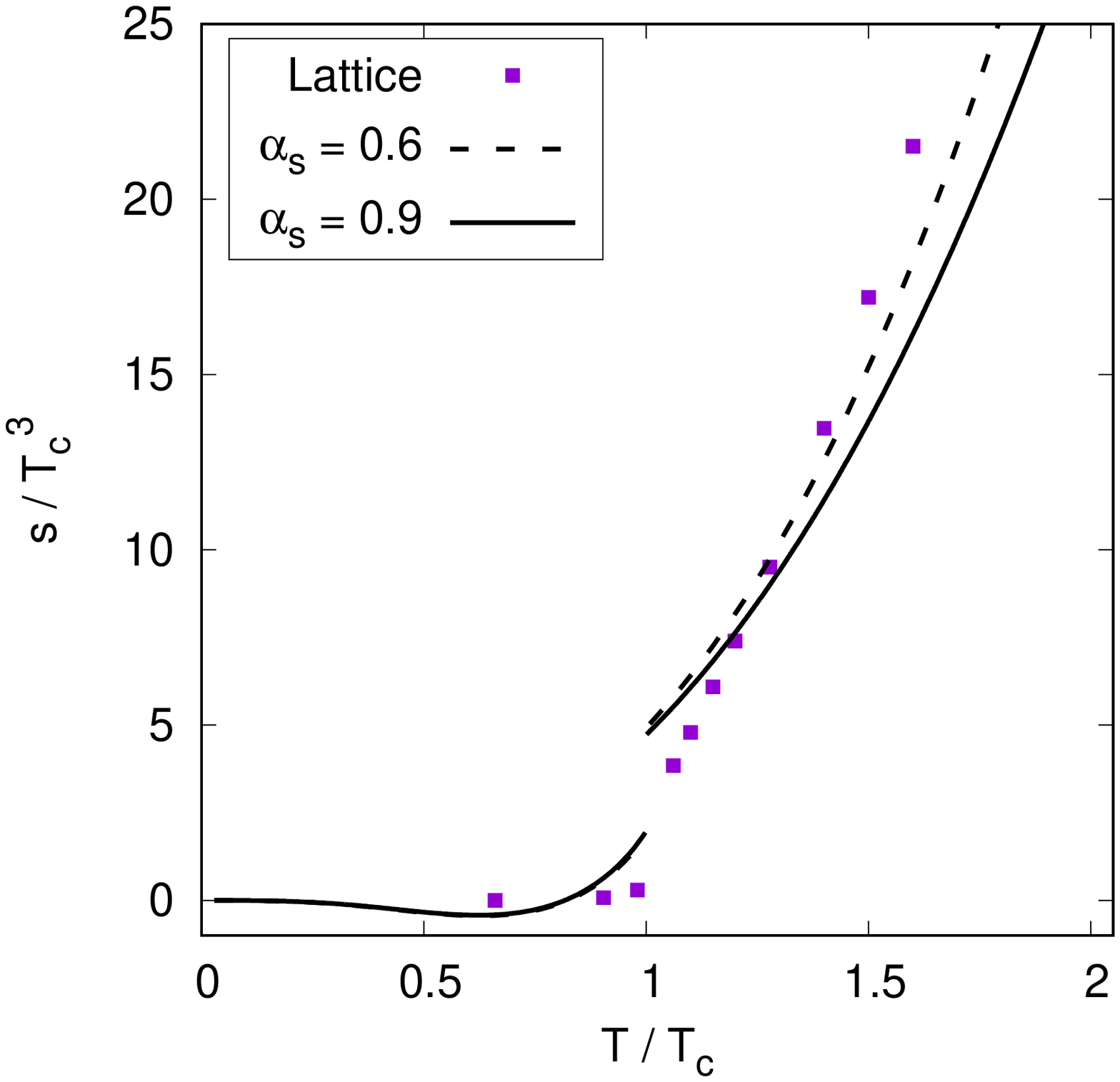}\hspace{-24mm}
\vspace{5pt}
\caption{Pressure (left) and entropy density (right) as functions of temperature computed in the GEP approximation, together with the lattice data of \cite{GP17}. Adimensionalized by the critical temperature $T_{c}$. For a discussion on the value of the coupling constant $\alpha_{s}$ see \cite{CS18}.}\label{fig:statefT}
\end{figure}
\vspace{5mm}
What are the consequences of the discontinuity found in $m(T)$ by the minimization of the GEP? In Fig.~\ref{fig:statefT} we display the pressure $p$ and entropy density $s$ computed in the GEP approximation,
\begin{equation}
p=-\left[\mc{F}_{G}(T,m(T))-\mc{F}_{G}(0,m_{0})\right]\ ,\qquad\qquad s=-\frac{d}{dT}\,\mc{F}_{G}(T,m(T))\ .
\end{equation}
While the pressure is a continuous function of the temperature, the entropy density has a discontinuity at $T=T_{c}$; this indicates that the system undergoes a first-order phase transition. Such a transition is well-known to occur in pure Yang-Mills theory: it is the deconfinement phase transition, by which the gluons\footnote{And, in full QCD, the quarks, although strictly speaking in this case there is no phase transition, but rather a crossover \cite{AEFK06,BBCD12,BBDD14}.} cease to be confined in color-singlet bound states and start behaving as free particles. Once again, we see that the existence of a link between the IR-massiveness of the gluon and the phenomenon of confinement is suggested by the results of the SME. With $m\approx 0.66$~GeV, $T_{c}$ would be predicted to be around $230$~MeV, not too far from the critical temperature found by lattice methods for pure Yang-Mills theory, $T_{c}\approx 270$~MeV \cite{SOBC14}. In passing, we note that the pressure and entropy density computed in the GEP approximation, when adimensionalized by the critical temperature $T_{c}$, are in fair agreement with the lattice data of \cite{GP17}.

The GEP analysis of pure Yang-Mills theory at $T\geq 0$ confirms that, below the critical temperature $T_{c}$, the Screened Massive Expansion is still expected to provide a better approximation of the infrared dynamics of the gauge sector compared to ordinary perturbation theory. At larger temperatures, on the other hand, the $T$-dependent gluon mass parameter can be understood to play essentially the same role as the ordinary thermal mass of massless particles, so that more specialized approaches such as the Hard Thermal Loop resummation \cite{BP92} might yield better results.

\subsection{The Landau-gauge gluon propagator and its poles at finite temperature}
\label{sec:glupropft}

In thermal field theory, the restriction of the imaginary time variable to the interval $[0,1/T]$ causes the breaking of the symmetry group SO(4) of $4$-dimensional Euclidean spacetime down to SO(3) -- that is, the group of ordinary spatial rotations. From a physical point of view, this is equivalent to the adoption of a reference frame with respect to which the state variables and the state functions of the thermodynamical system, such as the temperature and energy, are to be defined.

As a consequence of the breaking of SO(4) symmetry, at finite temperature the gluon propagator $\Delta_{\mu\nu}^{ab}$ cannot be expressed anymore in terms of just two scalar functions -- namely, its $4$-dimensionally transverse and longitudinal components --, but is instead determined by \textit{three} scalar components. Indeed, it can be shown \cite{KG06} that, in Fourier space, the most general expression for $\Delta_{\mu\nu}^{ab}$ at $T\geq 0$ is given by\footnote{We remark that, in Euclidean space, the $4$-dimensional transverse and longitudinal projectors are defined, respectively, as $t_{\mu\nu}(p)=\delta_{\mu\nu}-p_{\mu}p_{\nu}/p^{2}$ and $\ell_{\mu\nu}(p)=p_{\mu}p_{\nu}/p^{2}$.}
\begin{equation}\label{vjk219}
\Delta_{\mu\nu}^{ab}(p,T)=\left(\Delta_{T}(p,T)\,\mc{P}^{T}_{\mu\nu}(p)+\Delta_{L}(p,T)\,\mc{P}^{L}_{\mu\nu}(p)+\frac{\xi}{p^{2}}\,\ell_{\mu\nu}(p)\right)\delta^{ab}\ ,
\end{equation}
where $\mc{P}^{T}_{\mu\nu}(p)$ and $\mc{P}^{L}_{\mu\nu}(p)$ are \textit{3-dimensionally} transverse and longitudinal projectors,
\begin{equation}
\mc{P}^{T}_{\mu\nu}(p)=(1-\delta_{\mu4})(1-\delta_{\nu4})\left(\delta_{\mu\nu}-\frac{p_{\mu}p_{\nu}}{|\vec{p}|^{2}}\right)\ ,\qquad\quad \mc{P}^{L}_{\mu\nu}(p)=t_{\mu\nu}(p)-\mc{P}_{\mu\nu}^{T}(p)\ ,
\end{equation}
the fourth direction being that of the imaginary time variable, $x^{4}=\tau=it=ix^{0}$, and $\Delta_{T,L}(p,T)$ are the $3$-dimensionally transverse and longitudinal components of the gluon propagator. Because of SO(3) symmetry, the functions $\Delta_{T,L}$ do not depend on the direction of the three-dimensional vector $\vec{p}$, but only on its modulus $|\vec{p}|$,
\begin{equation}
\Delta_{T,L}(p,T)=\Delta_{T,L}(p^{4},|\vec{p}|,T)\ .
\end{equation}
That the $4$-dimensionally longitudinal component of $\Delta_{\mu\nu}^{ab}(p,T)$, $\xi/p^{2}$, is equal to its $T=0$ limit can be proved by exploiting the BRST invariance of the Faddeev-Popov Lagrangian, which still holds at finite temperature. During the rest of this section, we will use the terms transverse and longitudinal in the $3$-dimensional sense, unless otherwise specified.

Observe that, in order for SO(4) symmetry and the ordinary result to be recovered at zero temperature, we must have
\begin{equation}
\Delta_{T}(p,T=0)=\Delta_{L}(p,T=0)=\Delta(p)
\end{equation}
for the transverse and longitudinal propagators, where $\Delta(p)$ is the Euclidean ($4$-dimensionally transverse) propagator studied in Chpt.~\ref{chpt:sme}. A similar relation can be shown to hold \cite{KG06} when the spatial momentum $\vec{p}$ vanishes at $p^{4}\neq 0$:
\begin{equation}\label{smz839}
\Delta_{T}(p^{4},|\vec{p}|=0,T)=\Delta_{L}(p^{4},|\vec{p}|=0,T)\qquad\qquad(p^{4}\neq 0)\ .
\end{equation}
Eq.~\eqref{smz839} is easily understood to be a consequence of the fact that, at zero momentum, $\vec{p}=\vec{0}$, there is no spatial direction with respect to which the transverse and longitudinal components of the propagator $\Delta_{T,L}(p,T)$ can be distinguished.

Since the interval $[0,1/T]$ is bounded, the Fourier variable $p^{4}$ corresponding to the imaginary time $x^{4}=\tau$ is a discrete variable. Due to the boundary conditions of the thermal problem, which require the gluon field to be periodic in $\tau$ \cite{KG06}, $p^{4}$ takes on the values
\begin{equation}
p^{4}=\omega_{n}=2\pi nT\qquad\quad(n\in\Bbb{N})\ .
\end{equation}
The $\omega_{n}$'s are known as \textit{Matsubara frequencies}.\\

In \cite{SC21}, the Landau-gauge transverse and longitudinal components of the gluon propagator were computed to one loop at finite temperature in the Screened Massive Expansion, yielding
\begin{equation}\label{cin714}
\Delta_{T,L}(p,T)=\frac{Z_{T,L}(T)}{p^{2}[F(s(T))+F_{0}^{T,L}(T)+\pi_{T,L}(p,m(T),T)]}\ .
\end{equation}
In the above equation, $s(T)=p^{2}/m^{2}(T)$ is the adimensionalized Euclidean momentum, the gluon mass parameter $m^{2}=m^{2}(T)$ is taken to be a function of the temperature, $F(s)$ is the function already reported in Sec.~\ref{sec:glprsme}, $Z_{T,L}(T)$ and $F_{0}^{T,L}(T)$ are, respectively, multiplicative and additive temperature-dependent renormalization constants, and the functions $\pi_{T,L}(p,m,T)$, defined so that
\begin{equation}
\pi_{T,L}(p,m,T=0)=0\ ,
\end{equation}
contain the one-loop thermal corrections to the components of the propagator. The explicit form of $\pi_{T}(p,m,T)$ and $\pi_{L}(p,m,T)$ is reported in \cite{SC21} in terms of one-dimensional integrals which cannot be evaluated analytically.

The finite-temperature SME gluon propagator was then compared to the Landau-gauge lattice data of \cite{SOBC14} at zero Matsubara frequency ($n=0$). While the standard definition of the Screened Massive Expansion makes use of a single gluon mass parameter $m^{2}$ to rearrange the QCD perturbative series, in \cite{SC21} it was found that for $T>0$, and especially at high temperatures, it is not possible to reproduce the behavior of both the components of the gluon propagator by a single $T$-dependent value of $m^{2}$ in Eq.~\ref{cin714}. This is not totally unexpected, for the following reason. First of all, observe that the gluon mass term which is added and subtracted from the Faddeev-Popov action,
\begin{equation}\label{cin710}
\delta S=\frac{1}{2}\int \frac{d^{4}p}{(2\pi)^{4}}\ A_{\mu}^{a}(-p)\,m^{2}t^{\mu\nu}(p)\,A_{\nu}^{a}(p)\ ,
\end{equation}\cleannlnp
is $4$-dimensionally transverse, and therefore does not take into account possible differences in the behavior of the $3$-dimensionally transverse and longitudinal masses that can only arise at finite temperature. Second, the GEP analysis of the last section already showed that, at high temperatures, the minimum of the GEP is found at a linearly rising mass $m(T)\propto T$, which, as we noticed, resembles the thermal mass of a massless particle. For the vector bosons, it is well-known that the thermal masses associated to the transverse and the longitudinal components of the propagators \textit{do} have a different dependence on the temperature \cite{BP92}. Therefore, if the picture painted by the GEP analysis is qualitatively accurate, then the $\delta S$ in Eq.~\eqref{cin710} indeed will not be able to accurately reproduce the exact results at sufficiently high temperatures.

Nonetheless, the Screened Massive Expansion was still found to be able to provide a good semi-quantitative description of the gluon propagator at $T>0$, so long as the mass parameters in $\Delta_{T}(p,T)$ and $\Delta_{L}(p,T)$ are tuned separately as functions of the temperature. Of course, since using two different mass parameters is not allowed in the standard SME formalism, the results described in what follows are to be interpreted as estimates, rather than as first-principles calculations.
\vspace{5mm}
\begin{figure}[H]
\centering
\includegraphics[width=0.45\textwidth,angle=270]{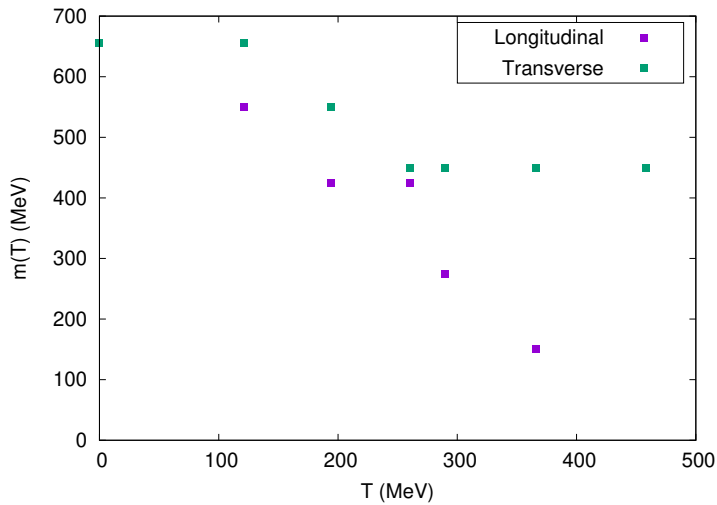}
\vspace{5pt}
\caption{Values of the gluon mass parameter which best fit the lattice data of \cite{SOBC14} for the transverse and the longitudinal components of the Landau-gauge Euclidean gluon propagator at zero Matsubara frequency, as a function of temperature.}\label{fig:glupropftmassT}
\end{figure}
\vspace{5mm}
The values of the gluon mass parameter $m^{2}$ which yield the best fit with the transverse and the longitudinal components of the lattice gluon propagator of \cite{SOBC14} at zero Matsubara frequency ($n=0$) are shown in Fig.~\ref{fig:glupropftmassT} as functions of the temperature. As we can see, while at low temperatures the transverse and the longitudinal mass remain sufficiently close to each other, as soon as the temperature approaches $T_{c}\approx270$~MeV their behavior starts to diverge radically. In particular, at high temperatures, the transverse gluon mass parameter is essentially constant, whereas the longitudinal one decreases with the temperature. We reiterate that, having been obtained by a non-standard procedure, the fitted values of the parameters should be regarded as effective values whose sole purpose is to reproduce the components of the propagator.
\newpage
\vspace*{5mm}
\begin{figure}[H]
\centering
\includegraphics[width=0.45\textwidth,angle=270]{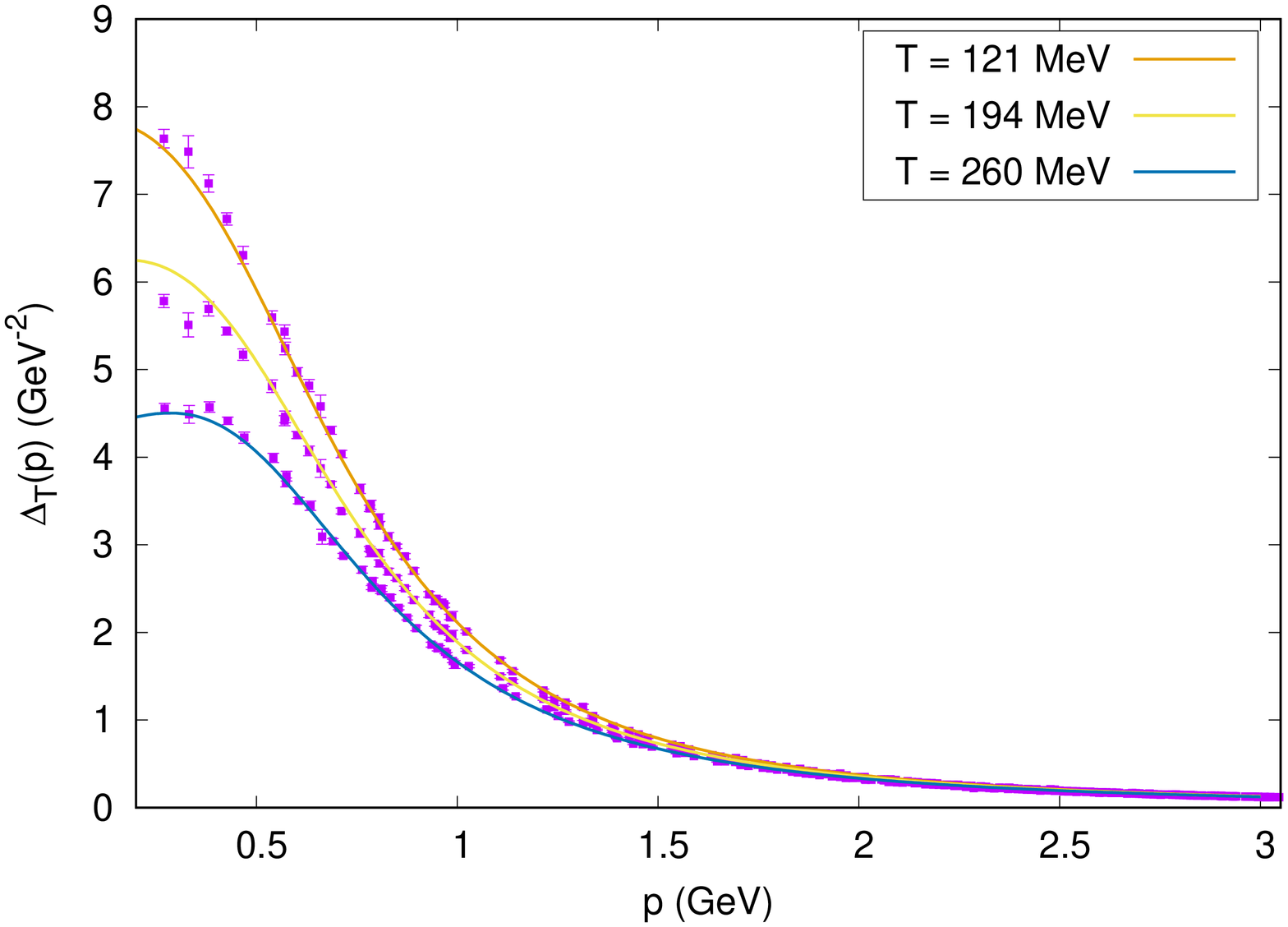}
\includegraphics[width=0.45\textwidth,angle=270]{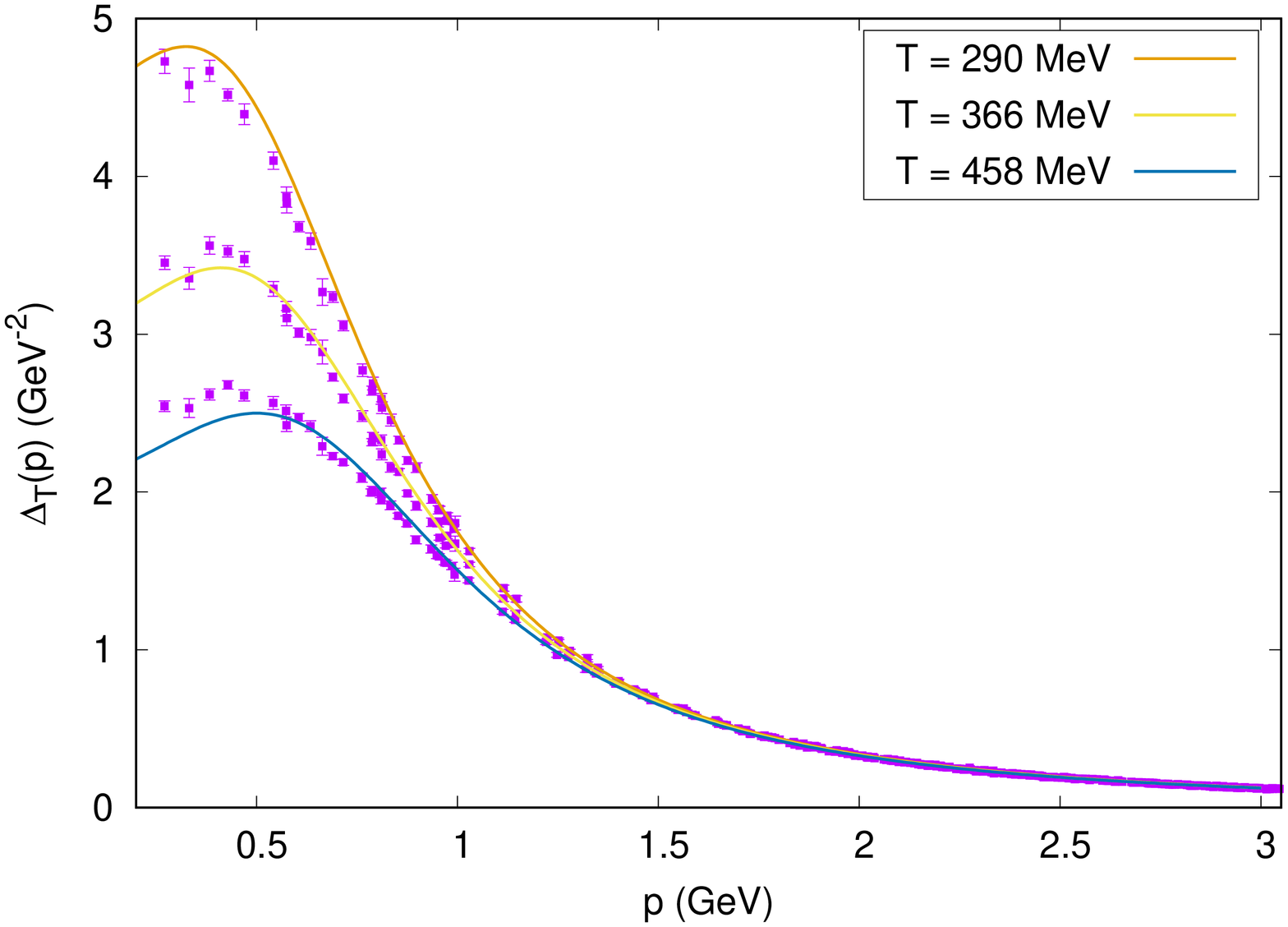}
\vspace{5pt}
\caption{Transverse component of the Landau-gauge Euclidean gluon propagator at zero Matsubara frequency ($n=0$) as a function of the modulus $|\vec{p}|$ of the $3$-dimensional momentum, for $T<T_{c}$ (top) and $T>T_{c}$ (bottom). Multiplicatively renormalized at $\mu_{0}=4$~GeV. Lattice data from \cite{SOBC14}.}\label{fig:glupropftt}
\end{figure}
\vspace{5mm}
In Fig.~\ref{fig:glupropftt} we display the transverse component of the Landau-gauge Euclidean gluon propagator at zero Matsubara frequency ($n=0$) as a function of the modulus $|\vec{p}|$ of the $3$-dimensional momentum for different values of the temperature. By tuning the gluon mass parameter as in Fig.~\ref{fig:glupropftmassT}, we see that the SME expression is able to give a good quantitative description of the transverse propagator over the whole momentum range $[0,3]$~GeV and for all the tested values of the temperature. When multiplicatively normalized at $\mu_{0}=4$~GeV as in the figure, $\Delta_{T}(p,T)$ monotonically decreases with the temperature at every fixed momentum $|\vec{p}|\leq3$~GeV. We remark that, in order to obtain the curves in Fig.~\ref{fig:glupropftt}, as well as those in the figures that follow, the values of the additive constants $F_{0}^{T,L}(T)$ contained in Eq.~\eqref{cin714} were also fitted at each temperature from the lattice data.

\begin{figure}[H]
\centering
\includegraphics[width=0.45\textwidth,angle=270]{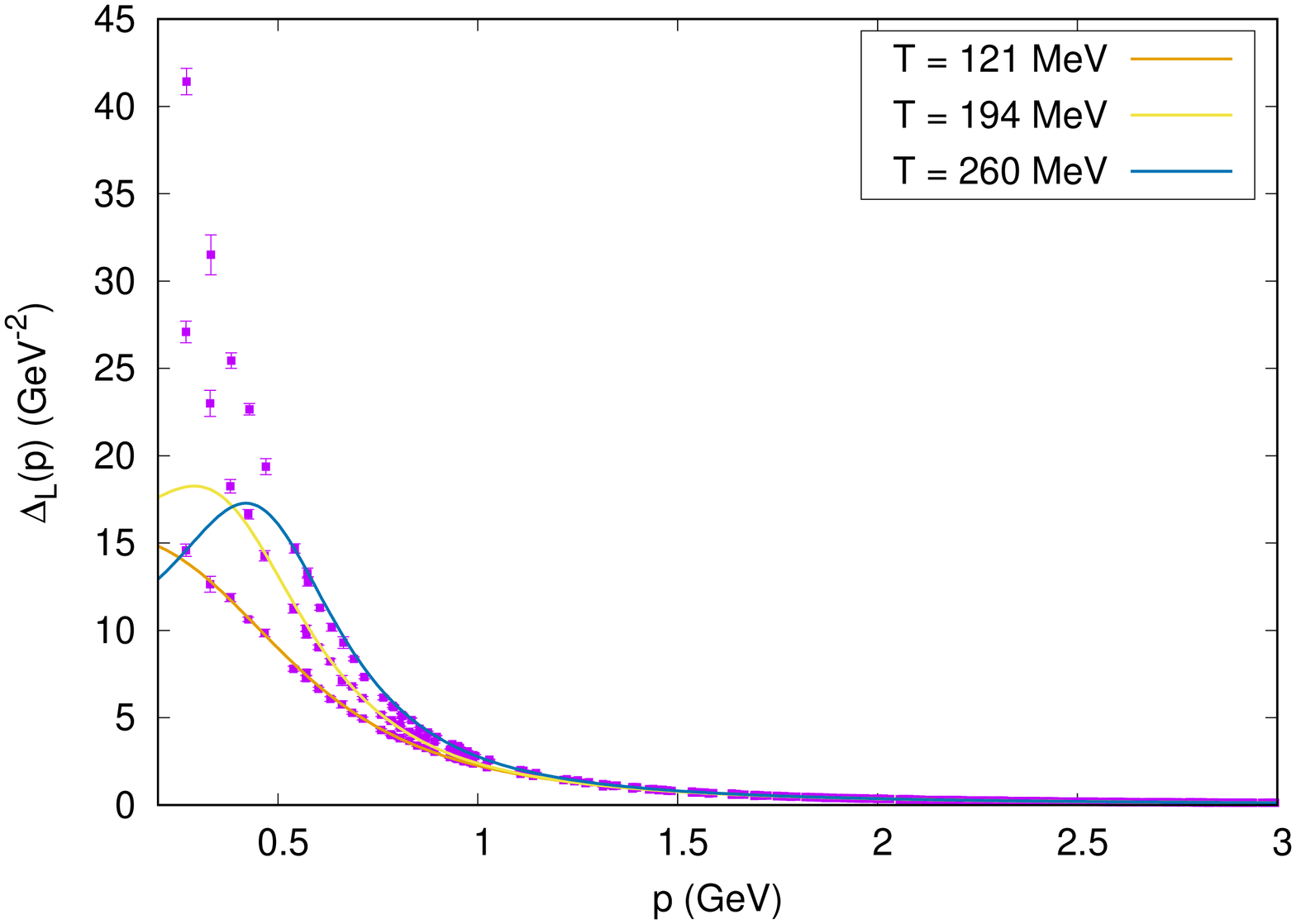}
\includegraphics[width=0.45\textwidth,angle=270]{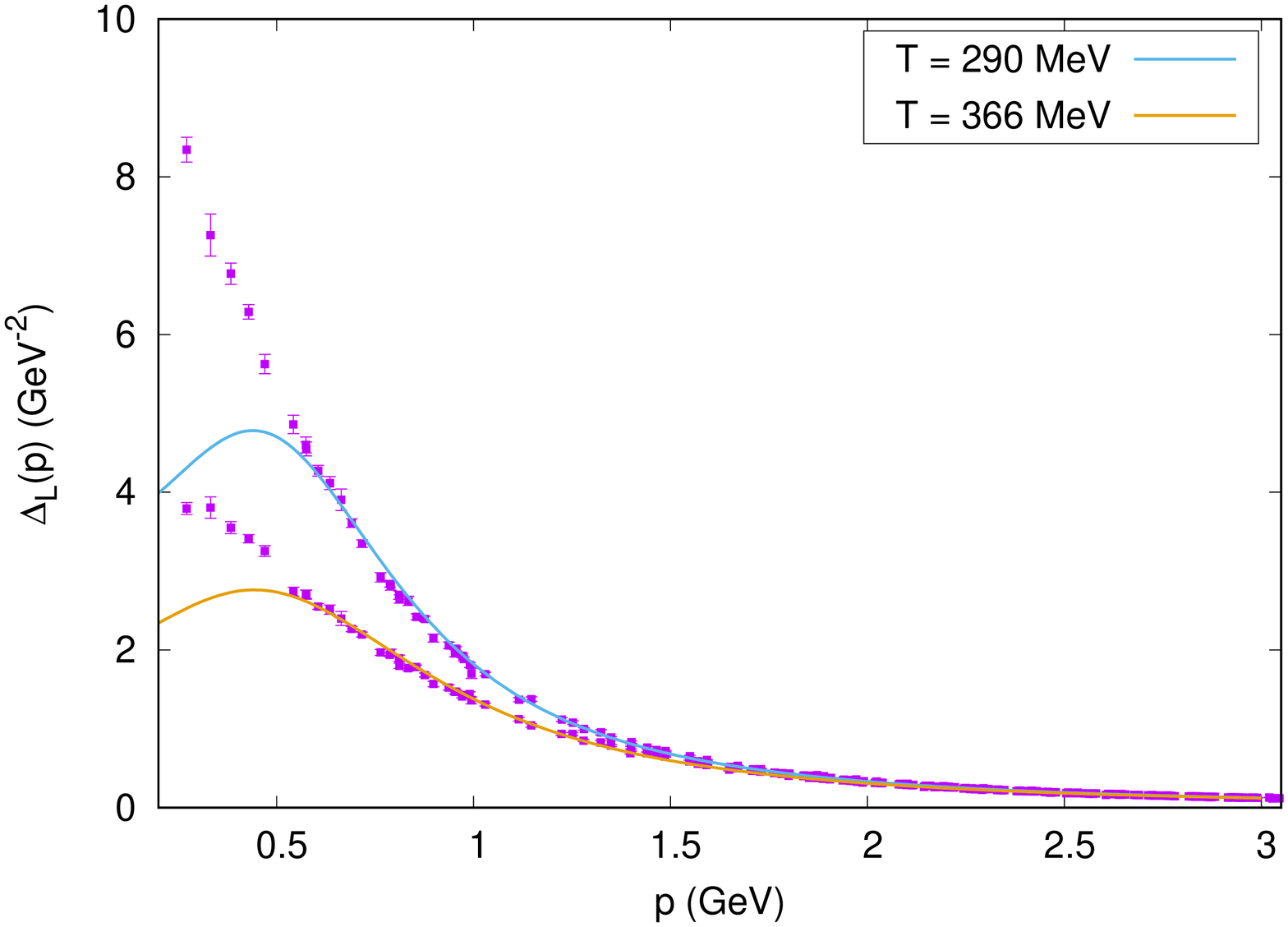}
\vspace{5pt}
\caption{Longitudinal component of the Landau-gauge Euclidean gluon propagator at zero Matsubara frequency ($n=0$) as a function of the modulus $|\vec{p}|$ of the $3$-dimensional momentum, for $T<T_{c}$ (top) and $T>T_{c}$ (bottom). Multiplicatively renormalized at $\mu_{0}=4$~GeV. Lattice data from \cite{SOBC14}.}\label{fig:glupropftl}
\end{figure}

In Fig.~\ref{fig:glupropftl} we display the longitudinal component of the Landau-gauge Euclidean gluon propagator at zero Matsubara frequency ($n=0$) as a function of the modulus $|\vec{p}|$ of the $3$-dimensional momentum. At variance with the transverse propagator, the longitudinal one -- when renormalized at $\mu_{0}=4$~GeV and at fixed momentum $|\vec{p}|$ -- is a non-monotonic function of the temperature. Specifically, $\Delta_{L}(p,T)$ first increases with the temperature, then attains a maximum at $T=T_{c}$, and finally it decreases at high temperatures just like its transverse counterpart. The critical temperature $T_{c}$ can indeed be defined as the point at which the longitudinal propagator changes its behavior. Interestingly, the change in behavior of $\Delta_{L}(p,T)$ at some fixed temperature is built into the SME propagator itself, as opposed to being an effect of our choice of free parameters \cite{SC21}.

As we can see, at low momenta and already for $T< T_{c}$, the SME longitudinal expression fails to reproduce the lattice data for all but the lowest value of temperature, $T=121$~MeV, the deviation from the lattice being larger around $T_{c}$. On the other hand, the agreement with the data is satisfactory for $|\vec{p}|\gtrsim0.5$~GeV.
\vspace{\fill}
\vspace{\fill}
\newpage
\vspace*{5mm}
\begin{figure}[H]
\centering
\includegraphics[width=0.45\textwidth,angle=270]{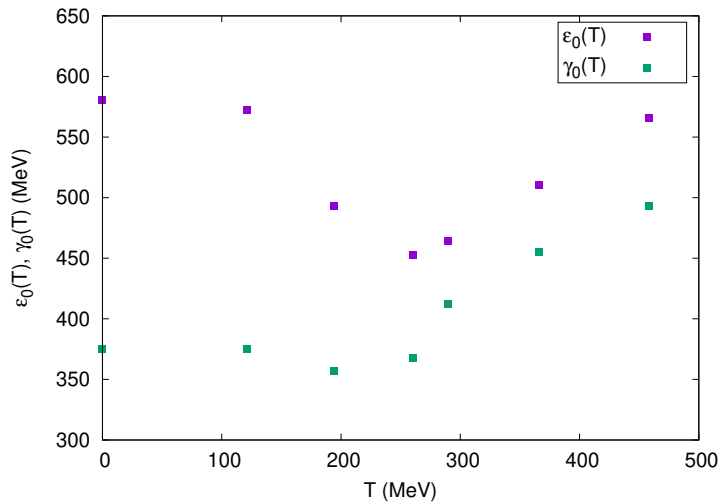}
\vspace{5pt}
\caption{Real ($\varepsilon_{0}(T)$) and imaginary part ($\gamma_{0}(T)$) of the gluon pole at vanishing spatial momentum $\vec{p}=0$ as functions of the temperature, obtained from the transverse component of the Landau-gauge gluon propagator.}\label{fig:glupropftpoles}
\end{figure}
\vspace{5mm}
Since the transverse component of the SME gluon propagator was found to be in good quantitative agreement with the lattice data down to vanishingly small spatial momenta, the results we just presented can be exploited to compute the position of the pole of the gluon propagator at $\vec{p}=0$ -- that is, the pole mass of the gluon -- as a function of the temperature\footnote{A similar calculation was carried out in \cite{SIR17b} under the assumption that the gluon mass parameter and additive renormalization constants can be approximated by their zero-temperature value at $T\neq 0$.}. Recall that, for $\vec{p}=0$ and $p^{4}\neq 0$, the two components of the gluon propagator coincide. Therefore, assuming that the determination of the gluon mass parameter $m(T)$ carried out by fitting the transverse propagator (Fig.~\ref{fig:glupropftmassT}) remains valid at non-zero Matsubara frequencies, what follows can in principle be interpreted to hold for both the components of the propagator.

By extending the imaginary-time component $p^{4}$ of the Euclidean momentum from the set of Matsubara frequencies $p^{4}\in2\pi T\, \Bbb{N}$ to the whole complex plane $p^{4}\in \Bbb{C}$, the equation $\Delta_{T}^{-1}(p^{4},\vec{p}=\vec{0},T)=0$ can be solved for $p^{4}=-i(\varepsilon_{0}(T)-i\gamma_{0}(T))$ at $m=m(T)$ -- the latter having been obtained by fitting the transverse data -- to yield the temperature-dependent mass $\varepsilon_{0}(T)$ and the zero-momentum damping factor $\gamma_{0}(T)$ of the gluon quasi-particles \cite{SC21}. The solutions are shown in Fig.~\ref{fig:glupropftpoles}. At $T=0$, the gluon poles are found at the position already reported in Sec.~\ref{sec:glghopt}, namely
\begin{equation}
\varepsilon_{0}(T=0)=\pm0.5810\ \text{GeV},\qquad\qquad\gamma_{0}(T=0)=\pm0.3751\ \text{GeV}\ .
\end{equation}
As the temperature increases, the mass $\varepsilon_{0}(T)$ first decreases to about $450$~MeV at $T=T_{c}$, and then starts to increase again, linearly with the temperature, for $T>T_{c}$. A similar pattern is followed by the damping rate, with some uncertainty on the temperature at which $\gamma_{0}(T)$ becomes an increasing function of the temperature\footnote{$\gamma_{0}(T)$ apparently starts to increase at a temperature $T<T_{c}$, rather than at $T=T_{c}$. However, this is most probably caused by the uncertainty that affects the underlying parameters, which is about $\pm50$~MeV for the mass parameter.} after having reached its minimum value of about $350$~MeV. The linear increase of the pole mass and zero-momentum damping rate found at high temperatures is typical of massless particles \cite{KG06}. An estimate of the full dispersion relations -- that is, of the gluon energy $\varepsilon_{|\vec{p}|}(T)$ and damping rate $\gamma_{|\vec{p}|}(T)$ as a function of the spatial momentum $|\vec{p}|$ -- at finite temperature can be found in \cite{SC21}.\\

The results presented in this section paint the picture of gluon quasi-particles which behave as massive below the critical temperature, and then become massless above $T=T_{c}$, where $\varepsilon_{0}(T)$ and $\gamma_{0}(T)$ can be interpreted, respectively, as a thermal mass and a thermal zero-momentum damping rate. Since $T_{c}\approx270$~MeV is the temperature at which the deconfinement phase transition occurs, we see that the SME predicts a strong correlation between dynamical mass generation and confinement in the gauge sector.

As for the accuracy of the Screened Massive Expansion at finite temperature, we saw that tuning the gluon mass parameter separately for the transverse and longitudinal components of the gluon propagator is able to provide an effective description of the dynamics of the gluons which is in good agreement with the lattice data at large spatial momenta, while being insufficient at small momenta as far as the longitudinal component is concerned. A proposal for an extension of the SME that would take into account the different behavior of the longitudinal and transverse gluon mass was advanced in \cite{SC21}.

\section{The Screened Massive Expansion of full QCD}
\renewcommand{\rightmark}{\thesection\ \ \ The Screened Massive Expansion of full QCD}
\label{sec:smeqk}

\subsection{Dynamical mass generation in the quark sector}
\label{sec:smeqkdmg}

It is a well-known fact that, due to the strong interactions, at low energies the quarks acquire a mass which for the lightest flavors is much larger than the quark mass parameter present in the Lagrangian. Such a phenomenon can be interpreted as a remnant of the violation of chiral symmetry that occurs in the limit in which the quarks are massless. In what follows, we give a brief introduction to this topic.\\

As we saw in Sec.~\ref{sec:operevres}, the non-vanishing of the quark condensate $\avg{\psibar\psi}$ triggers the dynamical generation of a mass for the massless quarks. $\avg{\psibar\psi}$ is a gauge-invariant quantity which, in the presence of chiral symmetry, would be forbidden to have a non-zero value. Indeed, if chiral transformations\footnote{We recall that the fifth gamma matrix $\gamma^{5}$ is defined as $\gamma^{5}=i\gamma^{0}\gamma^{1}\gamma^{2}\gamma^{3}$ and has vanishing anticommutation relations with the other gamma matrices, $\{\gamma^{\mu},\gamma^{5}\}=0$. Also, $\gamma^{5}=\mc{P}_{R}-\mc{P}_{L}$, where $\mc{P}_{L,R}$ are the projectors onto the left- and right-handed components of the Dirac field.},
\begin{equation}
\psi\to e^{i\alpha\gamma^{5}}\psi\ ,\qquad\qquad\psibar\to\psibar e^{i\alpha\gamma^{5}}\qquad\quad(\alpha\in\Bbb{R})\ ,
\end{equation}
were a symmetry of the vacuum, then the VEV of the quark operator $\psibar\psi$, which is not invariant under such transformations since
\begin{equation}
\psibar\psi=\psi_{L}^{\dagger}\psi_{R}+\psi_{R}^{\dagger}\psi_{L}\to\psibar e^{2i\alpha\gamma_{5}}\psi=e^{2i\alpha}\psi_{L}^{\dagger}\psi_{R}+e^{-2i\alpha}\psi_{R}^{\dagger}\psi_{L}\ ,
\end{equation}
would necessarily vanish. On the other hand, the fact that $\avg{\psibar\psi}\neq0$ signals the violation of chiral symmetry.

Symmetry under chiral transformations, which in the limit of massless quarks leave the classical quark Lagrangian invariant given that, for every quark flavor,
\begin{equation}
e^{i\alpha\gamma^{5}}\gamma^{\mu}e^{i\alpha\gamma^{5}}=\gamma^{\mu}e^{-i\alpha\gamma^{5}}e^{i\alpha\gamma^{5}}=\gamma^{\mu}\quad\quad\Longrightarrow\quad\quad\psibar e^{i\alpha\gamma^{5}}\,i\slashed{D}e^{i\alpha\gamma^{5}}\psi=\psibar\, i\slashed{D}\psi\ ,
\end{equation}
can be violated for a multitude of reasons. First of all, it may happen that the symmetry is \textit{anomalous}, in the sense that it cannot be realized at the level of the quantum theory. The\cleannlnp most renowned example of an anomalous chiral symmetry is the U(1) axial symmetry of full QCD, under which the up ($u$), down ($d$) and strange ($s$) quark fields -- taken to be massless -- transform as
\begin{equation}
u\to e^{i\alpha\gamma^{5}}u\ ,\qquad\qquad d\to e^{i\alpha\gamma^{5}}d\ ,\qquad\qquad s\to e^{i\alpha\gamma^{5}}s\ .
\end{equation}
The divergence of the corresponding current,
\begin{equation}
j^{\mu}_{5}=\overline{u}\gamma^{\mu}\gamma^{5}u+\overline{d}\gamma^{\mu}\gamma^{5}d+\overline{s}\gamma^{\mu}\gamma^{5}s\ ,
\end{equation}
instead of vanishing as would be suggested by the classical field equations, acquires a term which is quadratic in the gluon field-strength tensor,
\begin{equation}
\partial_{\mu}j^{\mu}_{5}=\frac{3N\alpha_{s}}{8\pi}\,\epsilon^{\mu\nu\sigma\tau}F^{a}_{\mu\nu}F^{a}_{\sigma\tau}\ ,
\end{equation}
implying that the symmetry is violated at the quantum level.

Second, it may happen that, while being a symmetry of the quantum theory, the vacuum state of the latter is not invariant under the chiral transformation. As a consequence, the symmetry is spontaneously broken in the vacuum. An example of a spontaneously broken chiral symmetry is the SU($n_{f}$) axial symmetry -- with $n_{f}$ the number of quarks --, which in the case $n_{f}=3$ acts on the up, down and quark fields as
\begin{equation}
\begin{pmatrix}u\\d\\s\end{pmatrix}\to e^{i\alpha^{A}T_{A}\gamma^{5}}\begin{pmatrix}u\\d\\s\end{pmatrix}\ ,
\end{equation}
where the $T_{A}$'s ($A=1,\dots,8$) are Gell-Mann matrices which generate the flavor SU(3) transformations. The currents associated to this symmetry,
\begin{equation}
j^{\mu}_{5A}=\overline{Q}\gamma^{\mu}T_{A}Q\ ,\qquad\qquad Q=\begin{pmatrix}u\\d\\s\end{pmatrix}\ ,
\end{equation}
are indeed conserved, $\partial_{\mu}j^{\mu}_{A}=0$; nonetheless, the vacuum state of QCD is not invariant under such transformations, resulting in chiral symmetry breaking (CSB) and in the existence of massless pseudo-scalar Goldstone bosons, namely -- in the limit of massless quarks -- the pions, kaons and lightest eta mesons.

Finally, chiral symmetry can be explicitly broken by the presence of quark mass terms $\psibar M \psi$ in the Lagrangian, which, as we saw earlier, are not invariant under chiral transformations. This is what happens in real-world QCD, where the pions, kaons and lightest eta mesons, instead of being massless as a consequence of CSB, posses masses that range from $135$~MeV to $548$~MeV.

Despite chiral symmetry not being an exact symmetry of full QCD, the mechanisms at play in chiral symmetry violation (either through anomalies or through chiral symmetry breaking) still cause an enhancement of the quark masses in the infrared. For the light quarks, this enhancement makes up for the majority of the mass with which they propagate at zero momentum; for instance, quarks with a Lagrangian mass of a few MeV turn out to propagate with masses $\approx300$-$400$~MeV in the $p\to 0$ limit. Such a phenomenon cannot be described in the framework of ordinary perturbation theory for two main reasons. First of all, the radiative pQCD corrections which contribute to the dressing of the quark masses are far too small to increase the values of the latter by orders of magnitude at energy scales in which the strong running coupling is well behaved. Second of all, the breakdown of ordinary perturbation theory in the infrared prevents us from computing the momentum-dependent quark masses at low energies, where their enhancement occurs.\newpage

The main objective of the following sections is to discuss how a shift of the quark Lagrangian similar to the one performed in Chpt.~\ref{chpt:sme} in the gluon sector of pure Yang-Mills theory can be exploited in full QCD to describe the phenomenon of dynamical mass generation in the quark sector, with a focus on the light quarks.

\subsection{The massive shift of the quark Lagrangian}

Recall that the quark Lagrangian $\mc{L}_{q}$ that appears in the Faddeev-Popov action of full QCD can be expressed in terms of the bare fields, coupling and quark mass as
\begin{equation}
\mc{L}_{q}=\psibar_{B}(i\slashed{\partial}-M_{B}+g_{B}\,\gamma^{\mu}A_{B\,\mu}^{a}T_{a})\psi_{B}\ .
\end{equation}
Given that, in what follows, we will not get into the specifics of renormalization -- which is discussed at length in \cite{CRBS21} --, we can assume that all the quantities in the above equation are renormalized and finite. In particular, we will denote the Lagrangian quark mass with a subscript $R$ and call it the \textit{renormalized mass}, in order to distinguish it from the mass developed by the quark in the infrared regime, whose scale is set by a parameter $M$ which we will refer to as the \textit{chiral mass}. The expression from which we will start for our study of the quark sector is then
\begin{equation}
\mc{L}_{q}=\psibar(i\slashed{\partial}-M_{R}+g\,\gamma^{\mu}A_{\mu}^{a}T_{a})\psi\ .
\end{equation}

In the previous section, we noted that the infrared enhancement of the light quark's mass caused by chiral symmetry violation cannot be described in ordinary perturbation theory, since the latter is unable to modify the value of any parameter by orders of magnitude via the radiative corrections. From a mathematical perspective, this translates to the fact that the scale $M$ of the infrared quark mass cannot be computed from the light quark's renormalized mass $M_{R}$ by truncating the ordinary perturbative series to any finite order in the coupling constant. Therefore, when setting up perturbation theory in the quark sector, we find ourself in a similar situation to that which occurs in the gluon sector.

If the aim of the perturbative expansion is to describe the low-energy dynamics of the light quarks, we should expect that a more accurate approximation of the exact results would be obtained by expanding around the infrared quark mass $M$, rather than around the renormalized value $M_{R}$. This can be achieved by splitting the quark Lagrangian $\mc{L}_{q}$ as
\begin{equation}\label{flz394}
\mc{L}_{q}=\mc{L}_{q,0}+\mc{L}_{q,\text{int}}
\end{equation}
with
\begin{equation}
\mc{L}_{q,0}=\psibar(i\slashed{\partial}-M)\psi\ ,\qquad\qquad \mc{L}_{q,\text{int}}=\psibar(g\slashed{A}^{a}T_{a}+M-M_{R})\psi\ ,
\end{equation}
and by using $\mc{L}_{q,0}$ as the order zero of the perturbative series and $\mc{L}_{q,\text{int}}$ as the interaction Lagrangian. Note that the split does not modify the quark Lagrangian $\mc{L}_{q}$ as a whole. It is understood that the expansion point for the gluon propagator must be chosen massive as in Chpt.~\ref{chpt:sme} in order to capture the correct infrared behavior of the gluons in addition to that of the quarks.

As a consequence of the split in Eq.~\eqref{flz394}, the zero-order quark propagator $S_{M}(p)$ reads
\begin{equation}
S_{M}(p)=\frac{i}{\slashed{p}-M}\ ,
\end{equation}
with the chiral mass $M$ replacing the renormalized mass $M_{R}$. Furthermore, two new quark two-point vertices arise in the interaction action:
\begin{equation}
S_{q,\text{int}}=-i\int d^{4}x\ \psibar(ig\slashed{A}^{a}T_{a}+\delta\Gamma_{q,1}+\delta\Gamma_{q,2})\psi\ ,
\end{equation}\cleannlnp
where
\begin{equation}\label{qmcts}
\delta\Gamma_{q,1}=iM,\qquad\qquad \delta\Gamma_{q,2}=-iM_{R}\ .
\end{equation}
The latter essentially play the same role as the gluon mass counterterm, and will thus be referred to as the quark mass counterterms.
\vspace{5mm}
\begin{figure}[H]
\centering
\includegraphics[width=0.85\textwidth]{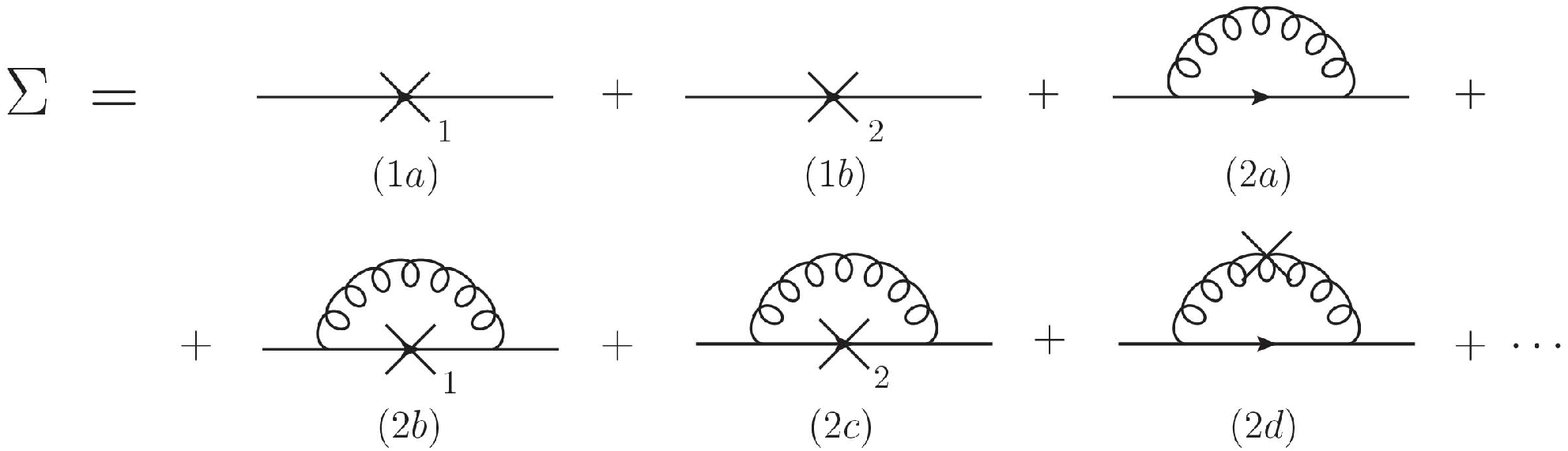}
\vspace{5pt}
\caption{Diagrams for the one-loop SME quark self-energy}\label{fig:quarkpoldiag}
\end{figure}
\vspace{5mm}
The dressed quark propagator $S(p)$ computed via Eq.~\eqref{flz394} can be expressed in terms of the quark self-energy $\Sigma(p)$ as
\begin{equation}\label{vsl592}
S(p)=\frac{i}{\slashed{p}-M-\Sigma(p)}\ .
\end{equation}
The quark mass counterterms in Eq.~\eqref{qmcts} contribute to $\Sigma(p)$ already at tree level via the diagrams (1a) and (1b) displayed in Fig.~\ref{fig:quarkpoldiag}; it is easy to see that the quark self-energy $\Sigma(p)$ can be put in the form
\begin{equation}\label{sigmaq1}
\Sigma(p)=-M+M_{R}+\Sigma^{(\text{loop})}(p)\ ,
\end{equation}
where the first two terms are provided by $\delta\Gamma_{q,1/2}$, whereas $\Sigma^{(\text{loop})}(p)$ is the contribution coming from the loops of the expansion. Since
\begin{equation}
S(p)=\frac{i}{\slashed{p}-M_{R}-\Sigma^{(\text{loop})}(p)}\ ,
\end{equation}
it is clear that, in the framework of the Screened Massive Expansion, the infrared enhancement of the quark mass, if any, will originate non-trivially from the loops, and not from the tree-level chiral mass present in Eq.~\eqref{vsl592}, just like in the gluon sector the gluon mass is generated by the loops and not by the tree-level gluon mass term $m^{2}$.

By defining functions $A(p^{2})$ and $B(p^{2})$ such that
\begin{align}\label{AandB}
\notag A(p^{2})&=1-\Sigma_{V}(p^{2})\ ,\\
B(p^{2})&=M_{R}+\Sigma_{S}(p^{2})\ ,
\end{align}
where $\Sigma_{V}(p^{2})$ and $\Sigma_{S}(p^{2})$ are, respectively, the vector and the scalar component of $\Sigma^{(\text{loop})}(p)$,
\begin{equation}
\Sigma^{(\text{loop})}(p)=\slashed{p}\ \Sigma_{V}(p^{2})+\Sigma_{S}(p^{2})\ ,
\end{equation}
the dressed quark propagator $S(p)$ can be put in the form
\begin{equation}\label{fvl402}
S(p)=\frac{iZ(p^{2})}{\slashed{p}-\mc{M}(p^{2})}\ ,
\end{equation}
where the functions $Z(p^{2})$ and $\mc{M}(p^{2})$, given by
\begin{equation}\label{ZandM}
Z(p^{2})=\frac{1}{A(p^{2})}\ ,\qquad\qquad \mc{M}(p^{2})=\frac{B(p^{2})}{A(p^{2})}\ ,
\end{equation}
are known respectively as the quark $Z$-function and mass function. By comparing Eq.~\eqref{fvl402} with the general form of the zero-order quark propagator, we see that $\mc{M}(p^{2})$ plays the role of the momentum-dependent mass of the quark. In particular, $\mc{M}(p^{2})$ determines the position of the poles of the quark propagator, given that
\begin{equation}
\frac{1}{\slashed{p}-\mc{M}(p^{2})}=\frac{\slashed{p}+\mc{M}(p^{2})}{p^{2}-\mc{M}^{2}(p^{2})}=\infty\qquad\Longleftrightarrow\qquad p^{2}=\mc{M}^{2}(p^{2}).
\end{equation}

In the zero-momentum limit, the quark mass function approaches the value
\begin{equation}
\mc{M}(0)=\frac{M_{R}+\Sigma_{S}(0)}{1-\Sigma_{V}(0)}\ .
\end{equation}
Therefore, if the components of the quark self-energy do not vanish at $p=0$, $\Sigma_{V}(0),\Sigma_{S}(0)\neq0$, the zero-momentum mass of the quark will be different from its renormalized mass, $\mc{M}(0)\neq M_{R}$. As we will see, this indeed turns out to be the case. In the framework of the Screened Massive Expansion, the energy scale for the difference between $\mc{M}(0)$ and $M_{R}$ will be set by $M$, and not by $M_{R}$, since as a consequence of the shift of the quark action it is the chiral mass $M$ that runs into the quark loops. By this mechanism, the infrared enhancement of the quark mass is made possible in the SME.

To end this section, we observe that -- at variance with ordinary pQCD -- the relation $\mc{M}(0)\neq M_{R}$ also holds within the SME in the chiral limit ($M_{R}=0$). In particular, the SME quark propagator can acquire a mass even when no mass term is present in the Lagrangian, as discussed in Secs.~\ref{sec:operevres} and~\ref{sec:smeqkdmg}. The chiral limit of the Screened Massive Expansion was investigated in \cite{Sir16a}, where the first results for full QCD were presented\footnote{We should mention that \cite{Sir16a} made use of lattice data for the quark mass function whose large errors -- customary in the context of unquenched calculations in the chiral limit -- only allowed for a qualitative comparison with the results of the SME.}.

\subsection{The quark propagator in the Landau gauge}

To compute the SME quark propagator, we follow the prescriptions laid down in Sec.~\ref{sec:genpropsme} regarding the number of mass counterterms to retain at a given order in perturbation theory. In Secs.~\ref{sec:ghprsme} and \ref{sec:glprsme} the ghost and gluon propagator were expanded to one loop and three vertices; in the quark sector we will do the same, working within the truncation scheme that in \cite{CRBS21} was termed the vertex-wise scheme. Additionally, we will present some results obtained in the so-called complex-conjugate scheme, to be illustrated later on. A third truncation scheme -- termed the minimalistic scheme -- was investigated in \cite{CRBS21} and will not be reported in what follows. The results presented in this section were obtained in the Landau gauge.

The diagrams with at most one loop and at most three vertices which contribute to the SME quark self-energy are displayed in Fig.~\ref{fig:quarkpoldiag}. In the figure, the crosses denoted with $1$ and $2$ are the quark mass counterterms $\delta\Gamma_{q,1}$ and $\delta\Gamma_{q,2}$, whereas the unlabeled cross is the gluon mass counterterm $\delta\Gamma_{\mu\nu}^{ab}$ -- Eq.~\eqref{glumct}. As we saw in Sec.~\ref{sec:smedef}, diagrams with one gluon mass counterterm can be computed from the corresponding uncrossed diagrams by differentiating the latter with respect to the gluon mass parameter. Similarly, since\newpage
\begin{equation}
\frac{i}{\slashed{p}-M}(i\lambda)\frac{i}{\slashed{p}-M}=-\lambda\frac{\partial}{\partial M}\,\frac{i}{\slashed{p}-M}\ ,
\end{equation}
where $\lambda=M,-M_{R}$, diagrams with one quark mass counterterm can be computed from the uncrossed diagrams by differentiating them with respect to the quark chiral mass $M$. In particular, we find that
\begin{align}
\notag\Sigma^{(2b)}(p)&=-M\frac{\partial}{\partial M}\,\Sigma^{(2a)}(p)\ ,\\
\Sigma^{(2c)}(p)&=M_{R}\frac{\partial}{\partial M}\,\Sigma^{(2a)}(p)\ ,\\
\notag\Sigma^{(2d)}(p)&=-m^{2}\frac{\partial}{\partial m^{2}}\,\Sigma^{(2a)}(p)\ ,
\end{align}
where $\Sigma^{(2a,2b,2c,2d)}(p)$ denote the contributions to $\Sigma(p)$ due to diagrams (2a) to (2d).

Diagram (2a) contains a mass divergence proportional to the quark chiral mass $M$ which has no counterpart in ordinary perturbation theory -- the mass divergences of the latter being proportional to $M_{R}$. Diagram (2b) cancels such a divergence by a mechanism analogous to the one discussed in Sec.~\ref{sec:glprsme} when addressing the gluon mass divergences. On the other hand, diagram (2c) contains the ordinary mass divergence $\propto M_{R}$, which only contributes to the renormalization of $M_{R}$ itself\footnote{More details on this can be found in \cite{CRBS21}.}. Finally, diagram (2d) is convergent.  Since in the Landau gauge none of the one-loop diagrams contain vector divergences -- that is, divergences proportional to the Dirac matrix $\slashed{p}$ --, the sum (2a+2b+2c+2d) only contains a divergence proportional to $M_{R}$, which is eliminated by renormalizing $M_{R}$.

As for the finite parts of diagrams (2b) and (2c), since
\begin{equation}
\Sigma^{(2b)}(p)+\Sigma^{(2c)}(p)=-(M-M_{R})\frac{\partial}{\partial M}\,\Sigma^{(2a)}(p)\ ,
\end{equation}
we see that for the light quarks -- $M_{R}\ll M$ -- we have $|\Sigma^{(2c)}(p)|\ll|\Sigma^{(2b)}(p)|$. Therefore, in what follows we will neglect\footnote{We checked that this does not substantially affect the results.} the finite part of $\Sigma^{(2c)}(p)$ \cite{CRBS21}.

An explicit calculation carried out in Euclidean space yields\footnote{In what follows we omit the subscripts $E$ from quantities defined in Euclidean space.}
\begin{equation}
\Sigma_{V}(p^{2})=\frac{\alpha_{s}}{3\pi}\ \sigma_{V}(p^{2})\ ,\qquad\Sigma_{S}(p^{2})=\frac{\alpha_{s}}{\pi}\ \sigma_{S}(p^{2})\ ,
\end{equation}
for the vector and the scalar component of the quark self-energy, where the functions $\sigma_{V}(p^{2})$ and $\sigma_{S}(p^{2})$ are reported in \cite{CRBS21}. In terms of these, the Euclidean quark $Z$- and mass function can be expressed as
\begin{equation}\label{gne049}
Z(p^{2})=\frac{1}{Z_{\psi}-\frac{\alpha_{s}}{3\pi}\ \sigma_{V}(p^{2})}\ ,\qquad \mc{M}(p^{2})=\frac{M_{R}+\frac{\alpha_{s}}{\pi}\,\sigma_{S}(p^{2})}{Z_{\psi}-\frac{\alpha_{s}}{3\pi}\ \sigma_{V}(p^{2})}\ ,
\end{equation}
where for completeness in the above equations we have reinstated the quark field renormalization factor $Z_{\psi}$. We remark that, because of the absence of vector divergences in the Landau-gauge one-loop quark self-energy, $Z_{\psi}$ is a finite quantity. By defining constants
\begin{equation}\label{h0k0def}
h_{0} =\frac{3\pi}{\alpha_{s}}\,Z_{\psi}\ ,\qquad\qquad k_{0}=\frac{\pi}{\alpha_{s}}\, M_{R}\ ,\qquad\qquad Z_{S}=\frac{3\pi}{\alpha_{s}}\ ,
\end{equation}
$Z(p^{2})$ and $\mc{M}(p^{2})$ can be put in the form
\begin{equation}\label{massk0h0}
Z(p^{2})=\frac{Z_{S}}{h_{0}-\sigma_{V}(p^{2})}\ ,\qquad\qquad\mc{M}(p^{2})=\frac{3[k_{0}+\sigma_{S}(p^{2})]}{h_{0}-\sigma_{V}(p^{2})}\ .
\end{equation}
The constant $Z_{S}$ can be fixed by renormalizing $Z(p^{2})$ in the MOM scheme,
\begin{equation}
Z(\mu^{2})=1
\end{equation}
at the initial renormalization scale $\mu$.
\vspace{5mm}
\begin{table}[H]
\setlength{\tabcolsep}{10pt}
\def\arraystretch{1.5}
\centering
\begin{tabular}{c||c|c|c}
$M_{\text{lat}}$ (MeV)&$M$ (MeV)&$h_{0}$&$k_{0}$ (MeV)\\
\hline
\hline
$18$&$268.0$&$2.656$&$-16.9$\\
$18^{\star}$&$197.6$&$2.051$&$6.8$\\
$36$&$228.7$&$2.418$&$11.5$\\
$54$&$221.4$&$2.577$&$40.0$\\
$72$&$238.4$&$2.977$&$70.1$\\
$90$&$249.0$&$3.207$&$102.5$
\end{tabular}
\caption{Parameters obtained by fitting the lattice data of \cite{KBLW05} for the Landau-gauge Euclidean quark mass function. The asterisked row was obtained by fixing $k_{0}=6.8$ and fitting the remaining parameters (see the text for details). $m=0.6557$~GeV.}\label{quarkfitvs}
\end{table}
\vspace{5mm}
\begin{table}[H]
\setlength{\tabcolsep}{10pt}
\def\arraystretch{1.5}
\centering
\begin{tabular}{c||c|c|c}
$M_{\text{lat}}$ (MeV)&$\alpha_{s}$&$M_{R}$ (MeV)&$P_{0}$ (MeV)\\
\hline
\hline
$18$&$2.605$&$-14.0$&$\pm387.4 \pm 180.9 i$\\
$18^{\star}$&$3.128$&$6.8$&$\pm349.2 \pm 193.1i$\\
$36$&$2.788$&$10.2$&$\pm371.7 \pm 185.4 i$\\
$54$&$2.663$&$33.9$&$\pm375.2 \pm 177.2 i$\\
$72$&$2.393$&$53.4$&$\pm392.9 \pm 167.6 i$\\
$90$&$2.261$&$73.8$&$\pm410.8 \pm 170.2 i$
\end{tabular}
\caption{Coupling constant, renormalized mass and quark poles corresponding to the parameters in Tab.~\ref{quarkfitvs}.}\label{quarkfitvsder}
\end{table}
\vspace{5mm}
In \cite{CRBS21}, the Landau-gauge Euclidean quark propagator computed in the Screened Massive Expansion was compared with the quenched lattice data of \cite{KBLW05} for quarks with masses $M_{\text{lat}}$ ranging from $18$~MeV to $271$~MeV \footnote{Roughly speaking, $M_{\text{lat}}$ is the mass that appears at tree level in the bare lattice quark propagator. See \cite{KBLW05} for more details.}. In what follows, we report our results up to $M_{\text{lat}}=90$~MeV, for which the condition $M_{R}\ll M$ is better fulfilled. The constants $h_{0}$ and $k_{0}$ that appear in Eq.~\eqref{massk0h0} were fitted from the data for the quark mass function $\mc{M}(p^{2})$ together with the chiral mass $M$ for each separate value of $M_{\text{lat}}$. As for the value of the gluon mass parameter $m^{2}$, we used $m=0.6557$~GeV, like in Sec.~\ref{sec:smeopt}. Albeit having been determined in pure Yang-Mills theory, this value is appropriate for comparisons with quenched lattice data, since the latter do not take into account the corrections to the gluon propagator due to the interactions with the quarks \cite{CRBS21}.

The outcome of the fit is reported in Tab.~\ref{quarkfitvs}. One may notice that the value of the constant $k_{0}$ obtained by fitting the $M_{\text{lat}}=18$~MeV dataset (first row) is negative, in contradiction with the definition given in Eq.~\eqref{h0k0def}, which for $M_{R},\alpha_{s}\geq 0$ would necessarily require $k_{0}\geq 0$. This issue was investigated in depth in \cite{CRBS21}, where it was noted that at high energies the $M_{\text{lat}}=18$~MeV data of \cite{KBLW05} are plagued by large oscillations -- presumably caused by discretization errors -- which make it very difficult to unambiguously extract a value of $k_{0}$ from the lattice results. As an alternative, in Tab.~\ref{quarkfitvs} we record a second determination of the $M_{\text{lat}}=18$~MeV parameters (second row, asterisked), which was obtained by fixing $k_{0}=6.8$~MeV \footnote{More details can be found in \cite{CRBS21} for the reasoning that leads to this (non-unique) choice.}, a value which -- modulo oscillations -- is still able to reproduce the lattice data with good precision.

In Tab.~\ref{quarkfitvsder}, we report the values of the coupling constant $\alpha_{s}$ and of the renormalized mass $M_{R}$ and the position of the poles of the quark propagator corresponding to the parameters in Tab.~\ref{quarkfitvs}. By Eqs.~\eqref{gne049}, \eqref{h0k0def} and \eqref{massk0h0}, $\alpha_{s}$ and $M_{R}$ can be computed from $h_{0}$ and $k_{0}$ via
\begin{equation}
\alpha_{s}=3\pi[h_{0}-\sigma_{V}(\mu^{2})]^{-1}\ ,\qquad\qquad M_{R}=\frac{3k_{0}}{h_{0}}\ ,
\end{equation}
where the renormalization scale $\mu$ was fixed to $4$~GeV. As discussed in \cite{CRBS21}, one should be careful when interpreting $\alpha_{s}$ with the actual value of the coupling constant at the scale $\mu$, mainly because the former is defined starting from the (renormalization of the) function $Z(p^{2})$, which to one loop -- as we will see in a moment -- is not well-behaved in the SME. The position of the quark poles is found by first solving the equation
\begin{equation}
p^{2}+\mc{M}^{2}(p^{2})=0
\end{equation}
in the complexified Euclidean space, and then converting the solutions to Minkowski space.
\vspace{\fill}
\begin{figure}[H]
\centering
\includegraphics[width=0.45\textwidth,angle=270]{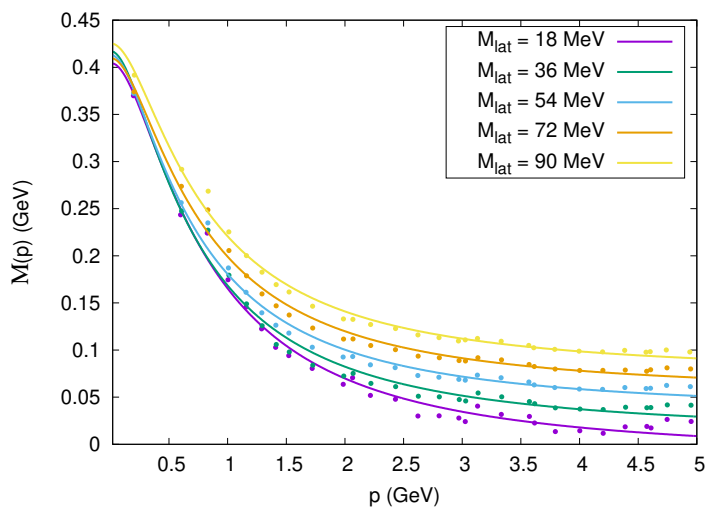}
\vspace{5pt}
\caption{Euclidean quark mass function in the Landau gauge ($\xi = 0$) for different values of the quark mass. Colored curves: SME with $h_{0}$, $k_{0}$ and $M$ fitted from the lattice data (Tab.~\ref{quarkfitvs}). Dots: lattice data from \cite{KBLW05}. $m=0.6557$~GeV.}\label{fig:quarkmassfunc}
\end{figure}
\vspace{\fill}
\newpage
In Fig.~\ref{fig:quarkmassfunc} we display the SME quark mass functions $\mc{M}(p^{2})$ computed by making use of the parameters in Tab.~\ref{quarkfitvs}, together with the lattice data of \cite{KBLW05}. As we can see, at zero momentum the quarks develop a mass $\mc{M}(0)\approx400$~MeV which is much larger than the UV limit of $\mc{M}(p^{2})$. While this infrared enhancement could not have been described within ordinary perturbation theory, the Screened Massive Expansion succeeds in reproducing the correct behavior of the mass function, thanks to the latter being dependent on two mass scales: a renormalized mass $M_{R}$, whose value is relevant to the UV regime, and a ``chiral'' mass $M$, which sets the scale for $\mc{M}(0)$. In doing so, it uses values of the coupling which are not too far from that of the SME running coupling at its maximum (Sec.~\ref{sec:smerg}). Notably, the value of the zero-momentum mass is nearly the same for all quarks with lattice masses up to $90$~MeV, confirming that in the deep infrared the origin of the quark masses is almost entirely accounted for by chiral symmetry violation. On the other hand, $M_{R}$ increases monotonically with $M_{\text{lat}}$ and is of the same order as the value of $\mc{M}(p^{2})$ evaluated at high energies. As for the chiral mass $M$, we find that its value also increases with $M_{\text{lat}}$, indicating that $M$ should be interpreted as a mass scale appropriate to the quark under examination, despite it yielding a value of $\mc{M}(0)$ which is independent from $M_{\text{lat}}$.

The quark poles calculated from the fitted mass functions are complex-conjugated (see Tab.~\ref{quarkfitvsder}), like in the gluon sector. This can be regarded both as evidence for quark confinement and as a hint that the quarks may be confined by a similar underlying mechanism to that which is at play for the gluons. The real and the imaginary part of the quark poles for $M_{\text{lat}}\leq 90$~MeV are found in the range $[349,411]$~MeV and $[170,193]$~MeV, respectively, with the former increasing with $M_{\text{lat}}$ and the latter slightly decreasing. In particular, we see that the energy scale for the poles is set by the chiral mass, rather than by $M_{R}$.
\vspace{5mm}
\begin{figure}[H]
\centering
\includegraphics[width=0.45\textwidth,angle=270]{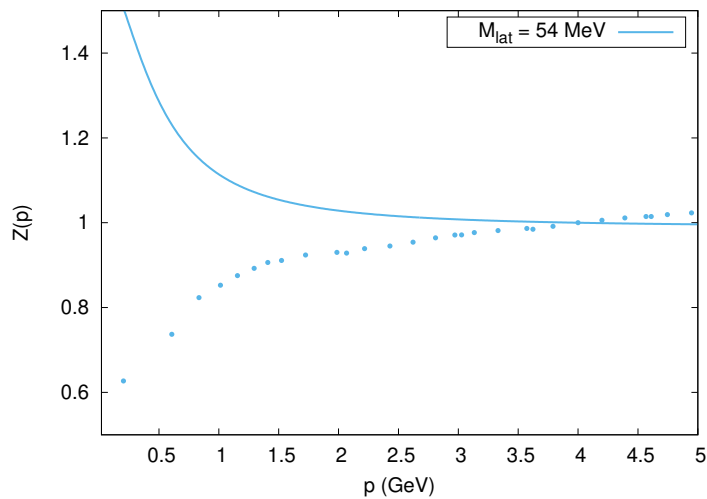}
\vspace{5pt}
\caption{Euclidean quark $Z$-function in the Landau gauge ($\xi = 0$) for $M_{\text{lat}}=54$~MeV. Blue curve: SME with $h_{0}$, $k_{0}$ and $M$ fitted from the lattice data (Tab.~\ref{quarkfitvs}). Dots: lattice data from \cite{KBLW05}. $m=0.6557$~GeV.}\label{fig:quarkzfunc}
\end{figure}
\vspace{5mm}
In Fig.~\ref{fig:quarkzfunc} we show the SME quark $Z$-function computed by the parameters in Tab.~\ref{quarkfitvs} for the single value $M_{\text{lat}}=54$~MeV, again with the lattice data of \cite{KBLW05}; as the results obtained for the other quark masses are similar to those in the figure, we will not display them in what follows. Clearly, the Screened Massive Expansion is not able to reproduce the correct behavior of $Z(p^{2})$ to one loop, yielding a function which decreases with momentum, instead of increasing as is found on the lattice. This is most probably due to the fact that -- as first argued by \cite{PTW14} -- the one-loop corrections to the vector component of the Landau-gauge quark self-energy are unusually small, even vanishing at $m^{2}=0$, so that the two-loop corrections must necessarily be taken into account. And indeed, in the framework of the Curci-Ferrari model, it was shown \cite{BGPR21} that the behavior of the $Z$-function can be fixed by going to two loops. Within the Screened Massive Expansion, similar knowledge is gained by computing the quark propagator in an alternative truncation scheme which in \cite{CRBS21} was termed the complex-conjugate (CC) scheme.

In the CC scheme, the one-loop quark self-energy is calculated by replacing the zero-order gluon propagator with the principal part of the dressed propagator in the internal gluon lines of the Feynman diagrams,
\begin{equation}\label{bla940}
\Delta_{m}(p^{2})=\frac{1}{p^{2}+m^{2}}\to\frac{1}{2\text{Re}\{R\}}\left[\frac{R}{p^{2}+p_{0}^{2}}+\frac{\overline{R}}{p^{2}+\overline{p_{0}^{2}}}\right]\ ,
\end{equation}
where $p_{0}^{2}$, $\overline{p_{0}^{2}}$, $R$ and $\overline{R}$ are, respectively, the poles and the residues of the optimized gluon propagator of Sec.~\ref{sec:glghopt}, and the normalization of the principal part is chosen so that the latter equals $1/p^{2}$ at high energies. As we saw in Sec.~\ref{sec:glghopt}, the principal part of the gluon propagator yields a good approximation of the full propagator provided that the former is multiplied by a constant. Since in the self-energy the zero-order gluon propagator always appears multiplied by the coupling constant $\alpha_{s}$, changing the normalization of the gluon propagator amounts to a redefinition of the coupling. Therefore, the replacement defined by Eq.~\eqref{bla940} allows us to compute the quark propagator in an approximation which takes into account the radiative corrections to the gluon propagator. While not being equivalent to a full two-loop calculation, the truncation provided by the CC scheme still implicitly includes contributions from the higher orders in the gluon perturbative series.
\vspace{5mm}
\begin{figure}[H]
\centering
\includegraphics[width=0.45\textwidth,angle=270]{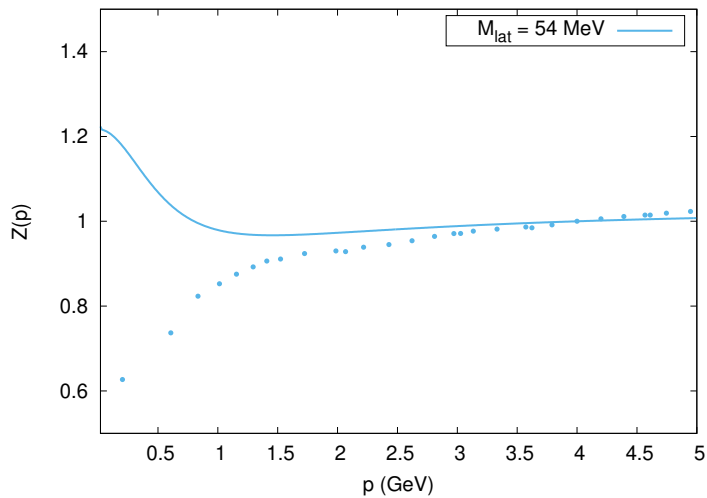}
\vspace{5pt}
\caption{Euclidean quark $Z$-function in the Landau gauge ($\xi = 0$) for $M_{\text{lat}}=54$~MeV. Blue curve: CC-scheme SME with $h_{0}$, $k_{0}$ and $M$ fitted from the lattice data. Dots: lattice data from \cite{KBLW05}.}\label{fig:quarkzfunccc}
\end{figure}
\vspace{5mm}
The SME $Z$-function computed in the CC scheme is shown in Fig.~\ref{fig:quarkzfunccc}, again for $M_{\text{lat}}=54$~MeV. As we can see, the higher-order corrections to the gluon propagator change the high-energy behavior of $Z(p^{2})$, turning it into an increasing function of momentum -- as it should be. Nonetheless, the quantitative agreement with the lattice data is still not good in the UV; moreover, at low energies, the CC-scheme $Z$-function still shows the incorrect behavior, indicating that a full two-loop calculation is required in order for the SME to be able to reproduce the lattice data. Further details on the CC scheme can be found in \cite{CRBS21}.

\subsection{Conclusions}

Despite not being as theoretically developed as in the gluon sector, lacking optimization procedures which would allow us to compute the parameters that appear in the quark propagator from first principles, the Screened Massive Expansion of full QCD still provides us with an accurate picture of dynamical mass generation in the quark sector. By shifting the expansion point for the quark perturbative series, the SME is able to incorporate a parameter $M$ -- the chiral mass -- into the expressions, that acts as the energy scale for the mass $\mc{M}(0)$ with which the quarks propagate at zero momentum.

In the framework of the Screened Massive Expansion, the infrared enhancement of the quark masses originates from the radiative corrections to the propagator, as expected for a truly dynamical phenomenon. The Euclidean mass functions $\mc{M}(p^{2})$ computed for the light quarks are found to be in good agreement with the quenched lattice data over a wide range of (lattice) masses, displaying saturation at $\mc{M}(0)\approx400$~MeV and a UV tail dominated by the value of the renormalized mass $M_{R}$. As a result of $\mc{M}(0)$ being much larger than $M_{R}$ for the light quarks, the latter have poles which, instead of being real and of the order of the renormalized mass, are complex-conjugate and of the order of $\mc{M}(0)$, giving evidence for quark confinement. These conclusions are expected to apply also to the chiral limit ($M_{R}=0$), although only preliminary results are available at present \cite{SIR16b}.

Due to the smallness of the $O(g^{2})$ corrections to the Landau-gauge self-energy, the standard one-loop truncation of the Screened Massive Expansion is inadequate for the calculation of the Landau-gauge quark $Z$-function. An alternative truncation of the quark propagator that takes into account the higher-order corrections to the gluon propagator can be shown to fix the high-energy qualitative behavior of $Z(p^{2})$, although neither the quantitative agreement with the lattice data, nor the low-energy limit of the $Z$-function substantially improved by employing such a method.

\chapter{The Dynamical Model}
\renewcommand{\leftmark}{\thechapter\ \ \ The Dynamical Model}
\renewcommand{\rightmark}{\thechapter\ \ \ The Dynamical Model}
\label{chpt:dynmod}

While the Screened Massive Expansion presented in Chpts.~\ref{chpt:sme} and \ref{chpt:smeapp} is able to account for dynamical mass generation in the gluon sector from first principles, the actual mechanism that triggers such a phenomenon is far from made clear by its formalism. Indeed, the shift that defines the SME -- motivated by the GEP favoring a massive gluon vacuum over the massless one of ordinary perturbation theory -- incorporates the gluon mass into the equations of QCD without explaining where the mass term in the Lagrangian originates from in the first place. A second drawback of the Screened Massive Expansion is the absence of a rigorous prescription for the truncation of its perturbative series at a given loop order. In Sec.~\ref{sec:genpropsme}, the minimum number of mass counterterms to include in the Feynman diagrams was fixed by the necessary requirement that the resulting Green functions be renormalizable. Their maximum number, on the other hand, was decided on principles of minimality which -- albeit certainly meaningful -- are perhaps not as strong as one would like.

The main objective of this chapter is to present a second perturbative framework for the computation of the Green functions of QCD at low energies, termed the Dynamical Model (DM). The Dynamical Model addresses both of the aforementioned issues moving from the hypothesis that a gauge-invariant, quadratic gluon condensate $\avg{(A^{h})^{2}}$ may be the root cause of DMG in the gauge sector. By making use of local composite operator (LCO) methods, it demonstrates that a mass term for the gluons is generated in the QCD Lagrangian as a consequence of the non-vanishing VEV of $(A^{h})^{2}$. The resulting Landau-gauge perturbative series has features that place it half way in between the Curci-Ferrari Model and the Screened Massive Expansion: with the former it has in common the explicit form of the one-loop gluon polarization and ghost self-energy, whereas with the latter it shares the fact that the gluon mass is generated beyond the tree level, from the loops of the expansion. At variance with the SME, the Dynamical Model can be formulated in such a way that the number of diagrams in the expansion is finite at fixed order in the coupling constant $g^{2}$.

The idea that DMG might be realized via gluon condensation is not at all new; as we saw in the \hyperref[chpt:intro]{Introduction}, the gluon mass has often been studied in the literature in terms of the dimension-$4$ condensate $\avg{F^{2}}$ \cite{Cor82,DL89,JA90,Lav91,LW96}. The possibility that a dimension-$2$ condensate of the form $\avg{A^{2}}$ \cite{CNZ99,BLLM01,GSZ01,GZ01} may contribute to mass generation (see Sec.~\ref{sec:operevres}), on the other hand, was explored only more recently, the main reason for this being that, since $A^{2}$ is not a gauge-invariant operator, the non-vanishing of its VEV could be interpreted as an indication that gauge symmetry is broken in the vacuum. Nonetheless, it was soon realized \cite{CDGL05,CDFG15} that a suitable generalization of the Landau-gauge operator $A^{2}$ could yield a gauge-invariant and BRST-invariant condensate with the right dimensions for playing the role of a gluon mass parameter $m^{2}$. Such a generalization was achieved by defining a gauge-invariant gluon field, customarily denoted by $A^{h}$ in the literature, which reduces to the ordinary gluon field $A$ in the Landau gauge.

The field $A^{h}$ as a tool to systematically investigate the infrared behavior of pure Yang-Mills theory was first introduced in the context of the Gribov-Zwanziger model \cite{CDFG15,CDFG16a,CDFG16b,CDPF17,CFPS17,CDGP18,MPPS19,DFPR19} with the objective of extending the framework of Gribov and Zwanziger to arbitrary covariant gauges while complying with BRST symmetry. The study of $A^{h}$ and its quadratic condensate in ordinary pure Yang-Mills theory -- that is, neglecting the issue of the Gribov copies --, on the other hand, was undertaken in \cite{CFGM16,CVPG18} and more recently in \cite{Dem20}, where the first preliminary expressions for the Landau-gauge gluon and the ghost propagators in the presence of a non-vanishing $\avg{(A^{h})^{2}}$ condensate were derived\footnote{See also \cite{DVRV22} for an application of the method to the study of the deconfinement phase transition.}.\\

In the present chapter we take up where \cite{Dem20} left off and discuss the generation of the infrared gluon mass within the framework of the Dynamical Model. As we will see, the inclusion of the $\avg{(A^{h})^{2}}$ condensate in the action of pure Yang-Mills theory causes the gluons to propagate as massive at tree level. The resulting Feynman diagrams contain internal gluon lines whose mass parameter $m^{2}\propto\avg{(A^{h})^{2}}$ sets the scale for the non-vanishing of the gluon polarization at zero momentum, yielding a propagator which -- as we will explicitly show in the Landau gauge -- saturates to a finite non-zero constant at $p=0$. In other words, a dynamical mass is generated for the gluons in the deep infrared within the Dynamical Model. The condensate $\avg{(A^{h})^{2}}$ itself -- and $m^{2}$ with it -- can be computed from first principles by solving a gap equation derived from a suitably defined effective potential.

Our results will be shown to be in good agreement with the lattice data of \cite{DOS16} over a wide range of momenta, once improved by Renormalization Group methods. The latter will also allow us to study the behavior of the Taylor running coupling computed in the Landau gauge, which in the Dynamical Model turns out to be free of Landau poles. The relevance of such a feature was already discussed in the framework of the Screened Massive Expansion, and does not need to be restated here.

This chapter is organized as follows. In Sec.~\ref{sec:dynmodcond} we will define the field $A^{h}$, derive an effective potential for its quadratic condensate and solve the corresponding gap equation. In the process of doing so, we will show that, on the shell of the gap equation, the Faddeev-Popov action $S_{\text{FP}}$ is dynamically equivalent to a second action $I$ whose explicit expression will be reported in Sec.~\ref{sec:dynmodpotcalc}. The action $I$ defines the Dynamical Model, which is the subject of Sec.~\ref{sec:dynmodprop}. In the latter, we will compute the Landau-gauge gluon and ghost propagators, perform their Renormalization Group improvement and compare our results with the lattice data of \cite{DOS16}. Finally, in Sec.~\ref{sec:dynmodconcl} we present our conclusions.

Most of the material presented in Sec.~\ref{sec:dynmodcond} and in Secs.~\ref{sec:dynmodrenfr}-\ref{sec:dynmodgpgse} is a review of the theory already developed in \cite{CFGM16,CVPG18,Dem20}. The main contribution of this thesis to the research on the Dynamical Model is the formulation of a new renormalization scheme -- the Dynamically Infrared-Safe (DIS) scheme, Sec.~\ref{sec:dynmodrgimp} -- in which the Landau-gauge propagators are well behaved and are found to reproduce well the lattice data (Sec.~\ref{sec:dynmodlat}). The results of Secs.~\ref{sec:dynmodrgimp} and \ref{sec:dynmodlat} are part of ongoing work in collaboration with D. Dudal, \textit{et al.}.
\newpage

\section{The BRST-invariant quadratic gluon condensate}
\renewcommand{\rightmark}{\thesection\ \ \ The BRST-invariant quadratic gluon condensate}
\label{sec:dynmodcond}

\subsection{A note on the conventions}

For the rest of this chapter, we will slightly change our formalism in order to conform to the conventions used in \cite{CFGM16,CVPG18,Dem20}. In the latter, like in Sec.~\ref{sec:cfrev}, all of the calculations are carried out in Euclidean space, where the Faddeev-Popov action $S_{\text{FP}}$ can be expressed as
\begin{equation}\label{gls502}
S_{\text{FP}}=\int d^{4}x\left(\frac{1}{4}F^{a}_{\mu\nu}F^{a\,\mu\nu}+\frac{\alpha}{2}\,B^{a}B^{a}+iB^{a}\partial^{\mu}A^{a}_{\mu}+\bar{c}^{a}\partial^{\mu}D_{\mu}c^{a}\right)\ ,
\end{equation}
with the Nakanishi-Lautrup field $B^{a}$ replaced by $iB^{a}$ and the ghost field $c^{a}$ replaced by $-c^{a}$. Because of these replacements, the BRST transformations under which $S_{\text{FP}}$ is invariant are given by
\begin{equation}
sA^{a}_{\mu}=-D_{\mu}c^{a}\ ,\qquad sc^{a}=\frac{g}{2}\,f^{abc}c^{b}c^{c}\ ,\qquad s\bar{c}^{a}=iB^{a}\ ,\qquad sB^{a}=0\ .
\end{equation}
In Eq.~\eqref{gls502} the symbol $\alpha$ is used in place of $\xi$ \footnote{The symbol $\xi$ will be reused for a new dynamical field, which will be defined in the next section.} to denote the gauge parameter.

As for $A^{U}$ -- that is, the field which is obtained by applying a gauge transformation $U$ to the gauge field $A$ --, we will adopt the definition
\begin{equation}\label{flm304}
A^{U}_{\mu}=U^{\dagger}\left(A_{\mu}+\frac{i}{g}\,\partial_{\mu}\right) U
\end{equation}
instead of one in which $U$ and $U^{\dagger}$ are interchanged. With this convention, the consecutive application of two gauge transformations $U_{1}$ and $U_{2}$ reads
\begin{equation}\label{vla941}
(A^{U_{1}})^{U_{2}}=A^{(U_{1}U_{2})}\ .
\end{equation}

\subsection[The field $A^{h}$ and its quadratic condensate]{The field $\boldsymbol{A^{h}}$ and its quadratic condensate}
\label{sec:brstinvcond}

Consider the functional $f_{A}[U]$ defined as \cite{CDGL05}
\begin{equation}
f_{A}[U]=\text{Tr}\left\{\int d^{4}x\ A^{U}\cdot A^{U}\right\}\ ,
\end{equation}
where the dot product denotes a contraction of the spacetime indices of the $A^{U}$'s and the trace is taken over the product of the SU(N) generators. A straightforward calculation \cite{CDGL05} shows that $f_{A}[U]$ is minimized by the transformation $U=h[A]$ which enforces the divergence equation
\begin{equation}\label{lsm391}
\partial^{\mu}A^{h[A]}_{\mu}=0\ .
\end{equation}
The resulting field $A^{h[A]}$, which for simplicity will also be denoted by $A^{h}$ in what follows, has the important property of being gauge invariant. This is a consequence of the identity
\begin{equation}
h[A^{U}]=U^{\dagger}h[A]\ ,
\end{equation}
which can be derived from Eqs.~\eqref{vla941} and \eqref{lsm391} by observing that
\begin{equation}
\partial^{\mu}A^{h[A]}_{\mu}=0\qquad\Longleftrightarrow\qquad \partial^{\mu}[(A^{U}_{\mu})^{U^{\dagger}h[A]}]=0\ ,
\end{equation}\cleannlnp
and implies the transformation law
\begin{equation}
A^{h[A]}\to (A^{U})^{h[A^{U}]}=A^{h[A]}\ .
\end{equation}

While the general solution of Eq.~\eqref{lsm391} is not known in closed form, an explicit perturbative expression for $A^{h}$ can still be obtained by defining a set of $A$-dependent fields $\xi^{a}[A]$ such that, with $\xi=\xi^{a}T_{a}$,
\begin{equation}
h[A]=e^{ig\xi[A]}\ .
\end{equation}
By plugging the latter into Eq.~\eqref{lsm391} and solving the equation order by order in $g$, one finds that \cite{CDGL05}
\begin{equation}
\xi[A]=\frac{\partial\cdot A}{\partial^{2}}+i\,\frac{g}{\partial^{2}}\left[\partial\cdot A,\frac{\partial\cdot A}{\partial^{2}}\right]+i\,\frac{g}{\partial^{2}}\left[A_{\mu },\partial^{\mu}\,\frac{\partial\cdot A}{\partial^{2}}\right]+\frac{i}{2}\frac{g}{\partial^{2}}\left[ \frac{\partial\cdot A}{\partial^{2}},\partial\cdot A\right]+\cdots\ ;
\end{equation}
it follows that the field $A^{h}$ can be expressed as
\begin{equation}\label{dkx392}
A^{h}_{\mu}=\left(\delta_{\mu\nu}-\frac{\partial_{\mu}\partial_{\nu}}{\partial^{2}}\right)\phi^{\nu}[A]\ ,
\end{equation}
where the vector field $\phi_{\mu}[A]$ is given by
\begin{equation}\label{vks941}
\phi_{\mu}[A] =A_{\mu}-ig\left[ \frac{\partial\cdot A}{\partial ^{2}},A_{\mu}\right]+\frac{ig}{2}\left[\frac{\partial\cdot A}{\partial ^{2}},\partial_{\mu}\frac{\partial\cdot A}{\partial ^{2}}\right]+\cdots\ .
\end{equation}
Eqs.~\eqref{dkx392} and \eqref{vks941} explicitly exhibit $A^{h}$ as a divergenceless version of the gauge field $A$, the difference $A^{h}-A$ being a non-local functional of the divergence $\partial\cdot A$. It is clear from the above expressions that $A^{h}$ and $A$ perturbatively coincide in the Landau gauge ($\alpha=0$, $\partial\cdot A=0$): if $\partial\cdot A=0$, then
\begin{equation}
\phi_{\mu}=A_{\mu}\quad\Longrightarrow\quad A_{\mu}^{h}=A_{\mu}\ .
\end{equation}

In the quantum context, the gauge invariance of $A^{h}$ translates into an invariance under BRST symmetry. Since $(A^{h})^{2}=(A^{h})^{a}\cdot (A^{h})^{a}$ is clearly also BRST invariant, its vacuum expectation value $\avg{(A^{h})^{2}}$ is not constrained to vanish in the covariant gauges under the assumption that BRST symmetry is not broken in the vacuum. In particular, it makes sense to ask whether such a condensate indeed exists and -- if it does -- what the consequences of its existence are on the vacuum structure of QCD.

The VEV of the quadratic operator $(A^{h})^{2}$ can be studied by making use of the local composite operator (LCO) formalism \cite{Ver95,VKVV01,DVS03}. Within the latter, a current $J$ is coupled to the operator $(A^{h})^{2}$ by introducing two extra terms in the Faddeev-Popov action,
\begin{equation}
S_{\text{FP}}\to S^{(1)}[J]\ ,
\end{equation}
where
\begin{equation}
S^{(1)}[J]=S_{\text{FP}}+\int d^{4}x\,\left(\frac{J}{2}\,(A^{h})^{2}-\frac{\zeta}{2}\,J^{2}\right)\ .
\end{equation}
In the above equation, the term quadratic in the current is introduced in order to ensure the renormalizability of the partition function $Z[J]$ \cite{VKVV01}, which, in Euclidean space, reads
\begin{equation}\label{hma945}
Z[J]=e^{-W[J]}=\int\mc{D}\mathscr{F}\ e^{-S^{(1)}[J]}\qquad\quad(\mc{D}\mathscr{F}=\mc{D}A\,\mc{D}\cbar\,\mc{D}c\,\mc{D}B)\ .
\end{equation}
$Z[J]$ can be differentiated with respect to the current $J$ to yield
\begin{equation}\label{dmd582}
\frac{\delta W}{\delta J}[J]=\frac{1}{2}\,\avg{(A^{h})^{2}}_{J}-\zeta J=\sigma[J]\ ;
\end{equation}
when evaluated at $J=0$, the last expression is equal to the BRST-invariant condensate $\sigma[J=0]=\frac{1}{2}\avg{(A^{h})^{2}}_{J=0}$.

An effective action $\Gamma[\sigma]$ for the condensate can be defined starting from $W[J]$. As in Sec.~\ref{sec:partfunct}, $\Gamma[\sigma]$ is obtained by first inverting the functional $\sigma[J]$ to a $J[\sigma]$, which we will simply denote by $J_{\sigma}$,
\begin{equation}
\frac{\delta W}{\delta J}[J_{\sigma}]=\frac{1}{2}\,\avg{(A^{h})^{2}}_{J_{\sigma}}-\zeta J_{\sigma}=\sigma\ ,
\end{equation}
and then by computing the Legendre transform of $W[J]$,
\begin{equation}\label{vkj293}
\Gamma[\sigma]=W[J_{\sigma}]-\int d^{4}x\ J_{\sigma}\,\sigma\ .
\end{equation}
The on-shell value of the BRST-invariant condensate is then obtained by differentiating $\Gamma[\sigma]$ with respect to $\sigma$,
\begin{equation}
\frac{\delta\Gamma}{\delta \sigma}[\sigma]=-J_{\sigma}\ ,
\end{equation}
and setting the derivative to zero:
\begin{equation}
\frac{\delta\Gamma}{\delta \sigma}[\sigma]=0\qquad\Longleftrightarrow\qquad \sigma=\frac{1}{2}\,\avg{(A^{h})^{2}}_{J=0}\ .
\end{equation}
The calculation of the effective action will be carried out explicitly in the next section.

\subsection{Calculation of the condensate's effective action}
\label{sec:dynmodpotcalc}

In order to compute the effective action $\Gamma[\sigma]$ -- Eq.~\eqref{vkj293} -- in a general covariant gauge, we first need to address two unusual properties of the action $S^{(1)}$ that appears in Eq.~\eqref{hma945}. These are the non-locality of the operator $A^{h}$ and the quadratic dependence of $S^{(1)}$ on both the current $J$ and the field $A^{h}$. Let us start from the first one.\\

As we saw in the last section, $A^{h}$ can be computed as a power series in the coupling $g$ whose zero-order term is the transverse projection $A^{T}$ of the field $A$,
\begin{equation}
A^{T}_{\mu}=\left(\delta_{\mu\nu}-\frac{\partial_{\mu}\partial_{\nu}}{\partial^{2}}\right)A^{\nu}\ ,
\end{equation}
and the remaining terms only depend on the divergence $\partial\cdot A$ of the gauge field. The presence of the operator $(\partial^{2})^{-1}$ both in the higher-order terms and in $A^{T}$ makes the series that defines $A^{h}$ highly non-local.

If we wish to localize the operator $A^{h}$, then a new dynamical field $\xi$ can be introduced in the FP action in such a way that $A^{h}$ will be expressed as a polynomial of infinite degree in $\xi$. This can be done as follows \cite{CVPG18}. First we insert a unity
\begin{equation}\label{dkx941}
1=\int\mc{D}F\ \delta(F)
\end{equation}
in the partition function $Z[J]$. Then we redefine $F$ to be a functional of a SU(N) algebra field $\xi=\xi^{a}T_{a}$, chosen so that $A^{h(\xi)}$ -- with $h(\xi)=e^{ig\xi}$~-- is divergenceless. In other words, we set
\begin{equation}
F=F[\xi]=\partial^{\mu}A_{\mu}^{h(\xi)}\ 
\end{equation}
and change variables of integration in Eq.~\eqref{dkx941} from $F$ to $\xi$,
\begin{equation}\label{fsz491}
1=\int\mc{D}\xi\ \text{det}\left(\frac{\delta F}{\delta \xi}\right)\ \delta(\partial^{\mu}A_{\mu}^{h(\xi)})\ .
\end{equation}
Under the sign of integral, $A^{h(\xi)}$ is clearly equal to the gauge-invariant field $A^{h}$ defined in the previous section. In particular, the $A^{h}$ which appears in the source term of $S^{(1)}$ can be replaced by $A^{h(\xi)}$ without spoiling the physical content of the partition function. In what follows, we will denote $h(\xi)$ simply by $h$, it being understood that the latter is $\xi$-dependent and equal to $e^{ig\xi}$.

With regard to the determinant and delta in Eq.~\eqref{fsz491}, these can be rewritten as a functional integral over new dynamical fields $\tau^{a}$ and $\eta^{a},\overline{\eta}^{a}$,
\begin{align}
\text{det}\left(\frac{\delta F}{\delta \xi}\right)\ \delta(\partial^{\mu}A_{\mu}^{h(\xi)})&=\text{det}\left(-\partial^{\mu}D_{\mu}(A^{h})\Lambda(\xi)\right)\ \delta(\partial^{\mu}A_{\mu}^{h(\xi)})=\\
\notag&=\mc{N}\int\mc{D}\tau\mc{D}\overline{\eta}\mc{D}\eta\ e^{-\Delta S_{1}}\ \text{det}\left(\Lambda(\xi)\right)\ .
\end{align}
In the above equation, $\mc{N}$ is an irrelevant constant, the action term $\Delta S_{1}$ reads
\begin{equation}\label{vmw940}
\Delta S_{1}=\int d^{4}x\, \left(\tau^{a}\partial^{\mu}A^{h,a}_{\mu}+ \overline{\eta}^a\partial^{\mu}D_{\mu}(A^h)\eta^{a} \right)\ ,
\end{equation}
where $D_{\mu}(A^{h})$ is the covariant derivative associated to the field $A^{h}$,
\begin{equation}
D_{\mu}^{ab}(A^{h})=\delta^{ab}\partial_{\mu}-gf^{abc}(A_{\mu}^{h})^{c}\ ,
\end{equation}
and $\Lambda(\xi)$ is defined as
\begin{equation}\label{eq:lambdadef}
\Lambda_{ab}(\xi)=\frac{2i}{g}\,\text{Tr}\left\{t_{a}\frac{\partial h^{\dagger}}{\partial \xi^{b}}h\right\}\ .
\end{equation}
The determinant $\text{det}(\Lambda(\xi))$ makes it first appearance in this thesis, having been neglected in \cite{CFGM16,CVPG18,Dem20} \footnote{\cite{CFGM16} also neglected the ghost term in Eq.~\eqref{vmw940}, which was introduced in \cite{CVPG18}.}. In Appendix~\ref{app:onedet}, we show that it decouples from the rest of the integral as long as the partition function is defined in dimensional regularization and the calculations are carried out perturbatively. Since we will be working under these hypotheses, the determinant will be dropped in what follows.

With the modifications we just discussed, the partition function $Z[J]$ reads
\begin{equation}\label{grt591}
Z[J]=\int\mc{D}\mathscr{F}\ e^{-S^{(2)}[J]}\ ,
\end{equation}
where the integration measure $\mc{D}\mathscr{F}$ is given by
\begin{equation}
\mc{D}\mathscr{F}=\mc{D}A\,\mc{D}\cbar\,\mc{D}c\,\mc{D}B\,\mc{D}\xi\,\mc{D}\tau\,\mc{D}\overline{\eta}\,\mc{D}\eta
\end{equation}
and in the action $S^{(2)}$,
\begin{equation}
S^{(2)}=S^{(1)}+\Delta S_{1}\ ,
\end{equation}
$A^{h}$ is to be expanded perturbatively,
\begin{equation}
A^{h}_{\mu}=A_{\mu}-\partial_{\mu}\xi+ig[A_{\mu},\xi]+\cdots\ .
\end{equation}\cleannlnp
Crucially, $S^{(2)}$ is invariant under an extended BRST symmetry which acts on the new fields as \cite{CVPG18}
\begin{equation}\label{gao586}
s \tau^a=s\overline{\eta}^a=s\eta^a=0\ ,\qquad\qquad sh=-igc^aT_{a}h\ .
\end{equation}
The latter translates into a corresponding transformation for the field $\xi$, which to lowest order reads
\begin{equation}
s\xi^{a}=-c^{a}+\frac{g}{2}f^{a}_{bc}\,c^{b}\xi^{c}+O(g^{2})\ .
\end{equation}
This extended BRST transformation, easily seen to be nilpotent since $s^{2}h=0$, was exploited in \cite{CVPG18} to prove the renormalizability of the theory defined by the action $S^{(2)}$.\\

The effective action $\Gamma[F]$ associated to the elementary fields $\mathscr{F}$ is usually computed by shifting the latter as $\mathscr{F}\to F+\delta\mathscr{F}$, so as to factorize an exponential of the form $\exp(-\int JF)$ from the partition function. In terms of the functional $W[J]$, such an exponential translates into the integral $-\int JF$ that appears in the definition of $\Gamma[F]$. In our partition function, Eq.~\eqref{grt591}, the shift cannot be performed in terms of the elementary fields due to the fact that the condensate $A^{h}$ not only appears quadratically in the source term, but is also a complicated function of both the gauge field $A$ and the algebra field $\xi$. In order to overcome this obstacle, a new dynamical field $\sigma$ can be introduced in the action in such a way that its Green functions coincide with those of the condensate \cite{VKVV01}. In the process of doing so, we will also get rid of the action term $\propto J^{2}$, which could potentially obstruct the calculation of the effective action.

Consider what happens if we insert a unity of the form
\begin{equation}
1=\mc{N}\int\mc{D}\sigma\ e^{-\Delta S_{2}}\ 
\end{equation}
in the partition function $Z[J]$, where the action term $\Delta S_{2}$ is given by
\begin{equation}
\Delta S_{2}=\frac{1}{2\zeta}\int d^{4}x\,\left(\sigma-\frac{1}{2}\,(A^{h})^{2}+\zeta J\right)^{2}\ .
\end{equation}
If we define a new action $S^{(3)}$ as
\begin{equation}
S^{(3)}=S^{(2)}+\Delta S_{2}\ ,
\end{equation}
then it is easy to see, by shifting $\sigma\to \sigma+\frac{1}{2}\,(A^{h})^{2}-\zeta J$ in the partition function, that
\begin{equation}
\avg{\sigma}_{J}\Big|_{S^{(3)}}=\sigma[J]\Big|_{S^{(2)}}\ ,
\end{equation}
where the left-hand-side average is computed with respect to the action $S^{(3)}$, whereas $\sigma[J]$ is the condensate given by Eq.~\eqref{dmd582} -- computed with respect to $S^{(2)}$. Since for any value of the current $J$ the VEV of the dynamical field $\sigma$ is equal to the BRST-invariant condensate, the latter can be studied by making use of the modified partition function
\begin{equation}
Z[J]=\int\mc{D}\mathscr{F}\ e^{-S^{(3)}[J]}\ ,
\end{equation}
where the action $S^{(3)}$ explicitly reads \cite{VKVV01}
\begin{equation}
S^{(3)}=S_{\text{FP}}+\Delta S_{1}+\int d^{4}x\ \left\{J\sigma+\frac{1}{2\zeta}\,\sigma^{2}-\frac{1}{2\zeta}\,\sigma (A^{h})^{2}+\frac{1}{8\zeta}\, [(A^{h})^{2}]^{2}\right\}\ .
\end{equation}\cleannlnp
As we can see, both the quadratic coupling of $A^{h}$ to $J$ and the quadratic current term $\propto J^{2}$ have disappeared from the new action. Instead, $S^{(3)}$ is linear in $J$, which is itself linearly coupled to the field $\sigma$.

Thanks to these features, the effective action can be computed by shifting $\sigma\to\Sigma+\delta\sigma$, where $\Sigma$ is the value of the condensate and $\delta\sigma$ quantifies the fluctuations of the new field around its VEV. Explicitly,
\begin{equation}\label{rks934}
Z[J]=e^{-W[J]}=e^{-\int d^{4}x\,J\Sigma}\int\mc{D}\mathscr{F}\ e^{-S^{(4)}}\ ,
\end{equation}
where the measure of integration $\mc{D}\mathscr{F}$ is given by
\begin{equation}
\mc{D}\mathscr{F}=\mc{D}A\,\mc{D}\cbar\,\mc{D}c\,\mc{D}B\,\mc{D}\xi\,\mc{D}\tau\,\mc{D}\overline{\eta}\,\mc{D}\eta\,\mc{D}\delta\sigma
\end{equation}
and the action $S^{(4)}$ reads
\begin{equation}\label{sma395}
S^{(4)}=S_{\text{FP}}+\Delta S_{1}+\int d^{4}x\ \left\{J\delta\sigma+\frac{1}{2\zeta}\,(\Sigma+\delta\sigma)^{2}-\frac{1}{2\zeta}\,(\Sigma+\delta\sigma) (A^{h})^{2}+\frac{1}{8\zeta}\, [(A^{h})^{2}]^{2}\right\}\ .
\end{equation}
After renaming $\Sigma\to \sigma$, from Eq.~\eqref{rks934} we obtain the following expression for the effective action $\Gamma[\sigma]$:
\begin{equation}\label{kfm394}
\Gamma[\sigma]=W[J_{\sigma}]-\int d^{4}x\ J_{\sigma}\sigma=\frac{1}{2\zeta}\int d^{4}x\ \sigma^{2}-\ln\int_{\avg{\delta\sigma}=0}\mc{D}\mathscr{F}\ e^{-I}\ ,
\end{equation}
where the action $I$ is given by \cite{Dem20}
\begin{align}\label{dynmodact}
I&=S_{\text{FP}}+\int d^{4}x\, \left(\tau^{a}\partial^{\mu}A^{h,a}_{\mu}+ \overline{\eta}^a\partial^{\mu}D_{\mu}(A^h)\eta^{a} \right)+\\
\notag&\qquad\quad+\int d^{4}x\ \left\{\frac{1}{2\zeta}\,(\delta\sigma)^{2}-\frac{1}{2\zeta}\,(\sigma+\delta\sigma) (A^{h})^{2}+\frac{1}{8\zeta}\, [(A^{h})^{2}]^{2}\right\}
\end{align}
and the linear terms $\propto\sigma\delta\sigma$ and $\propto J\delta\sigma$ have been replaced in Eq.~\eqref{sma395} by the requirement that the VEV $\avg{\delta\sigma}$ be equal to zero, so that $\sigma$ is the true value of the condensate.

Let us analyze the action $I$. First of all, we see that -- with $s\delta\sigma=0$ and by Eq.~\eqref{gao586}~-- $I$ is BRST invariant. Second, we see that in $I$ the fluctuation field $\delta\sigma$ has a quadratic term which we can interpret as the kinetic term for a perturbative expansion of $\Gamma[\sigma]$. The corresponding zero-order propagator $D_{(\delta\sigma)}(p^{2})$ reads
\begin{equation}
D_{(\delta\sigma)}(p^{2})=\zeta\ .
\end{equation}
Third, the field $\delta\sigma$ is coupled to the BRST-invariant gauge field $A^{h}$ via a cubic interaction $\propto \delta\sigma(A^{h})^{2}$. Moreover, as a result of our manipulations, a new interaction quartic in $A^{h}$ arises. Both of these terms are proportional to $\zeta^{-1}$, which can thus be regarded as a small parameter for the set-up of perturbation theory\footnote{Later on we will see that, by employing a procedure known as the reduction of couplings, $\zeta^{-1}$ can be taken to be proportional to $g^{2}$ to lowest order in perturbation theory, thus confirming this statement.}. We should remark that these interaction terms actually incorporate an infinite number of vertices that couple the gauge field $A$ to the fluctuation field $\delta\sigma$, to the algebra field $\xi$ and to itself, given that $A^{h}$ must be expanded in powers of $g\xi$. Finally, a quadratic term in $A^{h}$ is present in $I$, proportional to the VEV $\sigma$. Just like the cubic and quartic interactions, this term needs to be expanded in powers of $g$. To lowest order,
\begin{equation}
-\frac{\sigma}{2\zeta} (A^{h})^{2}=-\frac{\sigma}{2\zeta}\, (A-\partial\xi)^{2}+O(g/\zeta)\ .
\end{equation}\cleannlnp
We then see that the quadratic interaction yields a mass term for the gluon field,
\begin{equation}
\delta\mc{L}=\frac{m^{2}}{2}\,A^{2}\ ,
\end{equation}
where the gluon mass parameter $m^{2}$ is given by\footnote{Observe that $\sigma=\frac{1}{2}\avg{(A^{h})^{2}}$ does not necessarily imply $\sigma>0$, since the VEV needs to be renormalized and its value can become negative in the process. The renormalization of the condensate will be discussed in Sec.~\ref{sec:dynmodrenfr}.}
\begin{equation}
m^{2}=-\frac{\sigma}{\zeta}\ .
\end{equation}

As long as $\sigma$ is computed on-shell -- that is, as long as $\sigma$ solves the effective action equation $\delta\Gamma/\delta\sigma=0$~--, the Green functions of pure Yang-Mills theory can be evaluated by making use of the action $I$ instead of $S_{\text{FP}}$. This holds true due to the fact that setting the derivative of the effective action to zero is equivalent to computing the action at vanishing external current, so that $S_{\text{FP}}$ is recovered in the partition function\footnote{We remark that the modifications of the action which were performed in the present section -- being the result of insertions of unities in $Z[J]$ -- do not change neither the partition function, nor the physical content of the theory.}. In particular, if the effective action equation yields a non-zero value for $\sigma$, then a formulation of pure Yang-Mills theory that uses $I$ as its defining action will treat the gluons as massive at tree level from first principles, without changing the content of the FP action. The experience gained thanks to the Curci-Ferrari model and to the Screened Massive Expansion tells us that this should be sufficient to lead to dynamical mass generation in the gluon sector. We will come back to this subject in Sec.~\ref{sec:dynmodprop}.\\

Having laid out the technique for calculating the effective action associated to the BRST-invariant condensate, we can now proceed to derive an explicit expression for the effective potential $V(\sigma)$, defined as
\begin{equation}
V(\sigma)=\Gamma[\sigma]/\mc{V}_{4}\ ,
\end{equation}
where $\mc{V}_{4}$ is the $4$-dimensional Euclidean volume and $\Gamma[\sigma]$ is evaluated at constant $\sigma$. Since $\sigma$ is BRST invariant, the effective potential can be computed in the Landau gauge ($\alpha=0$) without any loss of generality.

As we saw in the last section, in the Landau gauge the fields $A^{h}$ and $A$ coincide. In particular, for $\alpha=0$ it is not necessary to localize the field $A^{h}$, so that the steps that led to the introduction of the fields $\tau$ and $\eta,\overline{\eta}$ can be skipped; moreover, $h$ can be everywhere set to $\one$. In other words, in the Landau gauge we can use the action
\begin{equation}
I_{L}=S_{\text{FP}}\big|_{\alpha=0}+\int d^{4}x\ \left\{\frac{1}{2\zeta}\,(\delta\sigma)^{2}-\frac{1}{2\zeta}\,(\sigma+\delta\sigma)\, A^{2}+\frac{1}{8\zeta}\, (A^{2})^{2}\right\}\ .
\end{equation}
By Eq.~\eqref{kfm394}, the effective potential will then be given by
\begin{equation}\label{bfg239}
V(\sigma)=\frac{1}{2\zeta}\,\sigma^{2}-\frac{1}{\mc{V}_{4}}\ln\int\mc{D}\mathscr{F}\ e^{-I_{L}}\ ,
\end{equation}
where
\begin{equation}
\mc{D}\mathscr{F}=\mc{D}A\,\mc{D}\cbar\,\mc{D}c\,\mc{D}B\,\mc{D}\delta\sigma\ .
\end{equation}

To lowest order in perturbation theory, the path integral in Eq.~\eqref{bfg239} is equal to the product of the determinants of the fields' zero-order propagators\footnote{Raised to the power of $1/2$ for the bosonic fields or to the power of $-1$ for the fermionic fields.}. All of these propagators are independent from $\sigma$, except for that of the gauge field -- $\Delta_{\sigma}$ in the following equations~--, which will thus yield the only non-constant contribution to $V(\sigma)$. Explicitly, to one loop, modulo $\sigma$-independent terms,
\begin{equation}
V(\sigma)=\frac{1}{2\zeta}\,\sigma^{2}-\frac{1}{\mc{V}_{4}}\ln[\text{det}(\Delta_{\sigma})^{1/2}]\ ,
\end{equation}
where, in momentum space,
\begin{equation}
(\Delta_{\sigma})_{\mu\nu}^{ab}(q)=\frac{1}{q^{2}-\sigma/\zeta}\,t_{\mu\nu}(p)\,\delta^{ab}\ .
\end{equation}
It follows that
\begin{equation}
V(\sigma)=\frac{\mu^{\epsilon}}{2\zeta}\,\sigma^{2}+\frac{(d-1)N_{A}}{2}\,\mu^{\epsilon}\int\frac{d^{d}q}{(2\pi)^{d}}\ln\left(q^{2}-\mu^{\epsilon}\frac{\sigma}{\zeta}\right)\ ,
\end{equation}
where $N_{A}=N^{2}-1$ is the dimension of the gauge group SU(N) and we have generalized to $d=4-\epsilon$ dimensions with an eye to the renormalization of $V(\sigma)$. In the above equation, $\mu$ is the scale introduced by dimensional regularization. The evaluation of the logarithmic integral yields the following regularized expression for the effective potential \cite{VKVV01,Dem20}:
\begin{equation}\label{vkd294}
V(\sigma)=\frac{\mu^{\epsilon}}{2\zeta}\,\sigma^{2}-\frac{3N_{A}}{64\pi^2\zeta^{2}}\,\sigma^{2}\left[\frac{2}{\epsilon}+\ln\left(-\frac{\overline{\mu}^2}{\mu^{\epsilon}\sigma/\zeta}\right)+\frac{5}{6}\right]\ ,
\end{equation}
where $\overline{\mu}=\sqrt{4\pi}e^{-\gamma_{E}/2}\mu$. It is clear from Eq.~\eqref{vkd294} where the need to introduce a new constant $\zeta$ comes from. Since the second term of the equation contains a divergence, we need a tunable -- and, more specifically, a renormalizable -- parameter in the first term in order to be able to absorb it. This is done by interpreting $\zeta$ as a bare parameter, $\zeta\to \zeta_{B}$, which is multiplicatively renormalized to a finite constant\footnote{We should mention that here we are not renormalizing the parameters in a systematic way. In Sec.~\ref{sec:dynmodrenfr}, we will give a slightly different definition of the renormalization factor $Z_{\zeta}$, which also takes into account the renormalization of the condensate $\avg{(A^{h})^{2}}$.},
\begin{equation}
\zeta_{B}=\mu^{-\epsilon}Z_{\zeta}\zeta\ .
\end{equation}
With these modifications, the effective potential reads \cite{VKVV01,Dem20}
\begin{equation}
V(\sigma)=\frac{\mu^{2\epsilon}}{2\zeta}(1-\delta Z_{\zeta})\,\sigma^{2}-\frac{3N_{A}}{64\pi^2\zeta^{2}}\,\mu^{2\epsilon}\sigma^{2}\left[\frac{2}{\epsilon}+\ln\left(-\frac{\overline{\mu}^2}{\mu^{\epsilon}\sigma/\zeta}\right)+\frac{5}{6}\right]\ ,
\end{equation}
where $\delta Z_{\zeta}=Z_{\zeta}-1$ and $Z_{\zeta}$ does not appear in the second term since the latter can be interpreted as a higher-order term. Setting
\begin{equation}
\delta Z_{\zeta}=-\frac{3N_{A}}{32\pi^{2}\zeta}\,\frac{2}{\epsilon}
\end{equation}
provides us with the effective potential renormalized in the $\overline{\text{MS}}$ scheme,
\begin{equation}\label{slf294}
V(\sigma)=\frac{\mu^{2\epsilon}}{2\zeta}\,\sigma^{2}-\frac{3N_{A}}{64\pi^2\zeta^{2}}\,\mu^{2\epsilon}\sigma^{2}\left[\ln\left(-\frac{\overline{\mu}^2}{\mu^{\epsilon}\sigma/\zeta}\right)+\frac{5}{6}\right]\ .
\end{equation}
In the next section we will rephrase $V(\sigma)$ in terms of the gluon mass parameter $m^{2}$ and show that its minimization leads to a non vanishing value of $m^{2}$.\newpage

\subsection{Dynamical mass generation: the gap equation}
\label{sec:dynmodgapeq}

Tracing back our steps to the definition of the gluon mass parameter, we see that in $d=4-\epsilon$ dimensions $m^{2}$ can be expressed in terms of the renormalized parameter $\zeta$ as
\begin{equation}\label{fjs294}
m^{2}=-\frac{\mu^{\epsilon}\sigma}{\zeta}\ ,
\end{equation}
where the factor of $\mu^{\epsilon}$ ensures that the right-hand side has the correct dimensions in the $\epsilon\to 0$ limit. A renormalized effective potential $V(m^{2})$ for the gluon mass parameter can then be derived from Eq.~\eqref{slf294}, yielding \cite{VKVV01,Dem20}
\begin{equation}\label{effpot}
V(m^2)=\zeta \frac{m^4}{2}-\frac{3N_{A}}{64\pi^2}\,m^4\left(\ln\frac{\overline{\mu}^2}{m^2}+\frac{5}{6}\right)\ .
\end{equation}
Both of the above equations have an explicit dependence on $\zeta$. Since the latter has no counterpart in the ordinary formulation of Yang-Mills theory, before trying to compute the on-shell value of the gluon mass parameter from the minimization of the effective potential we should find a way to fix its value from first principles.

To this extent, we observe that the fundamental quantities of Yang-Mills theory should not explicitly depend on $\zeta$, since the latter disappears from the equations in the limit of zero source $J$. The most obvious way to fix $\zeta$ would then be to require that the effective potential does not depend on it -- that is, to enforce the equation $dV/d\zeta=0$. Unfortunately, to the current order in perturbation theory, such a constraint does not yield meaningful results. In fact, not only does the vanishing of the derivative $dV/d\zeta$ give us no clue on the value of $\zeta$, but it also yields $m^{2}=0$, which is easily seen not to be a minimum of the effective potential: while $V(m^{2}=0)=0$, there exist values of $m^{2}$ for which $V(m^{2})$ is negative.

As an alternative, we could require that $\zeta$ be a function of the coupling constant $g^{2}$ alone -- so that $\zeta$ may be regarded as a quantity whose value is fixed order by order in perturbation theory by the interactions themselves \cite{Ver95,VKVV01,Dem20}. In other words, we may seek an expansion of $\zeta$ in the form
\begin{equation}\label{jso934}
\zeta(g^{2})=\frac{1}{g^{2}}\sum_{n}\ \zeta_{n}\,g^{2n}\ ,
\end{equation}
where the reason for having divided by $g^{2}$ in the above equation will become clear in a moment. If we want Eq.~\eqref{jso934} to be valid at all scales, then we should also require that the running of $\zeta$ be controlled by that of $g^{2}$ -- which in particular will lead to the constants $\zeta_{n}$ being scale-independent. These assumptions are at the core of the so-called Zimmermann reduction of couplings programme \cite{Zim85,HMTZ19}, which was proved to yield accurate results when applied in the context of the Gross-Neveu model \cite{VSV97}. In \cite{VKVV01,Gra03}, it was found that, in the $\overline{\text{MS}}$ scheme, the reduction yields
\begin{equation}
\zeta(g^2)=\frac{N_{A}}{g^2N}\frac{9}{13}+\frac{161}{52}\frac{N_{A}}{16\pi^2}+\cdots	 .
\end{equation}
Thus, to lowest order, $\zeta$ can be taken to be proportional to $g^{-2}$, with the first non-zero coefficients in Eq.~\eqref{jso934} given by
\begin{equation}
\zeta_{0}=\frac{9N_{A}}{13N} \ ,\qquad\qquad \zeta_{1}=\frac{161N_{A}}{52\cdot16\pi^{2}}\ .
\end{equation}\\

As a consequence of $\zeta$ being expressed as a power series in $g^{2}$, the mass parameter $m^{2}$ defined by Eq.~\eqref{fjs294} contains higher-order terms in the coupling constant \cite{VKVV01,Dem20},
\begin{equation}\label{erg324}
m^{2}=-\frac{\mu^{\epsilon}\sigma}{(\zeta_{0}/g^{2}+\zeta_{1})+O(g^{2})}=m_{0}^{2}\,\left(1-\frac{\zeta_{1}}{\zeta_{0}}\,g^{2}+O(g^{4})\right)\ ,
\end{equation}
where
\begin{equation}
m^{2}_{0}=-\frac{\mu^{\epsilon}g^{2}\sigma}{\zeta_{0}}\ .
\end{equation}
Since our calculation of the effective potential $V(m^{2})$ stops at one loop, we must be careful to separate the contributions coming from the different orders in $g^{2}$. An explicit calculation shows that, when expressed in terms of $m_{0}^{2}$, the effective potential reads \cite{VKVV01,Dem20}
\begin{equation}
V(m_{0}^{2})=\frac{9}{13}\frac{N_{A}}{N}\frac{m^{4}_{0}}{2g^{2}}-\frac{3N_{A}}{64\pi^2}\,m^{4}_{0}\left(\ln\frac{\overline{\mu}^{2}}{m^{2}_{0}}+\frac{113}{39}\right)+O(g^{2})\ ,
\end{equation}
where the higher-order terms come both from the $O(g^{4})$ corrections to $\zeta$ and $m^{2}$ and from the interaction terms in the Landau-gauge action $I_{L}$ that defines the partition function. In Sec.~\ref{sec:dynmodprop} we will see that, to lowest order in the coupling, only $m_{0}^{2}$ enters the expression of the gluon propagator computed in the presence of a non-vanishing condensate. For this reason, in what follows we will focus on the above expression for the effective potential, which allows us to compute the value of the mass parameter $m_{0}^{2}$ in terms of $g^{2}$ and of the renormalization scale $\overline{\mu}$.

The first derivative of $V(m^{2}_{0})$ with respect to $m^{2}_{0}$ reads
\begin{equation}
V^{\prime}(m_{0}^{2})=\frac{9}{13}\frac{N_{A}}{N}\frac{m^{2}_{0}}{g^{2}}-\frac{3N_{A}}{32\pi^2}\,m^{2}_{0}\left(\ln\frac{\overline{\mu}^{2}}{m^{2}_{0}}+\frac{187}{78}\right)\ .
\end{equation}
Equating the latter to zero yields the so-called \textit{gap equation}, which provides us with the on-shell value of the gluon mass parameter $m^{2}_{0}$ \cite{Dem20}:
\begin{equation}\label{erg325}
m^{2}_{0}=\frac{13Ng^{2}}{3\cdot32\pi^2}\,m^{2}_{0}\left(\ln\frac{\overline{\mu}^2}{m^{2}_{0}}+\frac{187}{78}\right)\ .
\end{equation}
The gap equation has two solutions. The first of these, $m^{2}_{0}=0$, corresponds to a vanishing potential,
\begin{equation}
V(m^{2}_{0}=0)=0\ .
\end{equation}
The second solution,
\begin{equation}\label{erg326}
m_{0}^{2}=\overline{\mu}^{2}\,\exp\left(\frac{187}{78}-\frac{3\cdot32\pi^{2}}{13Ng^{2}}\right)\ ,
\end{equation}
on the other hand, corresponds to a negative value of $V(m_{0}^{2})$,
\begin{equation}
V(m_{0}^{2})=-\frac{3N_{A}m^{4}_{0}}{128\pi^{2}}<0\ .
\end{equation}
Clearly, Eq.~\eqref{erg326} yields the global minimum of the effective potential.\\

The fact that the minimum of $V(m_{0}^{2})$ lies at a non-zero value of $m_{0}^{2}$ has far-reaching consequences. First of all, it proves that the gluon condensate $\avg{(A^{h})^{2}}$ does not vanish in pure Yang-Mills theory. We remark that, albeit having been derived in the Landau gauge, this result holds in every covariant gauge thanks to the BRST invariance of the operator $A^{h}$ and of its square. Second of all, as a consequence of $\avg{(A^{h})^{2}}\neq 0$, it shows that a mass term for the gluons is generated in every covariant gauge via the inclusion of the BRST-invariant condensate in the Faddeev-Popov Lagrangian. In the previous sections, this was achieved by consecutive transformations of the action $S_{\text{FP}}$ -- performed in such a way as to leave the contents of the theory unchanged -- which resulted in an action $I$ with respect to which the gluons propagate at tree level like massive particles. As we know by now, treating the gluons as massive at order zero in perturbation theory is expected to yield a dressed gluon propagator whose transverse component remains finite in the deep infrared. If this turns out to be the case also for the propagator computed by using the action $I$, then our findings will provide a strong indication that the occurrence of dynamical mass generation in the gluon sector may be triggered by the BRST-invariant condensate $\avg{(A^{h})^{2}}$. In the next section we will test our hypothesis by computing the DM propagators in the Landau gauge.

\section{Dynamical Model: the propagators in the Landau gauge}
\renewcommand{\rightmark}{\thesection\ \ \ Dynamical Model: the propagators in the Landau gauge}
\label{sec:dynmodprop}

Having seen how the BRST-invariant condensate $\avg{(A^{h})^{2}}$ can be included in the formalism of QCD, we are now in a position to give a precise definition of the Dynamical Model. The Dynamical Model (DM) is the reformulation of pure Yang-Mills theory that uses the action $I$ -- Eq.~\eqref{dynmodact} -- in place of the Faddeev-Popov action $S_{\text{FP}}$ to study the dynamics of the SU(N) gauge theories in a general covariant gauge. The equivalence between the Dynamical Model and the ordinary formulation of pure Yang-Mills theory was proved to hold in Sec.~\ref{sec:dynmodcond} provided that the gluon mass parameter $m^{2}$ -- equivalently, the gluon condensate~-- is computed on shell -- that is, on the solutions of the gap equation $V^{\prime}(m^{2})=0$.

In the present section, the dressed gluon and ghost propagators will be computed in the Landau gauge within the framework of the Dynamical Model both at fixed scale and by making use of the Renormalization Group. As we will see, in the limit of vanishing momenta the DM gluon propagator saturates to a finite constant. The latter will be shown to be proportional to the inverse of the gluon mass parameter $m^{2}$, implying that -- as far as the Landau gauge is concerned -- dynamical mass generation in the gluon sector can be accounted for by the non-vanishing of the BRST-invariant condensate $\avg{(A^{h})^{2}}$.

As the first step in our derivation of the propagators, we will start by discussing the renormalization of the Dynamical Model and by presenting its Feynman rules.

\subsection{Renormalization and Feynman rules}
\label{sec:dynmodrenfr}

In Sec.~\ref{sec:brstinvcond}, the source term for the BRST-invariant condensate $\avg{(A^{h})^{2}}$ was introduced disregarding the issue of renormalization. Since keeping track of the renormalization factors will be essential in what follows, let us take a step back and show how the parameters and fields of the Dynamical Model are to be renormalized.

When coupling the condensate to the external current $J$, two things need to be kept in mind. First of all, the BRST-invariant gauge field $A^{h}$ needs to be renormalized by making use of an appropriate renormalization factor $Z_{A^{h}}$. Second of all, even after having renormalized $A^{h}$, the VEV $\avg{(A^{h})^{a}(x)\cdot (A^{h})^{a}(x)}$ is still divergent due to the fact that, as a product of operators evaluated at the same spacetime point, the operator $(A^{h})^{2}$ is singular. This issue can be overcome by interpreting the partition function $Z[J]$ to be a functional of a renormalized current $J$ -- that is, by introducing a renormalization factor $Z_{J}$ for the current as well.\newpage

Starting from $Z_{A^{h}}$ and $Z_{J}$, one can define two mutually independent renormalization factors $Z_{2}$ and $Z_{\zeta}$ via the equations \cite{VKVV01}
\begin{gather}
J_{B}\,A^{h}_{B}\cdot A^{h}_{B}=(Z_{J}\,Z_{A^{h}})\,J\,A^{h}\cdot A^{h}=Z_{2}\,J\,A^{h}\cdot A^{h}\ ,\\
\zeta_{B}\,J_{B}^{2}=Z_{\zeta}\zeta\,\mu^{-\epsilon}\,J^{2}\ .\label{gba942}
\end{gather}
Note that, at variance with Sec.~\ref{sec:dynmodpotcalc}, the $Z_{\zeta}$ in Eq.~\eqref{gba942} does not renormalize the parameter $\zeta$ alone, but it also includes the renormalization of the current $J$. We will come back to this point at the end of the present section. $Z_{2}$ and $Z_{\zeta}$ are the renormalization factors that appear in the source term $\Delta S$ for the BRST-invariant potential. Explicitly,
\begin{equation}\label{vsl394}
\Delta S=\int d^{4}x\,\left(\frac{Z_{2}}{2}\,J\,A^{h}\cdot A^{h}-\mu^{-\epsilon}\frac{Z_{\zeta}\zeta}{2}\,J^{2}\right)\ .
\end{equation}
Differentiating the partition function $Z[J]$ with respect to $J$ while using Eq.~\eqref{vsl394} now yields the condensate $\sigma[J]$ in the form
\begin{equation}
\sigma[J]=\frac{\delta W}{\delta J}[J]=\frac{Z_{2}}{2}\avg{A^{h}\cdot A^{h}}_{J}-Z_{\zeta}\zeta\,\mu^{-\epsilon}\,J\ ,
\end{equation}
where the fields $A^{h}$ and $J$ are to be interpreted as renormalized quantities. In the above equation, the factors $Z_{2}$ and $Z_{\zeta}$ are needed to remove the divergences that arise from the product $A^{h}(x)\cdot A^{h}(x)$ and its VEV.

The fields $\xi$, $\tau$, $\eta$ and $\overline{\eta}$ that were introduced in Sec.~\ref{sec:dynmodpotcalc} with the purpose of localizing the operator $A^{h}$ also need to be renormalized. Since the main focus of what follows is computing the propagators in the Landau gauge -- within which, as discussed earlier, the localization step can be skipped --, we will go no further in discussing this topic. A complete treatment of the renormalization of the action $I$ in a general covariant gauge can be found in \cite{CVPG18}.\\

In the Landau gauge ($\alpha=0$) the DM action $I_{L}=I|_{\alpha=0}$ amended by taking $Z_{2}$ and $Z_{\zeta}$ into account can be explicitly computed to be
\begin{equation}
I_{L}=S_{\text{FP}}\big|_{\alpha=0}+\int d^{d}x\ \left\{\frac{\mu^{\epsilon}}{2Z_{\zeta}\zeta}\,(\delta\sigma)^{2}-\frac{\mu^{\epsilon}Z_{2}}{2Z_{\zeta}\zeta}\,(\sigma+\delta\sigma) A^{2}+\frac{\mu^{\epsilon}Z_{2}^{2}}{8Z_{\zeta}\zeta}\, (A^{2})^{2}\right\}\ .
\end{equation}
By rewriting the mass term as
\begin{equation}\label{fms393}
-\frac{\mu^{\epsilon}Z_{2}}{2Z_{\zeta}\zeta}\,\sigma A^{2}=-\frac{\mu^{\epsilon}}{2\zeta}\,\sigma A^{2}-\left(\frac{Z_{2}}{Z_{\zeta}}-1\right)\frac{\mu^{\epsilon}}{2\zeta}\,\sigma A^{2}\ ,
\end{equation}
a renormalized gluon mass parameter can be still defined in terms of the finite quantities $\sigma$ and $\zeta$ as
\begin{equation}
m^{2}=-\frac{\mu^{\epsilon}\sigma}{\zeta}\ ,
\end{equation}
so that the zero-order Euclidean gluon propagator $\Delta_{m\,\mu\nu}^{ab}(p)$ reads
\begin{equation}
\Delta_{m\,\mu\nu}^{ab}(p)=\frac{\delta^{ab}\,t_{\mu\nu}(p)}{p^{2}+m^{2}}\ .
\end{equation}
The second term in Eq.~\eqref{fms393}, on the other hand, provides a renormalization counterterm for the gluon mass operator: to lowest order,
\begin{equation}
\Delta\mc{L}_{\text{c.t.}}=(\delta Z_{2}-\delta Z_{\zeta})\,\frac{m^{2}}{2}\, A^{2}\qquad(\delta Z_{2,\zeta}=Z_{2,\zeta}-1)\ .
\end{equation}
As for the other terms in $I_{L}$, we see that -- as before -- the zero-order propagator $D_{(\delta\sigma)}(p^{2})$ of the fluctuation field $\delta\sigma$ can be expressed as
\begin{equation}
D_{(\delta\sigma)}(p^{2})=\zeta\mu^{-\epsilon}\ .
\end{equation}
Additionally, the analytical expressions for the cubic $\delta\sigma A^{2}$ and quartic $(A^{2})^{2}$ vertices, $\delta\Gamma_{\mu\nu}^{ab}$ \footnote{Not to be confused with the gluon mass counterterm of Chpts.~\ref{chpt:sme} and \ref{chpt:smeapp}.} and $\delta\Gamma_{\mu\nu\rho\sigma}^{abcd}$, respectively, are seen to be given by \cite{Dem20}
\begin{equation}
\delta\Gamma_{\mu\nu}^{ab}=\frac{\mu^\epsilon}{\zeta}\,\delta^{ab}\delta_{\mu\nu}\label{eq:AAsigma}
\end{equation}
and
\begin{equation}
\delta\Gamma_{\mu\nu\rho\sigma}^{abcd}=-\frac{\mu^\epsilon}{\zeta}\Big(\delta^{ab}\delta^{cd}\delta_{\mu\nu}\delta_{\rho\sigma}+\delta^{ac}\delta^{bd}\delta_{\mu\rho}\delta_{\nu\sigma}+\delta^{ad}\delta^{bc}\delta_{\mu\sigma}\delta_{\nu\rho}\Big)\ .\label{eq:AAAA}
\end{equation}
The renormalization factors $Z_{2}$ and $Z_{\zeta}$ contribute to these vertices and to the quadratic $\delta\sigma$ operator via appropriate renormalization counterterms, to be added to $\mc{L}_{\text{c.t.}}$. Finally, the remaining Feynman rules of the Dynamical Model are identical to those of ordinary pure Yang-Mills theory, and include the renormalization counterterms $Z_{A}$ and $Z_{g}$ for the gluon field and for the coupling constant.\\

To end this section, we point out that the condensate's effective potential $V(\sigma)$, computed to one loop while taking into account the appropriate renormalization factors, is given by
\begin{equation}
V(\sigma)=\frac{\mu^{2\epsilon}}{2Z_{\zeta}\zeta}\,\sigma^{2}+\frac{(d-1)N_{A}}{2}\mu^{\epsilon}\int\frac{d^{d}q}{(2\pi)^{d}}\,\ln\left(q^{2}-\mu^{\epsilon}\frac{\sigma}{\zeta}\right)\ ,
\end{equation}
where now $Z_{\zeta}$ is the renormalization factor defined by Eq.~\eqref{gba942}. Thus, while in Sec.~\ref{sec:dynmodpotcalc} $Z_{\zeta}$ was referred to $\zeta$ alone, we see that the renormalized expression of the potential does not change even when the correct renormalization factors are included. From $V(\sigma)$ we can derive a renormalized potential $V(m^{2})$ for the mass parameter $m^{2}$,
\begin{equation}
V(m^{2})=\frac{\zeta m^{4}}{2Z_{\zeta}}+\frac{(d-1)N_{A}}{2}\mu^{\epsilon}\int\frac{d^{d}q}{(2\pi)^{d}}\, \ln\left(q^{2}+m^{2}\right)
\end{equation}
whose first derivative $V^{\prime}(m^{2})$,
\begin{equation}\label{sdk450}
V^{\prime}(m^{2})=\frac{\zeta m^{2}}{Z_{\zeta}}+\frac{(d-1)N_{A}}{2}\mu^{\epsilon}\int\frac{d^{d}q}{(2\pi)^{d}}\frac{1}{q^{2}+m^{2}}\ ,
\end{equation}
once set to zero, provides us with the renormalized gap equation.

\subsection{The gluon and ghost propagators}
\label{sec:dynmodgpgse}

\begin{figure}[H]
\centering
\includegraphics[width=0.28\textwidth]{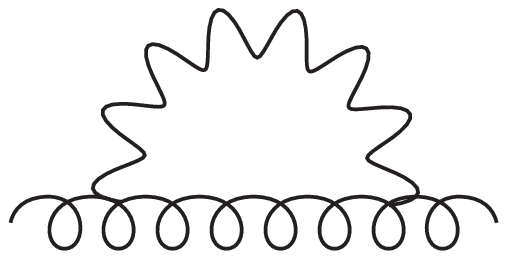}\hspace{2cm}\includegraphics[width=0.28\textwidth]{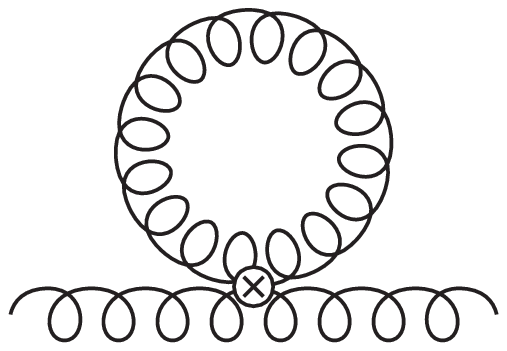}
\vspace{5pt}
\caption{Diagrams that contribute to the DM one-loop gluon polarization, arising from the new cubic and quartic vertices in the action $I_{L}$. The wiggly line represents the fluctuation field $\delta\sigma$'s propagator.}\label{fig:dynmodloops}
\end{figure}

Let us proceed to the calculation of the gluon and of the ghost propagators within the Dynamical Model. The Landau-gauge (inverse) gluon propagator computed in the framework of the DM can be expressed as \cite{Dem20}
\begin{equation}\label{vsf435}
[\Delta^{-1}(p)]_{\mu\nu}^{ab}=\delta^{ab}\left[Z_{A}\,p^{2}\,t_{\mu\nu}(p)+(1+\delta Z_{2}-\delta Z_{\zeta})\,m^{2}\,\delta_{\mu\nu}+\Pi_{\mu\nu}(p)\right]\ .
\end{equation}
In the above equation, $Z_{A}$ is the gluon field strength renormalization factor, the renormalization constants $\delta Z_{2}$ and $\delta Z_{\zeta}$ arise from the renormalization of the gluon mass operator $m^{2}A^{2}$, and $\Pi_{\mu\nu}(p)$ -- modulo color structure -- is the Landau-gauge gluon polarization. An explicit calculation shows that, to one loop,
\begin{align}\label{vsf436}
\Pi_{\mu\nu}(p)&=\Pi_{\mu\nu}^{(\text{CF})}(p)- \frac{4\cdot 2}{2!}\left(\frac{\mu^\epsilon}{2\zeta}\right)^2\frac{\zeta}{\mu^\epsilon}\int\frac{d^{d}q}{(2\pi)^{d}}\ \Delta_{m\,\mu\nu}(q)+\\
\notag&\quad+\frac{4\cdot 3}{4!}\frac{\mu^\epsilon}{\zeta}\left[\delta_{\mu\nu}N_{A}\int\frac{d^{d}q}{(2\pi)^{d}}\ \delta^{\sigma\tau}\Delta_{m\,\sigma\tau}(q)+2\int\frac{d^{d}q}{(2\pi)^{d}}\ \Delta_{m\,\mu\nu}(q)\right]\ .
\end{align}
Here $\Pi^{(\text{CF})}(p)$ -- given by diagrams (1), (2a) and (3a) in Fig.~\ref{fig:glupoldiag} -- is the gluon polarization computed in pure Yang-Mills theory by replacing the ordinary Landau-gauge massless zero-order gluon propagator by a massive one; as such, it is equal to the polarization computed within the Curci-Ferrari model -- hence the label CF. On the other hand, the second and third terms in Eq.~\eqref{vsf436} arise, respectively, from the interaction of the gluon field with the fluctuation field $\delta\sigma$ and from the self-interaction mediated by the new quartic $(A^{2})^{2}$ vertex. The corresponding diagrams are depicted in Fig.~\ref{fig:dynmodloops}.
\vspace{5mm}
\vspace{5mm}
It is easy to see that, in Eq.~\eqref{vsf436}, the second term inside the brackets cancels the second term in the first line. Therefore, the DM gluon polarization simplifies to
\begin{equation}\label{erv593}
\Pi_{\mu\nu}(p)=\Pi_{\mu\nu}^{(\text{CF})}(p)+\frac{(d-1)N_{A}}{2}\frac{\mu^\epsilon}{\zeta}\delta_{\mu\nu}\int\frac{d^{d}q}{(2\pi)^{d}}\frac{1}{q^2+m^2}\ .
\end{equation}
By plugging Eq.~\eqref{erv593} into Eq.~\eqref{vsf435} and observing that, to lowest order, the tree-level mass term in the latter can be rewritten as
\begin{equation}
(1+\delta Z_{2}-\delta Z_{\zeta})\,m^{2}=\delta Z_{2}\,m^{2}+Z_{\zeta}^{-1}\,m^{2}\ ,
\end{equation}
we find that the DM gluon propagator can be put in the form \cite{Dem20}
\begin{align}
\Delta^{-1}_{\mu\nu}(p)&=Z_{A}p^{2}\,t_{\mu\nu}(p)+\delta Z_{2}\,m^{2}\,\delta_{\mu\nu}+\Pi_{\mu\nu}^{(\text{CF})}(p)+\\
\notag&\quad+\delta_{\mu\nu}\left(Z_{\zeta}^{-1}\,m^{2}+\frac{(d-1)N_{A}}{2}\frac{\mu^\epsilon}{\zeta}\int\frac{d^{d}q}{(2\pi)^{d}}\frac{1}{q^2+m^2}\right)\ .
\end{align}

By comparing the last expression with Eq.~\eqref{sdk450}, we see that the terms in brackets are none other than the first derivative $V^{\prime}(m^{2})$ of the gluon mass parameter's effective potential, divided by $\zeta$. In particular, if $m^{2}$ solves the gap equation -- $V^{\prime}(m^{2})=0$ --, then these terms vanish and we are left with
\begin{equation}\label{vsf437}
\Delta(p^{2})=\frac{1}{Z_{A}p^{2}+\delta Z_{2}\,m^{2}+\Pi_{T}^{(\text{CF})}(p^{2})}
\end{equation}
for the transverse component $\Delta(p^{2})$ of the gluon propagator, where $\Pi_{T}^{(\text{CF})}(p^{2})$ is the transverse component of the Curci-Ferrari polarization. Since the equivalence between the Dynamical Model and ordinary pure Yang-Mills theory only holds on the solutions of $V^{\prime}(m^{2})=0$, we can take Eq.~\eqref{vsf437} as our final expression for $\Delta(p^{2})$.

Interestingly, the constant $\zeta$ disappears from the DM gluon propagator once the latter is computed on the shell of the gap equation. This could have been foreseen, given that such a parameter was not present in the Faddeev-Popov action in the first place. Nonetheless, an implicit dependence of the propagator on $\zeta$ still survives via the solutions of $V^{\prime}(m^{2})=0$. Within the reductions of coupling programme, this dependence on $\zeta$ is traded with a dependence on the coupling constant $g^{2}$ in a perturbative fashion -- see Sec.~\ref{sec:dynmodgapeq}. In particular, since both $\delta Z_{2}$ and $\Pi_{T}^{(\text{CF})}(p^{2})$ are already first-order in $g^{2}$, the mass parameter in Eq.~\eqref{vsf437} can be identified with the lowest-order $m_{0}^{2}$ defined in Sec.~\ref{sec:dynmodgapeq} -- namely
\begin{equation}
m_{0}^{2}=-\frac{\mu^{\epsilon}\,g^{2}Z_{2}\avg{A^{2}}}{2\zeta_{0}}\qquad\left(\zeta_{0}=\frac{9N_{A}}{13N}\right)\ .
\end{equation}
In what follows, we will not distinguish between $m^{2}$ and $m_{0}^{2}$, denoting by $m^{2}$ the value provided by the last equation.

A second observation we wish to make is that, again when enforcing $V^{\prime}(m^{2})=0$, the tree-level mass term $m^{2}$ is removed from $\Delta(p^{2})$. This is akin to what happens in the Screened Massive Expansion after the single-cross diagram -- Fig.~\ref{fig:gluctdiag}~-- is added to the polarization; similar cancellations can be shown to occur at every order in perturbation theory \cite{Dem20}. At variance with the SME, however, a renormalization counterterm of the form $\delta Z_{2}\,m^{2}$ is still left in $\Delta(p^{2})$. Such a feature is not accidental: the Curci-Ferrari polarization, in fact, contains a mass divergence (proportional to $m^{2}$) which needs to be absorbed into $\delta Z_{2}$ in order to obtain a finite result. Let us see explicitly how this works.

The one-loop Curci-Ferrari polarization was first computed in \cite{TW10}. Its transverse component reads
\begin{align}
\Pi_{T}^{(\text{CF})}(p^{2})&=-\frac{\lambda m^{2}}{6}\left(13s-\frac{9}{2}\right)\left(\frac{2}{\epsilon}+\ln\frac{\overline{\mu}^{2}}{m^{2}}\right)+\\
\notag&\quad\,-\frac{\lambda m^{2}}{24s^{2}} \bigg[\frac{242}{3} s^{3}-126 s^{2}+2s+(s^{2}-2)s^{3}\ln s+\\
\notag&\qquad\qquad\quad-2(s+1)^{3}(s^{2}-10s+1)\ln(s+1)+\\
\notag&\qquad\qquad\quad-s^{3/2}(s+4)^{3/2}(s^{2}-20s+12)\ln\left(\frac{\sqrt{s+4}-\sqrt{s}}{\sqrt{s+4}+\sqrt{s}}\right)\bigg]\ ,
\end{align}
where $s=p^{2}/m^{2}$, $\overline{\mu}=\sqrt{4\pi}e^{\gamma_{E}/2}\mu$ is the scale introduced by dimensional regularization and $\lambda$ is a normalized coupling constant defined as
\begin{equation}
\lambda=\frac{N\alpha_{s}}{4\pi}\ .
\end{equation}
The divergent part of $\Pi_{T}^{(\text{CF})}(p^{2})$ is given by
\begin{equation}\label{efr582}
[\Pi_{T}^{(\text{CF})}(p^{2})]_{\text{div.}}=-\lambda\left(\frac{13}{6}\,p^{2}-\frac{3}{4}\,m^{2}\right)\frac{2}{\epsilon}
\end{equation}
and contains two terms. The first one, proportional to $p^{2}$, is absorbed by the usual gluon field renormalization factor $Z_{A}=1+\delta Z_{A}$. The second one, proportional to $m^{2}$, can only be removed from the propagator if a corresponding counterterm is available. In the framework of the Dynamical Model, this counterterm is precisely $\delta Z_{2}$. In particular, from Eq.~\eqref{efr582}, we see that $\delta Z_{A}$ and $\delta Z_{2}$ must be chosen according to
\begin{equation}\label{eij485}
\delta Z_{A}=\frac{13\lambda}{6}\,\frac{2}{\epsilon}+\text{fin.}\ ,\qquad\qquad \delta Z_{2}=-\frac{3\lambda}{4}\,\frac{2}{\epsilon}+\text{fin.}\ ,
\end{equation}
where the finite terms depend on the renormalization scheme in which the propagator is defined. Note that $\delta Z_{A}$ contains the ordinary gluon field divergence -- see e.g. Sec.~\ref{sec:glprsme}. This happens because, just like the SME and the Curci-Ferrari model, the Dynamical Model as well does not modify the UV limit of pure Yang-Mills theory.

Once renormalized, the Landau-gauge DM gluon propagator reads
\begin{equation}
\Delta(p)=\frac{1}{p^{2}+\Pi_{T,R}^{(\text{CF})}(p^{2})}\ ,
\end{equation}
where $\Pi_{T,R}^{(\text{CF})}(p^{2})$ is the renormalized transverse Curci-Ferrari polarization. In the limit of vanishing momentum $p^{2}\to 0$, since
\begin{equation}\label{btm850}
\Pi^{(\text{CF})}_{T}(p^{2})\to\frac{3\lambda m^{2}}{4}\left(\frac{2}{\epsilon}+\frac{5}{6}+\ln\frac{\overline{\mu}^{2}}{m^{2}}\right)\neq 0\ ,
\end{equation}
the gluon propagator remains finite -- and is thus massive -- unless the finite terms of the renormalization constant $\delta Z_{2}$ are chosen to exactly eliminate the finite terms in $\Pi_{T,R}^{(\text{CF})}(p^{2})$. Of course, this will need to be avoided.\\

We now turn to the ghost sector. In the framework of the Dynamical Model, the ghost propagator can be expressed as
\begin{equation}\label{smk495}
\mc{G}(p^{2})=\frac{1}{Z_{c}\,p^{2}+\Sigma(p^{2})}\ ,
\end{equation}
where $Z_{c}$ is the ghost field renormalization constant and $\Sigma(p^{2})$ is the ghost self-energy. To one loop, since the corrections coming from the cubic $\delta\sigma A^{2}$ and quartic $(A^{2})^{2}$ vertices are higher-order, the Landau-gauge ghost self-energy diagrams are the same as those of ordinary Yang-Mills theory\footnote{Namely, the diagram on the left in Fig.~\eqref{fig:ghose}.}, albeit with massive zero-order propagators in their internal gluon lines. In other words, the DM ghost self-energy is equal to its Curci-Ferrari counterpart $\Sigma^{(\text{CF})}(p^{2})$ \cite{Dem20},
\begin{equation}
\Sigma(p^{2})=\Sigma^{(\text{CF})}(p^{2})\ .
\end{equation}
Like the gluon polarization, the one-loop Curci-Ferrari ghost self-energy was also first computed in \cite{TW10}, yielding
\begin{equation}
\Sigma^{(\text{CF})}(p^{2})=-\frac{3\lambda}{4}\,p^{2}\left(\frac{2}{\epsilon}+\ln\frac{\overline{\mu}^{2}}{m^{2}}\right)+\frac{\lambda}{4}\,p^{2}\left[\frac{(s+1)^{3}}{s^{2}}\,\ln(1+s)-s\ln s-\frac{1}{s}-5\right]\ .
\end{equation}
Since the divergent part of $\Sigma^{(\text{CF})}(p^{2})$ reads
\begin{equation}
[\Sigma^{(\text{CF})}(p^{2})]_{\text{div.}}=-\frac{3\lambda}{4}\,p^{2}\,\frac{2}{\epsilon}\ ,
\end{equation}
we see that the ghost field renormalization constant must be chosen according to
\begin{equation}\label{ref582}
\delta Z_{c}=\frac{3\lambda}{4}\,\frac{2}{\epsilon}+\text{fin.}\ .
\end{equation}
Again, the latter is equal to the ordinary divergence of pure Yang-Mills theory -- see e.g. Sec.~\ref{sec:ghprsme}.

In the limit of vanishing momenta $p^{2}\to 0$, the Landau-gauge ghost self-energy goes to zero like $p^{2}$. Explicitly,
\begin{equation}
\Sigma^{(\text{CF})}(p^{2})\to-\frac{3\lambda p^{2}}{4}\left(\frac{2}{\epsilon}+\frac{5}{6}+\ln\frac{\overline{\mu}^{2}}{m^{2}}\right)\ .\label{eq:sigmaghmu0}
\end{equation}
It follows from Eq.~\eqref{smk495} that the ghost propagator grows to infinity as $p^{2}\to 0$, confirming that within the Landau-gauge Dynamical Model the ghosts remain massless.

\subsection{RG improvement of the Dynamical Model in the Dynamically Infrared-Safe scheme}
\label{sec:dynmodrgimp}

In order to extend the validity of the Dynamical Model propagators to the widest possible range of momenta, the fixed-scale results described in the previous section can be improved by making use of Renormalization Group methods. In what follows, the propagators, the gluon mass parameter $m^{2}$ and the coupling $\lambda$ will be defined in the so-called Dynamically Infrared-Safe (DIS) scheme, which is presented for the first time in this thesis. The DIS scheme is just the MOM-Taylor scheme discussed in the context of the Screened Massive Expansion (Sec.~\ref{sec:smemomtay}), with an additional renormalization condition for the gluon mass parameter -- that is, for the renormalization factor $Z_{2}$.\\

Within the DIS scheme, the propagators are defined in the MOM scheme: denoting by $\Delta(p^{2};\mu^{2})$ and $\mc{G}(p^{2};\mu^{2})$, respectively, the gluon and ghost propagators renormalized at the scale $\mu$, we set
\begin{equation}
\Delta(\mu^{2};\mu^{2})=\frac{1}{\mu^{2}}\ ,\qquad\qquad\mc{G}(\mu^{2};\mu^{2})=\frac{1}{\mu^{2}}\ .
\end{equation}
From Eqs.~\eqref{vsf437} and \eqref{smk495}, it follows that in the DIS scheme the one-loop Landau-gauge DM field renormalization factors $Z_{A}$ and $Z_{c}$ are given by
\begin{equation}\label{vdi395}
Z_{A}=1-\delta Z_{2}\frac{m^{2}}{\mu^{2}}-\frac{\Pi^{(\text{CF})}_{T}(p^{2}=\mu^{2})}{\mu^{2}}\ ,\qquad\qquad Z_{c}=1-\frac{\Sigma^{(\text{CF})}(p^{2}=\mu^{2})}{\mu^{2}}\ ,
\end{equation}
where $\delta Z_{2}$ in the first of the above equations still needs to be fixed by appropriate renormalization conditions. The strong coupling, on the other hand, is defined in the Taylor scheme -- that is, by choosing the coupling renormalization factor $Z_{g}$ so that
\begin{equation}
Z_{g}Z_{A}^{1/2}Z_{c}=1\ .
\end{equation}
As we saw in Sec.~\ref{sec:smemomtay}, this is equivalent to having the beta function $\beta_{g}$ be equal to
\begin{equation}
\beta_{g}=\frac{g}{2}(\gamma_{A}+2\gamma_{c})\ ,
\end{equation}
where $\gamma_{A}$ and $\gamma_{c}$ are the gluon and ghost anomalous dimensions. A beta function $\beta_{\lambda}$ for the normalized coupling $\lambda=Ng^{2}/16\pi^{2}$ can be introduced as well, yielding
\begin{equation}
\beta_{\lambda}=\mu\frac{d\lambda}{d\mu}=\lambda(\gamma_{A}+2\gamma_{c})\ .
\end{equation}

With $Z_{A}$, $Z_{c}$ and $Z_{g}$ defined in the MOM-Taylor scheme, the only counterterm that remains to be fixed to complete the renormalization of the Landau-gauge Dynamical Model is $Z_{2}$. The latter, as we saw in the previous section, determines the renormalization of the gluon mass term in the propagator $\Delta(p^{2})$. Moreover, it enters the renormalization of the gluon mass parameter $m^{2}$ directly. In order to see this, recall that -- to lowest order in $g^{2}$~-- $m^{2}$ can be expressed in terms of the BRST-invariant condensate as
\begin{equation}
m^{2}=-\frac{g^{2}\mu^{\epsilon}}{2\zeta_{0}}\,Z_{2}\avg{A^{2}}\ .
\end{equation}
Substituting the coupling and the gluon field with their bare counterparts, we find that
\begin{equation}
m^{2}=-\frac{g^{2}\mu^{\epsilon}}{2\zeta_{0}}\,Z_{2}\avg{A^{2}}=-\frac{g_{B}^{2}\mu^{\epsilon}}{2\zeta_{0}}\,\frac{Z_{2}}{Z_{g}^{2}Z_{A}}\,\avg{A^{2}_{B}}\ .
\end{equation}
In particular, again to lowest order, the renormalization factor of $m^{2}$ is given by $Z_{2}/Z_{g}^{2}Z_{A}$. If the coupling is defined in the Taylor scheme, then such a factor can be rewritten as
\begin{equation}
\frac{Z_{2}}{Z_{g}^{2}Z_{A}}=\frac{Z_{2}Z_{c}^{2}}{Z_{g}^{2}Z_{A}Z_{c}^{2}}=Z_{2}Z_{c}^{2}\ .
\end{equation}
By introducing a gamma function $\gamma_{m^{2}}$ for the mass, such that
\begin{equation}
\mu\frac{dm^{2}}{d\mu}=\gamma_{m^{2}}\,m^{2}\ ,
\end{equation}
we see that in the Taylor scheme
\begin{equation}\label{fkm942}
\gamma_{m^{2}}=\gamma_{2}+2\gamma_{c}\ ,
\end{equation}
where $\gamma_{2}$ is the gamma function associated to the factor $Z_{2}$,
\begin{equation}
\gamma_{2}=\frac{\mu}{Z_{2}}\frac{dZ_{2}}{d\mu}\ .
\end{equation}

In order to fix $Z_{2}$, we first notice that, by Eqs.~\eqref{eij485} and \eqref{ref582},
\begin{equation}\label{ekf461}
(Z_{2}Z_{c})_{\text{div.}}=1\ .
\end{equation}
This relation was proved to hold to any perturbative order in \cite{DVS03}. Therefore, we could choose to set $Z_{2}=Z_{c}^{-1}$, extending the equality in Eq.~\eqref{ekf461} from the divergent terms to the full renormalization factors. However, this choice turns out to be disastrous when the MOM scheme is used for the propagators. To see this, observe that, by Eq.~\eqref{vdi395}, $Z_{2}=Z_{c}^{-1}$ would yield
\begin{equation}
\delta Z_{2}=-\delta Z_{c}=\frac{\Sigma^{(\text{CF})}(p^{2}=\mu^{2})}{\mu^{2}}\ .
\end{equation}
By expanding $\Sigma^{(\text{CF})}(\mu^{2})$ around $\mu^{2}=0$ -- Eq.~\eqref{eq:sigmaghmu0} --, we find that $\delta Z_{2}=-\delta Z_{c}$ would then imply the following asymptotic limit for the product $\delta Z_{2}\,m^{2}$ that appears in the gluon propagator:
\begin{equation}\label{fse592}
\delta Z_{2}\,m^{2}\to-\frac{3\lambda m^{2}}{4}\left(\frac{2}{\epsilon}+\frac{5}{6}+\ln\frac{\overline{\mu}^{2}}{m^{2}}\right)\ ,
\end{equation}
By Eq.~\eqref{btm850}, the latter is equal to $-\Pi_{T}^{(\text{CF})}(0)$. In other words, in the limit of vanishing renormalization scales, $\delta Z_{2}$ would kill the finite terms of the zero-momentum gluon polarization, which is precisely what we wanted to avoid in the context of the fixed-scale expansion.

The asymptotic behavior described by Eq.~\eqref{fse592} leads to a massless RG-improved gluon propagator, just like it does at fixed scale. To solve this issue, we can impose a slightly different renormalization condition on $\delta Z_{2}$: instead of setting $\delta Z_{2}=-\delta Z_{c}$, we define the former as
\begin{equation}\label{eq:dZ2dZc}
\delta Z_{2} = -\delta Z_{c}+\frac{5\lambda}{8}\ .
\end{equation}
The constant $5\lambda/8$, when multiplied by $m^{2}$, prevents the corresponding finite term in Eq.~\eqref{fse592} from entering $\delta Z_{2}$, so that $\Pi^{(\text{CF})}(0)$ is not renormalized to zero and the gluon propagator remains massive\footnote{We remark that the massiveness of the RG-improved gluon propagator does not depend on the specific coefficient of the constant term in Eq.~\eqref{eq:dZ2dZc}, as long as it is chosen different from zero. The factor of $5/8$ is the most natural choice, given that an opposite term is present in the low-energy expansion of $\delta Z_{c}$.}.\\

Eq.~\eqref{eq:dZ2dZc} -- together with the MOM condition for the propagator and the Taylor condition for the coupling -- completes the definition of the Dynamically Infrared-Safe scheme. We note that, since the derivative of $\lambda$ with respect to $\mu$ is $O(\lambda^{2})$, Eq.~\eqref{eq:dZ2dZc} implies that, to lowest order,
\begin{equation}
\gamma_{2}=-\gamma_{c}\ .
\end{equation}
In particular, by Eq.~\eqref{fkm942}, the gluon mass parameter's DIS anomalous dimension $\gamma_{m^{2}}$ reads
\begin{equation}
\gamma_{m^{2}}=\gamma_{c}\ .
\end{equation}
Thus, in the DIS scheme, the RG flow is determined by the beta and gamma functions
\begin{equation}\label{ksm358}
\beta_{\lambda}=\lambda(\gamma_{A}+2\gamma_{c})\ ,\qquad\qquad \gamma_{m^{2}}=\gamma_{c}\ ,
\end{equation}
where, to one loop,
\begin{align}\label{ksm359}
\gamma_{A}&=-\mu\frac{d}{d\mu}\left(\frac{\Pi^{(\text{CF})}_{T}(p^{2}=\mu^{2})}{\mu^{2}}+\frac{m^{2}}{\mu^{2}}\frac{\Sigma^{(\text{CF})}(p^{2}=\mu^{2})}{\mu^{2}}+\frac{5\lambda m^{2}}{8\mu^{2}}\right)\ ,\\
\gamma_{c}&=-\mu\frac{d}{d\mu}\left(\frac{\Sigma^{(\text{CF})}(p^{2}=\mu^{2})}{\mu^{2}}\right)\ .
\end{align}

An explicit calculation yields
\begin{equation}
\gamma_{A}\to-\frac{13\lambda}{3}\ ,\qquad\qquad\gamma_{c}\to -\frac{3\lambda}{2}\qquad\quad(\mu^{2}\to \infty)
\end{equation}
for the high-energy limit of the gluon and ghost anomalous dimensions. The latter are just the ordinary pQCD gamma functions: in the UV regime, the RG-improved gluon and ghost propagators will have the standard pQCD behavior. As for the running coupling and gluon mass parameter, we find that
\begin{align}
\beta_{\lambda}\to-\frac{22\lambda^{2}}{3}\ ,\qquad\qquad \gamma_{m^{2}}\to-\frac{3\lambda}{2}\qquad\quad(\mu^{2}\to \infty)\ ,
\end{align}
the first of which, again, is just the ordinary pQCD beta function, whereas the second yields a running $m^{2}(\mu^{2})$ which at large renormalization scales behaves as
\begin{equation}
m^{2}(\mu^{2})\sim [\lambda(\mu^{2})]^{\frac{9}{44}}\sim[\ln \mu^{2}]^{-\frac{9}{44}}\ ,
\end{equation}
thus decreasing as a negative rational power of the logarithm at high energies.\\

To end this section, we derive some useful expressions for the RG-improved gluon and ghost propagators\footnote{These can be shown to hold also in renormalization schemes which slightly differ from the DIS scheme -- see e.g. \cite{Dem20}.}. Let us start from the latter. Recall that the MOM-scheme RG-improved ghost propagator $\mc{G}(p^{2};\mu^{2}_{0})$ renormalized at the scale $\mu_{0}$ can be expressed in terms of the gluon anomalous dimension $\gamma_{c}$ as
\begin{equation}
\mc{G}\left(p^{2};\mu^{2}_{0}\right)=\frac{1}{p^{2}}\exp\left(\int^{p}_{\mu_{0}}\frac{d\mu}{\mu}\,\gamma_{c}\right)\ .
\end{equation}
Since in the DIS scheme $\gamma_{c}=\gamma_{m^{2}}$, the above equation can be rewritten as
\begin{align}\label{dln382}
\mc{G}\left(p^{2};\mu^{2}_{0}\right)&=\frac{1}{p^{2}}\exp\left(\int^{p}_{\mu_{0}}\frac{d\mu}{\mu}\,\gamma_{c}\right)=\frac{1}{p^{2}}\exp\left(\int^{p}_{\mu_{0}}\frac{d\mu}{\mu}\,\gamma_{m^{2}}\right)=\\
\notag&=\frac{1}{p^{2}}\exp\left(\int^{p}_{\mu_{0}}\frac{d\mu}{m^{2}}\,\frac{dm^{2}}{d\mu}\right)=\frac{1}{p^{2}}\frac{m^{2}(p^{2})}{m^{2}(\mu_{0}^{2})}\ .
\end{align}
We thus see that in the DIS scheme the RG-improved ghost propagator is equal to the running gluon mass parameter $m^{2}(p^{2})$ divided by $p^{2}$, normalized by the value of $m^{2}$ at the renormalization scale $\mu_{0}$. Similarly, since by Eq.~\eqref{ksm358} the gluon anomalous dimension $\gamma_{A}$ can be expressed as
\begin{equation}
\gamma_{A}=\frac{\beta_{\lambda}}{\lambda}-2\gamma_{m^{2}}\ ,
\end{equation}
the RG-improved DIS gluon propagator $\Delta(p^{2};\mu^{2}_{0})$ renormalized at the scale $\mu_{0}$ can be put in the form
\begin{align}\label{dln383}
\Delta\left(p^{2};\mu^{2}_{0}\right)&=\frac{1}{p^{2}}\exp\left(\int^{p}_{\mu_{0}}\frac{d\mu}{\mu}\,\gamma_{A}\right)=\frac{1}{p^{2}}\exp\left(\int^{p}_{\mu_{0}}\frac{d\mu}{\mu}\,\left[\frac{\beta_{\lambda}}{\lambda}-2\gamma_{m^{2}}\right]\right)\\
\notag&=\frac{1}{p^{2}}\exp\left(\int^{p}_{\mu_{0}}d\mu\,\left[\frac{1}{\lambda}\frac{d\lambda}{d\mu}-\frac{2}{m^{2}}\frac{dm^{2}}{d\mu}\right]\right)=\\
\notag&=\frac{1}{p^{2}}\frac{\lambda(p^{2})}{\lambda(\mu_{0}^{2})}\frac{m^{4}(\mu_{0}^{2})}{m^{4}(p^{2})}\ .
\end{align}

Eqs.~\eqref{dln382} and \eqref{dln383} allow us to study the infrared behavior of the RG-improved propagators analytically. Indeed, assuming that, as $p^{2}\to 0$, the running mass parameter $m^{2}(p^{2})$ saturates to a non-zero constant while the running coupling $\lambda(p^{2})$ goes to zero like $p^{2}$, the last equations tell us that $\Delta(p^{2};\mu_{0}^{2})$ also saturates to a non-zero constant, whereas $\mc{G}(p^{2};\mu_{0}^{2})$ diverges like $1/p^{2}$. The validity of these assumptions can be proved by solving the DIS RG equations for $\lambda$ and $m^{2}$ in the low-energy limit,
\begin{equation}
\mu\frac{d\lambda}{d\mu}\propto \frac{m^{2}\lambda^{2}}{\mu^{2}}\ ,\qquad\qquad \mu\frac{dm^{2}}{d\mu}\propto\lambda\mu^{2}\ ,
\end{equation}
which follow from the asymptotic behavior
\begin{equation}
\gamma_{A}\propto \frac{m^{2}\lambda}{\mu^{2}}\ ,\qquad\qquad \gamma_{c}\propto\frac{\mu^{2}\lambda}{m^{2}}\qquad\quad(\mu^{2}\to 0)\ .
\end{equation}
We thus conclude that in the deep infrared the RG-improved DIS propagators have the expected limits. In the next section, we will put our results to the test by comparing them with the lattice data.\newpage

\subsection{Comparison with the lattice data}
\label{sec:dynmodlat}

As we saw at the end of the previous section, the RG-improved Landau-gauge DM gluon and ghost propagators renormalized at the scale $\mu_{0}$ in the DIS scheme can be expressed as
\begin{equation}\label{vdk359}
\Delta(p^{2};\mu_{0}^{2})=\frac{1}{p^{2}}\frac{\lambda(p^{2})}{\lambda(\mu_{0}^{2})}\frac{m^{4}(\mu_{0}^{2})}{m^{4}(p^{2})}\ ,\qquad\qquad\mc{G}(p^{2};\mu_{0}^{2})=\frac{1}{p^{2}}\frac{m^{2}(p^{2})}{m^{2}(\mu_{0}^{2})}\ ,
\end{equation}
where $\lambda(p^{2})=N\alpha_{s}(p^{2})/4\pi$ is the running coupling and $m^{2}(p^{2})$ is the running gluon mass parameter. The latter can be computed by numerically integrating the Renormalization Group equations
\begin{equation}\label{vdk360}
\mu\frac{d\lambda}{d\mu}=\beta_{\lambda}\ ,\qquad\qquad\mu\frac{dm^{2}}{d\mu}=\gamma_{m^{2}}\,m^{2}\ ,
\end{equation}
with the beta and gamma functions $\beta_{\lambda}$ and $\gamma_{m^{2}}$ provided by Eqs.~\eqref{ksm358} and \eqref{ksm359}, starting from initial values $\lambda(\mu_{0}^{2})$ and $m^{2}(\mu_{0}^{2})$ at the renormalization scale $\mu_{0}^{2}$.

Within the Dynamical Model, $\lambda$ and $m^{2}$ are not independent parameters. On the contrary, they are related to one another via the gap equation $V^{\prime}(m^{2})=0$. Since the effective potential $V(m^{2})$ is RG invariant by definition, the order in which one solves the gap equation and the RG equations is irrelevant, as long as the choice of the renormalization scheme is consistent between the two.

In Sec.~\ref{sec:dynmodgapeq}, the gap equation was renormalized in the $\overline{\text{MS}}$ scheme and evaluated at fixed scale. To be consistent with the RG improvement of the propagators, in what follows we will instead use the RG-improved version of the potential -- namely,
\begin{equation}\label{LL6}
V(m^2)=\frac{9}{13} \frac{N_{A}}{N}\frac{m^4(\mu)}{2g^2(\mu)} \left(1+\beta_0\,\frac{g^2(\mu)}{16\pi^{2}}\ln\frac{m^2(\mu)}{\mu^2}\right)^{1+\gamma_0/\beta_0}\ ,
\end{equation}
where
\begin{equation}
\beta_{0}=\frac{11N}{3}\ ,\qquad\qquad\gamma_{0}=-\frac{3N}{2}
\end{equation}
are the one-loop coefficients of the $\overline{\text{MS}}$ beta function $\beta_{g}$ and gamma function $\gamma_{m^{2}}$. Eq.~\eqref{LL6} can be obtained by resumming $V(m^{2})$ to leading log within the $\overline{\text{MS}}$ scheme, as explained in \cite{Kas92}.

\begin{figure}[H]
\centering
\includegraphics[width=0.55\textwidth]{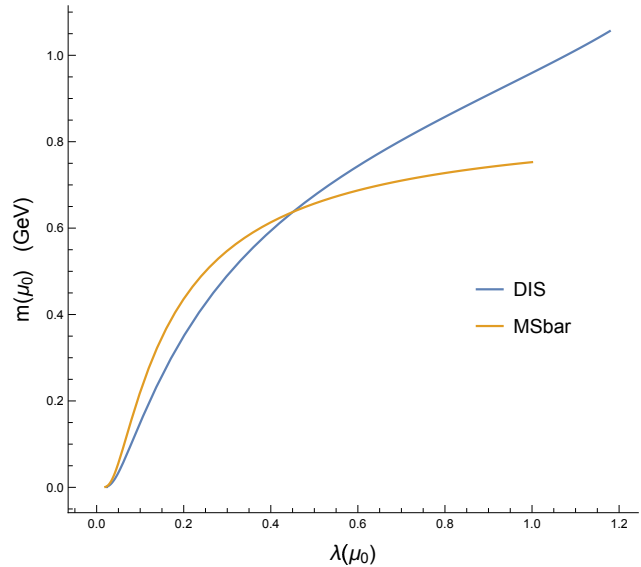}
\vspace{5pt}
\caption{Solutions of the RG-improved gap equation at the renormalization scale $\mu_{0}=1$~GeV. Blue curve: DIS scheme. Orange curve: $\overline{\text{MS}}$ scheme.}\label{fig:gapeqplot}
\end{figure}

To find the DIS solutions of $V^{\prime}(m^{2})=0$ starting from Eq.~\eqref{LL6}, one has to convert the coupling and mass parameter from one scheme to the other. This can be done by making use of the equations
\begin{equation}
m^{2}_{\text{DIS}}=\frac{Z_{m^{2},\overline{\text{MS}}}}{Z_{m^{2},\text{DIS}}}\,m^{2}_{\overline{\text{MS}}}\ ,\qquad\qquad
\lambda_{\text{DIS}}=\frac{Z_{g,\overline{\text{MS}}}^{2}}{Z_{g,\text{DIS}}^{2}}\,\lambda_{\overline{\text{MS}}}\ ,
\end{equation}
where $Z_{m^{2},\text{DIS}}$ and $Z_{g,\text{DIS}}$ (resp. $Z_{m^{2},\overline{\text{MS}}}$ and $Z_{g,\overline{\text{MS}}}$) are the mass and coupling renormalization factors evaluated in the DIS (resp. $\overline{\text{MS}}$) scheme. Explicitly, to one loop, these read
\begin{align}
m^{2}_{\text{DIS}}&=\left[1+\frac{5\lambda}{8}+(\delta Z_{c,\text{DIS}}-\delta Z_{c,\overline{\text{MS}}})\right]\,m^{2}_{\overline{\text{MS}}}\ ,\label{eq:massMStoDIS}\\
\lambda_{\text{DIS}}&=\Big[1+(\delta Z_{A,\text{DIS}}-\delta Z_{A,\overline{\text{MS}}})+2(\delta Z_{c,\text{DIS}}-\delta Z_{c,\overline{\text{MS}}})\Big]\,\lambda_{\overline{\text{MS}}}\ .\label{eq:lambdaMStoDIS}
\end{align}
The solutions of the RG-improved gap equation are displayed in Fig.~\ref{fig:gapeqplot} both in the DIS scheme and in the $\overline{\text{MS}}$ scheme.\newpage

\vspace*{5mm}
\vspace{5mm}
In what follows, the one-loop RG-improved Landau-gauge DM propagators $\Delta(p^{2};\mu_{0}^{2})$ and $\mc{G}(p^{2};\mu_{0}^{2})$ will be compared with the lattice data of \cite{DOS16}. The only free parameter of the comparison will be the value of the DIS coupling $\lambda(\mu_{0}^{2})$ at the initial renormalization scale $\mu_{0}$ -- taken to be equal to $1$~GeV \footnote{Since the lattice data of \cite{DOS16} were originally renormalized at $4$~GeV, a rescaling of the data was necessary to renormalize them at $1$~GeV.} --, the value of the DIS gluon mass parameter $m^{2}(\mu_{0}^{2})$ being calculated from the gap equation as a function of $\lambda(\mu_{0}^{2})$.

In order to fix $\lambda(\mu_{0}^{2})$, a combined fit of the gluon and ghost dressing functions -- $p^{2}\Delta(p^{2};\mu_{0}^{2})$ and $p^{2}\mc{G}(p^{2};\mu_{0}^{2})$, respectively -- was performed, with equal weights for both the functions. The fit yielded $\lambda(\mu_{0}^{2})=0.473$, corresponding  to $\alpha_{s}(\mu_{0}^{2})=1.981$ and to $m(\mu_{0}^{2})=0.655$~GeV by the DIS gap equation (Fig.~\ref{fig:gapeqplot}).

\begin{figure}[H]
\centering
\includegraphics[width=0.65\textwidth]{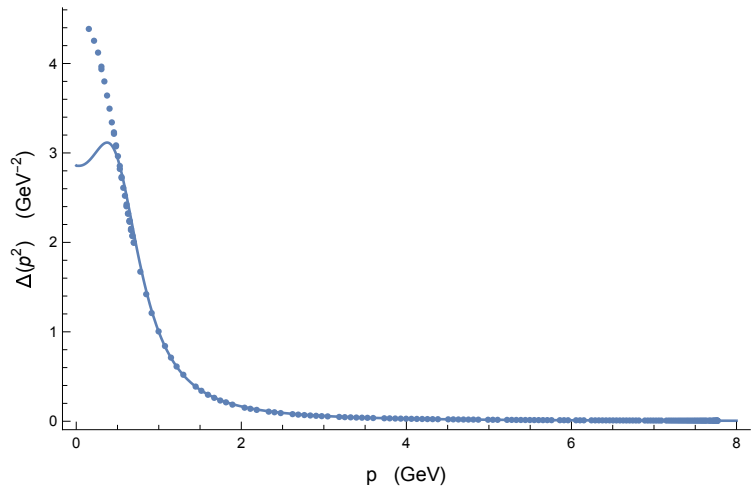}
\vspace{2pt}
\caption{Euclidean transverse gluon propagator in the Landau gauge ($\alpha=0$) renormalized at the scale $\mu_{0}=1$~GeV. Solid curve: one-loop RG-improved DM in the DIS scheme with the gluon mass parameter obtained from the gap equation; $\lambda(\mu_{0}^{2})=0.473$, $m(\mu_{0}^{2})=0.655$~GeV. Dots: lattice data from $\cite{DOS16}$.}\label{fig:dynmodglprop}
\end{figure}

\begin{figure}[H]
\centering
\includegraphics[width=0.65\textwidth]{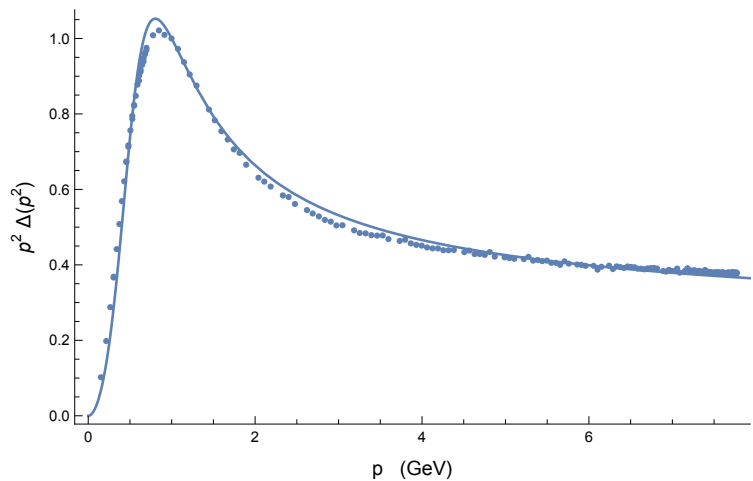}
\vspace{5pt}
\caption{Euclidean gluon dressing function in the Landau gauge ($\alpha=0$) renormalized at the scale $\mu_{0}=1$~GeV. As in Fig.~\ref{fig:dynmodglprop}.}\label{fig:dynmodgldress}
\end{figure}

The RG-improved one-loop Landau-gauge DM gluon propagator and dressing function are displayed, respectively, in Figs.~\ref{fig:dynmodglprop} and \ref{fig:dynmodgldress}, together with the lattice data of \cite{DOS16}. As we can see, the DIS functions show a very good agreement with the lattice over a wide range of momenta, extending from $p\approx0.5$~GeV up to $p\approx8$~GeV. In the infrared, as anticipated in the previous section, the gluon propagator saturates to a finite non-zero value, confirming that within the framework of the Landau-gauge Dynamical Model the gluon develops a mass. Nonetheless, the one loop approximation in which the propagator is computed is not able to reproduce the lattice results at energies lower than $\approx 0.5$~GeV. As discussed in Sec.~\ref{sec:smergopt} and reported likewise for the Screened Massive Expansion, this issue is common to massive perturbative truncations of Yang-Mills theory improved by the Renormalization Group, and is expected to be solved by going to higher order in perturbation theory \cite{GPRT19,CS20}.\newpage

\vspace*{5mm}
\vspace{5mm}
\begin{figure}[H]
\centering
\includegraphics[width=0.65\textwidth]{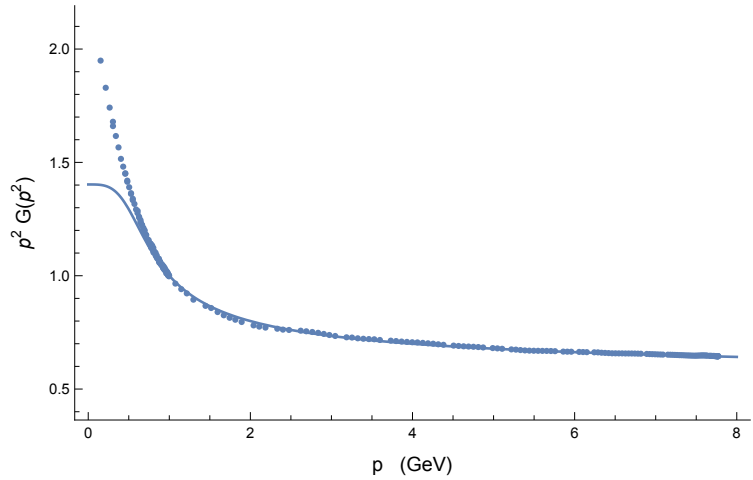}
\vspace{2pt}
\caption{Euclidean ghost dressing function in the Landau gauge ($\alpha=0$) renormalized at the scale $\mu_{0}=1$~GeV. As in Fig.~\ref{fig:dynmodglprop}.}\label{fig:dynmodghdress}
\end{figure}

\vspace{5mm}
In Fig.~\ref{fig:dynmodghdress} we display the RG-improved one-loop Landau-gauge DM ghost dressing function. At variance with the gluon sector, the agreement  between the DIS function and the lattice deteriorates below $p\approx1$~GeV. Nonetheless, the match between the two is very good at larger momenta -- up to $p\approx8$~GeV. In the limit of vanishing momenta, the dressing function $p^{2}\mc{G}(p^{2};\mu_{0}^{2})$ saturates to a constant, implying that $\mc{G}(p^{2};\mu_{0}^{2})$ grows to infinity like $1/p^{2}$. In other words, the RG-improved ghost propagator is massless, as expected.\newpage
\vspace{\fill}
\vspace{\fill}
\begin{figure}[H]
\centering
\includegraphics[width=0.65\textwidth]{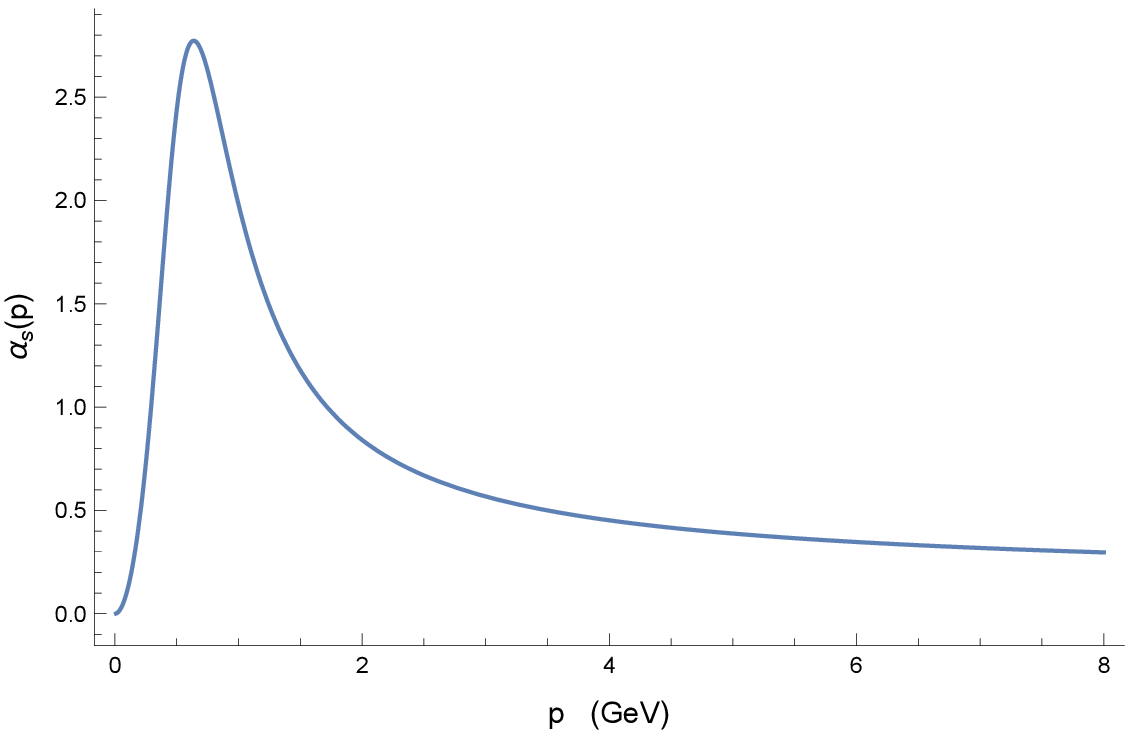}\\
\vspace{5mm}
\includegraphics[width=0.65\textwidth]{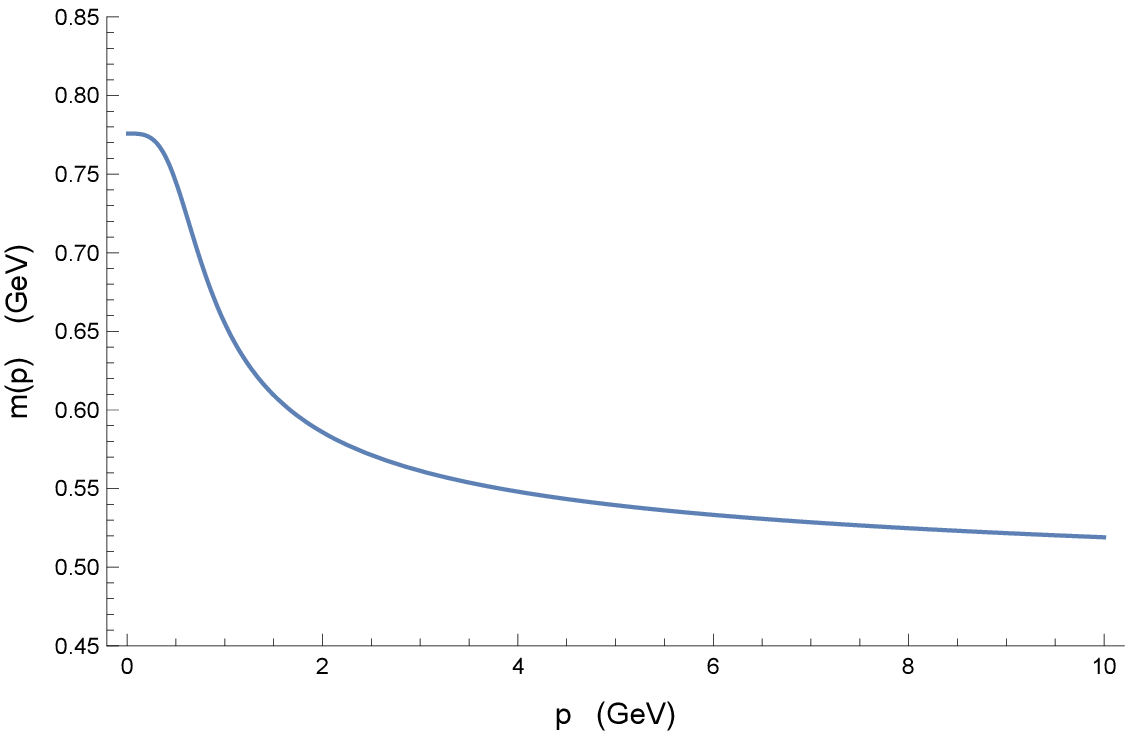}
\vspace{5pt}
\caption{DM running coupling (top) and gluon mass parameter (bottom) for the initial values $\lambda(\mu_{0}^{2})=0.473$ and $m(\mu_{0}^{2})=0.655$~GeV at $\mu_{0}=1$~GeV.}\label{fig:dynmodcoupmass}
\vspace{\fill}
\end{figure}
\newpage
In Fig.~\ref{fig:dynmodcoupmass} we show the DIS running coupling $\alpha_{s}(p^{2})=\frac{4\pi}{3}\lambda(p^{2})$ and gluon mass parameter $m(p^{2})$ obtained by integrating the RG equations with initial values $\lambda(\mu_{0}^{2})=0.473$ and $m(\mu_{0}^{2})=0.655$~GeV at $\mu_{0}=1$~GeV. Given that, within the DIS scheme, the coupling and propagators are defined, respectively, in the Taylor scheme and in the MOM scheme, the relation
\begin{equation}
\alpha_{s}(p^{2})=\alpha_{s}(\mu_{0}^{2})\,[p^{2}\Delta(p^{2};\mu_{0}^{2})][p^{2}\mc{G}(p^{2};\mu_{0}^{2})]^{2}
\end{equation}
still holds between $\alpha_{s}(p^{2})$ and $\Delta(p^{2};\mu_{0}^{2})$ and $\mc{G}(p^{2};\mu_{0}^{2})$, as can be easily seen from Eq.~\eqref{vdk359}. In particular, since the gluon propagator and the ghost dressing function saturate to a non-zero constant as $p^{2}\to 0$, $\alpha_{s}(p^{2})$ tends to zero like $p^{2}$ in the deep infrared. At intermediate energies, the running coupling first increases with the momentum -- attaining a maximum at $p\approx 0.64$~GeV, where $\alpha_{s}\approx2.77$ -- and then decreases again, showing the same behavior as the MOM-Taylor SME running coupling. In the UV, since the high-energy limit of the theory is not modified by the introduction of the condensate, $\alpha_{s}(p^{2})$ reduces to the ordinary one-loop pQCD running coupling.

The running gluon mass parameter $m^{2}(p^{2})$ is a decreasing function of momentum. For the fitted initial coupling $\lambda(\mu_{0}^{2})=0.473$, its saturation value at $p=0$ is found to be $m(0)\approx0.78$~GeV. At high energies, $m^{2}(p^{2})$ was already shown to have the behavior
\begin{equation}
m^{2}(p^{2})\sim\lambda^{\frac{9}{44}}(p^{2})\propto[\ln(p^{2})]^{-\frac{9}{44}}\ .
\end{equation}

\section{Conclusions}
\renewcommand{\rightmark}{\thesection\ \ \ Conclusions}
\label{sec:dynmodconcl}

In this chapter we showed that the non-vanishing of the BRST-invariant condensate $\sigma=\frac{1}{2}\avg{(A^{h})^{2}}$ in pure Yang-Mills theory can be proved in any covariant gauge by minimizing an effective potential $V(\sigma)$ derived by coupling the condensate to an external current $J$ in the Faddeev-Popov action $S_{\text{FP}}$. Subsequent manipulations of $S_{\text{FP}}$, performed with the double objective of localizing the field $A^{h}$ and of linearizing the source terms, led us to an action $I$ which, when computed on the shell of the gap equation $V^{\prime}(\sigma)=0$, is dynamically equivalent to the Faddeev-Popov action. $I$ defines the reformulation of pure Yang-Mills theory which we referred to as the Dynamical Model.

In the framework of the Dynamical Model, the gluons propagate as massive at tree level due to a mass term $\propto (A^{h})^{2}$ which appears in the action as a result of the non-vanishing of the condensate. The corresponding gluon mass parameter $m^{2}$ was found to be equal to $-\sigma/\zeta$, where $\zeta$ is a free parameter introduced in the formalism in order to renormalize the effective potential. By resorting to a procedure known as the reduction of couplings, the parameter $\zeta$ was expressed as a power series in the coupling constant $g^{2}$; given that, to lowest order, $\zeta$ was found to be proportional to the inverse coupling $g^{-2}$, the mass parameter $m^{2}$ turned out to be proportional to $\avg{g^{2}(A^{h})^{2}}$ plus higher-order corrections, which were tracked up to next-to-leading order as required for solving the gap equation.

In a general covariant gauge, the DM action contains a number of new fields. These are the field $\delta\sigma$ -- which quantifies the fluctuations of the operator $\frac{1}{2}(A^{h})^{2}$ around its vacuum expectation value -- and the fields $\tau^{a}$ and $\eta^{a}$, $\overline{\eta}^{a}$ -- which enforce the divergencelessness of the field $A^{h}$. $\tau^{a}$ and $\eta^{a}$, $\overline{\eta}^{a}$ can be neglected in the Landau gauge, where -- at least perturbatively -- $A^{h}$ coincides with the ordinary gluon field $A$, and as such already has zero divergence. As a result, the Landau-gauge DM action $I_{L}$ takes on a considerably simpler form, in which the $(A^{h})^{2}$ operator (and the corresponding condensate) is replaced by the ordinary quadratic gluon operator $A^{2}$ and the only new field is the fluctuation field $\delta\sigma$.

Following \cite{Dem20}, the one-loop DM gluon and ghost propagators were computed in the Landau gauge by making use of the action $I_{L}$. The latter contains two new interaction terms -- namely, a cubic interaction $\delta\sigma A^{2}$ and a quartic interaction $(A^{2})^{2}$ -- which are proportional to $\zeta^{-1}$. On the shell of the gap equation, the inclusion of the corresponding diagrams was shown to have the sole effect of removing the tree-level gluon mass term $m^{2}$ from the dressed gluon propagator, while leaving behind a counterterm $\delta Z_{2}$ required to renormalize the gluon polarization. Because of this, the one-loop Landau-gauge DM gluon and ghost propagators are the same as those computed within the Curci-Ferrari model, with the notable exception of the aforementioned lack of a tree-level mass term in the gluon propagator.

The Renormalization Group improvement of the DM propagators was performed in a renormalization scheme termed the Dynamically Infrared-Safe (DIS) scheme. In the DIS scheme, the propagators and coupling are defined, respectively, in the MOM and in the Taylor scheme, while the renormalization counterterm $\delta Z_{2}$ is chosen in such a way as to preserve the massiveness of the RG-improved gluon propagator at zero momentum. To one loop, the Landau-gauge DIS propagators can be expressed in terms of the running coupling $\lambda(p^{2})=\frac{N}{4\pi}\alpha_{s}(p^{2})$ and running gluon mass parameter $m^{2}(p^{2})$ alone. The former were compared to the lattice results of \cite{DOS16} and were found to be in very good agreement with the data over a wide range of momenta, extending from $p\approx 0.5$~GeV up to $p\approx 8$~GeV. Below $p\approx0.5$~GeV, the one-loop approximation is insufficient to reproduce the exact results; the agreement with the data is nonetheless expected to improve by going to higher order in perturbation theory.

The DIS running coupling was shown to have the behavior typical of Taylor couplings computed by massive perturbative methods: instead of developing an infrared Landau pole as in ordinary perturbation theory, it hits a maximum at $p\approx m$ and then decreases to zero like $p^{2}$ as $p\to 0$; in the UV, it reduces to the running coupling of ordinary pQCD. The running gluon mass parameter, on the other hand, strictly decreases with the momentum, reaching a saturation value $m(0)\approx0.78$~GeV and vanishing at high energies like a rational power of $\ln p^{2}$.\\

The Dynamical Model has the advantage of being a renormalizable framework within which -- by procedures such as the reduction of couplings -- the Green functions can be computed order-by-order in perturbation theory in terms of the strong coupling constant alone. This is made possible by the fact that the value of the condensate, and thus of the gluon mass parameter, can be computed by solving a gap equation which is itself built into the definition of the theory.

In the infrared, the Landau-gauge Dynamical Model reproduces the expected, non-perturbative behavior of pure Yang-Mills theory not only qualitatively -- by the saturation of the gluon propagator at zero momentum -- but also quantitatively, as demonstrated by a comparison with the lattice data. In the UV, where the effects of the gluon condensate are negligible, it reduces to ordinary perturbation theory.

The results presented in this chapter indicate that the non-vanishing of the BRST-invariant gluon condensate $\avg{(A^{h})^{2}}$ is a good candidate for explaining the mechanism by which dynamical mass generation occurs in the gluon sector of pure Yang-Mills theory.

\chapter{Conclusions and outlook}
\renewcommand{\leftmark}{\thechapter\ \ \ Conclusions and outlook}
\renewcommand{\rightmark}{\thechapter\ \ \ Conclusions and outlook}
\label{chpt:concl}

The breakdown of ordinary perturbation theory in the infrared regime of Quantum Chromodynamics, together with the occurrence of dynamical mass generation in the gluon sector, make it necessary to devise new analytical methods for studying the low-energy behavior of the strong interactions. In this thesis, two such methods -- the Screened Massive Expansion and the Dynamical Model -- have been presented within the framework of pure Yang-Mills theory, with applications of the former to full QCD.\\

The Screened Massive Expansion (SME) is a perturbative formulation of Quantum Chromodynamics that treats the transverse gluons as massive at tree level by performing a shift of the expansion point of the Faddeev-Popov gauge-fixed action. The shift is achieved by adding a mass term for the transverse gluons in the kinetic part of the FP Lagrangian and subtracting it back from its interaction part, so that the total action is left unchanged. As a result of the shift, the gluons propagate with a massive propagator at order zero in the perturbative series; moreover, a new interaction vertex -- quadratic in the gluon fields~-- arises in the Feynman rules of perturbation theory.

The presence of a gluon mass in the loops of the SME prevents the latter from developing infrared divergences and makes it possible for the gluon polarization not to vanish in the zero-momentum limit. Due to a cancellation which occurs between the tree-level mass inherited from the zero-order propagator and an opposite term contained in the polarization, the low-energy behavior of the transverse (dressed) gluon propagator is entirely determined by the interactions. At variance with ordinary pQCD, the one-loop SME gluon propagator is found to saturate to a finite constant -- and thus to develop a mass~-- at zero momentum, in agreement with the predictions of the lattice calculations. The ghost propagator, on the other hand, diverges in the infrared, thus remaining massless as expected. Both the propagators -- when evaluated in Euclidean space and at fixed scale~-- accurately reproduce the Landau-gauge pure Yang-Mills lattice data up to momenta of $\approx4$~GeV for the gluon, and of $\approx2$~GeV for the ghost.

By making use of optimization procedures based on principles of gauge invariance and of minimal sensitivity, the pure Yang-Mills fixed-scale SME propagators evaluated in a general covariant gauge can be made to depend -- modulo multiplicative renormalization~-- on the Landau-gauge value of the gluon mass parameter $m^{2}$ alone. In the gluon sector, this is accomplished by enforcing the gauge-parameter independence of the position of the (complex-conjugate) poles of the propagator -- as required by the Nielsen identities -- and by assuming that the phases of the residues at the poles are independent from the gauge parameter as well. The optimized propagators are found to be indistinguishable from those obtained from a full fit of the lattice data.

The domain of validity of the SME propagators can be extended to arbitrarily large\cleannlnp momenta by resorting to Renormalization Group (RG) methods. Within such a framework, the Taylor-scheme SME running coupling is free of Landau poles provided that the value of the coupling at the initial renormalization scale is not too large. At intermediate energies, the RG-improved and the optimized fixed-scale results can be made to match by choosing a suitable value of the initial coupling, leaving yet again $m^{2}$ as the only free parameter of the expansion. The corresponding propagators are found to be in good agreement with the lattice calculations over a wide range of momenta, extending from $\approx 0.7$~GeV up to $\approx8$~GeV. At low energies, the one-loop approximation is unable to capture the behavior of the lattice data. The discrepancy between the two is expected to be mitigated by going to higher order in perturbation theory.

The Screened Massive Expansion can be extended to finite temperatures $T\neq 0$ in order to study the thermal behavior of QCD. Within pure Yang-Mills theory, a one-loop evaluation of the (spatially) transverse and longitudinal gluon propagators is able to reproduce the lattice results only if the temperature dependence of the SME parameters is tuned separately for the two components. Doing so provides an effective description of the propagators at zero Matsubara frequency which is quite accurate in the transverse sector, but less so in the longitudinal one -- especially at low momenta and at high temperatures. The temperature dependence of the zero-(spatial-)momentum gluon poles -- that is, of the mass and damping factor of the gluon quasi-particles -- can be estimated by making use of the results obtained in the transverse sector. Both of them are found to decrease with $T$ below the critical temperature $T_{c}\approx270$~MeV corresponding to the deconfinement phase transition, and to increase roughly linearly with $T$ -- as is expected for massless particles~-- for $T>T_{c}$. Overall, we assess that at high temperatures the SME may be suboptimal as a perturbative method when compared to more refined approaches such as the Hard Thermal Loop resummation.

Within the quark sector of full QCD, a shift analogous to the one performed in the gluon sector can be used to study the infrared enhancing of the quark mass brought by the violation of chiral symmetry. The Landau-gauge Euclidean quark mass functions computed to one loop in the SME are found to be in very good agreement with the unquenched lattice results for light quarks of masses $\leq90$~MeV, displaying saturation values of about $\approx400$~MeV which are much larger than the quarks' renormalized masses. The one-loop SME quark $Z$-function, on the other hand, shows a decreasing behavior which conflicts with the lattice results. This might be a consequence of the unusually small one-loop corrections to the vector part of the Landau-gauge quark self-energy, which make it necessary to at least include the two-loop diagrams in the calculations. Evidence that the behavior of the $Z$-function can be fixed by going to higher order in perturbation theory can be provided within the framework of the SME by replacing the internal gluon lines of the one-loop self-energy diagrams with (the principal-part approximation of) the full dressed gluon propagator, thus taking into account the higher-order corrections to the latter. Doing so provides a $Z$-function which correctly increases with momentum at high energies, despite still showing the wrong behavior at low energies.\\

The Dynamical Model (DM) explores the possibility that the dynamical generation of an infrared mass for the gluons might be triggered by a non-vanishing condensate of dimension $2$ of the form $\langle (A^{h})^{2}\rangle$, where $A^{h}$ is a gauge- and BRST-invariant non-local version of the gluon field $A$. The former can be defined as the field obtained by applying a gauge transformation to $A$ in such a way as to make the latter divergenceless. When expanded in powers of the coupling constant, $A^{h}$ is found to be perturbatively equal to the transverse component of the gluon field, plus an infinite number of higher-order\cleannlnp terms which depend non-locally on the divergence $\partial\cdot A$.

The non-vanishing of the VEV $\langle (A^{h})^{2}\rangle$ within pure Yang-Mills theory can be studied by making use of Local Composite Operator methods. After coupling the operator $(A^{h})^{2}$ to an external current in the Faddeev-Popov action, successive transformations of the partition function lead to a BRST-invariant action $I$ in which the condensate $\langle (A^{h})^{2}\rangle$ appears multiplied by the operator $(A^{h})^{2}$ itself -- thus yielding a mass term for the gluons. An effective potential can then be derived which is found to have a minimum for a non-zero value of the condensate, showing that $\langle (A^{h})^{2}\rangle\neq 0$ in any covariant gauge.

When evaluated on the solutions of the gap equation -- that is, at vanishing first derivative of the effective potential --, the action $I$ is dynamically equivalent to the Faddeev-Popov action. The Green functions of pure Yang-Mills theory can thus be computed in any covariant gauge within the Dynamical Model, by making use of the action $I$. The DM incorporates the non-perturbative corrections brought by the condensate via a gluon mass parameter $m^{2}\propto-\langle (A^{h})^{2}\rangle$ and via a number of new vertices which also involve new bosonic and fermionic auxiliary fields. The derivation of the DM gluon and ghost propagators is the simplest in the Landau gauge, where $A^{h}=A$ and most of the auxiliary fields decouple. An explicit calculation shows that the one-loop Landau-gauge DM gluon polarization is equal to its Curci-Ferrari counterpart modulo new terms which cancel the tree-level gluon mass as soon as the gap equation is enforced. Similarly, the one-loop ghost self-energy is equal to the one computed within the Curci-Ferrari model. When renormalized in the Dynamically Infrared-Safe (DIS) scheme, the Euclidean RG-improved DM propagators are found to be in very good agreement with the Landau-gauge lattice data down to momenta $\approx0.5$~GeV. Thanks to the gap equation, this is achieved by making use of the strong coupling constant at the initial renormalization scale as the only free parameter of the expansion. The corresponding DIS running coupling is free of Landau poles, but still quite too large at intermediate energies for a one-loop approximation to be accurate down to vanishing scales; indeed, the approximation fails to reproduce the lattice data at momenta smaller than $\approx0.5$~GeV. The running gluon mass parameter is found to saturate to a finite constant in the zero momentum limit, and to vanish like a negative rational power of the logarithm at high energies.\\

The Screened Massive Expansion and the Dynamical Model paint a picture of the infrared regime of Quantum Chromodynamics which agrees with the Landau-gauge lattice calculations both qualitatively -- displaying dynamical mass generation in the gluon sector and a massless ghost propagator -- and quantitatively, within the limits of a one-loop approximation. The advantage they provide over other frameworks for studying the strong interactions at low energy is that of being first-principles, fully analytic methods, whose predictions can be improved systematically by evaluating the higher-order contributions to the QCD perturbative series. The way in which such predictions are made from first principles differs between the two frameworks. In the SME, the spurious free parameters of the expansion are fixed by requirements of gauge invariance and minimal sensitivity, which in the context of the Renormalization Group can then be exploited to express the strong running coupling as a function of the gluon mass parameter. In contrast, the Dynamical Model achieves the same goal -- although practically in the opposite sense, expressing the mass parameter in terms of the coupling -- thanks to a gap equation which is built into its very definition.

Both the SME and the DM use the ordinary Faddeev-Popov action as the starting point for the set-up of a modified perturbative expansion of QCD. This has three important consequences on the overall structure of the resulting frameworks. First of all, BRST invariance is retained as a symmetry: in the SME, this is true at the exact level, given that the total FP\cleannlnp action is not changed at all; in the DM, on the other hand, BRST invariance is achieved by a natural (nilpotent) extension of the standard BRST transformations to the new fields that appear in the action $I$. Second of all, the tree-level mass term which appears in the zero-order gluon propagators is removed from the corresponding dressed propagator: in the SME this happens thanks to an opposite gluon mass counterterm in the polarization, whereas in the DM the cancellation occurs as soon as the gap equation is enforced. In particular, in both cases the massiveness of the propagator can be specifically traced back to the loops of the expansion. Finally, none of the two frameworks addresses the issue of the Gribov copies. As we argued in the \hyperref[chpt:intro]{Introduction}, this can be justified on the account that the effects of the Gribov copies on the dynamics of the theory are expected to be less strong if the gluons acquire a mass.

Whereas the perturbative series of the SME contains an infinite number of crossed diagrams at each order in the coupling -- so that a criterion for its truncation has to be chosen explicitly by moving first and foremost from renormalizability \mbox{requirements --,} the opposite applies to the Dynamical Model. In practical terms, this means on the one hand that the SME and the DM radiative corrections turn out to be different when computed to fixed order in perturbation theory, and on the other hand that the gluon mass parameters $m^{2}$ defined within the two frameworks cannot be directly identified with one another. This last aspect is confirmed by the fact that the mass parameter of the SME is found not to run with the renormalization scale at one loop -- at variance with its DM counterpart, whose running is determined by that of the strong coupling constant and of the BRST-invariant condensate. Thus, while in the Dynamical Model the gluon mass parameter is interpreted in terms of the latter by its very definition, within the SME $m^{2}$ must be simply regarded as a dimensionful scale introduced in the formalism with the objective of providing the transverse gluons with a mass. Despite these differences, the results obtained at one loop within the two frameworks remain overall quite similar, and yield a value of $m\approx 0.65$~GeV (at $1$~GeV, as far as the DM is concerned) when the energy units of the theory are fixed by a comparison with the lattice data.\\

The achievements of the SME and of the DM in the two-point sector of pure Yang-Mills theory make the two approaches worthy of further research. As far as the Dynamical Model is concerned, work is already in progress to evaluate the propagators in an arbitrary covariant gauge so as to test whether the model is able to predict dynamical mass generation for the gluons beyond the Landau gauge. This is widely expected on the basis of the massiveness of the corresponding zero-order propagator, but needs to be confirmed by explicit calculations carried out in the context of specific renormalization schemes. Once in possession of the relevant analytic expressions, we will also be able to investigate whether the poles of the propagator are complex-conjugate and gauge-parameter independent, and to test the hypothesis -- advanced within the SME -- that the phases of its residues as well do not depend on the gauge.

In the framework of the Screened Massive Expansion, the gauge invariance of the gluon phases can in principle be used to make a determination of the gluon free parameters also within full QCD. Preliminary results suggest that this will be trickier to achieve than in pure Yang-Mills theory due to the larger number of parameters, which encompasses the chiral mass -- that is, the mass parameter that sets the scale for the quark mass function at low energies --, as well as the quark's renormalized mass and various renormalization constants. The latter might be fixed -- or at least put in relation to one another -- by enforcing the gauge invariance of the quark poles, which again holds because of the the Nielsen identities. One-loop expressions for the SME quark propagator in an arbitrary covariant gauge have already been derived with the objective of undertaking these analyses.\newpage

In the long run, it would be interesting to extend the Dynamical Model to full QCD and to use both the SME and the DM to evaluate higher-point Green functions such as the $3$-gluon, the ghost-gluon and the quark-gluon interactions vertices. Moreover, the calculations could be pushed to two loops in order to test whether the deep IR behavior of the RG-improved propagators and the overall behavior of the quark $Z$-function improve as expected when the higher-order corrections are included in the perturbative series. Our hope is that the results presented in this thesis and by previous works will generate further interest in the research on perturbative methods for probing the non-perturbative regime of Quantum Chromodynamics.

\appendix
\addappheadtotoc
\appendixpage
\titleformat{\chapter}[display]{\filright\normalfont\Huge\bfseries}{\chaptertitlename\ \thechapter}{15pt}{} 

\chapter{Canonical quantization of the Faddeev-Popov action}
\renewcommand{\leftmark}{\appendixname\ \thechapter: Canonical quantization of the Faddeev-Popov action}
\renewcommand{\rightmark}{\appendixname\ \thechapter: Canonical quantization of the Faddeev-Popov action}
\label{app:cqfp}
In Secs.~\ref{sec:fpquant} and \ref{sec:brst} we saw that in the presence of the Nakanishi-Lautrup field $B^{a}$ the Faddeev-Popov Lagrangian $\ELL_{\text{FP}}$ can be expressed as
\begin{align}
\ELL_{\text{FP}}&=-\frac{1}{2}\,\partial_{\mu}A_{\nu}^{a}\,\left(\partial^{\mu}A^{a\,\nu}-\partial^{\nu}A^{a\,\mu}\right)-gf^{a}_{bc}\,\partial_{\mu}A_{\nu}^{a}\,A^{b\,\mu}A^{c\,\nu}+\\
\notag &\quad\,-\frac{1}{4}\,g^{2}f^{a}_{bc}f^{a}_{de}\,A_{\mu}^{b}A_{\nu}^{c}A^{d\,\mu}A^{e\,\nu}+\frac{\xi}{2}\,B^{a}B^{a}-\partial^{\mu}B^{a}A_{\mu}^{a}+\\
\notag&\quad\,+\psibar(i\gamma^{\mu}\partial_{\mu}-M)\psi+g\,\psibar\gamma^{\mu}T_{a}\psi A_{\mu}^{a}+\\
\notag &\quad\,+\partial^{\mu}\cbar^{a}\partial_{\mu}c^{a}+gf^{a}_{bc}\,\partial^{\mu}\cbar^{a}A_{\mu}^{b}c^{c}\ .
\end{align}
Just like any Lagrangian, $\ELL_{\text{FP}}$ can be quantized by making use of the canonical formalism.

Within the canonical formalism, one associates conjugate momenta $\Pi$ to the time-derivatives of the fields $F$ using the formula
\begin{equation}\label{dgq945}
\Pi=\frac{\partial_{R}\mc{L}}{\partial \dot{F}}
\end{equation}
and then computes the Hamiltonian density $\mc{H}$ by performing a Legendre transform of the Lagrangian $\mc{L}$,
\begin{equation}\label{dkq485}
\mc{H}(F,\Pi)=\Pi\,\dot{F}(F,\Pi)-\mc{L}(F,\dot{F}(F,\Pi))\ ,
\end{equation}
where the functions $\dot{F}(F,\Pi)$ are obtained by inverting Eq.~\eqref{dgq945}. As quantum operators, the fields and their conjugate momenta are endowed with canonical (equal-time) anti/commutation relations of the form
\begin{equation}\label{dkq486}
[F(\vec{x},t),\Pi(\vec{y},t)]_{\mp}=i\,\delta(\vec{x}-\vec{y})\ ,
\end{equation}
where the upper (resp. lower) sign holds for bosonic (resp. fermionic) fields. The Heisenberg equations for the fields and their conjugate momenta are then obtained by taking the commutators of the former with the Hamiltonian operator $H=\int d^{3}x\ \mc{H}$,
\begin{equation}
\dot{F}=i[H,F]\ ,\qquad\qquad \dot{\Pi}=i[H,\Pi]\ .
\end{equation}

By applying these definitions to $\mc{L}_{\text{FP}}$, we find that the momenta $\Pi^{a\,\mu}$, $\Pi_{B}^{a}$, $\Pi_{c}^{a}$, $\Pi_{\cbar}^{a}$, $\Pi_{\psi}$ conjugate to the fields $A_{\mu}^{a}$, $B^{a}$, $c^{a}$, $\cbar^{a}$ and $\psi$, respectively, are given by
\begin{equation}\label{canmom}
\Pi^{a\,\mu}=F^{a\,\mu0}\ ,\qquad\Pi_{B}^{a}=-A_{0}^{a}\ ,\qquad \Pi_{c}^{a}=\partial_{0}\cbar^{a}\ ,\qquad\Pi_{\cbar}^{a}=-D_{0}c^{a}\ ,\qquad\Pi_{\psi}=i\psi^{\dagger}\ .
\end{equation}
\cleannlnp
We see that, formally, $\Pi^{a\,0}=0$. The term $\Pi^{a\,0}\dot{A}_{0}^{a}$ will thus vanish in Eq.~\eqref{dkq485}. Moreover, $\Pi_{B}^{a}=-A_{0}^{a}$. It follows that in the canonical formalism $A_{0}^{a}$ must not be treated as a field variable, but rather as the momentum conjugate to the field $B^{a}$ (modulo sign). From Eqs.~\eqref{dkq486} and \eqref{canmom} we can read out the following non-vanishing anti/commutation relations for the fields:
\begin{align}
&[A_{i}^{a}(\vec{x},t),F^{b\,j0}(\vec{y},t)]=i\delta^{ab}\delta_{i}^{j}\delta(\vec{x}-\vec{y})\ ,\qquad[A_{0}^{a}(\vec{x},t),B^{b}(\vec{y},t)]=i\delta^{ab}\delta(\vec{x}-\vec{y})\ ,\\
\notag &\{c^{a}(\vec{x},t),\partial_{0}\cbar^{b}(\vec{y},t)\}=i\delta^{ab}\delta(\vec{x}-\vec{y})\ ,\qquad\{\cbar^{a}(\vec{x},t),D_{0}c^{b}(\vec{y},t)\}=-i\delta^{ab}\delta(\vec{x}-\vec{y})\ ,\\
\notag &\qquad\qquad\qquad\qquad\qquad\quad\{\psi(\vec{x},t),\psi^{\dagger}(\vec{y},t)\}=\delta(\vec{x}-\vec{y})\,\one\ .
\end{align}

In order to derive the Faddeev-Popov Hamiltonian, we first note that, by Eq.~\eqref{canmom}, the time derivatives of the fields $A_{i}^{a}$, $c^{a}$ and $\cbar^{a}$ can be expressed in terms of the fields themselves and of their conjugate momenta as
\begin{align}\label{candermom}
\dot{A}^{a}_{i}=-\Pi^{a}_{i}+\partial_{i}A^{a}_{0}+gf^{a}_{bc}\,A_{i}^{b}A_{0}^{c}\ ,\qquad \dot{c}^{a}=-\Pi^{a}_{\cbar}-gf^{a}_{bc}\,A_{0}^{b}c^{c}\ ,\qquad\dot{\cbar}^{a}=\Pi^{a}_{c}\ .
\end{align}
Despite the lack of analogous relations for the time derivatives of the fields $\psi$ and $B^{a}$, the linearity of $\mc{L}_{\text{FP}}$ in $\dot{\psi}$ and $\dot{B}^{a}$, together with the result $\Pi_{\psi}=i\psi^{\dagger}$, $\Pi_{B}^{a}=-A_{0}^{a}$, allow us to compute $\mc{H}$ from Eq.~\eqref{candermom} alone. An explicit calculation that uses
\begin{align}
\mc{H}=\Pi^{a\,i}\dot{A}_{i}^{a}+\Pi_{B}^{a}\dot{B}^{a}+\Pi_{c}^{a}\dot{c}^{a}+\Pi_{\cbar}^{a}\dot{\cbar}^{a}+\Pi_{\psi}\dot{\psi}-\mc{L}_{\text{FP}}
\end{align}
as its starting point yields
\begin{align}
\mc{H}&=-\frac{1}{2}\,\Pi^{a\,i}\Pi^{a}_{i}+\Pi^{a\,i}\left(\partial_{i}A_{0}^{a}+gf^{a}_{bc}\,A_{b}^{i}A_{0}^{c}\right)+\frac{1}{4}\,F_{ij}^{a}F^{a\,ij}-\frac{\xi}{2}\,B^{a}B^{a}+\partial^{i}B^{a}A_{i}^{a}+\\
\notag&\quad\, +i\Pi_{\psi}\gamma^{0}\left(i\gamma^{i}D_{i}+g\gamma^{0}T^{a}A_{0}^{a}-M\right)\psi-\Pi_{c}^{a}\left(\Pi_{\cbar}^{a}+gf^{a}_{bc}\,A_{0}^{b}c^{c}\right)-\partial^{i}\cbar^{a}D_{i}c^{a}\ .
\end{align}
It can be checked that the Heisenberg equations for $A_{i}^{a}$, $c^{a}$ and $\cbar^{a}$ coincide with Eqs.~\eqref{candermom}, whereas the other Heisenberg equations can be rearranged so as to be formally identical to the field equations obtained by minimizing the Faddeev-Popov action $S_{\text{FP}}=\int d^{4}x\ \mc{L}_{\text{FP}}$.

\chapter{Perturbative decoupling of the determinant $\boldsymbol{det(\Lambda(\xi))}$ within dimensional regularization in the Dynamical Model}
\renewcommand{\leftmark}{\appendixname\ \thechapter: Perturbative decoupling of $\det(\Lambda(\xi))$ within dimreg in the DM}
\renewcommand{\rightmark}{\appendixname\ \thechapter: Perturbative decoupling of $\det(\Lambda(\xi))$ within dimreg in the DM}
\label{app:onedet}

In order to localize the BRST-invariant gluon field $A^{h}$, in Sec.~\ref{sec:dynmodpotcalc} we introduced a unity of the form
\begin{equation}\label{unityapp0}
1=\mathcal{N}\int \mathcal{D}\xi\,\mathcal{D}\tau\,\mathcal{D}\bar\eta\,\mathcal{D}\eta\,e^{-\Delta S_1}\det(\Lambda(\xi))
\end{equation}
in the partition function of the Dynamical Model. In this Appendix we show that the determinant $\det(\Lambda(\xi))$ does not perturbatively contribute to the $n$-point Green functions of the theory, as long as it is defined in dimensional regularization. As a consequence, when doing calculations in perturbation theory using dimensional regularization, the determinant can be suppressed by setting $\det(\Lambda(\xi))=1$.\\

In order to prove our statement, we first rewrite the determinant in terms of a functional integral over a new pair of ghost fields $(\lambda,\overline{\lambda})$,
\begin{equation}\label{eq:detlambda0}
\det(\Lambda(\xi))=\int\mathcal{D}\overline{\lambda}\mathcal{D}\lambda\ \exp\left\{-\int d^{d}x\ \overline{\lambda}^{a}\Lambda_{ab}(\xi)\lambda^{b}\right\}\ .
\end{equation}
Since perturbatively
\begin{equation}
\Lambda_{ab}(\xi)=\delta_{ab}-\frac{g}{2}\,f_{abc}\,\xi^{c}+\frac{g^{2}}{3!}\,f_{ace}f_{edb}\,\xi^{c}\xi^{d}+\cdots\ ,
\end{equation}
we may re-express Eq.~\eqref{eq:detlambda0} as
\begin{equation}
\det(\Lambda(\xi))=\int\mathcal{D}\overline{\lambda}\mathcal{D}\lambda\ e^{-(I_{0}+I_{1})}\ ,
\end{equation}
where the action terms $I_{0}$ and $I_{1}$ read
\begin{equation}
I_{0}=\int d^{d}x\ \overline{\lambda}^{a}\lambda^{a}\ ,\qquad\qquad I_{1}=\int d^{d}x\ \overline{\lambda}^{a}\Omega_{ab}(\xi)\lambda^{b}\ ,
\end{equation}
and $\Omega_{ab}(\xi)$ is given by
\begin{align}
\Omega_{ab}(\xi)=\Lambda_{ab}(\xi)-\delta_{ab}\ .
\end{align}
The action term $I_{1}$ contains the interactions between $(\lambda,\overline{\lambda})$ and $\xi$. The latter are quadratic in the ghost fields, with their $\xi$ dependence encoded in the function $\Omega_{ab}(\xi)$. $I_{0}$, on the other hand, contains the zero-order ghost propagator, which is easily seen to be $Q^{ab}(p)=\delta^{ab}$ in momentum space, or $Q^{ab}(x)=\delta^{ab}\delta(x)$ in coordinate space.

\begin{figure}[H]
\centering
\includegraphics[width=0.23\textwidth]{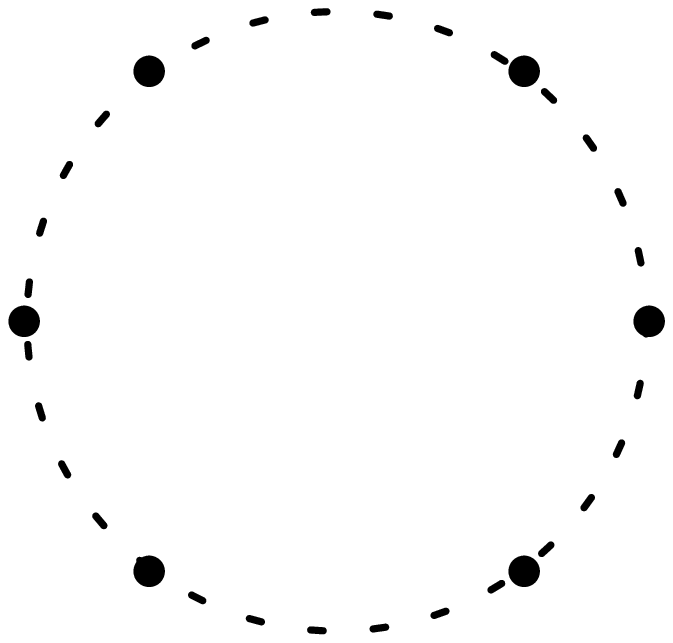}
\vspace{5pt}
\caption{Loop contributing to the ghost average in Eq.~\eqref{eq:avgn0conn} (example for $n=6$). The dashed line is the $(\lambda,\overline{\lambda})$ zero-order propagator.}
\label{fig:lambdaloop}
\end{figure}

Consider the vacuum expectation value $\langle\mathcal{O}\rangle$ of an operator $\mathcal{O}$ which does not depend on the newly-introduced fields $(\lambda,\overline{\lambda})$. This can be computed as
\begin{equation}\label{eq:lambdaovev}
\langle\mathcal{O}\rangle=\frac{\langle\mathcal{O}e^{-I_{1}}\rangle_{0}}{\langle e^{-I_{1}}\rangle_{0}}=\langle\mathcal{O}e^{-I_{1}}\rangle_{0,\text{conn.}}=\sum_{n=0}^{+\infty}\frac{(-1)^{n}}{n!}\langle\mathcal{O}I_{1}^{n}\rangle_{0,\text{conn.}}\ ,
\end{equation}
where the subscript 0 denotes that the average is to be taken with respect to the action $I_{0}$ plus any other $(\lambda,\overline{\lambda})$-independent term originally present in the full action of the theory. In $d$ dimensions, $\langle\mathcal{O}I_{1}^{n}\rangle_{0,\text{conn.}}$ explicitly reads
\begin{align}\label{eq:avgn0conn}
&\langle\mathcal{O}I_{1}^{n}\rangle_{0,\text{conn.}}=\int \prod_{i=1}^{n} d^{d}x_{i}\ \left\langle\mathcal{O}\prod_{j=1}^{n}\Omega_{a_{j}b_{j}}(\xi(x_{j}))\right\rangle_{00,\text{conn.}}\times\\
\notag&\qquad\qquad\qquad\qquad\times\left\langle\overline{\lambda}^{a_{1}}(x_{1})\lambda^{b_{1}}(x_{1})\cdots \overline{\lambda}^{a_{n}}(x_{n})\lambda^{b_{n}}(x_{n})\right\rangle_{\text{gh.},\text{conn.}}\ ,
\end{align}
where the subscript 00 denotes that the first average is to be taken with respect to the original action of the theory, whereas the subscript ``gh.'' denotes that the second average is to be taken with respect to the zero-order ghost action $I_{0}$. Diagrammatically, for each $n\geq 1$, the ghost average receives contributions from a single ghost loop, depicted in Fig.~\ref{fig:lambdaloop}. In coordinate space, suppressing the color structure, the diagram reads
\begin{equation}
(-1)(n-1)!\,\delta(x_{1}-x_{2})\cdots\delta(x_{n-1}-x_{n})\delta(x_{n}-x_{1})\ ,
\end{equation}
or, equivalently,
\begin{equation}
(-1)(n-1)!\,\delta(0)\int d^{d}x\ \prod_{i=1}^{n}\delta(x_{i}-x)\ ,
\end{equation}
where $\delta(0)$ is a Dirac delta in coordinate space,
\begin{equation}\label{eq:deltagh0}
\delta(0)=\int\frac{d^{d}q}{(2\pi)^{d}}\ 1\ .
\end{equation}
Therefore, for $n\geq 1$,
\begin{equation}
\langle\mathcal{O}I_{1}^{n}\rangle_{0,\text{conn.}}=(-1)(n-1)!\,\delta(0)\int d^{d}x\ \left\langle\mathcal{O}\text{Tr}\left\{\Omega^{n}(\xi(x))\right\}\right\rangle_{00,\text{conn.}}\ ,
\end{equation}
where $\Omega^{n}(\xi)$ is the matrix product of $n$ factors of $\Omega(\xi)$ and the trace is taken over the color indices.

In dimensional regularization, the integral in Eq.~\eqref{eq:deltagh0} vanishes (see e.g. \cite{Col84}). It follows that $\langle\mathcal{O}I_{1}^{n}\rangle_{0,\text{conn.}}=0$ for every $n\geq 1$, so that, going back to Eq.~\eqref{eq:lambdaovev},
\begin{equation}\label{eq:appafin}
\langle\mathcal{O}\rangle=\langle\mathcal{O}\rangle_{0}=\langle\mathcal{O}\rangle_{00}\ ,
\end{equation}
where to obtain $\langle\mathcal{O}\rangle_{00}$ we have integrated out the free ghost action $I_{0}$ from $\langle\mathcal{O}\rangle_{0}$. What Eq.~\eqref{eq:appafin} means is that the perturbative corrections to the vacuum expectation value $\langle\mathcal{O}\rangle$ due to the determinant $\det(\Lambda(\xi))$ vanish in dimensional regularization. Therefore, the vacuum expectation value of any operator $\mathcal{O}$ in the full theory can be computed by setting $\det(\Lambda(\xi))=1$ in its dimensionally regularized partition function.
\newpage
\vspace*{5mm}
\vspace{5mm}
One may have noticed that our proof -- besides dimensional regularization -- relies exclusively on the fact that $\Lambda(\xi)$ is equal to the unit matrix to lowest order in perturbation theory. The question arises, then, whether the proof is general enough to apply to the determinant of any such matrix. The answer is that, in general, it does not. Indeed, setting $\delta(0)=0$ in dimensional regularization is allowed if and only if the calculations can be carried out without spoiling the symmetries of the theory.

While Lorentz invariance is clearly preserved by the action in Eq.~\eqref{eq:detlambda0}, showing that  the latter does not violate the BRST invariance of the full action of the theory requires us to extend the symmetry to the ghost fields $\lambda$ and $\overline{\lambda}$. Indeed, a straightforward calculation starting from Eq.~\eqref{gao586} and from the definition of $\Lambda_{ab}(\xi)$ in Eq.~\eqref{eq:lambdadef} yields
\begin{equation}
s\Lambda_{ab}(\xi)=\Lambda_{ac}(\xi)\,\Psi^{c}_{b}(c,\xi)\ ,
\end{equation}
where
\begin{equation}
\Psi^{a}_{b}(c,\xi)=-\frac{\partial(s\xi^{a})}{\partial \xi^{b}}\ ,
\end{equation}
so that the new ghosts must have non-vanishing BRST transformations if the action in Eq.~\eqref{eq:detlambda0} is to be invariant. Since the BRST transformation does not act on the anti-ghost index of $\Lambda_{ab}(\xi)$, it is reasonable to define
\begin{align}\label{eq:ext2brst}
s\lambda^{a}=-\Psi^{a}_{b}(c,\xi)\,\lambda^{b}\ ,\qquad\qquad s\overline{\lambda}^{a}=0\ ,
\end{align}
where $s\lambda^{a}$ is chosen so that $s(\Lambda\lambda)=0$. The action in Eq.~\eqref{eq:detlambda0} -- and the full action of the theory together with it -- is invariant with respect to this extended BRST transformation. The nilpotency of the extended BRST operator is easily proved by observing that $s^{2}\Lambda_{ab}(\xi)=0$ -- which holds thanks to the nilpotency of $s$ on the fields $\xi$ and $c$ -- implies that
\begin{equation}
0=s^{2}\Lambda_{ab}=\Lambda_{ac}\left(\Psi^{c}_{d}\Psi^{d}_{b}+s\Psi_{b}^{c}\right)\ ,
\end{equation}
that is, $s\Psi=-\Psi^{2}$. When plugged into Eq.~\eqref{eq:ext2brst}, this relation ensures that $s^{2}\lambda^{a}=s^{2}\overline{\lambda}^{a}=0$.

\chapter{Published papers}
\renewcommand{\leftmark}{\appendixname\ \thechapter: Published papers}
\renewcommand{\rightmark}{\appendixname\ \thechapter: Published papers}
\label{app:published}
\clearpage
\thispagestyle{empty}
\phantomsection
\addtocounter{section}{1}
\addcontentsline{toc}{section}{\thesection\ \ \,G. Comitini and F. Siringo, Phys. Rev. D 97 (2018)}
\renewcommand{\rightmark}{\thesection\ \ \ G. Comitini and F. Siringo, Phys. Rev. D 97 (2018)}
\includepdf[pages=-, noautoscale=false, scale=0.86, offset=7.5mm 0, pagecommand={}]{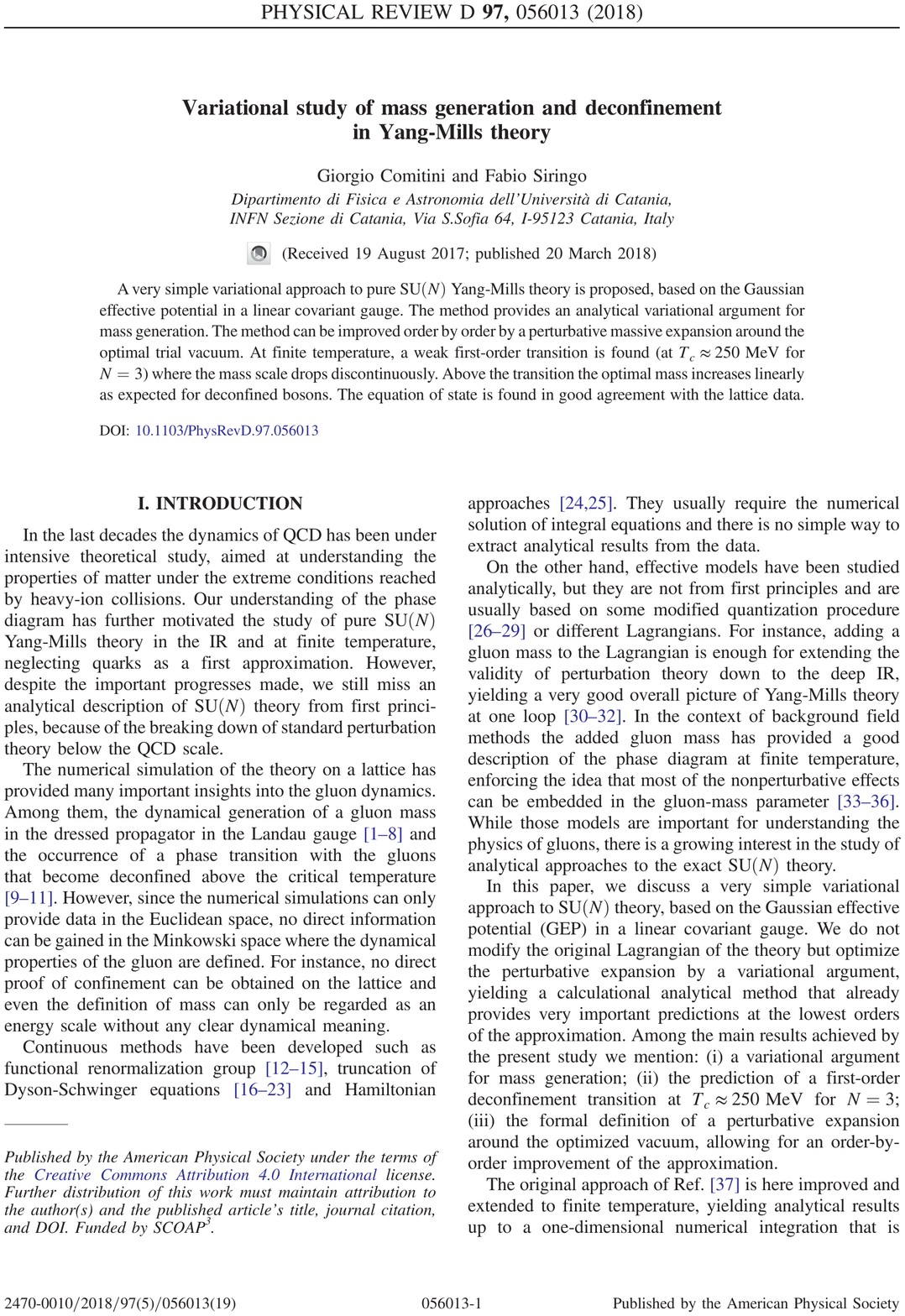}
\clearpage
\thispagestyle{empty}
\phantomsection
\addtocounter{section}{1}
\addcontentsline{toc}{section}{\thesection\ \ \,F. Siringo and G. Comitini, Phys. Rev. D 98 (2018)}
\renewcommand{\rightmark}{\thesection\ \ \ F. Siringo and G. Comitini, Phys. Rev. D 98 (2018)}
\includepdf[pages=-, noautoscale=false, scale=0.86, offset=7.5mm 0, pagecommand={}]{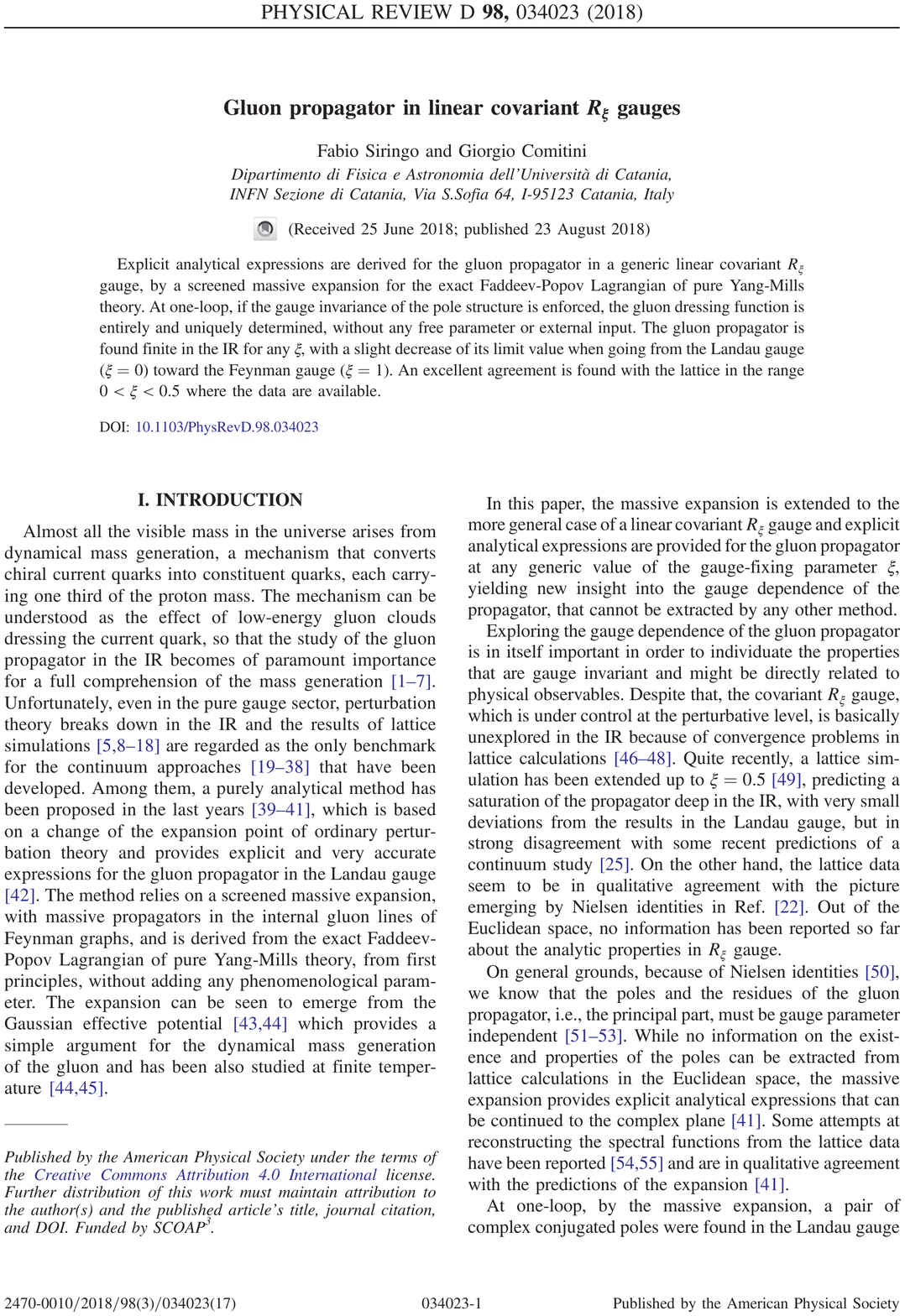}
\clearpage
\thispagestyle{empty}
\phantomsection
\addtocounter{section}{1}
\addcontentsline{toc}{section}{\thesection\ \ \,G. Comitini and F. Siringo, Phys. Rev. D 102 (2020)}
\renewcommand{\rightmark}{\thesection\ \ \ G. Comitini and F. Siringo, Phys. Rev. D 102 (2020)}
\includepdf[pages=-, noautoscale=false, scale=0.86, offset=7.5mm 0, pagecommand={}]{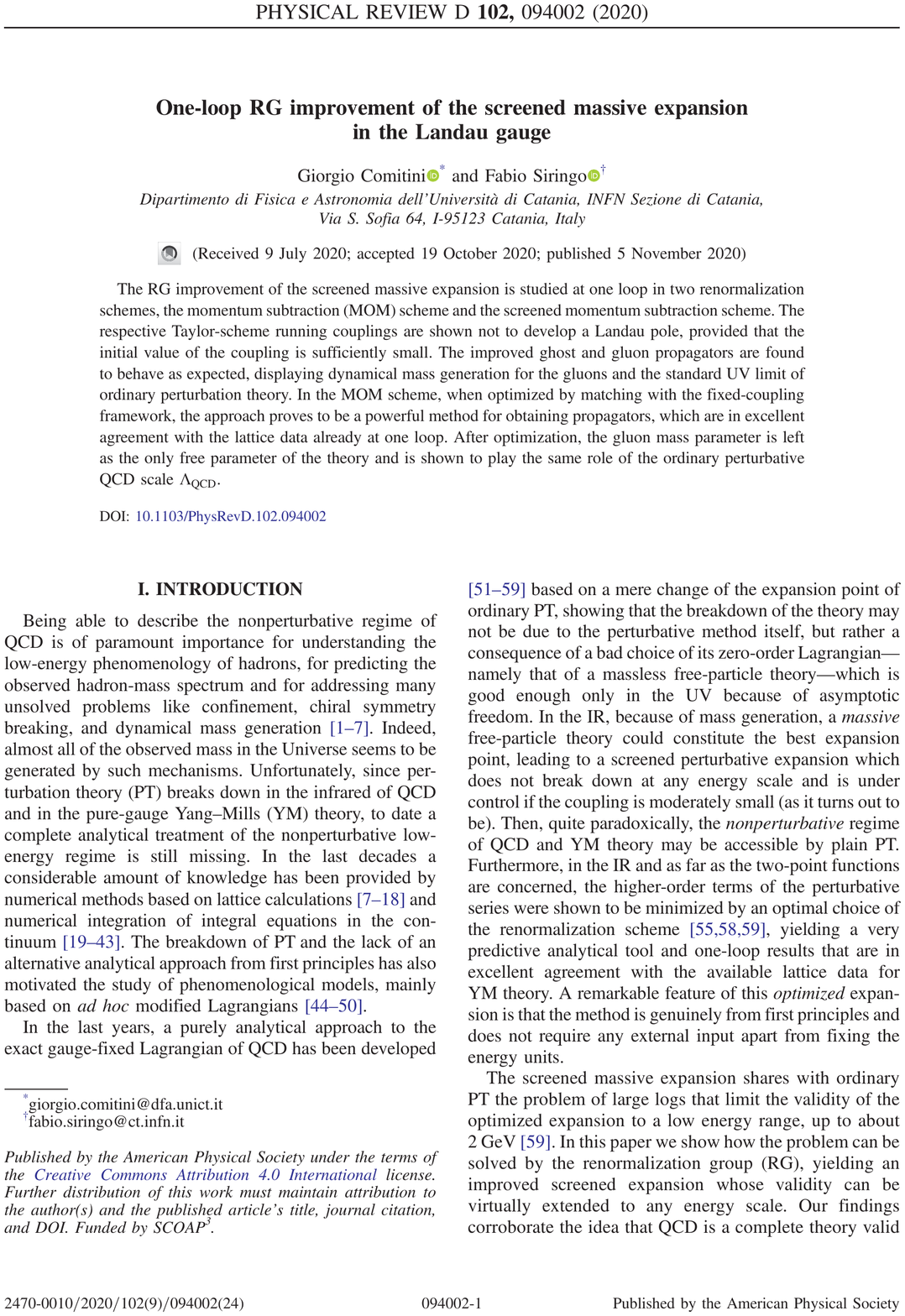}
\phantomsection
\addtocounter{section}{1}
\addcontentsline{toc}{section}{\thesection\ \ \,F. Siringo and G. Comitini, Phys. Rev. D 103 (2021)}
\renewcommand{\rightmark}{\thesection\ \ \ F. Siringo and G. Comitini, Phys. Rev. D 103 (2021)}
\includepdf[pages=-, noautoscale=false, scale=0.86, offset=7.5mm 0, pagecommand={}]{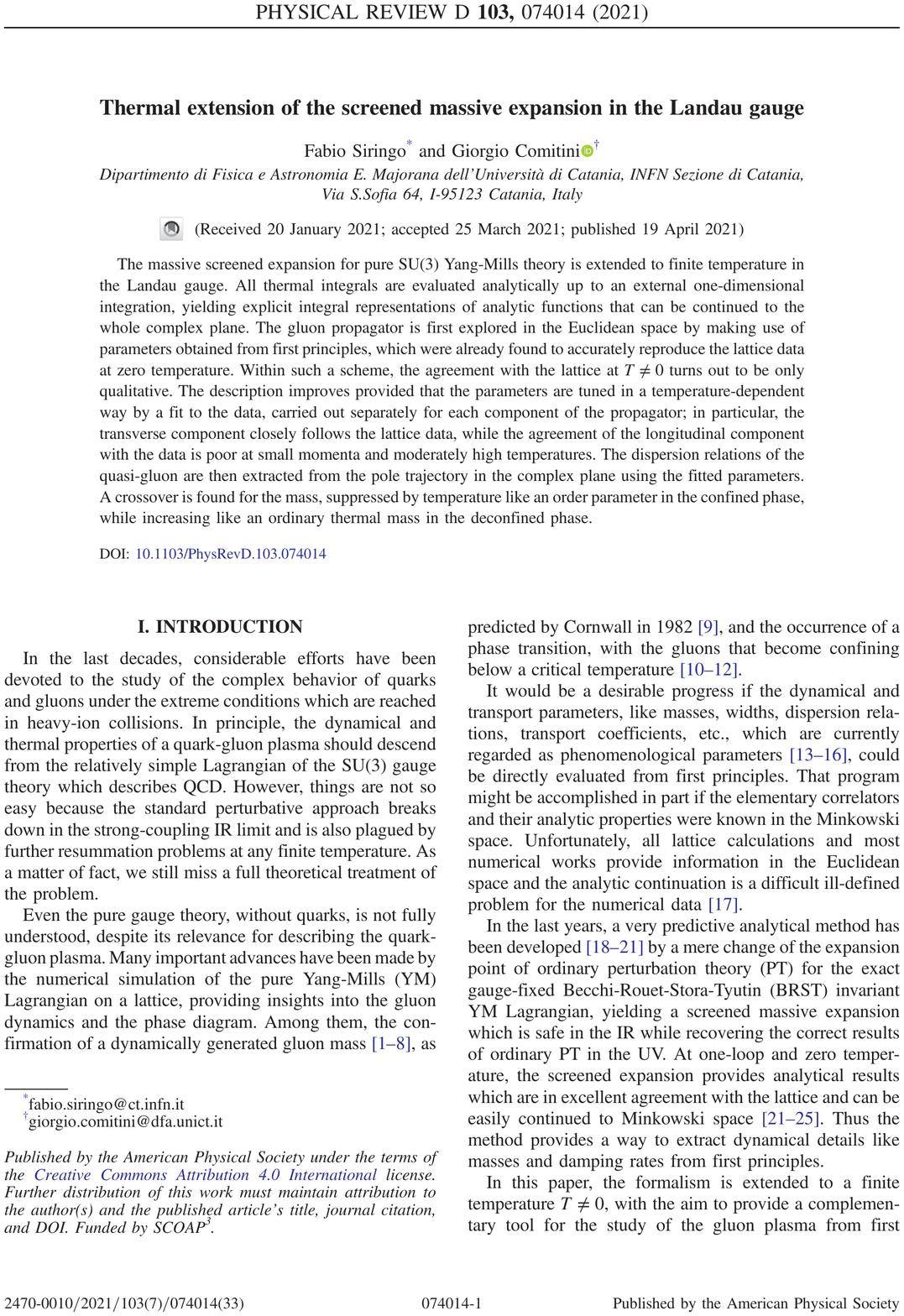}
\clearpage
\thispagestyle{empty}
\phantomsection
\addtocounter{section}{1}
\addcontentsline{toc}{section}{\thesection\ \ \,G. Comitini, D. Rizzo, M. Battello, and F. Siringo, Phys. Rev. D 104 (2021)}
\renewcommand{\rightmark}{\thesection\ \ \ G. C., D. Rizzo, M. Battello, and F. Siringo, Phys. Rev. D 104 (2021)}
\includepdf[pages=-, noautoscale=false, scale=0.86, offset=7.5mm 0, pagecommand={}]{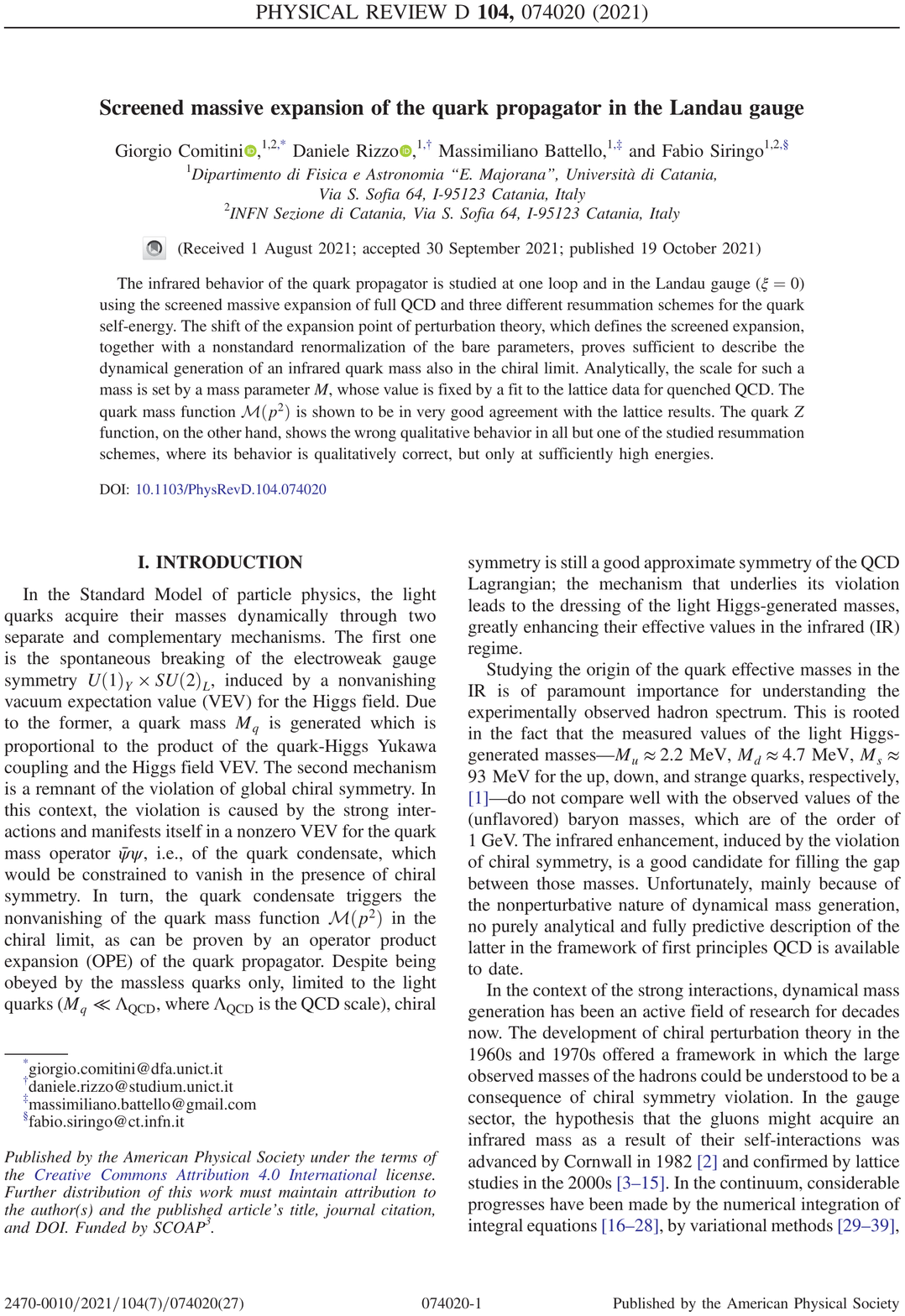}
\clearpage
\thispagestyle{empty}
\phantomsection
\addtocounter{section}{1}
\addcontentsline{toc}{section}{\thesection\ \ \,F. Siringo and G. Comitini, Phys. Rev. D 106 (2022)}
\renewcommand{\rightmark}{\thesection\ \ \ F. Siringo and G. Comitini, Phys. Rev. D 106 (2022)}
\includepdf[pages=-, noautoscale=false, scale=0.86, offset=7.5mm 0, pagecommand={}]{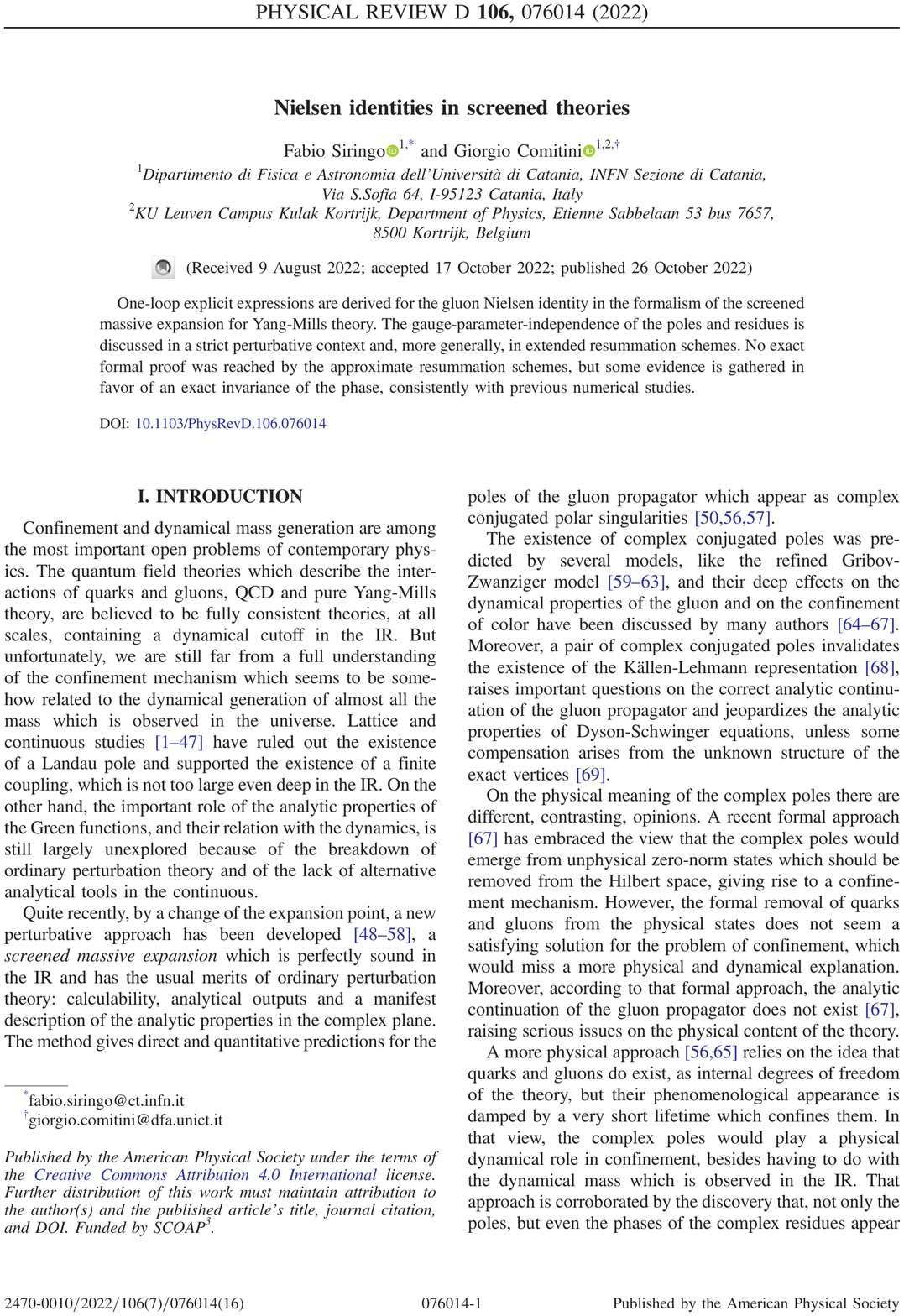}

\thispagestyle{empty}
\backmatter

\renewcommand{\leftmark}{Bibliography}
\renewcommand{\rightmark}{Bibliography}
\bibliography{biblio,chromobib}
\bibliographystyle{bibstyle}

\chapter{A note on the figures}
\renewcommand{\leftmark}{A note on the figures}
\renewcommand{\rightmark}{A note on the figures}

Most of the figures displayed in Chapters~\ref{chpt:sme} and~\ref{chpt:smeapp} were reproduced from the published papers already cited in the introductions to those chapters and within the text (see Appendix~\ref{app:published}).

\chapter{Acknowledgments}
\renewcommand{\leftmark}{Acknowledgments}
\renewcommand{\rightmark}{Acknowledgments}

Although it is a (bad) habit of mine not to write acknowledgments in theses, this one would feel especially incomplete if I didn't express gratitude to the people without whom I would never have made it this far.

First and foremost, I would like to thank my parents for their unwavering support throughout these last years of Ph.D., and long before them for that matter. It is truly hard for me to imagine what person I would be today without their kindness and affection.

Then, I would like to thank Fabio Siringo for the supervision and opportunities he provided me with since we started working on the Screened Massive Expansion. It feels like yesterday I entered his office seeking for a research topic for my Bachelor's thesis, and here we are now, years later, still looking for ways to make QCD work.

Furthermore, I would like to thank David Dudal both for his supervision on the Dynamical Model and for being a terrific host during the months spent in Kortrijk as part of the joint Ph.D. between Catania and Leuven. Despite the complications due to the pandemic, he managed to make me feel at home while being miles away.

Last, but absolutely not least, I need to express my gratitude to Cinzia for having had the patience to bear with me during these last few months. I am sorry I was not there at critical times, when you needed it the most. I will never be able to repay you enough for your understanding.

\end{document}